\documentclass[aps,prc,nofootinbib,superscriptaddress,onecolumn,10pt,floatfix]{revtex4-2}

\usepackage{spncci-recurrence1}
\usepackage{pmat}

\begin{document}

\title{Symplectic no-core configuration interaction framework for nuclear structure}

\author{Anna E. McCoy}
\email{amccoy@anl.gov}
\affiliation{Physics Division, Argonne National Laboratory, Argonne, Illinois 60439-4801, USA}

\affiliation{Department of Physics and Astronomy, University of Notre Dame, Notre Dame, Indiana 46556-5670, USA}

\author{Mark A. Caprio}
\affiliation{Department of Physics and Astronomy, University of Notre Dame, Notre Dame, Indiana 46556-5670, USA}
 
\author{Patrick J. Fasano}
\altaffiliation[Present address: ]{NextSilicon Inc., Minneapolis, Minnesota 55402-1572, USA}
\affiliation{Department of Physics and Astronomy, University of Notre Dame, Notre Dame, Indiana 46556-5670, USA}
 
 \author{Tom\'{a}\v{s} Dytrych}
\affiliation{Nuclear Physics Institute, Academy of Sciences of the Czech Republic, 250\,68 \v{R}e\v{z}, Czech Republic}
\affiliation{Department of Physics and Astronomy, Louisiana State University, Louisiana 70803-4001, USA}

\date{\today}

\begin{abstract}
We present the symplectic no-core configuration interaction (SpNCCI) framework, in which the nuclear many-body problem is solved a symmetry-adapted basis that explicitly encodes approximate symmetries associated with nuclear collectivity and deformation. 
In this framework, calculations are carried out in a basis organized into Sp(3,R) irreducible representations (irreps), each of which can be expressed as an infinite tower of U(3) irreps. 
In this framework, matrices of realistic relative two-body operators, such as the nuclear Hamiltonian, are computed directly in the Sp(3,R) many-body basis, obviating the need to expand all Sp(3,R) many-body states in, e.g., a U(3)-coupled configuration basis. 
Instead, many-body matrix elements are obtained via a recurrence relation that expresses a given matrix element in terms of matrix elements between basis states with fewer oscillator quanta. 
To use this recurrence method for computing matrix elements of relative two-body operators, we must first expand each operator into components of U(3) tensors.
To this end, we present a method for decomposing arbitrary operators into U(3) tensor components.
\end{abstract}

\maketitle

\section{Introduction}

Advancements in \emph{ab initio} nuclear structure methods have significantly extended the reach of \emph{ab initio} methods beyond light nuclei~\cite{prl-120-2018-152503-Morris,np-15-2019-428-Gysbers,prl-124-2020-232501-Yao,prl-126-2021-022501-Stroberg,np-18-2022-1196-Hu,fp-8-2020-1-Hergert,prl-126-2021-042502-Belley,prl-132-2024-182502-Belley,prc-110-2024-L011302-Hu}.
However, \emph{ab initio} methods still struggle to capture the relevant collective correlations necessary for describing highly deformed collective nuclei.  
Although signatures of collective behavior have been observed in \emph{ab initio} nuclear predictions of light and medium mass nuclei~\cite{plb-719-2013-179-Caprio,bjp-46-2019-445-Caprio,bjp-49-2022-057066-Caprio,epja-56-2020-120-Caprio,ijmpe-24-2015-1541002-Caprio,jpg-48-2021-075102-Zbikowski,plb-719-2013-179-Caprio,plb-856-2024-138870-McCoy,prc-91-2015-014310-Maris,prc-99-2019-029902-Maris-ERRATUM,prc-93-2016-051301-Stroberg,prl-111-2013-252501-Dytrych,prc-94-2016-011301-Jansen,prl-125-2020-102505-McCoy,prc-70-2004-054325-Pieper,npa-738-2004-357-Neff,jpcs-402-2012-012031-Maris,fbs-54-2013-1465-Yoshida,prl-117-2016-2225-Romero-Redondo,ps-91-2016-053002-Navratil,prl-134-2025-162503-Shen,prc-102-2020-051303R-Novario,prc-111-2025-044304-Sun,prx-15-2025-011028-Sun}, high-precision predictions of observables sensitive to collectivity often require the inclusion of multi-shell correlations, resulting in model spaces that are computationally inaccessible.  
Methods that have been applied to medium-mass and heavy nuclei are often limited to nuclei where deformation and collective effects are minimal~\cite{prl-118-2017-032502-Stroberg,rpp-77-2014-096302-Hagen}, while approaches that are more applicable to highly deformed nuclei are largely restricted to light nuclei~\cite{ppnp-69-2013-131-Barrett,prc-66-2002-044310-Pieper}.

Among these \emph{ab initio} methods is the no-core configuration interaction (NCCI) framework, which is also often referred to as the no-core shell model (NCSM)~\cite{ppnp-69-2013-131-Barrett}.  
In this framework energies and wave functions are obtained by solving the non-relativistic Schr\"odinger equation in a basis of antisymmetrized products of harmonic oscillator states (Slater determinants)~\cite{anp-9-1977-123-Whitehead,Brussard1977}, which are sometimes referred to as configurations.  
The relevant correlations necessary to describe highly collective bound states, including intruder states, can be captured in a sufficiently large model space~\cite{plb-856-2024-138870-McCoy}.
 However, due to combinatorial scaling in the size of the many-body basis with increasing mass number, the requisite model spaces often exceed current computational limits.  
 Accurate descriptions of observables that are sensitive to the tail of the wave function are particularly challenging~\cite{prc-86-2012-034312-Caprio,prc-90-2014-034305-Caprio,prc-104-2021-034319-Caprio,prc-105-2022-054301-Fasano,prc-105-2022-L061302-Caprio,prc-106-2022-034320-Caprio, prc-76-2007-064319-Pervin,npa-801-2008-21-Bogner,ijmpe-22-2013-1330016-Maris,rmp-87-2015-1067-Carlson,prc-93-2016-044331-Odell,ptrsla-382-2024-20230119-Roth}. 
While extensions of the NCCI framework that incorporate continuum effects through explicit inclusion of cluster states~\cite{ps-91-2016-053002-Navratil} can more readily build in the necessary long-range correlations and cluster effects, computational complexity results in additional limits on the basis size and mass number. 

However, the right choice of the many-body basis has the potential to significantly improve \emph{ab initio} descriptions of highly deformed or collective nuclei. 
One choice is to forego the advantages of working in a basis with definite rotational symmetry and instead carry out calculations in a deformed basis~\cite{
prc-102-2020-051303R-Novario,prc-110-2024-L011302-Hu,epja-61-2025-1-Scalesi}.   
Another choice is to attempt to identify physically adapted bases, i.e., bases which more closely resemble the nuclear wave function in terms of, e.g., long-range behavior \cite{
prc-86-2012-034312-Caprio,prc-105-2022-054301-Fasano,prc-99-2019-034321-Tichai,prc-103-2021-014321-Hoppe} or approximate symmetries~\cite{prl-111-2013-252501-Dytrych,jpg-35-2008-123101-Dytrych,cpc-207-2016-202-Dytrych,ppnp-89-2016-101-Launey,aarsscps-3-2018-17-McCoy,McCoy2018}.  
In such a basis, fewer basis states are required to capture the necessary correlations.  
The latter choice is the focus of this work. 

We describe a method for carrying out \emph{ab initio} many-body calculations in a basis which explicitly incorporates approximate symmetries of the nuclei.  Specifically, the basis incorporates symmetries that are tied to nuclear deformation and collective behavior, namely the symplectic symmetry $\grpsptr$ and Elliott's $\grpsu{3}$ symmetry~\cite{prsla-245-1958-128-Elliott,prsla-245-1958-562-Elliott, prsla-272-1963-557-Elliott, anp-1-67-1968-Harvey}.
This framework is referred to as the symplectic no-core configuration interaction (SpNCCI) framework. 

The development of a framework for carrying out realistic nuclear structure calculations in a symplectic basis has been ongoing since the 1970s. 
The symplectic model was first proposed by Rowe and Rosensteel as a microscopic framework for the nuclear collective model~\cite{prl-38-1977-10-Rosensteel,ap-126-1980-343-Rosensteel} that took into account both collective  and independent-particle degrees of freedom.  
The Lie algebra associated with the $\grpsptr$ symplectic symmetry is the smallest algebra which includes the quadrupole operator associated with nuclear shape,  the generators of orbital angular momentum, the kinetic energy and the harmonic oscillator Hamiltonian~\cite{prl-38-1977-10-Rosensteel,rpp-48-1985-1419-Rowe}.
Moreover, in its contraction limit, the generators of this algebra contract to the generators of giant monopole and quadrupole excitations~\cite{prl-47-1981-223-Rosensteel,plb-140-1984-155-Le_Blanc}. 
Thus it is not surprising that $\grpsptr$ has emerged as an approximate symmetry of nuclei which exhibit collective behavior~\cite{epj-229-2429-2020-Launey,fp-11-2023-1-Becker,jpg-35-2008-095101-Dytrych,jpg-35-2008-123101-Dytrych,ppnp-89-2016-101-Launey,prc-76-2007-014315-Dytrych,prc-105-2022-034306-Molchanov,prl-98-2007-162503-Dytrych,prl-111-2013-252501-Dytrych,prl-124-2020-042501-Dytrych,prl-125-2020-102505-McCoy,jpg-48-2021-075102-Zbikowski}.

Although $\grpsptr$ first gained interest because of its connection to a number of different collective models~\cite{ap-104-1977-134-Rosensteel,prl-38-1977-10-Rosensteel,ap-123-1979-36-Rosensteel,ap-126-1980-198-Rowe,ap-126-1980-343-Rosensteel}, for applications associated with the shell model or the NCCI framework, the most relevant sub-algebra is the one associated with Elliott's $\grpsu{3}$ symmetry.  
This algebra was the basis for Elliott's rotational model~\cite{prsla-245-1958-128-Elliott,prsla-245-1958-562-Elliott,prsla-272-1963-557-Elliott,prsla-302-1968-509-Elliott}, which was one of the first models to connect nuclear collective behavior to the shell model.  
Elliott's model provided a successful classification scheme for rotational bands in light nuclei~\cite{prsla-245-1958-128-Elliott,prsla-245-1958-562-Elliott,prsla-272-1963-557-Elliott,prsla-302-1968-509-Elliott,anp-1-67-1968-Harvey,prc-103-2021-L021302-Fu,Parikh1978,rmf-45-1999-86-Hirsch,Wilsdon1965} and was extended to heavier nuclei in the  pseudo-$\grpsu{3}$~\cite{npa-202-1973-433-Raju, npa-381-1982-1-Draayer, apny-156-1984-41-Draayer, prl-78-1997-436-Ginocchio} and proxy-$\grpsu{3}$~\cite{prc-95-2017-064325-Bonatsos,epja-53-2017-148-Bonatsos,epja-56-2020-239-Martinou,epja-57-2021-84-Martinou} models.

However, like most models restricted to a single valence shell, this framework was unable to accurately reproduce $E2$ transition strengths without introducing an effective charge.

In the symplectic model framework, the $\grpsptr$ generators, which include the kinetic energy and quadrupole operator, can excite particles across different oscillator shells.  
Thus, $\grpsptr$ provides a natural framework for a multi-shell extension of the Elliott model.  
Indeed, $\grpsptr$ model calculations demonstrated that $E2$ transitions strengths could be obtained without resorting to effective charges~\cite{npa-419-1984-1-Rosensteel}.
For this reason, an irreducible representation (irrep) of $\grpsptr$, which is the set of states \textemdash\, with different  numbers of oscillator quanta \textemdash\, connected by the $\grpsptr$ generators, is sometimes referred to as a collective subspace~\cite{npa-413-1984-215-Draayer}.  

Until recently, calculations in a symplectic basis~\cite{
prl-38-1977-10-Rosensteel,
npa-295-1978-34-Hecht,
ap-126-1980-343-Rosensteel,
plb-119-1982-249-Carvalho,
plb-124-1983-281-Draayer,
plb-125-1983-237-Draayer,
npa-414-1984-93-Park,
npa-413-1984-215-Draayer,npa-419-1984-1-Rosensteel,
rpp-48-1985-1419-Rowe,
npa-452-1986-240-Carvalho,
npa-452-1986-263-Le_Blanc, 
npa-454-1986-288-Vassanji,
Reske1984,npa-455-1986-315-Suzuki,ptp-77-1987-190-Suzuki,npa-465-1987-265-Carvalho,
prl-97-2006-202501-Rowe,
Escher1997,jmp-39-1998-5123-Escher} were largely restricted to calculations involving phenomenological interactions in a highly restricted basis consisting of a single $\grpsptr$ irrep.
While there were several efforts to carry out multi-irrep calculations~\cite{Reske1984,npa-455-1986-315-Suzuki,ptp-77-1987-190-Suzuki,npa-413-1984-215-Draayer}, these methods were limited at the time by computational resources.  One of these approaches, proposed by Reske in the mid 1980s~\cite{Reske1984}, evaluated matrix elements in a symplectic basis via a recurrence.  
The seeds of the recurrence, which were a small subset of the many-body matrix elements,  could be computed by expanding the symplectic states in terms of harmonic oscillator configurations.  
The method presented in this work builds upon the approach pioneered by Reske.

It wasn't until recently, with the development of the symmetry-adapted no-core shell model (SA-NCSM)~\cite{cpc-207-2016-202-Dytrych,ppnp-89-2016-101-Launey,epj-229-2429-2020-Launey}, that \emph{ab initio} calculations in a symplectic basis became computationally feasible.  
In this framework, calculations are carried out in a basis spanned by $\grpu{3}$-coupled many-body configurations.  
These coupled configurations are linear combinations, having definite Elliott $\grpsu{3}$ symmetry and definite $\grpu{1}$ symmetry (definite number of oscillator quanta), of the Slater determinant configurations typically used in the NCCI framework. 
Many-body states with definite $\grpsptr$ symmetry can then be obtained as linear combinations of the $\grpu{3}$-coupled configurations. 
In the SA-NCSM, this linear combination is obtained by diagonalizing a special operator related to the $\grpsptr$ Casimir operator that commutes with all of the generators of $\grpsptr$.  
Thus, eigenstates of this operator will necessarily have definite $\grpsptr$ symmetry. 
Although one could, in principle, expand the symplectic states by diagonalizing the same operator in the NCCI basis~\cite{Dytrych2001}, each symplectic state is a linear combinations of \emph{many} different Slater determinants.
By fully expanding the symplectic states in an NCCI basis, one loses many of the advantages of working in a symmetry-adapted basis.  
In contrast, expanding in an $\grpu{3}$-coupled basis allows for the use of group theoretical machinery~\cite{Wybourne1974,tmp-13-1972-1171-Klimyk,rmp-7-1975-153-Klimyk,ptrsla-277-1975-545-Butler,McCoy2018,Escher1997}  to simplify many-body calculations.

Although the SA-NCSM has been successfully applied to light and select medium-mass nuclei using symmetry based truncations of the many-body basis~\cite{prl-111-2013-252501-Dytrych,ppnp-67-2012-516-Draayer,ppnp-89-2016-101-Launey,epj-229-2429-2020-Launey,arnps-71-2021-253-Launey,prc-108-2023-054303-Sargsyan,cpc-280-2022-108476-Mercenne,prl-128-2022-202503-Sargsyan,prc-108-2023-024304-Heller,arxiv:2501.00682-Becker,plb-866-2025-139563-Sargsyan,psotancp-2038-2018-020004-Launey}, this method of constructing the symplectic basis does limit the degree to which such truncations can be applied. 
Within the many-body basis, there can be many different $\grpsptr$ irreps with the same quantum numbers. 
Using the above technique, one can only identify the subspace spanned by all of the $\grpsptr$ irreps with the same quantum numbers and thus cannot resolve individual irreps. 
However, the ability to  truncate by individual $\grpsptr$ irreps  has the potential to reduce the necessary basis size by orders of magnitude and significantly extend the already impressive results obtained with the SA-NCSM.

In this work, we propose an alternative method for constructing states with definite $\grpsptr$ symmetry, which allows for more fine-grained truncations of the many-body basis. 
In the SpNCCI framework, physical applications of which may be found in Refs.~\cite{aarsscps-3-2018-17-McCoy,McCoy2018,prl-125-2020-102505-McCoy,Herko2024}, the basis is obtained by first explicitly expanding a select set of special $\grpsptr$ states in terms of $\grpu{3}$-coupled configurations.  
Specifically, for each $\grpsptr$ irrep, a single state (namely, a ``highest weight state'') belonging to the unique $\grpu{3}$ irrep with fewest oscillator quanta in the $\grpsptr$ irrep is explicitly constructed.  
We refer to this $\grpu{3}$ irrep with fewest oscillator quanta as the lowest grade irrep (LGI).   
The remaining states within an $\grpsptr$ irrep are then obtained by repeatedly acting on this special state in the LGI with the operators forming the $\grpsptr$ algebra.

To avoid the overhead of explicitly constructing the basis, we use variants of the recurrence method initially put forth by Reske~\cite{Reske1984}.
In the SpNCCI framework, matrix elements of operators, such as the Hamiltonian, are constructed via a recurrence relation over matrix elements, similar to the recurrence derived by Reske~\cite{Reske1984}.  
Here, the seeds of the recurrence for each pair of $\grpsptr$ irreps are matrix elements between the corresponding pair of special states belonging to the LGIs that are explicitly expanded in terms of the $\grpsu{3}$-coupled configurations. 
Recurrence relations are presented which permit evaluation of the matrix elements of a two-body nuclear Hamiltonian.
These recurrence relations provide, more generally, for the evaluation of matrix elements of an arbitrary intrinsic (Galilean-invariant) two-body operator, and thus for the evaluation of, \textit{e.g.}, electromagnetic moment and transition observables.  
Recurrence relations may also be derived for calculating generic laboratory-frame one- and two-body densities~\cite{Herko2024}.

When working in a basis with definite symmetry, powerful group theoretical machinery can be used to calculate matrix elements if the operator of interest is a tensor with respect to the symmetry group.  
Although realistic Hamiltonians, and other operators of interest, are not tensors with respect to $\grpsptr$ or $\grpsu{3}$, they can be expressed as linear combinations of components of $\grpsu{3}$ tensors.
Building on the efforts of, e.g., Hecht, Suzuki and Reske~\cite{npa-455-1986-315-Suzuki,ptp-77-1987-190-Suzuki,Reske1984}, we present a systematic way of decomposing arbitrary operators, which have no particular symmetry properties with respect to $\grpsu{3}$ or $\grpsptr$, in such a way.

In this article we first review Elliott's $\grpsu{3}$ symmetry (Section~\ref{section:u3}). In Section~\ref{section:u3-many-body} we describe the construction of a $\grpu{3}$ many-body basis and how matrix elements can be evaluated in such a basis while taking advantage of group theoretical machinery. 
We then sketch the salient features of $\grpsptr$ irreps (Section~\ref{section:sp3r}) before presenting the method used in the SpNCCI framework for constructing an $\grpsptr$ many-body basis (Section~\ref{subsection:sp3r-many-body}).  
In Section~\ref{section:recurrence}, we then derive the recurrence relations used to carry out calculations in this $\grpsptr$ many-body basis and provide details on how to evaluate the seed matrix elements in a $\grpu{3}$ many-body basis.   
Details and identities used in deriving these relations are provided in the appendices.  In Appendix~\ref{appendix:upcoupling} we outline an algorithm for decomposing operators which are not good $\grpsu{3}$ tensors in terms of components of $\grpsu{3}$ tensors.   
We then provide expressions for generic coupled commutators of two or three $\grpsu{3}$ tensors in Appendix~\ref{sec:app-commutator} and specialize to the specific commutators used in this work in  Appendix~\ref{sec:app-coupled-commutator-unit-tensors}. 
Preliminary expressions for the recurrence relations presented in this article were reported in Ref.~\cite{McCoy2018}.

\section{
Elliott's $\grpsu{3}$
\label{section:u3}
}

Much of the formalism used to develop the SpNCCI framework hinges on the fact that the states within $\grpsptr$ irreps are organized into irreps of Elliott's $\grpsu{3}$.  
This enables the use of powerful and well established group theoretical tools for calculations in an $\grpsu{3}$ basis~\cite{rmp-34-1962-813-Moshinsky,np-62-1965-1-Hecht, jmp-6-1965-142-Kaufman,jmp-6-1965-1584-Derome,jmp-7-1966-612-Derome,jmp-8-1967-714-Derome,jmp-8-1967-63-Resnikoff,npa-111-1968-681-Vergados,jmp-13-1972-1985-Biedenharn,tmp-13-1972-1171-Klimyk,ptrsla-277-1975-545-Butler,rmp-7-1975-153-Klimyk,jmp-14-1973-1904-Draayer,cpc-5-1973-405-Akiyama,Wybourne1974,npa-244-1975-365-Hecht, apny-95-1975-139-Moshinsky,jmp-19-1978-1513-Millener, jmp-23-1982-2022-OReilly,Hecht1987, jpa-23-1990-407-Hecht, Escher1997,
Chen2002,cpc-159-2004-121-Bahri,jpa-41-2008-065206-Rowe,McCoy2018,
privcom-Rowe-2016,
cpc-244-2019-442-Langr, cpc-269-2021-108137-Dytrych,arxiv-Herko}.\footnote{   
While many of these tools could, in principle, be generalized to $\grpsptr$, the applicability of these tools would be limited  in actual calculations by the lack of available $\grpsptr$ coupling coefficients.
In contrast, there are several codes available for computing $\grpsu{3}$ coupling and recoupling coefficients~\cite{cpc-5-1973-405-Akiyama,cpc-269-2021-108137-Dytrych,arxiv-Herko}}
Thus, before diving into the description of $\grpsptr$ irreps, we will first briefly review a few key features of Elliott's $\grpsu{3}$ algebra, the labeling scheme used in this work for $\grpu{3}$ irreps, the connection between Elliott's $\grpsu{3}$ framework and the shell model, and a few useful definitions pertaining to, e.g., Racah reduction formulas and $\grpsu{3}$ coupling coefficients.  

The connection between $\grpsu{3}$, collective behavior, and the shell model arose from the realization that, within a single oscillator shell, the quadrupole operator
\begin{equation}
Q_{2M}=\sqrt{\frac{16\pi}{5}}\sum_{s=1}^{A}\frac{\textbf{x}_s^2}{b^2}Y_{2M}(\hat{\textbf{x}}_s)
\label{eq:Q_mass}
\end{equation}
 has the same matrix elements as the so-called algebraic quadrupole operator 
\begin{equation}
\mathcal{Q}_{2M}=\sqrt{\frac{4\pi}{5}}\sum_{n=1}^{A}\left[\frac{\textbf{x}_s^2}{b^2}Y_{2\mu}(\hat{\textbf{x}}_s)+b^2\hbar^2\textbf{k}_s^2Y_{2\mu}(\hat{\textbf{p}}_s)\right],
\label{eq:Q_alg}
\end{equation}
where  $\textbf{x}_{s}$ is the position of the $s$-th particle and the  wave vector $\vec{k}_s$ is related to the momentum vector by $\vec{p}_{s}=\hbar \vec{k}_{s}$. The parameter $b$ is the harmonic oscillator length parameter given by $b=\hbar/(m_N\Omega)$, where $m_N$ is the nucleon mass and $\Omega$ is the harmonic oscillator frequency. 
The  five components of the quadrupole tensor operator, $\mathcal{Q}_{2M}$, when taken together with the orbital angular momentum operators $L_{1M}$, which generate the group of rotations in three dimensions $\grpso{3}$, form an $\grpsu{3}$ algebra known as Elliott's $\grpsu{3}$~\cite{prsla-245-1958-128-Elliott}.  
Representations of this $\grpsu{3}$ algebra provided a natural framework for describing nuclear deformation and rotation in a shell model context.  This framework became known as Elliott's rotational model~\cite{prsla-245-1958-128-Elliott,prsla-245-1958-562-Elliott,prsla-272-1963-557-Elliott,prsla-302-1968-509-Elliott}. 

In Elliott's rotational model, rotational dynamics are governed by an isotropic harmonic oscillator potential plus a residual quadruple-quadrupole interaction, i.e., $H=H_0-\chi\mathcal{Q}_2\cdot\mathcal{Q}_2$.  
In this framework the rotational intrinsic state has definite $\grpsu{3}$ symmetry, labeled by quantum numbers $(\lambda\mu)$~\cite{prsla-245-1958-128-Elliott,prsla-272-1963-557-Elliott,ps-91-2016-033003-Rowe}. 
In a large quantum number contraction limit, these labels correspond approximately to the deformation parameters $\beta$ and $\gamma$ by~\cite{zpa-329-1988-33-Castanos}
\begin{equation}
\begin{split}
\beta^2\propto (\lambda^2+\lambda\mu+\mu^2+3\lambda+3\mu+3)\\
\gamma=\tan^{-1}\left(\frac{\sqrt{3}(\mu+1)}{2\lambda+\mu+3}\right).
\end{split}
\end{equation}
In addition to the $\grpsu{3}$ labels, the intrinsic state is also labeled by $K$, which is the projection\footnote{
A basis for an $\grpsu{3}$ irrep is obtained by carrying out Peierls-Yoccoz angular momentum projection~\cite{ppsla-70-1957-381-Peierls}
from one extremal state of the irrep ($\lambda\mu)$~--- either the highest-weight or lowest-weight state.
} of the orbital angular momentum of the state on the symmetry axis of the co-rotating frame.
That is, the intrinsic state is denoted by $\tket{(\lambda\mu)K}$. 
The intrinsic state projects onto states with definite orbital angular momentum $L$ and projection $M$.  
The resulting state is denoted by $\tket{(\lambda\mu)KLM}$.  
To see the connection to the rotor model, we can re-express Elliott's Hamiltonian in terms of $\vec{L}^2$  and the quadratic $\grpsu{3}$ Casimir operator  as $H=H_0-\chi(6C_{\grpsu{3}}-3\vec{L}^2)$.  
Since the eigenvalues of the Casimir depend only on $(\lambda\mu)$,  the energy of each eigenstate $\tket{(\lambda\mu)KLM}$ is given by 
\begin{equation}
E=E_0(\lambda\mu)+aL(L+1).
\end{equation} 
In systems with $S=0$ (and thus $J=L$), the energies associated with these states are the characteristic energies of a rotational band~\cite{Bohr1998:v1,Rowe2010}.  
When both $\lambda$ and $\mu$ are non-zero (and thus $\gamma\ne0$), the intrinsic state can have different projections $K$ onto the symmetry axis,  which result in different $K$ bands.  
The possible projections $K$ for an intrinsic state with $\grpsu{3}$ symmetry  $(\lambda\mu)$ are given by 
\begin{equation}
K=\min(\lambda,\mu),\, \min(\lambda,\mu)-2,\dots,1\text{ or }0,
\end{equation}
and the possible $L$ of the states within each band are given by~\cite{prsla-245-1958-128-Elliott}
\begin{equation}
L=
\begin{cases}
K, K+1,\dots,K+\max(\lambda,\mu)&K\ne 0\\
\max(\lambda,\mu),\,\max(\lambda,\mu)-2,\dots, 1\text{ or }0,& K=0.
\end{cases}
\end{equation}

The set of states $\{\ket{(\lambda\mu)KLM}\}$ spans an irreducible representation (irrep)\footnote{
Mathematically, an irrep is a space together with a set of operators representing the action of the group generators on that space.  
Thus it would be more correct to say that the states $\ket{(\lambda\mu)KLM}$ form a basis for the carrier space of the irrep.  
But, like many physicists, we forgo the mathematically correct terminology where the distinction is unnecessary, and refer to the space as the irrep.
In other physics applications, the operators alone may be referred to as the irrep when the space can be deduced from context~\cite{Casten2000}.
} of $\grpsu{3}$ with quantum numbers $(\lambda\mu)$.  
Within a given $\grpsu{3}$ irrep $(\lambda\mu)$ each set of states with fixed $K$ and $L$ form an irrep of $\grpso{3}$, the group of rotations in three dimensions.   Since each $\grpsu{3}$ irrep is organized into $\grpso{3}$ irreps  $L$,  we say that the basis of the irrep reduces  the group chain $\grpsu{3}\supset\grpso{3}$.  

States with same $L$ but different $K$ are not, in general, orthogonal.  
However, an orthogonal basis can be readily constructed for the irrep~\cite{npa-111-1968-681-Vergados,jmp-41-2000-6544-Rowe,cpc-159-2004-121-Bahri,jpa-41-2008-065206-Rowe,jmp-14-1973-1904-Draayer}. 
The states of the orthogonal basis for the irrep are labeled by $\ket{(\lambda\mu)\kappa LM}$, where $\kappa$ is an index (with values $\kappa=1,\dots,\kappa_{\max}$) that distinguishes between different $\grpso{3}$ irreps with the same label $L$.  
This index is often referred to as an inner multiplicity index or a branching multiplicity index, and $\kappa_{\max}$ is the multiplicity of $L$ in $(\lambda\mu)$.\footnote{The maximum value of $\kappa$ for given  $L$  is~\cite{rmp-21-1949-494-Racah,jmp-6-1965-142-Kaufman}
\begin{equation}
\kappa_{\max}=\max\left(0,\floor*{\frac{\lambda+\mu+2-L}{2}}\right)
-\max\left(0,\floor*{\frac{\lambda+1+L}{2}}\right)
-\max\left(0,\floor*{\frac{\mu+1-L}{2}}\right),
\end{equation}
where $\floor*{x}$ denotes the integer part of $x$. 
}  To summarize, the labeling scheme for a basis of an $\grpsu{3}$ irrep which reduces the orbital angular momentum group chain is:
\begin{equation}
\begin{array}{ccccc}
\grpsu{3} & \supset & \grpso{3}  \\
 (\lambda,\mu) & \kappa & L M \\
\end{array}.
\label{su3_chain}
\end{equation}

The significance of Elliott's $\grpsu{3}$ in a shell model context arises from the fact that the three-dimensional harmonic oscillator Hamiltonian commutes with all of the generators of a $\grpu{3}$ group, which contains Elliott's $\grpsu{3}$ as a sub-group.   
The generators of this $\grpu{3}$ group are given by
\begin{equation}
C_{ij}=\frac12\sum_s^{A}(c_{si}^\dagger c_{sj}+c_{sj}c^\dagger_{si}),\label{su3_cartesian_generators}
\end{equation}
where 
$c_{si}^{\dagger}=\sqrt{\frac{1}{2}}\left(b^{-1}x_{si}-ibk_{si}\right)$ and $c_{si}=\sqrt{\frac{1}{2}}\left(b^{-1}x_{si}+ibk_{si}\right)$
are the harmonic oscillator ladder operators which create or annihilate oscillator quanta in the $i$-th direction. 

The $C_{ij}$ operators with $i\ne j$ are the $\grpu{3}$ ladder operators, which ladder between states with the same total number of oscillator quanta, but with differing numbers of oscillator quanta in the Cartesian directions.   
For example, $C_{12}$ annihilates an oscillator quantum in the $y$ direction and creates a quantum in the $x$ direction.  
The weight operators $C_{ii}$ do not connect different basis states, but rather provide information about the state upon which they act.  Specifically, the eigenvalue of $C_{ii}$ corresponds to the number of quanta in $i$-th direction.  

The $\grpu{3}$ irreps are labeled by $\omega=[\omega_1\omega_2\omega_3]$, where each $\omega_i$ correspond to the number of quanta in the $i$-th direction.\footnote{
The labels correspond to the labels of an extremal weight, which must satisfy $C_{ij}\ket{\omega}=0$ for all $i>j$.
}  
The three labels $\omega_i$ must satisfy  $\omega_1\ge\omega_2\ge\omega_3\ge0$.  
An equivalent labeling is given in terms of the total number of oscillator quanta $N=\omega_1+\omega_2+\omega_3$ and the Elliott $\grpsu{3}$ labels $(\lambda_\omega\mu_\omega)$, where $\lambda_\omega=\omega_1-\omega_2$, $\mu_\omega=\omega_2-\omega_3$.  That is, $\omega$  can equivalently be written as $\omega= N_\omega(\lambda_\omega\mu_\omega)$.  

The product of two irreps, i.e., a Kronecker product, will not, in general, be irreducible.  
However, it can be broken up into a direct sum of irreps, i.e.,  $\omega_1\times\omega_2=\oplus_{f}\omega_f$.  
The particular irreps which appear in the Kronecker product can be deduced using, e.g., the Littlewood rules~\cite{jmp-19-1978-720-Braunschweig} or the O'Reilly formula~\cite{jmp-23-1982-2022-OReilly}.  
The relation between the basis for an irrep in the Kronecker product $\omega_f$ and the bases of the irreps $\omega_1$ and $\omega_2$ is given by a unitary transformation, where the matrix elements of the transformation are given by $\grpsu{3}$ coupling coefficients~\cite{npa-170-1971-34-Hecht}:
\begin{equation}
\ket{\rho \omega_3\kappa_3L_3M_3}
=\sum_{\substack{\kappa_1 L_1M_1\\\kappa_2L_2M_2}}
\cg{(\lambda_{\omega_1}\mu_{\omega_1})}{\kappa_1 L_1M_1}
{(\lambda_{\omega_2}\mu_{\omega_2})}{\kappa_2L_2M_2}
{(\lambda_{\omega_3}\mu_{\omega_3})}{\kappa_3L_3M_3}_{\rho}
\ket{\omega_1\kappa_1L_1M_1}\,
\ket{\omega_2\kappa_2L_2M_2}.
\end{equation}
The index $\rho$ distinguishes different $\grpu{3}$ irreps which arise in the Kronecker product with the same labels $\omega_3$.   Note that $N_{\omega_3}=N_{\omega_1}+N_{\omega_2}$ is implicit in the expression above. 

\subsection{$\grpsu{3}$ tensors\label{subsection:su3_tensors}}
For calculations in a basis with definite $\grpsu{3}$  symmetry, it is useful to express the generators as components of an $\grpsu{3}$ irreducible tensor, in order to make use of, e.g., the Wigner-Eckart theorem~\cite{Wybourne1974,tmp-13-1972-1171-Klimyk,rmp-7-1975-153-Klimyk} and Racah reduction formulas~\cite{ptrsla-277-1975-545-Butler,jmp-39-1998-5123-Escher,McCoy2018,Escher1997,privcom-Rowe-2016} to simplify the calculations. 
An $\grpsu{3}$ tensor $T^{(\lambda_t\mu_t)}_{\kappa_t L_tM_t}$ is a set of operators which form a basis for an $\grpsu{3}$ irrep and thus transform  under the adjoint action of the generators of the group as 
\begin{equation}
[C_{LM},T^{(\lambda_t\mu_t)}_{\kappa_tL_tM_t}]
=\sum_{\kappa_t'L_t'M_t'}\braket{(\lambda_t\mu_t)\kappa_t'L_t'M_t'|C_{LM}|(\lambda_t\mu_t)\kappa_tL_tM_t}T^{(\lambda_t\mu_t)}_{\kappa_t'L_t'M_t'},
\label{eq:su3_tensor_def}
\end{equation}
where $C_{2M}=\mathcal{Q}_{2M}/\sqrt{3}$ and $C_{1M}=L_{1M}$.  
Note that it follows that the coupled product of two tensors, given by 
\begin{equation}
[T^{(\lambda_t\mu_t)}\times S^{(\lambda_s\mu_s)}]^{\rho_0(\lambda_0\mu_0)}_{\kappa_0L_0M_0}\\
=\sum_{\substack{\kappa_tL_tM_t\\\kappa_s L_sM_s}}
\cg{(\lambda_t\mu_t)}{\kappa_tL_tM_t}{(\lambda_s\mu_s)}{\kappa_sL_sM_s}{(\lambda_0\mu_0)}{\kappa_0L_0M_0}_{\rho_0}
T^{(\lambda_t\mu_t)}_{\kappa_t L_tM_t}
S^{(\lambda_s\mu_s)}_{\kappa_s L_sM_s},
\label{eq:coupled-tensors}
\end{equation}
will again be a tensor.  
That is, the coupled product will also satisfy the commutation relation \eqref{eq:su3_tensor_def}.  
Similarly, as discussed in Appendix~\ref{sec:app-commutator}, the coupled commutator of two tensor operators will also be a tensor. 
However, the adjoint of a tensor does not generally transform tensorially~\cite{Wybourne1974}, i.e., as a standard basis of an $\grpsu{3}$ irrep.  
The adjoint, however, is related component-wise by a conjugation relation to the covariant adjoint~\cite{RoweWood2010} tensor $\tilde{T}^{(\mu_t\lambda_t)}$, which does transform as a standard basis for an irrep of $\grpsu{3}$ with quantum numbers $\widetilde{(\lambda_t\mu_t)}=(\mu_t\lambda_t)$.  This relation is given by
\begin{equation}
\tilde{T}^{(\mu_t\lambda_t)}_{\kappa_t'\tilde{L}_t\tilde{M}_t}  
=\sum_{\kappa_t\kappa_t'}\Phi_{\kappa_t\kappa_t'}[(\mu_t\lambda_t)\tilde{L}_t\tilde{M}_t]
\big(T^{(\lambda_t\mu_t)}_{\kappa_tL_tM_t}\big)^\dagger,
\label{su3_tensor_conj}
\end{equation}
where $\Phi_{\kappa_t\kappa_t'}[(\mu_t\lambda_t)\tilde{L}_t\tilde{M}_t]$ is a $\kappa_{t,\max}\times\kappa_{t,\max}$ conjugation matrix. Because $\grpso{3}$ irreps are self conjugate, $\tilde{L}_t=L_t$.  However, under the conjugation relation, the sign on the projection flips ($\tilde{M}_t=-M_t$).  
Under the  Condon-Shortley convention, the conjugation matrix is consistent with the standard coupling coefficients and is given in terms of the identity-coupled coupling coefficient as
\begin{equation}
\Phi_{\kappa'\kappa}[(\lambda\mu) L M]=\sqrt{\dim(\lambda\mu)}\cg{(\lambda\mu)}{\kappa L M}{(\mu\lambda)}{\kappa' L\,- M}{(0,0)}{00}.
\label{eq:phi_def}
\end{equation}
When $\lambda\ne\mu$ or  $\kappa_{\max}=1$, the conjugation matrix  simplifies to the identity matrix times a phase factor given by 
\begin{equation}
\Phi[(\lambda\mu)L M]=(-)^{\lambda+\mu+L+M}.
\end{equation}
The relationship in \eqref{su3_tensor_conj} can be inverted using the conjugation matrix identity 
\begin{equation}
\sum_{\kappa'} \Phi_{\kappa\kappa'}[(\mu\lambda)\tilde{L}\tilde{M}]\Phi_{\kappa'\kappa''}[(\lambda\mu)LM]=\delta_{\kappa,\kappa''}.
\end{equation}

To rewrite the $\grpu{3}$ generators in terms of $\grpsu{3}$ tensors, we first note that the harmonic oscillator ladder operators can be related to components of $\grpsu{3}$ irreducible tensors with tensor character $(1,0)$ and $(0,1)$, respectively, as\footnote{See, e.g., Appendix~F of Ref.~\cite{jpg-47-2020-122001-Caprio} for more details.}
\begin{equation}
\begin{aligned}
c^{\dagger(1,0)}_{1\pm 1}&=\mp\frac{1}{\sqrt{2}}\left(c^{\dagger}_1\pm ic^{\dagger}_2\right),
&c^{\dagger(1,0)}_{10}&=c^{\dagger}_3,\\
c^{(0,1)}_{1\pm 1}&=\mp\frac{1}{\sqrt{2}}\left(c_1\pm ic_2\right),
&c^{(0,1)}_{10}&=c_3.
\end{aligned}
\label{eq:boson-tensors}
\end{equation}
The $\grpsu{3}$ generators can then  be expressed in terms of $\grpsu{3}$-coupled products of these as 
\begin{align}
C_{LM}^{(1,1)}&=\sqrt{2}\sum_{s=1}^{A}\left[c_s^{\dagger(1,0)}\times c_s^{(0,1)}\right]_{LM}^{(1,1)}
\label{su3_tensor_generators}
\intertext{and}
H_{00}^{(0,0)}&=\frac{\sqrt{3}}{2}\sum_{s=1}^{A}\Big(
\left[c_s^{\dagger(1,0)}\times c_s^{(0,1)}\right]_{00}^{(0,0)}+ \left[c_s^{(0,1)}\times c_s^{\dagger(1,0)}\right]_{00}^{(0,0)}
\Big)
\label{u3_tensor_generators_H}.
\end{align}
The eight components of $C^{(1,1)}$  are the generators of $\grpsu{3}$, 
\begin{equation}
C^{(1,1)}_{2M}=\frac{1}{\sqrt{3}}\mathcal{Q}_{2M}\qquad C^{(1,1)}_{1M}=L_M.\label{CQL_rel}
\end{equation}  
The tensor $H^{(0,0)}$ is the spherical shell model Hamiltonian.  
This is more apparent if we use the coupled commutator  $[c_s^{(0,1)}, c_s^{\dagger(1,0)}]_{00}^{(0,0)}=\frac{3}{2}$, to rewrite this operator as 
\begin{equation}
H_{00}^{(0,0)}=\sqrt{3}\sum_{s=1}^{A}\left[c_s^{\dagger(1,0)}\times c_s^{(0,1)}\right]_{00}^{(0,0)}+\frac{3}{2}A\label{su3_tensor_generators}.
\end{equation}
Note that the first term is just the oscillator number operator expressed as an $\grpsu{3}$ tensor:
\begin{equation}
N^{(0,0)}=\sqrt{3}\sum_{s=1}^{A}\left[c_s^{\dagger(1,0)}\times c_s^{(0,1)}\right]_{00}^{(0,0)}.
\end{equation}

\subsection{Reduced matrix elements of $\grpsu{3}$ tensors}
Now that the generators are expressed as $\grpsu{3}$ tensors, we can use the Wigner-Eckart theorem to  calculate the matrix elements of these tensor operators.  
For a general $\grpsu{3}$ tensor $T^{(\lambda_0\mu_0)}$, the Wigner-Eckart theorem is given by\footnote{As a notational convenience, we will often use $\grpu{3}$ labels $\omega$ in expressions for, e.g.,  $\grpsu{3}$  coupling and recoupling coefficients, even though only the $\grpsu{3}$ labels $(\lambda_\omega\mu_\omega)$ are relevant for evaluating such coefficients. } 
\begin{equation}
\me{\varrho'\omega'\kappa'L'M'}{T^{(\lambda_0,\mu_0)}_{\kappa_0L_0M_0}}{\varrho\omega\kappa LM}
=\sum_{\rho_0}
\cg{\omega}{\kappa LM}{(\lambda_0,\mu_0)}{\kappa_0L_0M_0}{\omega'}{\kappa'L'M'}_{\rho_0}
\rme{\varrho'\omega'}{T^{(\lambda_0,\mu_0)}}{\varrho\omega}_{\rho_0},\label{U3_Wigner}
\end{equation}
where $\varrho'$ and $\varrho$ represent any additional labels necessary to uniquely identify the irreps beyond the $\grpu{3}$ labels $\omega'$ and $\omega$, respectively. 
The first factor on the right hand side in \eqref{U3_Wigner} is an $\grpsu{3}$ coupling coefficient, and the second is an $\grpsu{3}$ reduced matrix element (RME), which is independent of all of the branching quantum numbers ($\kappa LM$) associated with the subgroup $\grpso{3}$.  In contrast to the $\grpsu{2}$ Wigner-Eckart theorem, in which the matrix element is given by a single product, of a unique RME with a unique $\grpsu{2}$ coupling coefficient, the RME and $\grpsu{3}$ coupling coefficient appearing here in the $\grpsu{3}$ Wigner-Eckart theorem both depend upon the outer multiplicity index $\rho_0$, and a summation over this index is entailed.  

Using the Wigner-Eckart theorem, calculating matrix elements for a tensor operator simplifies to calculating the RMEs, after which the RMEs can be combined with different coupling coefficients to obtain all of the matrix elements. 

In particular, for the generators of $\grpu{3}$, the matrix elements between members of an $\grpu{3}$ irrep $\omega$ can be obtained from the RMEs\footnote{The RME of $C^{(1,1)}$ is derived by noting that the $\grpsu{3}$ Casimir operator is given by $C_{\grpsu{3}}=C^{(1,1)}\cdot C^{(1,1)}$~\cite{ps-91-2016-033003-Rowe}, and thus the sign of the matrix element is undetermined.  We follow the convention of Rowe~\cite{ps-91-2016-033003-Rowe} in fixing the sign such that the $\grpso{3}$ RME of $L_{1M}$ is always positive.  That is, we fix the sign of the RME to be the sign of the $\grpsu{3}$ reduced coupling coefficient  $((\lambda\mu)\kappa L (1,1)1||(\lambda\mu)\kappa L)$.}
\begin{equation}
\rme{\omega'}{C^{(1,1)}}{\omega}=\sqrt{\frac{4}{3}}(\lambda_\omega^2+\lambda_\omega\mu_\omega+\mu_\omega^2+3\lambda_\omega+3\mu_\omega)^{\frac12}\delta_{\omega,\omega'}
\label{Crme}
\end{equation}
 and 
 \begin{equation}
\rme{\omega'}{H^{(0,0)}}{\omega}=N_\omega\delta_{\omega,\omega'},
\label{Hrme}
\end{equation}
by applying the  appropriate coupling coefficients to obtain the matrix elements for $H_{00}$, $\mathcal{Q}_{2M}$ and $L_{1M}$. 

The RMEs of a \emph{coupled} product of two $\grpsu{3}$ tensors can be expressed as a sum of products of the RMEs of the individual tensor operators, using Racah's reduction formulas.  
There are two cases to consider.  
In the first case, the two tensor operators $T^{(\lambda_t\mu_t)}$ and $S^{(\lambda_s\mu_s)}$ act on the same Hilbert space.  The RME of the coupled product of these tensors is given by~\cite{ptrsla-277-1975-545-Butler,McCoy2018,privcom-Rowe-2012}
\begin{multline}
\rme{\varrho'\omega'}
{\big[T^{(\lambda_t\mu_t)}\times S^{(\lambda_s\mu_s)}\big]^{\rho_0(\lambda_0\mu_0)}}
{\varrho\omega}_{\rho}\\
=\sum_{\bar{\omega}\bar{\varrho}\rho_0'\rho_t\rho_s}
\Phi_{\rho_0\rho_0'}[(\lambda_t\mu_t)(\lambda_s\mu_s)(\lambda_0\mu_0)]
U\big(\omega(\lambda_s\mu_s)\omega'(\lambda_t\mu_t);\bar{\omega}\rho_s\rho_t(\lambda_0\mu_0)\rho_0'\rho\big)\\
\rme{\varrho'\omega'}
{T^{(\lambda_t\mu_t)}}
{\bar{\varrho}\bar{\omega}}_{\rho_t}
\rme{\bar{\varrho}\bar{\omega}}
{S^{(\lambda_s\mu_s)}}
{\varrho\omega}_{\rho_s},
\label{su3-Racah1}
\end{multline}
where $\Phi[\dots]$ is an $\grpsu{3}$ phase matrix~\cite{Escher1997} associated with interchanging the coupling order of two $\grpsu{3}$ irreps, and $U[\dots]$ is a unitary Racah coefficient~\cite{np-62-1965-1-Hecht}, which relates different coupling orders of three $\grpsu{3}$ irreps.

In the second case, where  $T^{(\lambda_t\mu_t)}$ and $S^{(\lambda_s\mu_s)}$ act on different Hilbert spaces, then the Racah reduction formula is given by~\cite{ptrsla-277-1975-545-Butler,Escher1997,jmp-39-1998-5123-Escher,McCoy2018}
\begin{multline}
\rme{[\varrho_1'\omega_1',\varrho_2'\omega_2']\rho'\omega'}
{\big[T^{(\lambda_t\mu_t)}\times S^{(\lambda_s\mu_s)}]\big]^{\rho_0(\lambda_0\mu_0)}}
{[\varrho_1\omega_1,\varrho_2\omega_2]\rho\omega}_{\rho_0'}\\
=\sum_{\rho_s\rho_t}
\left[\begin{array}{cccc}
\omega_1&\omega_2&\omega&\rho\\
(\lambda_t\mu_t)&(\lambda_s\mu_s)&(\lambda_0\mu_0)&\rho_0\\
\omega_1'&\omega_2'&\omega'&\rho'\\
\rho_t&\rho_s&\rho_0'
\end{array}
\right]
\rme{\varrho_1'\omega_1'}{T^{(\lambda_t\mu_t)}}{\varrho_1\omega_1}_{\rho_t}\,
\rme{\varrho_2'\omega_2'}{S^{(\lambda_s\mu_s)}}{\varrho_2\omega_2}_{\rho_s}.
\label{su3-Racah2}
\end{multline}

In this case, the coefficient in brakets is a unitary 9-$(\lambda\mu)$ symbol~\cite{jmp-19-1978-1513-Millener} (analogous to the unitary 9-$j$ symbol). 
 
\section{
Calculations in a $\grpu{3}$ many-body basis
\label{section:u3-many-body}
}In this section, we describe a computational framework for carrying out calculations in a many-body basis which has definite $\grpu{3}$ symmetry.  The basis used in this framework consists of $\grpu{3}$-coupled many-body configurations, which are distributions of particles over single particle states which have been coupled together to have definite $\grpu{3}$ symmetry~\cite{pr-12-1974-201-Anyas-Weiss,npa-244-1975-365-Hecht,npa-897-2013-109-Luo,cpc-207-2016-202-Dytrych}.  The resulting coupled configurations are linear combinations of Slater determinants.  Constructing the basis in this way enables the use of the well established second quantization techniques for computing matrix elements for  a fermionic many-body system. 

The coupled configurations may have additional symmetries depending on whether protons and neutrons as treated as distinct particle types (proton-neutron scheme)  or as different states of the same particle type under isospin symmetry (isospin scheme).  In both schemes, the $\grpu{3}$-coupled configurations, by construction, also have definite total spin $S$.  Thus the coupled configurations can be organized into irreps of $\grpu{3}\times\grpsuS$, where $\grpsuS$ is the group generated by the spin operators $\{S_0,S_\pm\}$.  Within each irrep, the $S$ is coupled to the different orbital angular momenta $L$ appearing in the $\grpu{3}$ irrep $\omega$ to obtain states with definite $J$, which is associated with the group $\grpsuJ$ generated by the angular momentum operators $\{J_0,J_\pm\}$.\footnote{For later reference, we note that the problem of finding the $J$ arising in the Kronecker product $L\times S$, by the standard triangle inequality, can be described in the context of the group $\grpso{3}\times\grpsuS$ as finding the branching to its subgroup $\grpsuJ$.}  Thus, the coupled configurations reduce the group chain 
\begin{equation}
\begin{array}{ccccc}
\omega&\kappa&L\\
\mathrm{U}(3)&\supset&\mathrm{SO}(3)\\
&&\times&\supset&\grpsuJ\\
&&\grpsuS&&J\\
&&S\\
\end{array}.
\label{su3_so3_su2_chain}
\end{equation}

In the isospin scheme, it is natural to construct coupled configurations which also have good total isospin $T$.  
The resulting configurations then form irreps of $\grpsuT$, which is the group generated by the isospin operators $\{T_0,T_\pm\}$.  
In addition to $\grpsuT$, there is another symmetry which plays a key role in constructing the $\grpu{3}$ many-body basis in the isospin scheme, namely, the Wigner $\grpsu{4}$ symmetry~\cite{pr-51-1937-106-Wigner}.  
The $\grpsu{4}$ group associated with this symmetry is generated by the $\grpsuS$ and $\grpsuT$ generators, along with the Gamow-Teller operators~\cite{jmp-10-1969-1571-Hecht}.  
When taken together with a particle number operator, these operators generate a $\grpu{4}$ group, which, as discussed below, plays a key role in enforcing antisymmetry within each shell of a coupled configuration. 

Generically, irreps of $\grpu{4}$ are labeled by  $[{\bf f}]=[f_1f_2f_3f_4]$, where $f_i$ is the length of the $i$th row of a four-rowed Young tableau~\cite{trsa-239-1943-305-Littlewood}.  
The rows of the tableau correspond to the four possible combinations of spin and isospin projections ($m_s=\pm1/2$ and $m_t=\pm1/2$). 
The length of each row corresponds to the number of particles with a given pair of projections ($m_sm_t$) for the highest weight state of the irrep, which is the unique state that is annihilated by all of the $\grpu{4}$ ladder operators ($S_+$, $T_+$, etc.).   These irrep labels are related to those of the subgroup $\grpsu{4}$ by $[{\boldsymbol \tau}]=[f_1-f_2,f_2-f_3,f_3-f_4]$. 

Since $\grpsuS\times\grpsuT$ is a subgroup of $\grpsu{4}$ [and thus of $\grpu{4}$], the isospin and spin of the configuration is obtained by branching from total $[{\bf f}]$ to $S$ and $T$~\cite{jmp-10-1969-1571-Hecht}.  
As there can be more than one $\grpsuS\times\grpsuT$ irrep appearing in $[{\bf f}]$ with same labels $ST$, we distinguish different irreps with the same labels $ST$ by a multiplicity index $\beta$ which has maximum value $\beta_{\max}$.    
Thus, to obtain coupled configurations with definite spin and isospin, we construct coupled configurations that reduce the group chain 

\begin{equation}
\begin{array}{ccccc}
\omega&\kappa&L\\
\mathrm{U}(3)&\supset&\mathrm{SO}(3)\\
&&\times&\supset&\grpsuJ\\
\grpu{4}&\supset&\grpsuS\times\grpsuT&&J\\
\left[ {\bf f}\right]&\beta&ST
\end{array}.
\label{su3_so3_su4_chain}
\end{equation}
Note, however, that this is not the only coupled-configuration basis which can be constructed with good total spin and isospin symmetry in the isospin scheme. 

For example, in earlier $\grpu{3}$ shell model applications~\cite{pr-12-1974-201-Anyas-Weiss,npa-244-1975-365-Hecht}, the $\grpu{4}$ symmetry was enforced only within each shell to ensure antisymmetry.  
The coupled configuration in these works had good total spin and isospin symmetry, but not good total $\grpu{4}$ symmetry.  
Here we choose to construct a basis which does have good total $\grpu{4}$ symmetry.

In the following, we describe how the $\grpu{3}$ many-body basis is constructed for either scheme (Section~\ref{subsection:su3-many-body}).  We then briefly lay out the computational scheme for computing matrix elements in this basis (Section~\ref{subsection:su3basis-me}), which is based on the computational schemes of Ref.~\cite{cpc-83-1994-59-Bahri,cpc-207-2016-202-Dytrych}.

\subsection{Constructing many-body states\label{subsection:su3-many-body}}

In principle, a many-body state with definite $\grpsu{3}$ symmetry can be obtained by coupling the single particle states, following the $\grpu{3}$ and $\grpsuS$ [or $\grpu{4}$] coupling rules. 
The single particle basis used in constructing the $\grpu{3}$ many-body basis are harmonic oscillator eigenstate combined with spin (and isospin).
In the proton-neutron scheme, the single particle states are  denoted by $\ket{\chi Nlsjm}$, where $\chi$ indicates the species ($\pi$ for protons or $\nu$ for neutrons), $N$ is the principal quantum number associated with the harmonic oscillator shell, $l$ is the orbital angular momentum, $s$ is the spin, and $j$ and $m$ are the angular momentum and its projection, respectively.  
The spin label $s$ is often suppressed, as $s=1/2$ for protons and neutrons.
In an isospin scheme, the states are labeled by isospin $t$ and its projection $m_t$ rather than $\chi$.  However, note that  $\chi$ can be readily related to $m_t$,  e.g, $m_t=1/2$ can be identified with protons ($\pi$) and $m_t=-1/2$ with neutrons ($\nu$).\footnote{None of the results presented in this work depend on the choice to identify, e.g., protons by $m_t=1/2$ rather than $m_t=-1/2$.}
For a single particle, the $\grpu{4}$ irrep is unique and is labeled by $[{\bf f}]=[1]\equiv[1,0,0,0]$.  
Thus, in the isospin scheme, we denote the single particle state by $\ket{[{\bf f}]Nlsjmtm_t}$.

In both the proton-neutron and isospin schemes, the harmonic oscillator states have definite $\grpu{3}$.  
Specifically, the $\grpu{3}$ symmetry is labeled by $\omega=N_\omega(N,0)$, where $N_\omega=N+\frac32$ is the oscillator energy of the state.
The set of harmonic oscillator states $\tket{Nlm_l}$ with fixed $N$ form an basis for a $\grpu{3}$ irrep that reduces the group chain $\grpu{3}\supset\grpso{3}$.\footnote{Note that these $\grpu{3}$ irreps are more technically irreps of a projective representation of $\grpu{3}$ which is a double cover of $\grpu{3}$~\cite{Borel2000}.  Analogous to the relationship between $\grpsu{2}$ and $\grpso{3}$, the projective group has the same underlying algebra,  but the irrep labels of this group may take on half-integer values.    Similarly, the $\grpsptr$ many-body basis described in Section~\ref{subsection:sp3r-many-body} actually reduces the group chain of the metaplectic group $\mathrm{Mp}(3)$, which is a two-fold covering group of $\grpsptr$ with the same algebra as $\grpsptr$.  The projective realization of $\grpu{3}$ is a subgroup of this metaplectic group.  Again, the only relevant consequence is that the irrep labels may have half-integer values.}  
Thus, after taking into account the spin $s$ (which couples with $l$ to obtain states with good angular momentum $j$), we have single particle states, denoted by $\tket{Nlsjm}$.
Specifically, these states form a basis which reduces the group chain~\eqref{su3_so3_su2_chain}. 
Similarly, in the isospin scheme, the single particle states form bases for irreps of $\grpsu{3}\times\grpu{4}$ labeled by $\omega[1]$.  
These states form a basis which reduces the group chain~\eqref{su3_so3_su4_chain}.

However, when constructing a many-body state in this way, care must be taken to ensure that the coupled products of the single particles states lie in the antisymmetric space of $A$ particles. To avoid constructing many-body states which lie outside the antisymmetric space, we use a group theoretical relation between the permutation symmetry group $S_n$ for $n$ particles and the $\grpu{\Lambda}$ symmetry associated with each the distribution of particles over the $\Lambda$ possible single particle states within an oscillator shell.  
The number of single particle states $\Lambda=\Omega N_\Sigma$ is the  product of the number of spatial states $\Omega=(N+1)(N+2)/2$ and the number of spin (or spin and isospin) substates $N_\Sigma$, where  $N_\sigma=2$ in the proton-neutron scheme and $N_\Sigma=4$ in the isospin scheme.  The relevant relation between $S_n$ and $\grpu{\Lambda}$ is that the Young tableau labeling the $S_n$ symmetry of $n$ particles will have the same shape as the Young tableau which describes the $\grpu{\Lambda}$ symmetry associated with the distribution of those particles over the available single particle states.  
Because a fully antisymmetric product state has permutation symmetry $[1^n]$, the corresponding $\grpu{\Lambda}$ irrep will also be labeled by $[1^n]$.  It then follows that antisymmetric irreps of $\grpu{3}\times\grpsuS$ or $\grpu{3}\times\grpu{4}$ that could be obtained by coupling together single particle states within a single oscillator shell must lie within this $\grpu{\Lambda}$ irrep. 
Thus, the problem of constructing the  antisymmetrized coupled products of single particle states within each shell is equivalent to constructing the irreps of $\grpu{3}\times\grpsuS$ or $\grpu{3}\times\grpu{4}$ which belong to the $\grpu{\Lambda}$ irrep $[1^n]$.   
The full many-body state can then be obtained by coupling together antisymmetric irreps from each shell. 

To identify the antisymmeteric $\grpu{3}\times\grpu{N_\Sigma}$ irreps, we start by imposing the $\grpu{\Lambda}$ symmetry $[1^n]$.  
We then factorize the spatial degrees of freedom from the spin (or spin and isospin) degrees of freedom, which gives rise to a direct product group $\grpu{\Omega}\times \grpu{N_\Sigma}$, which is a subgroup of $\grpu{\Lambda}$.  
Irreps of this product group are labeled by the two Young tableau, $[{\bf f}^\Omega]$ and $[{\bf f}^\Sigma]$, which label the irreps of  $\grpu{\Omega}$  and $\grpu{N_\Sigma}$, respectively.  
It can be shown that if an irrep of this product group  belongs to the $[1^n]$ irrep of $\grpu{\Lambda}$, then $[{\bf f}^\Omega]$ and $[{\bf f}^\Sigma]$ must be conjugates under permutation symmetry~\cite{Chen2002},\footnote{Conjugation under permutation symmetry corresponds to exchange the rows and columns of the Young diagram.}  i.e., $[{\bf f}^\Omega]=[{\bf \tilde{f}}^\Sigma]$, where the tilde indicates conjugation. 
In the proton-neutron scheme, the $\grpu{2}$ irrep can alternatively be labeled by the number of particles ($\sum_if_i$) along with the irrep label $S=(f^\Sigma_1-f^\Sigma_2)/2$ for its $\grpsuS$ subgroup. 
 In the isospin scheme, $[{\bf f}^\Sigma]= [{\bf f}]$. 
The set of $\grpu{3}$ irreps appearing in an $[{\bf f}^\Omega]$ irrep can be obtained via $\grpu{\Omega}\rightarrow\grpu{3}$ branching rules~\cite{cpc-56-1989-279-Draayer}.\footnote{A code that generates the set of $\omega$  that arise within a given $[{\bf f}^\Omega]$ irrep is available~\cite{cpc-244-2019-442-Langr}.}
 Thus, identifying antisymmetric $\grpu{3}\times\grpu{N_\Sigma}$ irreps within a single oscillator shell is equivalent to constructing states which reduce the group chain 
\begin{equation}
\begin{array}{ccccc}
&&[{\bf \tilde{f}}^\Sigma]&\alpha&\omega\\
&& \grpu{\Omega}&\supset& \grpu{3}\\
\grpu{\Lambda}[S_n]&\supset& \times \\
\,[1^n]&&\grpu{N_\Sigma}.\\
&&[{\bf f}^\Sigma]
\end{array}
\label{eq:single_shell_irreps}
\end{equation}
where $\alpha$ is a branching multiplicity index distinguishing different $\grpu{3}$ irreps in $[{\bf\tilde{f}}^\Sigma]$ with the same label $\omega$. 
To simplify notation in the remainder of this work, we will suppress redundant labels, such as $[{\bf \tilde{f}}^\Sigma]$.  In the proton-neutron scheme, we will also suppress the redundant $\grpu{2}$ label $[{\bf f}^\Sigma]$, as this label can be deduced from the labels $[1^n]$ and $S$.

To construct a basis for a $\grpu{3}\times\grpu{N_\Sigma}$ irrep which reduces the group chain in \eqref{eq:single_shell_irreps}, we first construct the  highest weight state, which is the unique state in the irrep which is simultaneously annihilated by the raising operators of the generators of each the groups in the chain~\cite{cpc-83-1994-59-Bahri}.  This state can be obtained by solving for the simultaneous null space of all of the raising operators in a Slater determinant basis spanned by single particle states within the given oscillator shell.  Once this highest weight state is constructed, other states in the $\grpu{3}\times\grpu{N_\Sigma}$ irrep can  be obtained using the $\grpu{3}$ and $\grpu{N_\Sigma}$  lowering operators. 

The full many-body state is then obtained by successively coupling together $\grpu{3}\times\grpu{N_\Sigma}$ irreps from different shells. 
In the proton-neutron scheme, an $\grpu{3}\times\grpsuS$ many-body state is  obtained by first coupling  together irreps for protons and neutrons separately to obtain states with  definite total $\omega_\pi S_\pi$ and $\omega_\nu S_\nu$, respectively, and then coupling together the proton and neutron irreps.  That is, the many-body states are given by 
\begin{multline}
\ket{\varrho \omega S\kappa L JM}=
\Bigg[
\left[
	\left[
		\left[
			\ket{\Gamma_{1\pi}}
			\times
			\ket{\Gamma_{2\pi}}
		\right]^{\rho_{12\pi}\omega_{12\pi} S_{12\pi}}
		\times\ket{\Gamma_{3\pi}}
	\right]^{\Gamma_{12,3\pi}}
\times \dots
\right]^{\rho_\pi\omega_\pi S_\pi}\\
\times
\left[
	\left[
		\left[
			\ket{\Gamma_{1\nu}}
			\times
			\ket{\Gamma_{2\nu}}
		\right]^{\rho_{12\nu}\omega_{12\nu}S_{12\nu}}
		\times\ket{\Gamma_{3\nu}}
	\right]^{\omega_{12,3\nu}S_{12,3\nu}}
\times \dots
\right]^{\rho_\nu\omega_\nu S_\nu}
\Bigg]^{\rho\omega S}_{\kappa LJM},
\label{eq:su3pn_scheme}
\end{multline}
where $\Gamma_{i} = [1^{n_i}]\alpha_i \omega_i S_i$ labels the irrep in the $i$th oscillator shell. On the left hand side, $\varrho$ is a composite of all of the additional quantum numbers that describe the product state, beyond the total symmetry labels. In particular, $\varrho$ includes the labels $\Gamma_i$ associated with each oscillator shell and the labels associated with the intermediate couplings, e.g., $\rho_{12}\omega_{12}S_{12}$.   

In the isospin scheme, the many-body state is similarly obtained by coupling together irreps from different shells according to $\grpu{3}$ and $\grpu{4}$ coupling rules, and then branching the final $\grpu{3}\times\grpu{4}$ irrep to obtain states with definite isospin $T$ as well as orbital angular momentum, spin and angular momentum.  That is, the many-body state is given by   
\begin{equation}
\ket{\varrho \omega [{\bf f]} \beta S T \kappa L JM}=
\Bigg[
	\left[
		\left[
			\ket{\Gamma_{1}}
			\times
			\ket{\Gamma_{2}}
		\right]^{\rho_{12}\omega_{12} \delta_{12}[{\bf f}_{12}]}
		\times\ket{\Gamma_{3}}
	\right]^{\rho_{12,3}\omega_{12,3}\delta_{12,3}[{\bf f}_{12,3}]}
\times \dots
\Bigg]^{\rho\omega \delta [{\bf f}]\beta ST}_{\kappa LJM},
\label{eq:su3st_scheme}
\end{equation}
where now $\Gamma_i= [1^{n_i}][{\bf f}_i]\alpha_i \omega_i $.  The label $\varrho$ again represents all of the additional quantum numbers for the state, including the $\Gamma_i$,  the outer-multiplicity indices ($\delta$'s) distinguishing different $\grpu{4}$ irreps which can arise in the coupled product of two $\grpu{4}$ irreps, and the branching multiplicities ($\beta$'s).

The method for carrying out calculations in a symplectic many-body method described in this work is largely independent of chosen coupling scheme.  
Thus, for the remainder of this work we will use $\Sigma$ to denote scheme specific spin quantum numbers, e.g., $S_\pi S_\nu$, or $[f]\beta TM_T$, and only point out instances where the distinction may be important.   
That is, each many-body state will be labeled by $\ket{\varrho \omega \kappa L\Sigma S JM}$.

\subsection{Computing matrix elements in a $\grpu{3}$ many-body basis\label{subsection:su3basis-me}}

To carry out calculations in the $\grpu{3}$ many-body basis described above, we write the operators in terms of coupled products of fermionic creation and annihilation operators which have definite $\grpsu{3}\times\grpsu{N_\Sigma}$ symmetry.  Note, the group $\grpsu{N_\Sigma}$ is the subgroup of $\grpu{N_\Sigma}$ obtained by removing the particle number operator from the underlying Lie algebra.  In the proton-neutron scheme, this subgroup is just $\grpsuS$.  In the isospin scheme, this subgroup is $\grpsu{4}$, which contains $\grpsuS\times\grpsuT$ as a subgroup.  We label irreps of $\grpsu{4}$ by $[{\bf \tau}]=[f_1-f_2, f_2-f_3,f_3-f_4]$.  
Obtaining such products of the creation and annihilation operators is straightforward, as the single-particle creation and annihilation operators are themselves $\grpsu{3}\times\grpsu{N_\Sigma}$ tensors.

In the proton-neutron scheme, the fermion creation operators can be expressed as components of $\grpsu{3}\times\grpsu{2}$ tensors as 
\begin{equation}
a^{\dagger(N,0)}_{\chi ljm}= a^{\dagger}_{\chi Nljm}\qquad \tilde{a}^{(0,N)}_{\chi ljm}=\tilde{\Phi}[(N,0)lsjm]a_{\chi Nlj-m},
\end{equation}
where $\tilde{\Phi}[(N,0)lsjm]$ is the conjugation phase factor for the group chain~\eqref{su3_so3_su2_chain}.\footnote{\label{fn:sp_ho_conjugation_phase_pn}As with the $\grpu{3}$ [see \eqref{eq:phi_def}], we follow the convention that the conjugation phase factor is given by the phase of the identity-coupled coupling coefficient. 
 The conjugation phase factor for the single-particle harmonic oscillator basis is thus given by 
\begin{displaymath}
\Phi[(N,0)lsjm]=\sign\ccg
{(N,0)}{l s}{ jm}
{(0,N)}{l s}{j\,- m}
{(0,0)}{00}{00}
 = (-)^N(-)^{j+m}.
\end{displaymath}
Note that for $\grpsu{3}$ irreps with $\lambda=0$ or $\mu=0$, the branching to $\grpso{3}$ is always multiplicity free.}
One can construct coupled products of these operators which are $\grpsu{3}\times\grpsuS$ tensors using the coupling coefficients of the group chain~\eqref{su3_so3_su2_chain}.  These coupling coefficient can be factored into a product of an $\grpsu{3}$ reduced coupling coefficient,  a unitary 9-$j$ symbol, and a Clebsch-Gordon coefficient as ~\cite{Chen2002,McCoy2018}: 
\begin{equation}
\ccg
{(\lambda_1\mu_1)}
{\kappa_1 L_1 S_1} 
{J_1M_1}
{(\lambda_2\mu_2)}
{\kappa_2 L_2 S_2}
{J_2M_2}
{(\lambda\mu)}{\kappa L S}
{JM}_\rho
= \rcc{(\lambda_1\mu_1)}{\kappa_1 L_1}{(\lambda_2\mu_2)}{\kappa_2 L_2}{(\lambda\mu)}{\kappa L}_{\rho}
\uninej{L_1}{S_1}{J_1}
{L_2}{S_2}{J_2}
{L}{S}{J}
(J_1M_2;J_2M_2|JM).
\label{eq:u3lsjm_cc}
\end{equation}
 
The coupled products of these operators can be obtained by coupling together the creation and annihilation operators following a similar coupling scheme as that used in constructing the many-body states \eqref{eq:su3pn_scheme}. That is, we first couple together of the creation and annihilation operators which create or annihilate a particle in the $i$th oscillator shell to have definite $\grpsu{3}\times\grpsu{2}$ tensor character
\begin{equation}
\mathcal{A}^{(\lambda_{i}\mu_{i})S_{i}}_\chi=\left[a^{\dagger(N_i,0)}_\chi\times a^{\dagger(N_i,0)}_\chi\times..\times\tilde{a}^{(0,N_i)}_\chi\times\tilde{a}^{(0,N_i)}_\chi\right]^{(\lambda_{i}\mu_{i})S_{i}}.
\end{equation}
Then, for protons and neutrons separate, we couple the operators from each shell together.  Finally we coupled the proton operator to the neutron operator to obtain
\begin{multline}
\mathcal{O}^{\rho_0(\lambda_0\mu_0)S_0}\\
=
\Bigg[
\left[
	\left[
	[
		\mathcal{A}_\pi^{(\lambda_{1\pi}\mu_{1\pi})S_{1\pi}}\times \mathcal{A}_\pi^{(\lambda_{2\pi}\mu_{2})S_{2\pi}}
	]^{\rho_{12\pi}(\lambda_{12\pi}\mu_{12\pi})S_{12\pi}}
	\times 
	\mathcal{A}_\pi^{(\lambda_{03}\mu_{03})S_{03}}
	\right]^{\rho_{12,3\pi}(\lambda_{12,3\pi}\mu_{12,3\pi})S_{12,3\pi}}
	\times \dots
\right]^{\rho_\pi\omega_{0\pi}S_{0\pi}}\qquad\qquad\\
\times 
\left[
	\left[
	[
		\mathcal{A}_\nu^{(\lambda_{1\nu}\mu_{1\nu})S_{1\nu}}\times \mathcal{A}_\nu^{(\lambda_{2\nu}\mu_{2})S_{2\nu}}
	]^{\rho_{12\nu}(\lambda_{12\nu}\mu_{12\nu})S_{12\nu}}
	\times 
	\mathcal{A}_\pi^{(\lambda_{03}\mu_{03})S_{03}}
	\right]^{\rho_{12,3\nu}(\lambda_{12,3\nu}\mu_{12,3\nu})S_{12,3\nu}}\times\dots
\right]^{\rho_\pi\omega_{0\pi}S_{0\pi}}
\Bigg]^{\rho_0(\delta_0)\omega_0\Sigma_0S_0}.
\end{multline}

To evaluate operator matrix elements, we compute RMEs of the coupled creation and annihilation operators $\trme{\varrho'\Sigma' \omega' S'}{\mathcal{O}^{\rho_0(\lambda_0\mu_0)S_0}}{\varrho\Sigma \omega S}_{\rho_0'}$.  These RMEs can be evaluated through repeated application of Racah factorization to rewrite the RME of $\mathcal{O}^{\rho_0(\lambda_0\mu_0)S_0}$ in terms of sums of products of recoupling coefficients and RMEs $\trme{\Gamma_{i\chi}'}{\mathcal{A}^{(\lambda_i\mu_i)S_{i}}_\chi}{\Gamma_{i\chi}}$.  
For this group chain \eqref{su3_so3_su2_chain}, Racah's factorization for operators acting on different spaces (i.e., different shells) is given by~\cite{McCoy2018} 
\begin{multline}
\rme{[\varrho_1'\omega_1'S_1,\varrho_2'\omega_2'S_2']\rho'\omega'S'}
{\big[T^{(\lambda_t\mu_t)S_t}\times S^{(\lambda_s\mu_s)S_s}\big]^{\rho_0(\lambda_0\mu_0)S_0}}
{[\varrho_1\omega_1S_1,\varrho_2\omega_2S_2]\rho\omega}_{\rho_0'}\\
=\sum_{\rho_s\rho_t}
\left[\begin{array}{cccc}
\omega_1&\omega_2&\omega&\rho\\
(\lambda_t\mu_t)&(\lambda_s\mu_s)&(\lambda_0\mu_0)&\rho_0\\
\omega_1'&\omega_2'&\omega'&\rho'\\
\rho_t&\rho_s&\rho_0'
\end{array}
\right]
\uninej{S_1}{S_2}{S}{S_t}{S_s}{S_0}{S_1'}{S_2'}{S'}\\
\times \rme{\varrho_1'\omega_1'S_1'}{T^{(\lambda_t\mu_t)S_t}}{\varrho_1\omega_1S_1}_{\rho_t}\,
\rme{\varrho_2'\omega_2'S_2'}{S^{(\lambda_s\mu_s)S_s}}{\varrho_2\omega_2S_2}_{\rho_s}.
\label{su3su2-Racah2}
\end{multline}

To obtain the single shell RMEs $\trme{\Gamma_{i\chi}'}{\mathcal{A}^{(\lambda_i\mu_i)S_{i}}_\chi}{\Gamma_{i\chi}}$, we first compute the matrix element of one of the components of the tensor operator $\mathcal{A}^{(\lambda_i\mu_i)S_{i}}_\chi$ between the highest weight states of the irreps $\ket{\Gamma_i}$ and $\ket{\Gamma_i'}$.
Because these highest weight states are explicitly expanded in terms of Slater determinants when constructing the $\grpu{3}$ many-body basis, standard shell model techniques for computing matrix elements of second quantized operators can be used to compute this matrix element.
The RME is then obtained from the matrix element using the Wigner-Eckart theorem.

Similarly, in the isospin scheme, the fermionic creation and annihilation operators can be expressed as components of $\grpu{3}\times\grpu{4}$ tensors as 
\begin{equation} 
a^{\dagger(N,0)[1]}_{lsjmtm_t}= a^{\dagger}_{Nlsjmtm_t} \qquad \tilde{a}^{(0,N)[01]}_{ljmm_t}=
\tilde{\Phi}\big[(N,0)[1]lsjmtm_t\big]a_{Nlj,-m,-m_t}.
\end{equation} 
where $\tilde{\Phi}\big[[1](N,0)lsjmtm_t\big]$ is the conjugation phase factor for the group chain \eqref{su3_so3_su4_chain}.\footnote{Again, we follow the convention that the  conjugation phase factor is given by the sign of the identity-coupled coupling coefficient for the group chain \eqref{su3_so3_su4_chain}.
The conjugation phase, given by the identity-coupled coupling coefficient, is then
\begin{equation}
\tilde{\Phi}\big[[1](N,0)lsjmtm_t\big]=(-)^N(-)^{j+m}\phi([1])(-)^{t+m_t},
\end{equation}  
where $\phi([1])=\sign\trcc{[1]}{s t}{[01]}{s t}{[0]}{00}$.
Note that coupled products involving the $\grpsu{4}$ irrep [1] always have $\delta_{\max}=1$.  Similarly, for this irrep, $\beta_{\max}=1$.  
}  
These creation and annihilation operators can be operators can be coupled to form good $\grpsu{3}\times\grpsu{4}$ tensors using the $\grpsu{3}\times\grpsu{4}$ coupling coefficients, which can be expressed in terms of a product of coupling and reduced coupling coefficients as 
\begin{multline}
\left(\begin{array}{cc}
(\lambda_1\mu_1)[{\boldsymbol \tau}_1] & (\lambda_2\mu_2)[{\boldsymbol \tau}_2]\\
\kappa_1 L_1 \beta_1S_1 T_1& \kappa_2 L_2 \beta_2S_2T_2\\
 J_1M_1 &J_2M_2
\end{array}
\right|
\left.
\begin{array}{c}
(\lambda\mu)[{\boldsymbol \tau}]\\
\kappa L \beta ST\\ JM
\end{array}
\right)_{\rho,\delta}
= 
\rcc{(\lambda_1\mu_1)}{\kappa_1 L_1}{(\lambda_2\mu_2)}{\kappa_2 L_2}{(\lambda\mu)}{\kappa L}_{\rho}
\uninej{L_1}{S_1}{J_1}
{L_2}{S_2}{J_2}
{L}{S}{J}
(J_1M_1;J_2M_2|JM)\\
\times
\rcc
{[{\bf \tau}_1]}{\beta_1S_1T_1}
{[{\bf \tau}_2]}{\beta_2S_2T_2}
{[{\bf \tau}]}{\beta ST}_\delta
(T_1M_1;T_2M_2|TM).
\label{eq:su3su4cc}
\end{multline}
The factors on the right hand side now additionally include an $\grpsu{4}$ reduced coupling coefficient and an isospin Clebsch-Gordan coefficient. 
Just as in the proton-neutron scheme, we first couple the operators acting on the same oscillator shell together 
\begin{equation}
\mathcal{A}^{(\lambda_{i}\mu_{i})[{\boldsymbol \tau}_i]}=\left[a^{\dagger(N_i,0)}\times a^{\dagger(N_i,0)}\times..\times\tilde{a}^{(0,N_i)}\times\tilde{a}^{(0,N_i)}\right]^{(\lambda_{i}\mu_{i})[\boldsymbol{\tau}_i]}
\end{equation}
and then couple together the products of operators acting in different shells to obtain the total operator

\begin{multline}
\mathcal{O}^{\rho_0(\lambda_0\mu_0)\delta_0[\boldsymbol{\tau}_0]}\\=
\left[\left[\left[
\mathcal{A}^{(\lambda_{1}\mu_{1})[\boldsymbol{\tau}_1]}
\times
\mathcal{A}^{(\lambda_{2}\mu_{2})[\boldsymbol{\tau}_2]}
\right]^{\rho_{12}(\lambda_{12}\mu_{12})\delta_{12}[\boldsymbol{\tau}_{12}]}
\times
\mathcal{A}^{(\lambda_{3}\mu_{3})[\tau_3]}
\right]^{\rho_{12,3}(\lambda_{12,3}\mu_{12,3})\delta_{12,3}[\boldsymbol{\tau}_{12,3}]}\times\dots
\right]^{\rho_0(\lambda_0\mu_0)\delta_0[\boldsymbol{\tau}_0]}.
\end{multline}
The reduced matrix elements of this operator are then obtained following the same procedure as in the proton-neutron scheme.  The Racah factorization is now given by\footnote{This expression can be obtained from the expressions for Racah's factorization for a generic group chain given in, e.g., Ref.~\cite{McCoy2018}.}
\begin{multline}
\rme{[\varrho_1'\omega_1'[{\bf f}_1'],\varrho_2'\omega_2'[{\bf f}_2']]\rho'\omega'\delta'[{\bf f}']}
{\big[T^{(\lambda_t\mu_t)[{\boldsymbol \tau}_t]}\times S^{(\lambda_s\mu_s)[{\boldsymbol\tau }_s]}]\big]^{\rho_0(\lambda_0\mu_0)\delta_0[{\boldsymbol \tau}_0]}}
{[\varrho_1\omega_1[{\bf f}_1]S_1,\varrho_2\omega_2[{\bf f}_2]S_2]\rho\omega\delta [{\bf f}]}_{\rho_0',\delta_0'}\\
=\sum_{\delta_s\delta_t\rho_s\rho_t}
\left[\begin{array}{cccc}
\omega_1&\omega_2&\omega&\rho\\
(\lambda_t\mu_t)&(\lambda_s\mu_s)&(\lambda_0\mu_0)&\rho_0\\
\omega_1'&\omega_2'&\omega'&\rho'\\
\rho_t&\rho_s&\rho_0'
\end{array}
\right]
\left[
\begin{array}{cccc}
[{\mathbf f}_1]&[{\mathbf f}_2]&[{\mathbf f}]&\delta\\{}[{\boldsymbol \tau}_t]&[{\boldsymbol \tau}_s]&[{\boldsymbol \tau}_0]&\delta_0\\{}
[{\mathbf f}_1']&[{\mathbf f}_2']&[{\mathbf f}']&\delta'\\
\delta_t&\delta_s&\delta_0'
\end{array}
\right]\\
\times \rme{\varrho_1'\omega_1'[{\mathbf f}_1']}{T^{(\lambda_t\mu_t)[{\boldsymbol \tau}_t]}}{\varrho_1\omega_1[{\mathbf f}_1]}_{\rho_t,\delta_t}\,
\rme{\varrho_2'\omega_2'[{\mathbf f}_2']}{S^{(\lambda_s\mu_s)[{\boldsymbol \tau}_s]}}{\varrho_2\omega_2[{\mathbf f}_2]}_{\rho_s,\delta_s},
\label{su3su2-Racah2}
\end{multline}
where the $[f]$ and $[\tau]$ dependent factor is an $\grpsu{4}$ unitary 9-$[\tau]$ symbol~\cite{Chen2002} analogous to the 9-$J$ symbol.  Note that while we use the $\grpu{4}$ labels of the bra and ket in the 9-$[\tau]$ symbol, only the related $\grpsu{4}$ labels are relevant for evaluating the symbol.
 \section{
$\grpsptr$ irreducible representations
\label{section:sp3r}
}

In this section, we provide a brief review of $\grpsptr$ irreducible representations which will be used in constructing a symplectic many-body basis in Section~\ref{subsection:sp3r-many-body}.  Specifically, we discuss different realizations of the $\grpsptr$ generators which are most relevant in the context of collective dynamics or calculations in a harmonic oscillator basis (Section~\ref{subsection:sp3r_alg}).  We then provide an overview of the construction of a basis for an $\grpsptr$ irrep using vector coherent state (VCS) theory (Sections~\ref{subsection:sp3r_irrep}), and analytic expression for the RMEs of the $\grpsptr$ generators (\ref{subsection:sp3r-generator-rme}).  

\subsection{Realizations of $\grpsptr$ generators \label{subsection:sp3r_alg}}
The algebra of $\grpsptr$ can be written in terms of different sets of generators, which correspond to different sub-models or sub-algebras.  
One such realization, which is related to a number of collective models, is given by expressing all 21 generators in terms of one-body, bilinear products of position and momentum coordinates $x_{si}$ and $p_{si}$~\cite{rpp-48-1985-1419-Rowe}, as
\begin{equation}
\begin{aligned}
{Q}_{ij}&=\sum_{s=1}^{A}{x}_{si}{x}_{sj}
\qquad&{P}_{ij}&=\sum_{s=1}^{A}(x_{si}p_{sj}+p_{si}x_{sj})\\
 L_{ij}&=\sum_{s=1}^{A}(x_{si}p_{sj}-x_{sj}p_{si})
 \qquad&K_{ij}&=\sum_{s=1}^{A}p_{si}p_{sj}, \label{sp_generators_canonical}
\end{aligned}
\end{equation}
where $Q_{ij}$ is the Cartesian quadrupole moment tensor,  $P_{ij}$ and  $L_{ij}$ are generators of deformation and rotations, and  $K_{ij}$ is the quadrupole flow tensor.  

For calculations in a harmonic oscillator many-body basis, it is useful to write generators in terms of the oscillator ladder operators $c^\dagger_{si}$ and $c_{sj}$.  In terms of these operators, the $\grpsptr$ generators are given by
\begin{equation}
\begin{aligned}
A_{ij}&=\sum_s^{A} c^\dagger_{si}c^\dagger_{sj}\\
B_{ij}&=\sum_s^{A}c_{si}c_{sj}\\
C_{ij}&=\frac12\sum_s^{A}(c_{si}^\dagger c_{sj}+c_{sj}c^\dagger_{si}).
\label{sp_generators_boson}
\end{aligned}
\end{equation}

In this form, the generators can be grouped into the generators $C_{ij}$ of $\grpu{3}$ given in \eqref{su3_cartesian_generators} and symplectic raising ($A_{ij}$) and lowering ($B_{ij}$) operators, which create and annihilate two oscillator quanta, respectively.  
That is, $A_{ij}$ and $B_{ij}$ ladder between states with $N_\omega$ differing by two.  

Just as with $\grpsu{3}$,  when working in a harmonic oscillator many-body basis, it is useful to express the symplectic generators as $\grpsu{3}$ irreducible tensor operators in order to make use of  group-theoretical tools to simplify the calculations. 
Using the expressions for the oscillator ladder operators as $\grpu{3}$ tensors, as given in \eqref{eq:boson-tensors}, the symplectic generators in  \eqref{sp_generators_boson} can be rearranged into $\grpsu{3}$ tensor operators as~\cite{rosensteel1992:sp3r-tensors-gtssnp91,prc-65-2002-054309-Escher}
\begin{equation}
\begin{split}
A_{LM}^{(2,0)}
&=\frac{1}{\sqrt{2}}\sum_{s=1}^{A}\left[c_s^{\dagger(1,0)}\times c_s^{\dagger(1,0)}\right]_{LM}^{(2,0)}\\
B_{LM}^{(0,2)}
&=\frac{1}{\sqrt{2}}\sum_{s=1}^{A}\left[{c}_s^{(0,1)}\times c_s^{(0,1)}\right]_{LM}^{(0,2)}\\
C_{LM}^{(1,1)}
&=\sqrt{2}\sum_{s=1}^{A}\left[c_s^{\dagger(1,0)}\times c_s^{(0,1)}\right]_{LM}^{(1,1)}\\
H_{00}^{(0,0)}
&=\frac{\sqrt{3}}{2}\sum_{s=1}^{A}\Big(
\left[c_s^{\dagger(1,0)}\times c_s^{(0,1)}\right]_{00}^{(0,0)}+ \left[c_s^{(0,1)}\times c_s^{\dagger(1,0)}\right]_{00}^{(0,0)}
\Big)
\label{tensor_generators}.
\end{split}
\end{equation}
The six components of $A^{(2,0)}$ and the six components of $B^{(0,2)}$ are the symplectic raising and lowering operators, which are associated with the giant monopole ($L=0$) and quadrupole $(L=2)$ excitations, the eight components of the $\grpsu{3}$ tensor $C^{(1,1)}$ ($L=1,2$) and the spherical shell model Hamiltonian $H^{(0,0)}$ ($L=0$)~\cite{rpp-48-1985-1419-Rowe}.  

From the realization of the generators in terms of position and momentum in \eqref{sp_generators_canonical}, it can be shown that several operators which play an important role in nuclear calculations  can be expressed as linear combinations of components of these four $\grpsu{3}$ tensors in \eqref{tensor_generators}. Namely, the kinetic energy operator, the quadrupole operator, and the monopole operator $M_{00}=\sum_i{r_i^2}$ used to calculate $E0$ transitions and root-mean-square radii.  The expressions for these operators in terms of the $\grpsu{3}$ form of the $\grpsptr$ generators are obtained using the relations between the different realization of the $\grpsptr$ generators in \eqref{sp_generators_canonical}-\eqref{tensor_generators}.
The resulting expressions are: 
\begin{equation}
\begin{split}
T_{00}&=b^{-2}\left(H^{(0,0)}-\sqrt{\frac32}A^{(2,0)}_{00}-\sqrt{\frac32}B_{00}^{(0,2)}\right)\\
Q_{2M}&=b^2\sqrt{\frac{15}{16\pi}}\left(C^{(1,1)}_{2M}+A^{(2,0)}_{2M}+B^{(0,2)}_{2M}\right)\\
M_{00}&=b^2\left(H_{00}^{(0,0)}+\sqrt{\frac32}A^{(2,0)}_{00}+\sqrt{\frac32}B_{00}^{(0,2)}\right).
\end{split}
\end{equation}

\subsection{Basis for an $\grpsptr$ irrep
\label{subsection:sp3r_irrep}
}
In this section, we describe how an irrep of $\grpsptr$ is constructed and define the labeling scheme for $\grpsptr$ irreps used in this work.  
Unlike simpler groups like $\grpsu{3}$ or $\grpso{3}$, the group $\grpsptr$ is non-compact.
Thus, unitary irreps of $\grpsptr$ are infinite dimensional.
Fortunately, the $\grpsptr$ irreps relevant for the SpNCCI framework are discrete representations.   
Each such $\grpsptr$ irrep is composed of an infinite tower of $\grpu{3}$ irreps.  
Within each $\grpsptr$ irrep,  the $\grpu{3}$ irreps are partially ordered, or graded, by $N_\omega$, which corresponds to the oscillator energy (eigenvalue of $H^{(0,0)}$) associated with the irrep.  
There  exists a unique $\grpu{3}$ irrep $\sigma=N_\sigma(\lambda_\sigma\mu_\sigma)$ with lowest grade (lowest oscillator energy), for which all states within the irrep are annihilated by the lowering operator ($B^{(0,2)}\ket{\sigma}=0$).  
We refer to this as the \emph{lowest grade irrep} (LGI).\footnote{In other works~\cite{prl-98-2007-162503-Dytrych,prc-76-2007-014315-Dytrych,jmp-39-1998-5123-Escher,ppnp-89-2016-101-Launey,prc-76-2007-014315-Dytrych,jpg-35-2008-095101-Dytrych,jpg-35-2008-123101-Dytrych,epj-229-2429-2020-Launey,plb-727-2013-511-Dreyfuss,prc-95-2017-044312-Dreyfuss}, the LGI is referred to as a symplectic bandhead.}
\begin{figure}
\begin{center}
\includegraphics[width=0.8\hsize]{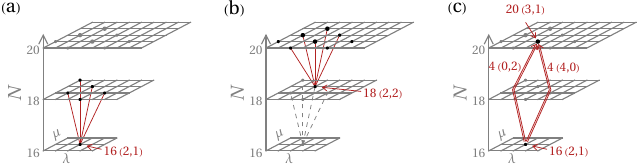}
\end{center}
\caption[Generating $\grpsptr$ irreps.]{$\grpu{3}$ irreps in the $\grpsptr$ irrep $N_\sigma(\lambda_\sigma\mu_\sigma)=16(2,1)$.  
Each irrep, with labels $(\lambda\mu)$, is indicated as a dot on the grid formed by different values of $\lambda$ and $\mu$. The size of the dot corresponds to the multiplicity $\upsilon_{\max}$ of that $\grpu{3}$ irrep in the $\grpsptr$ irrep. 
The irreps obtained by (a) acting with a single raising operator on the LGI, (b) acting on an excited $\grpu{3}$ irrep with a single raising operator and (c) acting on the LGI with a raising polynomial with $N_n=4$ are shown. 
Action of the raising operator is indicated by the red lines connecting the irreps.  
\label{fig:sp_laddering}}
\end{figure}

If the LGI of the $\grpsptr$ irrep is known, then the remainder of the irrep can be obtained through repeated action of the symplectic raising operator on the LGI.  
To see how this works, we take, as an example, an $\grpsptr$ irrep with LGI $\sigma=16(2,1)$, which appears in the many-body basis of $\isotope[8]{Be}$.  
Acting on $\sigma$ with the symplectic raising operator $A^{(2,0)}$ will ladder to a set of $\grpu{3}$ irreps $\omega$ with different $\grpsu{3}$ labels $(\lambda_\omega\mu_\omega)$, which arise in the Kronecker product  $(\lambda_\sigma\mu_\sigma)\times (2,0)$. 
Since $A^{(2,0)}$ carries two oscillator quanta, all of the irreps $\omega$ will have $N_\omega=N_\sigma+2$.  The possible irreps $\omega$ obtained by laddering with $A^{(2,0)}$ are shown in Fig.~\ref{fig:sp_laddering}(a).  In this figure, each grid corresponds to a space with a given grade, or number of oscillator quanta $N$.  The points on the grid represent the $\grpsu{3}$ irreps with that grade.  
The red lines connecting point in different $N$ spaces indicate the action of the symplectic raising operator $A^{(2,0)}$.   
As shown in Fig.~\ref{fig:sp_laddering}(a), the raising operator acting on $\sigma=16(2,1)$ ladders to the set of  $\grpu{3}$ irreps  $\{\omega\}=\{18(3,0), 18(1,1), 18(4,1), 18(2,2), 18(0,4)\}$.  

If we then act with the raising operator on one of the resulting $\grpu{3}$ irreps $\omega$, e.g., $\omega=18(2,2)$, we now obtain a set of $\grpu{3}$ irreps $\omega'$, with  $N_\omega'=N_\sigma+4$, and  $(\lambda_\omega',\mu_\omega')$  arising in the Kronecker product $(\lambda_\omega,\mu_\omega)\times(2,0)$.  
The $\grpu{3}$ irreps obtained by laddering with $A^{(2,0)}$ from $\omega=18(2,2)$ are shown in Fig.~\ref{fig:sp_laddering}(b).  
However, note that not all of the $\grpsu{3}$ irreps which appear in the product of $(\lambda_\omega\mu_\omega)\times(2,0)$ appear on the $N=20$ grid, as only the irreps which belong to the $\grpsptr$ irrep $\sigma=16(2,1)$ are shown.
For example, if we act on the $\grpu{3}$ irrep $18(2,2)$ which arises in the $\grpsptr$ irrep shown, the $\grpu{3}$ irrep (0,4) is in the product $(2,2)\times(2,0)$, but the $\grpu{3}$ irrep $20(0,4)$ does not belong to the $\grpsptr$ irrep.  

The reason for this is that the $\grpu{3}$ irreps which belong to the $\grpsptr$ irrep are only those which are related to the LGI by \emph{symmetric} coupled products of $\grpsptr$ raising operators.  
 To see why, consider what would happen if instead of first coupling $A^{(2,0)}$ onto $\sigma$ to get $\omega$ and then coupling $A^{(2,0)}$ onto $\omega$ to get $\omega'$, i.e., $[A^{(2,0)}\times[A^{(2,0)}\times\ket{\sigma}]^{\omega}]^{\omega'}$,  we change the coupling order and first couple the two symplectic raising operators together to obtain a raising polynomial with U(3) character $n=N_n(\lambda_n,\mu_n)$.  The raising polynomial is then coupled to the LGI $[[A^{(2,0)}\times A^{(2,0)}]^n\times\ket{\sigma}]^{\rho\omega'}$.\footnote{
Note that because the symplectic raising and lowering operators actually form a $\grpu{3}$ tensor, the tensors should really be labeled as $A^{2(2,0)}$ and the coupled product $A^{2(2,0)}\times A^{2(2,0)}$ results in $\grpu{3}$ tensors, not just $\grpsu{3}$ tensors. 
However, for notational convenience, we treat the $N_n=2$ label on the symplectic raising operator as implicit.}   
Because the raising operators $A^{(2,0)}_{LM}$ commute, only the symmetrically coupled raising operators, i.e., those with even $\lambda_n$ and $\mu_n$, are allowed \cite{ap-126-1980-343-Rosensteel}.  
Thus the two raising operators can only couple to $n=4(4,0)$ and $4(0,2)$. 
 The irrep $20(0,4)$ does not appear in the Kronecker product of $n\times\sigma$ for either of these $n$, and thus does not appear in the $\grpsptr$ irrep $\sigma$.

In a given $\grpsptr$ irrep, there can be more than one $\grpu{3}$ irrep with the same quantum numbers appearing in the $\grpsptr$ irrep. 
This multiplicity is indicated in Fig.~\ref{fig:sp_laddering} by the size of the dots representing the $\grpu{3}$ irreps.  
One example is provided by the irrep labeled by $20(3,1)$ highlighted in Fig.~\ref{fig:sp_laddering}(c).  
Such  irreps can be obtained by laddering with the coupled product of raising operators with $n=4(4,0)$ or $n=4(0,2)$.  
When there is more than one $\grpu{3}$ irrep with the same labels obtained by acting on the LGI with the raising polynomials, the $\grpu{3}$ irreps thus obtained will not, in general, be orthogonal.  To construct an orthogonal basis for the $\grpsptr$ irrep, we need to identify an orthogonal basis for each $\omega$ subspace spanned by the $\grpu{3}$ irreps with the same label $\omega$.
In the remainder of this section, we will describe a method for generating a set of orthogonal $\grpu{3}$ irreps which span the $\grpsptr$ irrep.

To generate the set of U(3) irreps constituting the $\grpsptr$ irrep at each specific grade $N_\omega$, we first construct the set of raising polynomial $P^n_{\alpha_n}(A)=\mathcal{N}\left[A^{(2,0)}\times A^{(2,0)}\times...\times A^{(2,0)}\right]^{n}_{\alpha_n}$ with $N_n=N_\omega-N_\sigma$, and $\mathcal{N}$ is a normalization coefficient.   
 These raising polynomials are obtained by successively coupling the $\grpsptr$ raising operators $A^{(2,0)}$  together  using the relation~\cite{npa-455-1986-315-Suzuki}
\begin{equation}
P^{n}_{\alpha_{n}}(A)=\sum_{\bar{n}}\frac{2}{N_{n}}(n||b^{\dagger(2,0)}||\bar{n})[A^{(20)}\times P^{\bar{n}}(A)]^{n}_{\alpha_{n}}.
\label{raising_polynomial_recurrence} \end{equation}
The factor $(n||b^{\dagger(2,0)}||n')$ is the RME of the $\grpsu{3}$ tensor $b^{\dagger(2,0)}$, whose components are the generators of a boson (Heisenberg) algebra  in a Cartesian oscillator basis~\cite{jmp-24-1983-2461-Rosensteel}.  This RME is given by 
\begin{widetext}
\begin{equation}
\begin{aligned}
(n'||b^{\dagger(2,0)}||n)=&\left(\frac{(n_1+4)(n_1-n_2+2)(n_1-n_3+3)}{2(n_1-n_2+3)(n_1-n_3+4)}\right)^{\frac12}\delta_{n'_1,n_1+2}\delta_{n_2',n_2}\delta_{n_3',n_3}\\\\
&+\left(\frac{(n_2+3)(n_1-n_2)(n_2-n_3+2)}{2(n_1-n_2-1)(n_2-n_3+3)}\right)^{\frac12}\delta_{n'_1,n_1}\delta_{n_2',n_2+2}\delta_{n_3',n_3}\\\\
&+\left(\frac{(n_3+2)(n_2-n_3)(n_1-n_3+1)}{2(n_1-n_3)(n_2-n_3-1)}\right)^{\frac12}\delta_{n_1',n_1}\delta_{n_2',n_2}\delta_{n_3',n_3+2}.
\label{n_boson}
\end{aligned}
\end{equation}
\end{widetext}
Once we have the set of raising polynomials, we generate a set of $\grpu{3}$ irreps with grade $N_\omega$ by acting on the LGI with these raising polynomials to obtain
\begin{equation}
\ket{\sigma n\rho\omega\kappa LM}=\left[P^n(A)\times\ket{\sigma}\right]^{\rho\omega}_{\kappa LM}.
\label{polynomial_on_lgi}
\end{equation}
However,  as noted above, states forming the $\grpu{3}$ irreps with the same label $\omega$ but different $n$ and $\rho$ are not, in general,  orthogonal. 
To obtain an orthogonal sets of states that span the $\omega$ subspaces, we take linear combinations of the non-orthogonal states using a so-called $K$ matrix, which is a matrix $(K_\sigma^\omega)$ whose matrix elements are the projections of $\ket{\sigma n\rho\omega\kappa LM}$ onto $\ket{\sigma\upsilon\omega\kappa LM}$.  The index $\upsilon$ is a branching multiplicity index distinguishing different $\grpu{3}$ irreps with the same labels arising in the irrep $\sigma$.  The branching multiplicity $\upsilon_{\max}$ corresponds to the column rank of $K$.  
The states forming orthogonal bases for the $\grpu{3}$ irreps are given in terms of the inverse of the $K$-matrix as
\begin{equation}
\ket{\sigma \upsilon\omega\kappa LM}=\sum_{n\rho}(K_\sigma^\omega)_{\upsilon[n\rho]}^{-1}\ket{\sigma n\rho\omega\kappa LM}.\label{orthobasis}
\end{equation}

Analytic expression for these matrix elements have been derived using vector coherent state theory~\cite{jmp-25-1984-2662-Rowe,jmp-36-1995-1520-Rowe,Hecht1987}.  We will not go into detail of the derivation of the $K$ matrix (see, e.g., Refs.~\cite{jmp-25-1984-2662-Rowe,jmp-36-1995-1520-Rowe,Hecht1987,ps-91-2016-033003-Rowe}), but a summary of expressions necessary for carrying out calculations in the SpNCCI framework is given here. 

The $K$ matrix can be computed indirectly from the matrix $KK^\dagger$.  
A recurrence relation for the matrix elements of $KK^\dagger$ is given by
\begin{multline}
(KK^\dagger)_{n_1'\rho_1'n_2'\rho_2'}^{\omega'}\\
=\frac{2}{N_\omega'-N_\sigma}\sum_{\omega n_1\rho_1n_2\rho_2}\big(\Omega(n_1'\omega')-\Omega(n_1\omega)\big)(KK^\dagger)_{n_1\rho_1 n_2\rho_2}^\omega(\sigma n_1'\rho_1'\omega'||b^{\dagger(2,0)}||\sigma n_1\rho_1\omega)(\sigma n_2'\rho_2'\omega'||b^{\dagger(2,0)}||\sigma n_2\rho_2\omega),
\label{K2matrix}
\end{multline}
where the coefficient $\Omega(n\omega)$ is given by
\begin{equation}
\Omega(n\omega)=\frac14\sum_{j=1}^3\left[2\omega_j^2-n_j^2+8(\omega_j-n_j)-2j(2\omega_j-n_j)\right].\label{Omega}
\end{equation}
The RMEs in \eqref{K2matrix} are those of the  $b^{\dagger(2,0)}$ boson operator \eqref{n_boson} in a $\grpu{6}$ basis which is related to the $\grpsptr$ basis by VCS theory. This RME is given by 
\begin{equation}
(\sigma n'\rho'\omega'||b^{\dagger(2,0)}||\sigma n\rho\omega)
=(-)^{\omega+\omega'}
U\big((20)(\lambda_n\mu_n)(\lambda_{\omega'},\mu_{\omega'})(\lambda_\sigma,\mu_\sigma);(\lambda_{n'}\mu_{n'})\_\rho';(\lambda_\omega\mu_\omega)\rho\_\big)
(n'||b^{\dagger(2,0)}||n).
\label{adagger}
\end{equation}
Here the coefficient $U(\dots)$ is a unitary $\grpsu{3}$ Racah coefficient~\cite{np-62-1965-1-Hecht}, and $(n'||b^{\dagger(2,0)}||n)$ is given by \eqref{n_boson}. 
The $K$ matrix is then extracted from the matrix $KK^\dagger$ by unitary transforming $KK^\dagger$ to the form $KK^\dagger=UDU^\dagger$, where $D$ is a diagonal matrix with eigenvalues $k_{ii}^2$ along the diagonal.  
Because the matrix $KK^\dagger$ is non-negative by construction, the eigenvalues are real and non-negative. 
Thus we can write the $K$ matrix as $K_{ij}=U_{ij}k_{ii}$.  

If we are considering a nuclear system with mass number $A>6$, then the maximum value for $\upsilon$ corresponds to the number of $n\rho$ pairs.  That is, $\upsilon_{\max}=\mathrm{num}(n\rho)$.  
However,  if  $A<6$, then the set of states $\ket{\sigma n\rho\omega}$ can form an overcomplete basis for the $\grpsptr$ irrep.   
This overcompleteness manifests as zero eigenvalues of the $K$ matrix and thus $\upsilon_{\max}<\mathrm{num}(n\rho)$. In this case, the inverse of $K$ appearing in \eqref{orthobasis} is the \emph{right} inverse of $K$.

\subsection{Matrix elements of the symplectic generators
\label{subsection:sp3r-generator-rme}
}
VCS theory also provides analytic expressions for the RMEs of the symplectic ladder operators~\cite{jmp-25-1984-2662-Rowe}.
The RME of the symplectic raising operator $A^{(2,0)}$, derived using VCS theory, is given by 
\begin{equation}
\rme{\sigma \upsilon'\omega'}{A^{(2,0)}}{\sigma \upsilon\omega}
=\sum_{\substack{n'\rho'\\n\rho}}
(K_\sigma^{\omega'})_{\upsilon',[n'\rho']}
(\sigma n'\rho'\omega'||b^\dagger||\sigma n\rho\omega) 
(K_\sigma^\omega)^{-1}_{[n\rho],\upsilon}
\label{Arme},
\end{equation}
where the reduced matrix element of the U(6) boson operator $b^{\dagger(2,0)}$ is given by \eqref{adagger} and  the $K$ matrix is obtained by the recurrence \eqref{K2matrix}.

The RME of the lowering operator $B^{(0,2)}$ is then obtained by making use of the fact that  $(A^{(2,0)}_{LM})^\dagger=(-1)^{M}B^{(0,2)}_{L\,-M}$.  Thus, the RME of the symplectic lowering operator $B^{(0,2)}$ is related to the RME of the raising operator  $A^{(2,0)}$  by
\begin{equation}
\rme{\sigma \upsilon'\omega'}{B^{(0,2)}}{\sigma \upsilon\omega}
=(-1)^{\omega-\omega'}
\sqrt{\frac{\mathrm{dim}(\omega)}{\mathrm{dim}(\omega')}}
\rme{\sigma \upsilon\omega}{A^{(2,0)}}{\sigma\upsilon'\omega'}. 
\label{AB_relation}
\end{equation}

 \section{Symplectic many-body basis}
\label{subsection:sp3r-many-body}

In this section, we define the $\grpsptr$ many-body basis for the SpNCCI framework.   
The $\grpsptr$ basis can be viewed as a unitary transformation of the harmonic oscillator many-body basis to a basis which has definite $\grpsptr$ symmetry.  
That is, the $\grpsptr$ many-body basis is an expression of the Hilbert space for the many-nucleon system as a  sum of  subspaces that form irreps of $\grpsptr$, or more specifically, irreps of $\grpsptr\times\grpsu{N_\Sigma}$.  
Because of the close connection between the generators of $\grpsptr$ and collective phenomena, these $\grpsptr$ irreps are sometimes referred to as collective subspaces. 

The symplectic many-body basis is constructed by first expanding the LGI of each irrep in terms of $\grpu{3}$ many-body states and then laddering with the symplectic raising operator (as described in Section~\ref{section:sp3r}) to generate the remainder of the states.  
Because $\grpsptr$ is non-compact, there is no highest grade $\grpu{3}$ irrep, i.e., there is no $\grpu{3}$ irrep that will be annihilated by the raising operator $A^{(2,0)}$.  
Thus, in principle, each $\grpsptr$ irrep is infinite dimensional. 
However, in practice, the SpNCCI basis is truncated to finite size, by, e.g., restricting the irreps such that $N_{\omega}$ is below some chosen cutoff.  One option is to apply the same truncation used in an $\Nmax$ truncated NCCI basis, in which the basis include only states with number of excitation quanta $\Nex$ up to some chosen maximum number of quanta $\Nmax$ (i.e., $\Nex\le\Nmax$). In the symplectic basis, the number of excitation quanta associated with a given $\grpu{3}$ irrep  is given by $N_{\omega,\mathrm{ex}}=N_{\omega}-N_{\sigma,0}$, where $N_{\sigma,0}$ is the lowest $N_\sigma$ arising in the many-body space.  That is, $N_{\omega,\mathrm{ex}}$ is identical to $\Nex$ and measures the number of oscillator quanta relative to the lowest Pauli-allowed filling of oscillator shells for the given nucleus.  If all $\grpsptr$ irreps, truncated to include only those $\grpu{3}$ irreps with $N_{\omega,\mathrm{ex}}\le\Nmax$ are included in the basis, then the $\grpsptr$ basis would span the same space as the $\Nmax$ truncated NCCI basis.

For many calculations, the dynamics we are interested in are those in the center-of-mass rest frame.  Thus we restrict the symplectic basis to include only those states with zero center of mass motion, i.e., center-of-mass free (CMF) states.  
Because the $\grpu{3}$ generators commute with the center-of-mass number operator, we are guaranteed that $\grpu{3}$ irreps preserve the exact factorization of center-of-mass inherent in the harmonic oscillator basis.  
Thus it is always possible to construct LGIs that are CMF~\cite{zp-158-1960-284-Kretzschmar, np-21-1960-508-Verhaar, npa-170-1971-34-Hecht, npa-255-1975-315-Millener,npa-897-2013-109-Luo}.
As discussed below, the remainder of the CMF basis can then be obtained by laddering with \emph{intrinsic} symplectic raising operators which do not act the center-of-mass coordinate and thus preserve the CMF nature of the states.  

Each LGI  is identified as an $\grpu{3}$ irrep satisfying the criteria that it is annihilated by the symplectic lowering operator $B^{(0,2)}$ and the center-of-mass number operator $N_\cm$.  
That is, we are looking for linear combination
\begin{equation}
\tket{\gamma\sigma\Sigma S}=b^{\Sigma S}_{[\gamma\sigma],[\varrho\omega]}\tket{\varrho\omega\Sigma S}
\label{eq:lgi_expansion}
\end{equation}
 of the $\grpu{3}$ many-body irreps defined in Section~\ref{section:u3-many-body}  such that 
\begin{equation}
B^{(0,2)}\ket{\gamma\sigma \Sigma S}=\sum_{\varrho\omega} b^{\Sigma S}_{[\gamma\sigma],[\varrho\omega]}B^{(0,2)}\ket{\varrho\omega \Sigma S }=0
\end{equation}
and 
\begin{equation}
N^{(0,0)}_{\cm,00}\ket{\gamma\sigma\Sigma S}=\sum_\varrho b^{\sigma S}_{(\gamma\sigma),\varrho}N^{(0,0)}_{\cm}\ket{\varrho\sigma \Sigma S}=0, 
\end{equation}
where $\gamma$ is an index distinguishing different LGIs sharing the same $\omega \Sigma S$ labels.  Recall that $\varrho$ [see \eqref{eq:su3pn_scheme} and \eqref{eq:su3st_scheme}] is a composite of all of the additional quantum numbers labeling the $\grpu{3}$ many-body irrep. 
In other words, we want to solve for the simultaneous null space of the symplectic lowering operator and the center-of-mass number operator. 
Formulating this as a matrix null space problem, we solve for the null space of a matrix whose matrix elements are the RMEs $\trme{\varrho'\omega' \Sigma S}{B^{(0,2)}}{\varrho\sigma\Sigma S}$ and $\trme{\varrho'\sigma \Sigma S}{N_{\cm}^{(0,0)}}{\varrho\sigma\Sigma S}$.  The resulting matrix has the form shown in Fig.~\ref{fig:lgi_matrix}.  
We note that the null vectors obtained from the null solver are not unique.  Any orthogonal linear combination of these null vectors would provide an equally valid basis for the null space and therefore an equally valid expansion for a set of LGIs. 
\begin{figure}
\[
\begin{pmat}[{...}]
\rme{1\sigma \Sigma S}
{N^{(0,0)}_\mathrm{cm}}
 {1 \Sigma \sigma S}
 &\rme{1\sigma \Sigma S}
{N^{(0,0)}_\mathrm{cm}}
 {2 \Sigma \sigma S}
 &\cdots
 &\rme{1\sigma \Sigma S}
 {N^{(0,0)}_\mathrm{cm}}
 {\varrho_\mathrm{max} \sigma \Sigma S}\cr
 \rme{2\sigma \Sigma S}
{N^{(0,0)}_\mathrm{cm}}
 {1 \Sigma \sigma S}
 &\rme{2\sigma \Sigma S}
{N^{(0,0)}_\mathrm{cm}}
 {2 \Sigma \sigma S}
 &\cdots
 &\rme{2\sigma \Sigma S}
 {N^{(0,0)}_\mathrm{cm}}
 {\varrho_\mathrm{max} \sigma \Sigma S}\cr
 \vdots&&\vdots\cr
\rme{\varrho_{\mathrm{max}}\sigma \Sigma S} 
{N^{(0,0)}_\mathrm{cm}}
{1 \sigma \Sigma S}
&\rme{\varrho_{\mathrm{max}}\sigma \Sigma S}
{N^{(0,0)}_\mathrm{cm}}
{2 \sigma \Sigma S}
&\cdots
&\rme{\varrho_{\max}\sigma \Sigma S}
{N^{(0,0)}_\mathrm{cm}}
{\varrho_{\max} \sigma \Sigma S}\cr\-
\rme{1\omega_1 \Sigma S}
 {B^{(0,2)}}
 {1 \sigma S} 
 & \rme{1\omega_1 \Sigma S}
 {B^{(0,2)}}
 {2 \sigma S} 
 &\cdots 
 &\rme{1\omega_1 S}
 {B^{(0,2)}}
 {\varrho_\mathrm{max} \sigma S}\cr
\rme{2\omega_1 \Sigma S}
 {B^{(0,2)}}
 {1 \sigma S} 
 & \rme{2\omega_1 \Sigma S}
 {B^{(0,2)}}
 {2 \sigma S} 
 &\cdots 
 &\rme{2\omega_1 S}
 {B^{(0,2)}}
 {\varrho_\mathrm{max} \sigma S}\cr
 \vdots&&\vdots\cr
 \rme{\varrho'_\mathrm{max}\omega_1 \Sigma S}
 {B^{(0,2)}}
 {1 \sigma\Sigma S}
 & \rme{\varrho'_\mathrm{max}\omega_1 \Sigma S}
 {B^{(0,2)}}
 {2 \sigma\Sigma S}
 &\cdots 
 &\rme{\varrho'_{\mathrm{max}}\omega_1\Sigma S}
 {B^{(0,2)}}
 {\varrho_{\mathrm{max}} \sigma \Sigma S}\cr\cr
\rme{1\omega_2 \Sigma S}
 {B^{(0,2)}}
 {1 \sigma S} 
 & \rme{1\omega_2 \Sigma S}
 {B^{(0,2)}}
 {2 \sigma S} 
 &\cdots 
 &\rme{1\omega_2 S}
 {B^{(0,2)}}
 {\varrho_\mathrm{max} \sigma S}\cr
 \rme{2\omega_2 \Sigma S}
 {B^{(0,2)}}
 {1 \sigma S} 
 & \rme{2\omega_2 \Sigma S}
 {B^{(0,2)}}
 {2 \sigma S} 
 &\cdots 
 &\rme{2\omega_2 S}
 {B^{(0,2)}}
 {\varrho_\mathrm{max} \sigma S}\cr
 \vdots&&\vdots\cr
 \rme{\varrho'_\mathrm{max}\omega_2 \Sigma S}
 {B^{(0,2)}}
 {1 \sigma\Sigma S}
 & \rme{\varrho'_\mathrm{max}\omega_2 \Sigma S}
 {B^{(0,2)}}
 {2 \sigma\Sigma S}
 &\cdots 
 &\rme{\varrho'_{\mathrm{max}}\omega'\Sigma S}
 {B^{(0,2)}}
 {\varrho_{\mathrm{max}} \sigma \Sigma S}\cr
  \vdots&&\vdots\cr
\end{pmat}
\]
 \caption{
\label{fig:lgi_matrix}
 Form of the matrix used to obtain the expansion of the CMF LGIs with labels $\sigma\Sigma S$ in terms of SU(3) many-body irreps.  Elements are given by the RMEs of the operators $B^{(0,2)}$ and $N^{(0,0)}_{\cm}$ in an $\grpsu{3}$ many-body basis.  The dashed line delimits the sub-matrices corresponding to the RMEs of $B^{(0,2)}$ and $N_{\cm}^{(0,0)}$.   
Note that because we are solving for LGIs with fixed $\sigma\Sigma S$, we need only construct the matrix for the set of ket $\grpu{3}$ irreps belonging to the $\grpu{3}$ many-body basis which have the same labels $\sigma\Sigma S$.  Different irreps are distinguished by the multiplicity index $\varrho=1,..,\varrho_{\mathrm{max}}$.    
The possible final (bra) irreps differ for the two operators.
 Since $N_{\cm}^{(0,0)}$ is an $\grpu{3}$ scalar operator, the same set of irreps appear in the bras as in the kets of the RMEs, sharing the labels $N_{\omega}$ and $(\lambda_\omega\mu_\omega)$.  
 For $B^{(0,2)}$, the possible bra irreps are those with two fewer quanta $(N_{\omega_i}=N_{\omega}-2)$ and $(\lambda_{\omega_i}\mu_{\omega_i})$ arising in the Kronecker product $(\lambda_\omega\mu_\omega)\times(0,2)$.  
 Because the operators conserve total spin, isospin and $\grpsu{4}$, this matrix is block diagonal in $\Sigma S$. 
 }
 \end{figure}

Once the LGI is constructed, the remainder of the basis states are obtained by acting on the LGI with the symplectic raising polynomials $P^n(A)$ as in \eqref{polynomial_on_lgi}-\eqref{orthobasis}.   
To ensure the symplectic basis is CMF, the raising polynomial applied is constructed in terms of the \emph{intrinsic} symplectic raising operators $A^{\prime (2,0)}$, which act only on the intrinsic coordinates of the state and thus leave the center of mass unchanged.  
To obtain the intrinsic symplectic generators, we can rewrite the generators in \eqref{tensor_generators} in terms of  intrinsic harmonic oscillator ladder operators
\begin{equation}
\begin{aligned}
c_s^{\prime\dagger(1,0)}&=c_s^{\dagger(1,0)}-\frac{1}{\sqrt{A}}c_{\cm}^{\dagger(1,0)},
&\quad
&c_{\cm}^{\dagger(1,0)}=\frac{1}{\sqrt{A}}\sum_t^A c_t^{\dagger(1,0)}\\
c_s^{\prime(0,1)}&=c_s^{(0,1)}-\frac{1}{\sqrt{A}}c_{\cm}^{(0,1)}, 
&\quad 
& c_{\cm}^{(0,1)}=\frac{1}{\sqrt{A}}\sum_t^Ac_t^{(0,1)},
\end{aligned}
\label{boson_intrinsic}
\end{equation}
where the primes on the operators indicate that these operators are intrinsic.
The intrinsic symplectic generators, can then be obtained by making the substitutions $c_s^{\dagger(1,0)}\rightarrow c_s^{\prime\dagger(1,0)}$ and $c_s^{(0,1)}\rightarrow c_s^{\prime(0,1)}$ in \eqref{tensor_generators}.  It can then be shown\footnote{See, e.g., Appendix~A.4 of Ref.~\cite{jpg-47-2020-122001-Caprio} for more details.} that the resulting intrinsic generators can be obtained by subtracting off a center of mass contribution from the symplectic generators in \eqref{tensor_generators}
\begin{equation}
\begin{aligned}
A^{\prime(2,0)}&=A^{(2,0)}-A_{\cm}^{(2,0)}, 
&\qquad\qquad 
A_{\cm}^{(2,0)}&=\frac{1}{\sqrt{2}}\left[c_{\cm}^{\dagger(1,0)}\times c_{\cm}^{\dagger(1,0)}\right]^{(2,0)}\\ 
B^{\prime(0,2)}&=B^{(0,2)}-B_{\cm}^{(0,2)}, 
&\qquad\qquad
B_{\cm}^{(0,2)}&=\frac{1}{\sqrt{2}}\left[c_{\cm}^{(0,1)}\times c_{\cm}^{(0,1)}\right]^{(0,2)}\\ 
C^{\prime(1,1)}&=C^{(1,1)}-C_{\cm}^{(1,1)},
&\qquad\qquad
C_{\cm}^{(1,1)}&=\sqrt{2}\left[c_{\cm}^{\dagger(1,0)}\times c_{\cm}^{(0,1)}\right]^{(1,1)}\\ 
H^{\prime(0,0)}&=H^{(0,0)}-H_{\cm}^{(0,0)},
&\qquad
H_{\cm}^{(0,0)}&=\sqrt{3}\left[c_{\cm}^{\dagger(1,0)}\times c_{\cm}^{(0,1)}\right]^{(0,0)}+\frac{3}{2}.
\label{intrinsic-sp3r-gen}
\end{aligned}
\end{equation}
Note that  intrinsic harmonic oscillator hamiltonian can be written in terms of the intrinsic number operator  $N^{\prime(0,0)}$ as $H^{\prime(0,0)}=N^{\prime(0,0)}+\frac32(A-1)$.

The intrinsic symplectic many-body states are obtained as in \eqref{orthobasis}, but with the raising polynomial given by coupled products of the intrinsic symplectic raising operator $A^{\prime(2,0)}$.  The resulting $\grpu{3}$ irrep is then branched to states with definite orbital angular momentum, which when coupled with the spin $S$ of the LGI give many-body basis states with definite angular momentum $J$.  That is, a symplectic many-body state is given by 
\begin{equation}
\ket{\gamma \sigma \upsilon \omega\kappa L\Sigma SJM}=\sum_{n\rho}(K^\omega_\sigma)^{-1}_{v,[n\rho]}\left[P^n(A^{\prime(2,0)})\times\ket{\gamma\sigma \Sigma S}\right]^{\rho\omega}_{\kappa LJM}.
\label{polynomial_expansion_manybody_basis}
\end{equation}
 \section{Calculating matrix elements of relative operators in an $\grpsptr$ basis}
\label{section:recurrence}

To carry out calculations in a symplectic many-body basis, we need to compute the matrix elements of operators.  
To accomplish this, one could explicitly expand each $\grpsptr$ many-body state in the $\grpu{3}$ many-body basis and then compute the matrix elements in the $\grpu{3}$ basis, as discussed in Section~\ref{subsection:su3basis-me}.  
However, in this work, we propose an alternative method, which eliminates the need to expand every $\grpsptr$ state in the $\grpu{3}$ many-body basis. 
Here, we expand only the LGIs in the $\grpu{3}$ many-body basis, to compute the RMEs of the operators between LGIs.  
The remaining RMEs are then computed via recurrence relations for which the RMEs between the LGI act as the seeds.  
The recurrence relations, derived in this section, depend only on $\grpu{3}$ and $\grpsu{2}$ coupling and recoupling coefficients, $K$-matrices, and the $\grpsptr$ generator matrix elements~\eqref{Arme}-\eqref{AB_relation}. 

In order to make use of this recurrence method to evaluate RMEs of an operator, we first need to write this operator in terms of $\grpsu{3}\times\grpsu{2}$ tensors, in the proton-neutron scheme, or $\grpsu{3}\times\grpsu{4}$ tensors, in the isospin scheme.
Although an operator will not, in general, be a good tensor of this type, we can expand any operator in terms of components of such tensors.  
In the NCCI framework, an operator is typically defined by their matrix elements in the harmonic oscillator basis.  
As shown in Appendix~\ref{appendix:upcoupling}, if we take these matrix elements as a starting point, we can expand this operator in terms of a set of special tensor operators, known as unit tensors or basis operators~\cite{ptrsla-277-1975-545-Butler}.  
In general, unit tensors are defined with respect to a specific basis, as having a single non-zero RME in that basis. 

In this work, we expand intrinsic operators, such as the two-body Hamiltonian, in terms of unit tensors.
A two-body intrinsic operator acts only the relative coordinate degrees of freedom $(x_i-x_j)/\sqrt{2}$ (and the spins of the two associated nucleons).
However, because we are computing the matrix elements of this operator in an antisymmeterized basis, we can treat the operator is if it acts only on the relative coordinate $x_{\rel}=(x_1-x_2)/\sqrt{2}$ (and the spins of these first two nucleons).  
Thus, we decompose a relative operator in terms of relative unit tensors which are defined with respect to the harmonic oscillator basis for the space in which these states are defined, namely, for $x_{\rel}$ (and the spins of the two nucleons). 

The states in this relative harmonic oscillator basis are labeled by $\ket{ NL \Sigma SJ M}$, where $N$ is the number of relative oscillator quanta, $L$ is the relative orbital angular momentum of the two particles, $S$ is the spin of the two-particle system and $J$ is the relative angular momentum, obtained by coupling $L\times S \rightarrow J$.
The label $\Sigma$ represents all additional labels necessary to distinguish the state.
In the proton-neutron scheme, $\Sigma$ corresponds to the species of the two particles ($\pi\pi$, $\pi\nu$ or $\nu\nu$), and, in the isospin scheme, $\Sigma$ corresponds to the $\grpu{4}$ label, isospin and isospin projection of the two particles ($[f]TM_T$).
Like the single particle harmonic oscillator basis states discussed in Section~\ref{section:u3-many-body}, the relative harmonic oscillator basis states have definite $\grpsu{3}$ symmetry, again given by $\omega = [N+\frac32](N,0)$.  

Let us first consider the proton-neutron scheme.  With respect to the relative harmonic oscillator basis just defined, $\grpsu{3}\times\grpsuS$ unit tensors are defined by having RMEs
\begin{equation}
\rme{N'\Sigma'S'}{\mathcal{U}^{(\lambda_0,\mu_0)S_0}(\bar{N}'\bar{\Sigma}'\bar{S}',\bar{N}\bar{\Sigma}\bar{S})}{N\Sigma S}
=\delta_{\bar{N},N}\delta_{\bar{S},S}\delta_{\bar{\Sigma},\Sigma}
\delta_{\bar{N}',N'}\delta_{\bar{S}',S'}\delta_{\bar{\Sigma}',\Sigma'}.
\label{eq:relative_unit_tensors_u3s}
\end{equation}

If the operator is already a good $\grpsu{3}\times\grpsuS$ tensor $V^{(\lambda_0\mu_0)S_0}$,  then the operator can be written in terms of these unit tensors as 
\begin{equation}
V^{(\lambda_0\mu_0)S_0}
=\sum_{\substack{
\bar{N}'\bar{S}'\\
\bar{N}\bar{S}
}}
\rme{\bar{N}'\bar{\Sigma}'\bar{S}'}{V^{(\lambda_0\mu_0)S_0}}{\bar{N}\bar{\Sigma}\bar{S}}\,
\scrU^{(\lambda_0\mu_0)S_0}(\bar{N}'\bar{\Sigma}'\bar{S}',\bar{N}\bar{\Sigma}\bar{S}),
\label{tensor_rel_unit_tensor_expansion}
\end{equation}
where the expansion coefficient $\trme{\bar{N}'\bar{\Sigma}'\bar{S}'}{V^{(\lambda_0\mu_0)S_0}}{\bar{N}\bar{\Sigma}\bar{S}}$  is just the $\grpsu{3}\times\grpsuS$ RME of the tensor operator in the relative harmonic oscillator basis.
If, however, the intrinsic operator is not an $\grpsu{3}\times\grpsuS$ tensor, but it is defined by its $\grpsuJ$-reduced matrix elements $\trme{\bar{N}'\bar{L}'\bar{\Sigma}'\bar{S}'\bar{J}'}{V_{J_0}}{\bar{N}\bar{L}\bar{\Sigma}\bar{S}\bar{J}}$, then we can expand this operator in terms of \emph{components} of these relative $\grpsu{3}\times\grpsuS$ unit tensors as 
\begin{equation}
V_{J_0 M_0}=
\sum_{\substack{(\lambda_0\mu_0)\kappa_0\\L_0S_0}}
\sum_{\substack{
\bar{N}'\bar{\Sigma}'\bar{S}'\\
\bar{N}\bar{\Sigma}\bar{S}
}}\mathcal{V}^{(\lambda_0\mu_0)S_0}_{\kappa_0L_0J_0}
\scrU^{(\lambda_0\mu_0)S_0}_{\kappa_0L_0J_0M_0}(\bar{N}'\bar{\Sigma}'\bar{S}',\bar{N}\bar{\Sigma}\bar{S})
\label{eq:suJ-su3suS-unit-tensors}
\end{equation}
where the expansion coefficient $\mathcal{V}^{(\lambda_0\mu_0)S_0}_{\kappa_0L_0J_0}$ is given by
\begin{multline}
\mathcal{V}^{(\lambda_0\mu_0)S_0}_{\kappa_0L_0J_0}\\
=
\sum_{\substack{
\bar{L}'\bar{J}'
\bar{L}\bar{J}
}}
(-)^{\bar{N}}
\frac{\hat{\bar{J}}'\hat{S}_0}{\hat{J}_0\hat{\bar{S}}'}
\left[\frac{\dim(\lambda_0\mu_0)}{\dim(\bar{N}',0)}\right]^{\frac12}
\rme{\bar{N}'\bar{L}'\bar{\Sigma}'\bar{S}'\bar{J}'}
{V_{J_0}}
{\bar{N}\bar{L}\bar{\Sigma}\bar{S}\bar{J}}
\rcc{(\bar{N}',0)}{\bar{L}'}{(0,\bar{N})}{\bar{L}}{(\lambda_0\mu_0)}{\kappa_0L_0}
\uninej{\bar{L}'}{\bar{S}'}{\bar{J}'}{\bar{L}}{\bar{S}}{\bar{J}}{L_0}{S_0}{J_0}.
\label{eq:suJ-su3suS-unit-tensors-coef}
\end{multline}

With the operator expanded in terms of unit tensors, the matrix elements of the operator are given by
\begin{multline}
\me{\gamma'\sigma'\upsilon'\omega'\Sigma'S'\kappa'L'J'M'}
{V_{J_0M_0}}
{\gamma\sigma\upsilon\omega \Sigma S\kappa LJM}\\
=
\sum_{\substack{(\lambda_0\mu_0)\kappa_0\\L_0S_0}}
\sum_{\substack{\bar{N}'\bar{\Sigma'}\bar{S}'\\\bar{N}\bar{\Sigma}\bar{S}}}
\mathcal{V}^{(\lambda_0\mu_0)S_0}_{\kappa_0L_0J_0}
(JM;J_0M_0|J'M')
\uninej{L}{S}{J}{L_0}{S_0}{J_0}{L}{S}{J}\qquad\qquad\qquad\\
\times
\sum_{\rho_0}
\rcc{\omega}{\kappa L}{(\lambda_0\mu_0)}{\kappa_0L_0}{\omega'}{\kappa'L'}_{\rho_0}
\rme{\gamma'\sigma'\upsilon'\omega'\Sigma'S'}
{\mathcal{U}^{(\lambda_0,\mu_0)S_0}(\bar{N}'\bar{\Sigma}'\bar{S}',\bar{N}\bar{\Sigma}\bar{S})}
{\gamma\sigma\upsilon\omega \Sigma S}_{\rho_0}.
\label{many-body-mes-operator}
\end{multline}
Here we have used the Wigner-Eckart theorem for the group chain \eqref{su3_so3_su2_chain} to write the matrix element of the relative unit tensor in terms of the coupling coefficient for this chain, given by~\eqref{eq:u3lsjm_cc}, and $\grpu{3}\times\grpsuS$ reduced matrix elements of the relative unit tensor in the symplectic basis. 
Thus, computing the matrix elements of the operator reduces to calculating the $\grpsu{3}\times\grpsuS$ RMEs of relative unit tensors.  

In the isospin scheme, we would further decompose the operators in terms of  $\grpsu{3}\times\grpsu{4}$  unit tensors, which would enable the use of, e.g., the $\grpsu{4}$ Wigner-Eckart theorem.  
In this case, we would decompose the operators in terms of $\grpsu{3}\times\grpsu{4}$ relative unit tensors defined as having RMEs 
\begin{equation}
\rme{N'[{\boldsymbol\tau}']}{\mathcal{U}^{(\lambda_0\mu_0)[{\boldsymbol \tau}_0]}(\bar{N}'[\bar{\boldsymbol\tau}'],\bar{N}[\bar{\boldsymbol \tau}])}{N[{\boldsymbol \tau}]}
=
\delta_{\bar{N},N}\delta_{[\bar{\boldsymbol \tau}],[{\boldsymbol \tau}]}
\delta_{\bar{N}',N'}\delta_{[\bar{\boldsymbol \tau}'],[{\boldsymbol \tau}']},
\label{eq:u3u4_unit_tensor_def}
\end{equation}
where $[{\boldsymbol \tau}]$ is the $\grpsu{4}$ label associated with the two-particle state, which is $[{\boldsymbol\tau}]=[2]$ if $N$ is odd and $[{\boldsymbol\tau}]=[1^2]$ if $N$ is even. The $\grpsu{4}$ tensor label $[{\bf \tau_0}]$  arises in the $\grpsu{4}$ Kronecker product $[\bar{\boldsymbol\tau}']\times[\tilde{\bar{\boldsymbol \tau}}]$.   In this case, the $\grpsu{4}$ Kronecker product is always multiplicity free, i.e., the outer multiplicity is $\delta_{\mathrm{max}}=1$. Note that the components  $\mathcal{U}^{(\lambda_0\mu_0)[{\boldsymbol \tau}_0]}_{S_0T_0}(\bar{N}'[\bar{\boldsymbol\tau}'],\bar{N}[\bar{\boldsymbol \tau}])$ of this tensor form $\grpsuS\times\grpsuT$ tensors.

In the following, we derive a recurrence relation for computing these RMEs of  operators in the symplectic many-body basis. 
Although we will specifically derive the recurrence relation for relative unit tensors in this work, the general recurrence scheme described here can also be applied to the RMEs of other classes of operators, including generic one- and two-body operators~\cite{Reske1984,Escher1997,Herko2024}.
Thus, we shall carry out the derivation of the recurrence relations in terms of a generic $\grpsu{3}$ tensor $\mathcal{O}^{(\lambda_0\mu_0)}$, for as long as this is practical, before specializing to relative unit tensors. 

The recurrence relation generates reduced matrix elements of the operator between symplectic states, in terms of reduced matrix elements of unit tensors, between symplectic states with lower grade, i.e., lower $N_{\omega}$.  
The recurrence is derived by first ``peeling off" a symplectic raising operator from the symplectic state and then commuting the raising operator through the operator $\mathcal{O}^{(\lambda_0\mu_0)}$.  The raising operator then acts as a lowering operator on the bra state.  
The result is an expression for an RME between a pair of SU(3) irreps $\omega$ and $\omega'$ in terms of RMEs of the operator between irreps with two fewer quanta.

Schematically, a $\grpu{3}$ irrep in an $\grpsptr$ irrep with a given value for $N_\omega$, which we will denote $\ket{N_\omega}$, can be expressed in terms of a $\grpu{3}$-coupled product of $a$ intrinsic symplectic raising operators  $A^{\prime(2,0)}$, which form a raising polynomial carrying $N_n=2a$ oscillator quanta, acting on the LGI $\tket{N_\sigma}$, as in \eqref{polynomial_on_lgi}:
\begin{equation*}
\ket{N_\omega}=\underbrace{[A^{\prime(2,0)}\times A^{\prime(2,0)}\times..A^{\prime(2,0)}]^n}_{a}\ket{N_\sigma}.
\end{equation*}
We can peel off one of the factors of $A^{\prime(2,0)}$ and recouple the remaining $A^{\prime(2,0)}$ to form a new raising polynomial with two fewer oscillator quanta [$N_n'=2(a-1)$]:
\begin{equation*}
 \ket{N_\omega}=\big[A^{\prime(2,0)}\times\underbrace{[A^{\prime(2,0)}\times A^{\prime(2,0)}\times..A^{\prime(2,0)}]}_{a-1}\big]\ket{N_\sigma}
\end{equation*}
This new raising polynomial acts on $\tket{N_\sigma}$, generating $\grpu{3}$ irreps with two fewer quanta than $\ket{N_\omega}$.  Thus, we now have a relation between a $\grpu{3}$ irrep $\tket{N_\omega}$ and  irreps with with two fewer oscillator quanta $\tket{N_\omega-2}$:
\begin{equation*}
  \ket{N_\omega}=A^{\prime(2,0)}\ket{N_\omega-2}.
\end{equation*}
 
We then use this relationship between states to derive an expression relating  the matrix elements between states with $N_\omega$ and $N_{\omega'}$ oscillator quanta to matrix elements between states with $N_\omega-2$ and $N_{\omega'}-2$ oscillator quanta.  
We begin by peeling off a symplectic raising operator from the ket and then commuting the raising operator $A^{\prime(2,0)}$ through the operator, so that we can act with $A^{(2,0)}$ on the bra:
\begin{align*}
\rme{N_{\omega'}}{\mathcal{O}^{(\lambda_0\mu_0)}}{N_\omega}&=\rme{N_{\omega'}}{\mathcal{O}^{(\lambda_0\mu_0)}A^{\prime(2,0)}}{N_\omega-2}\\
&=\rme{N_{\omega'}}{A^{\prime(2,0)}\mathcal{O}^{(\lambda_0\mu_0)}}{N_\omega-2}+\rme{N_{\omega'}}{[\mathcal{O}^{(\lambda_0\mu_0)},A^{\prime(2,0)}]}{N_\omega-2}.
\end{align*}

The action of $A^{(2,0)}$ on the bra is that of its covariant adjoint  $B^{(0,2)}$.  
Thus, acting with $A^{(2,0)}$ on the bra lowers the number of oscillator quanta of the bra by two: 
\begin{align*}
\rme{N_{\omega'}}{\mathcal{O}^{(\lambda_0\mu_0)}}{N_\omega}
&=\rme{N_{\omega'}}{A^{\prime(2,0)}}{N_{\omega'}-2}\rme{N_{\omega'}-2}{\mathcal{O}^{(\lambda_0\mu_0)}}{N_\omega-2}\nonumber\\
&\quad+\rme{N_{\omega'}}{[\mathcal{O}^{(\lambda_0\mu_0)},A^{\prime(2,0)}]}{N_\omega-2}.
\end{align*}
The RMEs of $A^{\prime(2,0)}$ are known analytically. 
Thus it only remains to evaluate the commutator of the operator with the symplectic raising operator.  

In a few special cases, the commutator has a particular convenient form.  In the case in which the operator is a relative unit tensor, as we shall see below, the commutator can itself be expressed as a linear combination of relative unit tensors.  
The expression above reduces to 
\begin{equation*}
\rme{N_{\omega'}}{\mathcal{U}^{(\lambda_0\mu_0)}}{N_\omega}=\sum_i\mathcal{C}_{1,i}\rme{N_{\omega'}-2}{\mathcal{U}^{(\lambda_0\mu_0)}_i}{N_\omega-2}+\sum_j\mathcal{C}_{2,j}\rme{N_{\omega'}}{\mathcal{U}^{(\lambda_0\mu_0)}_j}{N_\omega-2},
\end{equation*}
where $\mathcal{C}_1$ and $\mathcal{C}_{2,i}$ can be analytically expressed in terms of RMEs of the symplectic generators and known $\grpsu{3}$ and $\grpsu{2}$ recoupling coefficients.
Alternatively, if the operator is an elementary one-body operator $[a^\dagger a]^{(\lambda_0\mu_0)}$ or an elementary two-body operator $[a^\dagger a^\dagger aa]^{(\lambda_0\mu_0)}$,
a recurrence scheme of this form can similarly be applied, as shown in Refs.~\cite{Reske1984,Escher1997,Herko2024}. 
In any of these cases, the recurrence relation thus obtained expresses the RME of an operator in terms of RMEs of operators similar in form to the original operator, but between states with fewer oscillator quanta.
Such a recurrence relation can be used to generate the RMEs of such operators between all states belonging to two different $\grpsptr$ irreps. 
The seeds of the recurrence are the RMEs between $\grpu{3}$ irreps for which we cannot peel off any additional raising operators, i.e., RMEs between the LGIs of the two $\grpsptr$ irreps.

Since any one- or two-body operator can be expressed in terms of creation and annihilation operators, this recurrence method can be used to compute the RMEs of any one- or two-body operator, not just relative operators.
The advantage of expanding relative operators in terms of relative unit tensors, rather than elementary two-body operators $[a^\dagger a^\dagger aa]^{(\lambda_0\mu_0)}$, is that significantly fewer terms need to be evaluated. 

The recurrence as described above assumes we can continually peel off raising operators from the ket until we reach the LGI.  
This will not be the case if the difference in grade $N_n$ between $\omega$ and $\sigma$ ($N_n=N_\omega-N_\sigma$) is larger than $N_{n'}$.  
In this case, the action of the symplectic raising operator will annihilate the bra before we have recursed down to $\sigma$.  
However, we may use conjugation relations to relate the RME of interest to RMEs $\trme{N_\omega}{(\mathcal{O}^{(\lambda_0\mu_0)})^\dagger}{N_{\omega}'}$, of the adjoint of the operator, which do satisfy this constraint.
For example, for the relative unit tensors, the relevant conjugation relation is given by 
\begin{multline}
  \rme{\gamma'\sigma'\upsilon'\omega'\Sigma'S'}{\mathcal{U}^{(\lambda_0\mu_0)S_0}(\bar{N}'\bar{\Sigma}'\bar{S}',\bar{N}\bar{\Sigma}\bar{S})}{\gamma\sigma\upsilon\omega \Sigma S}_{\rho_0} \\ 
  = (-1)^{\omega'+\omega+\bar{N}'+\bar{N}+S'+S+\bar{S}'+\bar{S}}
\frac{\hat{S}\hat{\bar{S}}'}{\hat{S}'\hat{\bar{S}}}
  \left[
  \frac{\dim(\bar{N}',0)\dim(\omega)}{\dim(\bar{ N},0)\dim(\omega')}
  \right]^{\frac12}
  \rme{\gamma\sigma\upsilon\omega\Sigma S}
  {\mathcal{U}^{\widetilde{(\lambda_0\mu_0})S_0}(\bar{N}\bar{\Sigma}\bar{S},\, \bar{N}'\bar{\Sigma}'\bar{S}')}
  {\gamma'\sigma'\upsilon'\omega'\Sigma'S'}_{\rho_0}.
\end{multline}

In the following we derive the recurrence relations specifically for calculating RMEs of relative unit tensors.  
Since only the seeds of the recurrence relation derived in this section depends on whether the unit tensors have $\grpsuS$ or $\grpu{4}$ tensor character, we will only derive the recurrence for $\grpsu{3}\times\grpsuS$ relative unit tensors, as appropriate to the proton-neutron scheme, and simply note where the calculations of the many-body matrix elements may differ for the isospin scheme. 
The key steps  are as follow: 
\begin{itemize}
\item Peel off a symplectic raising operator (Section~\ref{subsection:peelA}).
\item Commute the raising operator through the unit tensor (Section~\ref{subsection:commuting}). 
\item Evaluate the commutator of $A^{\prime(2,0)}$ and $\scrU^{(\lambda_0\mu_0)}$ (Section~\ref{subsection:evaluate:AUcom}).
\end{itemize}
Once we have the expression for the recurrence (Section~\ref{subsection:recurrence_relations}), all that remains is to compute the seeds, i.e., compute the RMEs of the unit tensors between LGI (Section~\ref{subsection:seeds}).

\subsection{ Peeling off a symplectic raising operator\label{subsection:peelA}}

To peel off a symplectic raising operator from the ket of the RME appearing in~\eqref{many-body-mes-operator},
we start from the expression \eqref{polynomial_expansion_manybody_basis} for the basis states in terms of the symplectic raising polynomial.
A symplectic raising operator can be peeled off from the raising polynomial using the recurrence relation 
 \eqref{raising_polynomial_recurrence}, which constructs $P^n(A^\prime)$ as a sum over $\grpu{3}$ quantum numbers $n_1$ of products of $\left[A^{\prime(2,0)}\times
    P^{n_1}(A^{\prime})\right]^n$. 
Substituting \eqref{raising_polynomial_recurrence} into \eqref{polynomial_expansion_manybody_basis}, we obtain the expression 
\begin{equation}
\ket{\sigma\upsilon\omega\kappa
  L\Sigma SJM}=\sum_{n\rho}(K_\sigma^\omega)_{\upsilon[n\rho]}^{-1}\sum_{n_1}\frac{2}{N_n}(n||b^{\dagger(2,0)}||n_1)
  \left[\left[A^{\prime(2,0)}\times
    P^{n_1}(A^\prime)\right]^n\times\ket{\sigma\Sigma S}\right]^{\rho\omega}_{\kappa
  LJM}.
  \end{equation}
  The raising polynomial $P^{n_1}(A^\prime)$ [which carries two fewer quanta than the original raising polynomial $P^n(A^\prime)$] can then be coupled directly onto the LGI by changing the coupling order of the raising operator, raising polynomial and LGI and inverting \eqref{polynomial_expansion_manybody_basis}.  
  The resulting expression is
\begin{equation}
\ket{\sigma\upsilon\omega\kappa L\Sigma SJM}
=\sum_{\omega_1\upsilon_1\rho}\Xi(\upsilon_1\omega_1;\upsilon\omega)\Big[
A^{\prime(2,0)}
	\times\ket{\sigma\upsilon_1\omega_1\Sigma S}
\Big]^{\rho\omega}_{\kappa LJM},
  \label{cfp1}
\end{equation}
where 
\begin{equation}
\Xi(\upsilon_1\omega_1;\upsilon\omega)
=\sum_{n\rho n_1\rho_1}(K_\sigma^\omega)_{\upsilon[n\rho]}^{-1}\frac{2}{N_n}(n||b^{\dagger(2,0)}||n_1)(K_\sigma^{\omega_1})_{[n_1\rho_1],\upsilon_1}
 U\big[(2,0)n_1\omega\sigma;n\_\rho;\omega_1\rho_1\_\big].
\end{equation}

We can now use the above expression to peel off a raising operator from the ket of the RME.  
For simplicity, we will restrict the following discussion to operators $\mathcal{O}^{(\lambda_0\mu_0) S_0 }$ which are $\grpsu{3}\times\grpsuS$ tensors, since we can always decompose any operator in terms of $\grpsu{3}\times\grpsuS$ unit tensors.  
To do this, we must first re-express the RME as the overlap of a bra with the coupled action of the operator on the ket\footnote{See, e.g., Appendix~A.5 of Ref.~\cite{RoweWood2010} for the analogous relation in the case of $\grpsu{2}$ RMEs.}
\begin{equation}
  \label{eq:relative_unit_tensor_u3s_rme_overlap}
 \rme{\gamma'\sigma'\upsilon'\omega'\Sigma'S'}{\mathcal{O}^{(\lambda_0\mu_0) S_0 }}{\gamma\sigma\upsilon\omega \Sigma S}_{\rho_0}
 =
 	\bra{\gamma'\sigma'\upsilon'\omega'\Sigma'S'}
	\left[
 		\overleftarrow{
\mathcal{O}^{(\lambda_0\mu_0) S_0 }
			\times
			\ket{\sigma\upsilon\omega\Sigma S}
			}
	 \right]^{\rho_0\omega'S'},
\end{equation}
where the arrow over the coupled action of the unit tensor on the ket indicates right-to-left coupling order rather than the standard left-to-right coupling order.  In order to peel off the raising operator, we have to first reverse the coupling order of the operator and ket using the interchange phase matrix $\Phi[\dots]$, and then we can then peel off a symplectic raising operator using the expression above.  
The expression for the RME of the unit tensor becomes
\begin{equation}
\begin{split}
 &\rme{\gamma'\sigma'\upsilon'\omega'\Sigma'S'}{\mathcal{O}^{(\lambda_0\mu_0) S_0 }}{\gamma\sigma\upsilon\omega \Sigma S}_{\rho_0}\\
 &=\sum_{\rho_0'}\Phi_{\rho_0\rho_0'}[\omega(\lambda_0\mu_0)\omega']
\sum_{\omega_1\upsilon_1}
	\Xi(\upsilon_1\omega_1;\upsilon\omega)
 	\bra{\gamma'\sigma'\upsilon'\omega'\Sigma'S'}
	\left[
 		\mathcal{O}^{(\lambda_0\mu_0) S_0 }
		\times \left[
			A^{\prime(2,0)}
			\times
			\ket{\sigma\upsilon_1\omega_1\Sigma S}
		\right]^{\omega S}
	 \right]^{\rho_0'\omega'S'}.
\end{split}
\end{equation}

\subsection{Commuting through the symplectic raising operator\label{subsection:commuting}}

In order to commute the raising operator through the unit tensor, we must first recouple the raising operator to the unit tensor using an $\grpsu{3}$ recoupling coefficient: 
\begin{equation}
\begin{split}
 &\rme{\gamma'\sigma'\upsilon'\omega'\Sigma'S'}{\mathcal{O}^{(\lambda_0\mu_0) S_0}}{\gamma\sigma\upsilon\omega \Sigma S}_{\rho_0}\\
 &=\sum_{\rho_0'}\Phi_{\rho_0\rho_0'}[\omega(\lambda_0\mu_0)\omega']
\sum_{\omega_1\upsilon_1}\Xi(\upsilon_1\omega_1;\upsilon\omega)\sum_{(\lambda_0'\mu_0')\rho_0''}
U[(\lambda_0\mu_0)(2,0)\omega'\omega_1(\lambda_0'\mu_0')\_\,\rho_0''\omega\_\,\rho_0']\\
 	&\times \bra{\gamma'\sigma'\upsilon'\omega'\Sigma'S'}\left[
		\left[
 			\mathcal{O}^{(\lambda_0\mu_0) S_0 }
			\times A^{\prime(2,0)}
		\right]^{(\lambda_0'\mu_0')S_0}
		\times
		\ket{\sigma\upsilon_1\omega_1\Sigma S}
	 \right]^{\rho_0''\omega'S'}.
\end{split}
\end{equation}
We can then commute the raising operator through, thereby introducing a coupled commutator [defined in  \eqref{eqn:coupled-commutator-su3-defn}]. Using the expression for an $\grpsu{3}$ coupled commutator in terms of $\grpsu{3}$ coupled products \eqref{eqn:coupled-commutator-su3-difference}, we obtain
\begin{equation}
\begin{split}
 &\rme{\gamma'\sigma'\upsilon'\omega'\Sigma'S'}{\mathcal{O}^{(\lambda_0\mu_0) S_0}}{\gamma\sigma\upsilon\omega \Sigma S}_{\rho_0}\\
 &=\sum_{\rho_0'}\Phi_{\rho_0\rho_0'}[\omega(\lambda_0\mu_0)\omega']
\sum_{\omega_1\upsilon_1}\Xi(\upsilon_1\omega_1;\upsilon\omega)\sum_{(\lambda_0'\mu_0')\rho_0''}
U[(\lambda_0\mu_0)(2,0)\omega'\omega_1(\lambda_0'\mu_0')\_\,\rho_0''\omega\_\,\rho_0']\\
 	&\bra{\gamma'\sigma'\upsilon'\omega'\Sigma'S'}
	\Bigg[
		\Big(\big[
 			A^{\prime(2,0)}
			\times 
			\mathcal{O}^{(\lambda_0\mu_0) S_0 }
		\big]^{(\lambda_0'\mu_0')S_0}\\
		&\qquad\qquad\qquad\qquad+
		\Phi[(\lambda_0\mu_0)(2,0)(\lambda_0'\mu_0')] \big[
 			\mathcal{O}^{(\lambda_0\mu_0) S_0 },\,
			A^{\prime(2,0)}
		\big]^{(\lambda_0'\mu_0')S_0}
		\Big)
		\times
		\ket{\sigma\upsilon_1\omega_1\Sigma S}
	 \Bigg]^{\rho_0''\omega'S'}.
\end{split}
\end{equation}
Note, the phase factor $\Phi[(\lambda_0\mu_0)(2,0)(\lambda_0'\mu_0')]$ appearing in
the second term evaluates to $+1$.  We can then re-express the overlap
(which now involves the coupled commutator) and rewrite the overlaps as RMEs to
obtain an expression for the RME of $\mathcal{O}^{(\lambda_0\mu_0)S_0}$ in terms of RMEs where the ket has
two fewer quanta:
\begin{equation}
\begin{split}
  &\rme
  {\gamma'\sigma'\upsilon'\omega'\Sigma'S'}
  {\mathcal{O}^{(\lambda_0\mu_0) S_0}}
  {\gamma\sigma\upsilon\omega \Sigma S}_{\rho_0}\\
 &=\sum_{\rho_0'}\Phi_{\rho_0\rho_0'}[\omega(\lambda_0\mu_0)\omega']
\sum_{\omega_1\upsilon_1}
\Xi(\upsilon_1\omega_1;\upsilon\omega)\sum_{(\lambda_0'\mu_0')\rho_0''}
U[(\lambda_0\mu_0)(2,0)\omega'\omega_1(\lambda_0'\mu_0')\_\,\rho_0''\omega\_\,\rho_0']
\sum_{\rho_{00}}
\Phi_{\rho_0''\rho_{00}}[(\lambda_0'\mu_0')\omega_1\omega']\\
 	&\qquad\times\Big\{
	\rme{\gamma'\sigma'\upsilon'\omega'\Sigma'S'}{
		\left[
			A^{\prime(2,0)}\times
 			\mathcal{O}^{(\lambda_0\mu_0) S_0 }
		\right]^{(\lambda_0'\mu_0')S_0}
		}
		{\sigma\upsilon_1\omega_1\Sigma S}_{\rho_{00}}\\
	 &\qquad\qquad +
 	\rme{\gamma'\sigma'\upsilon'\omega'\Sigma'S'}
    {
		\left[
 			\mathcal{O}^{(\lambda_0\mu_0) S_0},
			A^{\prime(2,0)}
		\right]^{(\lambda_0'\mu_00)S_0}
		}
		{\sigma\upsilon_1\omega_1\Sigma S}_{\rho_{00}}\Big\}.
\end{split}
\label{eq:unit_tensor_mb_rme1_generic}
\end{equation}
What remains is to derive the expressions for evaluating these RMEs.  
The first RME on the right hand side can be evaluated using the Racah reduction formula \eqref{su3-Racah1}. 
To evaluate the second RME, we need to first evaluate the coupled commutator, which depends on the particular choice of operator.

For the remainder of this article we specialize to the case of interest for this work, namely, relative unit tensors. Thus
$\mathcal{O}^{(\lambda_0\mu_0)S_0}=\mathcal{U}^{(\lambda_0\mu_0)S_0}(\bar{N}'\bar{\Sigma}'\bar{S},\bar{N}\bar{\Sigma}\bar{S})$,and~\eqref{eq:unit_tensor_mb_rme1_generic} becomes
\begin{equation}
\begin{split}
  &\rme
  {\gamma'\sigma'\upsilon'\omega'\Sigma'S'}
  {\mathcal{U}^{(\lambda_0\mu_0)S_0}(\bar{N}'\bar{\Sigma}'\bar{S},\bar{N}\bar{\Sigma}\bar{S})}
  {\gamma\sigma\upsilon\omega \Sigma S}_{\rho_0}\\
 &=\sum_{\rho_0'}\Phi_{\rho_0\rho_0'}[\omega(\lambda_0\mu_0)\omega']
\sum_{\omega_1\upsilon_1}
\Xi(\upsilon_1\omega_1;\upsilon\omega)\sum_{(\lambda_0'\mu_0')\rho_0''}
U[(\lambda_0\mu_0)(2,0)\omega'\omega_1(\lambda_0'\mu_0')\_\,\rho_0''\omega\_\,\rho_0']
\sum_{\rho_{00}}
\Phi_{\rho_0''\rho_{00}}[(\lambda_0'\mu_0')\omega_1\omega']\\
 	&\qquad\times\Big\{
	\rme{\gamma'\sigma'\upsilon'\omega'\Sigma'S'}{
		\left[
			A^{\prime(2,0)}\times
 			\mathcal{U}^{(\lambda_0\mu_0)S_0}(\bar{N}'\bar{\Sigma}'\bar{S},\bar{N}\bar{\Sigma}\bar{S})
		\right]^{(\lambda_0'\mu_0')S_0}
		}
		{\sigma\upsilon_1\omega_1\Sigma S}_{\rho_{00}}\\
	 &\qquad\qquad +
 	\rme{\gamma'\sigma'\upsilon'\omega'\Sigma'S'}
    {
		\left[
 			\mathcal{U}^{(\lambda_0\mu_0)S_0}(\bar{N}'\bar{\Sigma}'\bar{S},\bar{N}\bar{\Sigma}\bar{S}),
			A^{\prime(2,0)}
		\right]^{(\lambda_0'\mu_00)S_0}
		}
		{\sigma\upsilon_1\omega_1\Sigma S}_{\rho_{00}}\Big\}.
\end{split}
\label{eq:unit_tensor_mb_rme1}
\end{equation}

\subsection{Evaluating the coupled commutator of $A^{\prime(2,0)}$ and $\mathcal{U}^{(\lambda_0\mu_0)S_0}$\label{subsection:evaluate:AUcom}}

To evaluate the second RME appearing on the right hand side of~\eqref{eq:unit_tensor_mb_rme1}, involving the coupled commutator, we first wish to rewrite the coupled commutator in terms of unit tensors.
To do so, we expand $A^{\prime(2,0)}$ in terms of relative unit tensors, which reexpresses the coupled commutator in terms of the coupled commutators of two relative unit tensors.
This commutator, in turn, can be expanded as of a sum of relative unit tensors. 

To accomplish these steps, we use the trick that the intrinsic symplectic raising operator $A^{\prime(2,0)}$ can be rewritten in terms of oscillator ladder operators $\zeta_m^\dagger$ defined with respect to the Jacobi coordinate $\xi_m$ and its conjugate momentum $\eta_m$.  The Jacobi coordinate and momentum are defined as
\begin{equation}
\xi_m=\frac{1}{\sqrt{m(m+1)}}\left[\sum_{l=1}^m x_l -m x_{m+1}\right] \qquad 
\eta_m=\frac{1}{\sqrt{m(m+1)}}\left[\sum_{l=1}^m k_l -m k_{m+1}\right].
\end{equation}
The operator $\zeta^\dagger$ is related to $c^\dagger$ by the same orthogonal transformation matrix as that relating single particle and Jacobi coordinates ($\xi_m=\sum_l^AO_{ml}x_m$).   The symplectic raising operator is given in terms of these Jacobi-coordinate oscillator ladder operators  by~\cite{Reske1984}
\begin{equation}
\begin{split}
A_{LM}^{\prime(2,0)}=\frac{1}{\sqrt{2}}\sum_{s=1}^{A-1}\left[\zeta_s^{\dagger(1,0)}\times \zeta_s^{\dagger(1,0)}\right]_{LM}^{(2,0)}.
\label{A_jacobi}
\end{split}
\end{equation}
Note that the sum is over only the first $A-1$ Jacobi coordinates and thus does not include the center-of-mass coordinate $\xi_A=x_{\cm}=A^{-1/2}\sum_i^Ax_i$. 

Since the relative unit tensor is only a function of the first Jacobi coordinate, only the first term in the sum in \eqref{A_jacobi} will have a non-zero commutator with the relative unit tensor.  Thus, 
\begin{equation}
[\scrU^{(\lambda_0\mu_0) S_0}(\bar{N}'\bar{\Sigma}'\bar{S}',\bar{N}\bar{\Sigma}\bar{S}),\,A^{\prime(2,0)}]^{(\lambda_0'\mu_0')S_0}
= \big[\scrU^{(\lambda_0\mu_0) S_0 }(\bar{N}'\bar{\Sigma}'\bar{S}',\bar{N}\bar{\Sigma}\bar{S}),\,[\zeta_1^{\dagger(1,0)}\times \zeta_1^{\dagger(1,0)}]^{(2,0)}\big]^{(\lambda_0'\mu_0')S_0}.
\end{equation}
Since $[\zeta_1^{\dagger(1,0)}\times \zeta_1^{\dagger(1,0)}]^{(2,0)}$ depends only on the first Jacobi coordinate, we can expand this operator in terms of relative unit tensors using \eqref{tensor_rel_unit_tensor_expansion}. The RME of this operator in the relative harmonic oscillator basis is that of the symplectic raising operator, which is given by 
\begin{equation}
\rme{\bar{N}'_a\bar{\Sigma}'_a\bar{S}'_a}{A^{\prime(2,0)}}{\bar{N}_a\bar{\Sigma}_a\bar{S}_a}=\sqrt{\dim(\bar{N}_a,0)}\delta_{\bar{N}_a',\bar{N}_a+2}\delta_{\bar{\Sigma}'_a,\bar{\Sigma}_a}\delta_{\bar{S}_a'\bar{S}_a}.
\end{equation}
Thus, the coupled commutator of $A^{\prime(2,0)}$ and a relative unit tensor reduces to
\begin{multline}
[\scrU^{(\lambda_0\mu_0) S_0}(\bar{N}'\bar{\Sigma}'\bar{S}',\bar{N}\bar{\Sigma}\bar{S}),\,A^{\prime(2,0)}]^{(\lambda_0'\mu_0')S_0}\\
= \sum_{\bar{N}_a\bar{\Sigma}_a\bar{S}_a}\sqrt{\dim(\bar{N}_a,0)}[\scrU^{(\lambda_0\mu_0) S_0}(\bar{N}'\bar{\Sigma}'\bar{S}',\bar{N}\bar{\Sigma}\bar{S}),\scrU^{(2,0)0}(\bar{N}_a+2\,\bar{\Sigma}_a\bar{S}_a,\bar{N}_a\bar{\Sigma}_a\bar{S}_a)]^{(\lambda_0'\mu_0')S_0}.
\label{eq:AU_commutator}
\end{multline} 
The coupled commutator of two relative unit tensors appearing on the right hand side of \eqref{eq:AU_commutator} may in turn be expanded as a linear combination of relative unit tensors, given by \eqref{eq:unit_tensor_commutator}.

\subsection{The RME recurrence relations\label{subsection:recurrence_relations}}
To obtain the recurrence relation we now combine the three steps listed above in Sections~\ref{subsection:peelA}-\ref{subsection:evaluate:AUcom}.  We start with the expression for the RME of a unit tensor in \eqref{eq:unit_tensor_mb_rme1}.  To evaluate the first term, we apply Racah's reduction formula \eqref{su3-Racah1}.  To evaluate the second term, we substitute the expression \eqref{eq:AU_commutator} for the commutator of $A^{\prime(2,0)}$ with $\scrU$ into the second term of \eqref{eq:unit_tensor_mb_rme1}, and then re-express the coupled commutator as a linear combination of unit tensors using \eqref{eq:unit_tensor_commutator}.  After applying various orthogonality and symmetry properties of the Racah coefficients~
 \cite{jmp-7-1966-612-Derome,
 np-62-1965-1-Hecht,
 jmp-8-1967-714-Derome,
 jmp-8-1967-63-Resnikoff,
 npa-111-1968-681-Vergados,
 npa-129-1969-647-Draayer,
 jmp-14-1973-1904-Draayer,
 ptrsla-277-1975-545-Butler,
  npa-244-1975-365-Hecht,
 jmp-19-1978-1513-Millener,
npa-356-1981-146-Hecht,
 jmp-24-1983-785-Hecht,
 jpa-23-1990-Hecht, 
 jmp-41-2000-6544-Rowe}
 to simplify the expression and eliminate some of the sums over outer multiplicity indices, we obtain recurrence relation for computing RMEs of relative unit tensors in the symplectic many-body basis.  The resulting recurrence relation (for $N_{n'}>N_n$) is given by (here we underline the RMEs of unit tensors to highlight that the recurrence is over such RMEs)
\begin{equation}
\begin{split}
  &\underline{\rme{\gamma'\sigma'\upsilon'\omega'\Sigma'S'}
  {\scrU^{(\lambda_0\mu_0) S_0 }(\bar{N}'\bar{\Sigma}'\bar{S}',\bar{N}\bar{\Sigma}\bar{S})}
  {\gamma\sigma\upsilon\omega \Sigma S}_{\rho_0}}\\[2ex]
  &=\sum_{\omega_1\upsilon_1}
  \Bigg\{
    \frac{2}{N_\omega - N_\sigma}
    \sum_{\substack{n\rho\\n_1\rho_1}}
      (K_\sigma^\omega)_{\upsilon[n\rho]}^{-1}
      (n\Vert{}a^\dagger\Vert{}n_1)
      U[(2,0)n_1 \omega \omega_\sigma;n\_\rho;\omega_1\rho_1\_]
      (K_\sigma^{\omega_1})_{[n_1\rho_1],\upsilon_1}
  \Bigg\}\\
&\times\Bigg\{\sum_{\omega_2\upsilon_2}
    (-1)^{\omega_2+\omega'}
    \sum_{\rho_0'}
    \Big(
      \rme{\sigma'\upsilon'\omega'}
      {A^{\prime(2,0)}}
      {\sigma'\upsilon_2\omega_2}
      U[(2,0) \omega_1 \omega' (\lambda_0\mu_0); \omega \_\, \rho_0; \omega_2 \rho_0'\_\,]\\
  &\hspace{7cm}
  \times
     \underline{ \rme{\gamma'\sigma'\upsilon_2\omega_2\Sigma'S'}
      {\mathcal{U}^{(\lambda_0\mu_0)S_0}(\bar{N}'\bar{\Sigma}'\bar{S}',\bar{N}\bar{\Sigma}\bar{S})}
      {\gamma\sigma\upsilon_1\omega_1\Sigma S}_{\rho_0'}}
    \Big)\\\\
&\qquad+
  (-1)^{\omega+\omega_1}(-1)^{\omega_0+\omega_0'}
  \sum_{\rho_0'\omega_0'}
  \Bigg(
\left[\frac{\dim(\lambda_0'\mu_0')\dim(\bar{N},0)}{\dim(\lambda_0\mu_0)}\right]^{\frac12}\\
&\qquad\qquad\times
      U[\omega_1 (2,0) \omega' (\lambda_0\mu_0); \omega\_\,\rho_0;(\lambda_0'\mu_0')\_\,\rho_0']
      U[(\bar{N}',0) (0,\bar{N}) (\lambda_0'\mu_0') (2,0);(\lambda_0\mu_0)\_\,\_\,;(0,\bar{N}-2)\_\,\_\,]\\
&\hspace{7cm}
\underline{\rme{\gamma'\sigma'\upsilon'\omega' \Sigma'S'}
{\mathcal{U}^{(\lambda_0'\mu_0')S_0}(\bar{N}'\bar{\Sigma}'\bar{S}',\bar{N}-2\,\bar{\Sigma}\bar{S})}
{\gamma\sigma\upsilon_1\omega_1\Sigma S}_{\rho_0'}}
\Bigg)\\\\
&\qquad-
  (-1)^{\omega+\omega_1}(-1)^{\omega_0+\omega_0'}
  \sum_{\rho_0'\omega_0'}
  \Bigg(\left[\frac{\dim(\bar{N}',0)^2\dim(\lambda_0'\mu_0')}{\dim(2,0)\dim(\bar{N},0)}\right]^{\frac12}\\
&\qquad\qquad\times
      U[\omega_1 (2,0) \omega' (\lambda_0\mu_0); \omega\_\,\rho_0;(\lambda_0',\mu_0')\_\,\rho_0']\,
      U[(\bar{N}'+2,0)(0,\bar{N}') (\lambda_0'\mu_0') (\lambda_0\mu_0);(2,0)\_\,\_\,;(0,\bar{N})\_\,\_\,]\\
  &\hspace{7cm}
  \underline{
  \rme{\gamma'\sigma'\upsilon'\omega' \Sigma'S'}
  {\mathcal{U}^{(\lambda_0'\mu_0') S_0}(\bar{N}'+2\,\bar{\Sigma}'\bar{S}',\bar{N}\bar{\Sigma}\bar{S})}
  {\gamma\sigma\upsilon_1\omega_1\Sigma S}_{\rho_0'}}
    \Bigg)
\Bigg\}.
\end{split}
\label{eq:recurrence_final}
\end{equation}
In the summation over $\omega_1$ and $\omega_2$, note that $N_{\omega_1}=N_\omega-2$ and $N_{\omega_2}=N_{\omega'}-2$.  

The recurrence relation above does not depend on $\Sigma S$.   Thus, in the isospin scheme, if the operators are decomposed in terms of $\grpsu{3}\times\grpsu{4}$ unit tensors, the same recurrence relation can be used to compute the RMEs of a $\grpsu{3}\times\grpsu{4}$ unit tensor by simply replacing each of the $\grpsuS$-reduced matrix elements with a corresponding $\grpsu{4}$ reduced matrix elements.  For example, substituting  
\begin{equation}
\begin{split}
&\rme{\gamma'\sigma'\upsilon'\omega'\Sigma'S'}
  {\scrU^{(\lambda_0\mu_0) S_0 }(\bar{N}'\bar{\Sigma}'\bar{S}',\bar{N}\bar{\Sigma}\bar{S})}
  {\gamma\sigma\upsilon\omega \Sigma S}_{\rho_0}\\
  &\qquad \longrightarrow 
  \rme{\gamma'\sigma'\upsilon'\omega'[{\boldsymbol f}']}
  {\scrU^{(\lambda_0\mu_0) [{\boldsymbol \tau}_0] }(\bar{N}'[\bar{\boldsymbol \tau}'],\bar{N}[\bar{\boldsymbol \tau}])}
  {\gamma\sigma\upsilon\omega [{\boldsymbol f}]}_{\rho_0,\delta_0}.
\end{split}
\end{equation}

\subsection{Calculating the seeds\label{subsection:seeds}}
The seeds for the recurrence, which are the RMEs of unit tensors between LGIs, can be obtained by first calculating the RMEs of the relative unit tensors in the $\grpu{3}$ many-body basis  and then transforming the RMEs to the $\grpsptr$ many-body basis.  Recall that the unitary transformation from the $\grpu{3}$ many-body basis to the $\grpsptr$ many-body basis is given by the null vectors of $B^{(0,2)}$ and $N^{(0,0)}_\cm$ in \eqref{eq:lgi_expansion}.  

To compute the RMEs of the unit tensors in the $\grpu{3}$ many-body basis, we need to express the relative unit tensors in terms of $\grpu{3}$ creation and annihilation operators  as discussed in Section~\ref{subsection:su3basis-me}.  
In order to do so, we need to first obtain an expression for the relative unit tensor in the two-body product basis. 
It is convenient to first write the relative unit tensors in terms of unit tensors defined with respect to the two-body product basis, which are defined, in the proton-neutron scheme, by having RME
\begin{multline}
\rme{
[N_1'N_2']\omega \Sigma S}{
\mathcal{U}^{\bar{\rho}_0(\lambda_0,\mu_0)S_0}\big([\bar{N}_1'\bar{N}_2']\bar{\omega}'\bar{\Sigma}'\bar{S}',[\bar{N}_1\bar{N}_2]\bar{\omega}\bar{\Sigma}\bar{S}\big)
}{[N_1N_2]\omega \Sigma S
}_{\rho_0}\\
=
\delta_{N_1',\bar{N}_1'}\delta_{N_2',\bar{N}_2'}\delta_{\omega',\bar{\omega}'}\delta_{\Sigma',\bar{\Sigma}'}\delta_{S',\bar{S}'}
\delta_{N_1,\bar{N}_1}\delta_{N_2,\bar{N}_2}\delta_{\omega,\bar{\omega}}\delta_{\Sigma,\bar{\Sigma}}\delta_{S,\bar{S}}\delta_{\bar{\rho}_0,\rho_0}.\nonumber
\end{multline}
A relative unit tensor can be written in terms of these two-body unit tensors as 
\begin{multline}
\mathcal{U}^{(\lambda_0,\mu_0)S_0}(\bar{N}'\bar{\Sigma}'\bar{S}',\bar{N}\bar{\Sigma}\bar{S})\\
=\sum_{\substack{\bar{N}_1'\bar{N}_2'\bar{\omega}'\bar{\Sigma}'\bar{S}'\\\bar{N}_1\bar{N}_2\bar{\omega}\bar{\Sigma}\bar{S}\rho_0}}
\rme{[\bar{N}_1'\bar{N}_2']\bar{\omega}'\bar{\Sigma}'\bar{S}'}
{\mathcal{U}^{(\lambda_0,\mu_0)S_0}(\bar{N}'\bar{\Sigma}'\bar{S}',\bar{N}\bar{\Sigma}\bar{S})}
{[\bar{N}_1\bar{N}_2]\bar{\omega}\bar{\Sigma} \bar{S}}_{\rho_0}
\mathcal{U}^{\rho_0(\lambda_0,\mu_0)S_0}
\big(
[\bar{N}_1'\bar{N}_2']\bar{\omega}'\bar{\Sigma}'\bar{S}',
[\bar{N}_1\bar{N}_2]\bar{\omega}\bar{\Sigma}\bar{S}
\big). 
\label{eq:rel-unit-in-tb-unit}
\end{multline}
The  reduced matrix element of the relative unit tensor in the two-body harmonic oscillator basis is obtained by upgrading the relative unit tensor to an operator on the combined space defined on the relative and center-of-mass coordinates $x_{\rel}$ and $x_{\cm}=(x_1+x_2)/\sqrt{2}$.  
The relation between the bases is given by the Talmi-Moshinsky transformation for an $\grpsu{3}$ coupled basis~\cite{jmp-6-1965-142-Kaufman,npa-318-1979-1-Hecht,rmf-9-1960-181-moshinsky}:
\begin{multline}
\rme{[\bar{N}_1'\bar{N}_2']\bar{\omega}'\bar{\Sigma}'\bar{S}'}
{\mathcal{U}^{(\lambda_0,\mu_0)S_0}(\bar{N}'\bar{\Sigma}'\bar{S}',\bar{N}\bar{\Sigma}\bar{S})}
{[\bar{N}_1\bar{N}_2]\bar{\omega} \bar{\Sigma}\bar{S}}_{\rho_0}\\
=\sum_{N_\cm}
2\frac{(-)^{\bar{N}+\bar{N}'+\bar{\omega}+\bar{\omega}'}}
{\sqrt{(1+\delta_{\bar{N}_1,\bar{N}_2})(1+\delta_{\bar{N}_1'\bar{N}_2'})}}
\left(
\begin{array}{c}
\bar{N}_{1}\,\bar{N}_{2}\\
\bar{\omega}
\end{array}
\right|
\left.\begin{array}{c}
\bar{N}\, \bar{N}_\cm\\
\bar{\omega}
\end{array}
\right)
\left(
\begin{array}{c}
\bar{N}_{1}'\,\bar{N}_{2}'\\
\bar{\omega}'
\end{array}
\right|
\left.\begin{array}{c}
\bar{N}' \,\bar{N}_\cm\\
\bar{\omega}
\end{array}
\right)\\
\times 
U\big((\bar{N}_\cm,0)(\bar{N},0)\bar{\omega}'(\lambda_0,\mu_0);\bar{\omega}\_\,\rho_0;(\bar{N}',0)\big).
\label{eq:rel-tb-rme}
\end{multline}
Here $(-)^\omega=(-)^{\lambda_\omega+\mu_\omega}$. 
 The Talmi-Moshinsky bracket is given by a Wigner $d$-function evaluated at $\pi/2$, as
\begin{equation}
\left(
\begin{array}{c}
\bar{N}_{1}\bar{N}_{2}\\
\bar{\omega}
\end{array}
\right|
\left.\begin{array}{c}
\bar{N} \bar{N}_\cm\\
\bar{\omega}
\end{array}
\right)=d^{\frac{\lambda_{\bar{\omega}}}{2}}_{\frac{(\bar{N}_1-\bar{N}_2)}{2},\frac{(\bar{N}-\bar{N}_\cm)}{2}}\left(\frac{\pi}{2}\right)\label{MTB}. 
\end{equation}
Expressions for the Wigner $d$ function evaluated at $\pi/2$ are given in, e.g., Ref~\cite{Edmonds1960}.

The two-body unit tensor operator can be expressed in terms of $\grpsu{3}\times\grpsuS$ coupled creation and annihilation operators as~\cite{McCoy2018}
\begin{multline}
\mathcal{U}^{\rho_0(\lambda_0,\mu_0)S_0}\Big([\bar{N}_1'\bar{N}_2']\bar{\omega}'\bar{S}',[\bar{N}_1\bar{N}_2]\bar{\omega}\bar{S}\Big)\\
=\sum_{\bar{\rho}_0}(-1)^{\lambda_0+\mu_0+\bar{\omega}+\bar{\omega}'}\Phi_{\rho_0\bar{\rho}_0}[\bar{\omega}(\lambda_0,\mu_0)\bar{\omega}']
\sqrt{\frac{\dim(\bar{\omega}')}{\dim(\lambda_0,\mu_0)}}\frac{\hat{\bar{S}}'}{\hat{S}_0}\\\times
\left[
[a_{\chi_1}^{\dagger(\bar{N}_1',0)}\times a_{\chi_2}^{\dagger(\bar{N}_2',0)}]^{\bar{\omega}' \bar{S}'}
\times
[\tilde{a}_{\chi_2}^{(0,\bar{N}_2)}\times \tilde{a}_{\chi_1}^{(0,\bar{N}_1)}]^{\tilde{\bar{\omega}}\bar{S}}
\right]^{\bar{\rho}_0(\lambda_0,\mu_0)S_0}.
\label{eq:tb-unit-second-quantized-pn}
\end{multline}
Combining the second quantized expression for the two-body unit tensor \eqref{eq:tb-unit-second-quantized-pn} with the expansion of the relative unit tensor in terms of two-body unit tensors \eqref{eq:rel-unit-in-tb-unit}, we obtain an expression for the relative unit tensor in terms of $\grpsu{3}$-coupled creation and annihilation operators:  
\begin{equation}
\begin{aligned}
&\mathcal{U}^{(\lambda_0,\mu_0)S_0}(\bar{N}'\bar{S}',\bar{N}\bar{S})\\
&=\sum_{\substack{\bar{N}_1'\bar{N}_2'\bar{\omega}'\bar{S}'\\\bar{N}_1\bar{N}_2\bar{\omega}\bar{S}\rho_0}N_{\cm}}
\frac{2}
{\sqrt{(1+\delta_{\bar{N}_1,\bar{N}_2})(1+\delta_{\bar{N}_1'\bar{N}_2'})}}
\left(
\begin{array}{c}
\bar{N}_{1}\,\bar{N}_{2}\\
\bar{\omega}
\end{array}
\right|
\left.\begin{array}{c}
\bar{N}\, \bar{N}_\cm\\
\bar{\omega}
\end{array}
\right)
\left(
\begin{array}{c}
\bar{N}_{1}'\,\bar{N}_{2}'\\
\bar{\omega}'
\end{array}
\right|
\left.\begin{array}{c}
\bar{N}' \,\bar{N}_\cm\\
\bar{\omega}
\end{array}
\right)\\
&\qquad\quad\times
U\big((\bar{N}_\cm,0)(\bar{N},0)\bar{\omega}'(\lambda_0,\mu_0);\bar{\omega}\_\,\rho_0;(\bar{N}',0)\big)\\
&\qquad\qquad\times
\sum_{\bar{\rho}_0}
\Phi_{\rho_0\bar{\rho}_0}[\bar{\omega}(\lambda_0,\mu_0)\bar{\omega}']
\sqrt{\frac{\dim(\bar{\omega}')}{\dim(\lambda_0,\mu_0)}}\frac{\hat{\bar{S}}'}{\hat{S}_0}
\left[
[a_{\chi_1}^{\dagger(\bar{N}_1',0)}\times a_{\chi_2}^{\dagger(\bar{N}_2',0)}]^{\bar{\omega}' \bar{S}'}
\times
[\tilde{a}_{\chi_2}^{(0,\bar{N}_2)}\times \tilde{a}_{\chi_1}^{(0,\bar{N}_1)}]^{\tilde{\bar{\omega}}\bar{S}}
\right]^{\bar{\rho}_0(\lambda_0,\mu_0)S_0}.
\label{eq:rel-unit-second-quantized-pn}
\end{aligned}
\end{equation}
Note that in simplifying the expression above, we make use of the fact that  $(-)^{\lambda_0+\mu_0+\bar{N}+\bar{N}'}=1$.
The seed RMEs between LGIs of these operators can then be computed by first expanding the LGIs in terms of $\grpu{3}$ many-body states \eqref{eq:lgi_expansion}, and then computing the RMEs of these relative unit tensors in the $\grpu{3}$ many-body basis as discussed in Section~\ref{subsection:su3basis-me}.

In the isospin scheme, the relative unit tensor \eqref{eq:u3u4_unit_tensor_def} can be expressed in terms of $\grpsu{3}\times\grpsu{4}$ coupled creation and annihilation operators as 

\begin{equation}
\begin{aligned}
&\mathcal{U}^{\rho_0(\lambda_0,\mu_0)[{\bf \tau}_0]}\Big([\bar{N}'[\bar{\boldsymbol\tau}'],\bar{N}[\bar{\boldsymbol\tau}]\Big)\\
&=\sum_{\substack{\bar{N}_1'\bar{N}_2'\bar{\omega}'[\bar{\boldsymbol \tau}']\\\bar{N}_1\bar{N}_2\bar{\omega}[\bar{\boldsymbol \tau}]\rho_0}N_{\cm}}
\frac{2}
{\sqrt{(1+\delta_{\bar{N}_1,\bar{N}_2})(1+\delta_{\bar{N}_1'\bar{N}_2'})}}
\left(
\begin{array}{c}
\bar{N}_{1}\,\bar{N}_{2}\\
\bar{\omega}
\end{array}
\right|
\left.\begin{array}{c}
\bar{N}\, \bar{N}_\cm\\
\bar{\omega}
\end{array}
\right)
\left(
\begin{array}{c}
\bar{N}_{1}'\,\bar{N}_{2}'\\
\bar{\omega}'
\end{array}
\right|
\left.\begin{array}{c}
\bar{N}' \,\bar{N}_\cm\\
\bar{\omega}
\end{array}
\right)\\
&\qquad\quad\times
U\big((\bar{N}_\cm,0)(\bar{N},0)\bar{\omega}'(\lambda_0,\mu_0);\bar{\omega}\_\,\rho_0;(\bar{N}',0)\big)\\
&\qquad\qquad\times\sum_{\bar{\rho}_0}\Phi_{\rho_0\bar{\rho}_0}[\bar{\omega}(\lambda_0,\mu_0)\bar{\omega}']
\check{\Phi}\big[[\bar{\boldsymbol\tau}][{\boldsymbol \tau}_0][\bar{\boldsymbol\tau}']\big]]\cev{\Phi}\big[[{\boldsymbol \tau}_0][\bar{\boldsymbol\tau}][\bar{\boldsymbol\tau}']
\sqrt{\frac{\dim(\bar{\omega}')\dim([\bar{\boldsymbol \tau}'])}{\dim(\lambda_0,\mu_0)\dim([{\boldsymbol \tau}_0])}}\\
&\qquad\qquad\quad\times
\left[
[a^{\dagger(\bar{N}_1',0)}\times a^{\dagger(\bar{N}_2',0)}]^{\bar{\omega}' [\bar{\boldsymbol\tau}']}
\times
[\tilde{a}^{(0,\bar{N}_2)}\times \tilde{a}^{(0,\bar{N}_1)}]^{\tilde{\bar{\omega}}[\tilde{\bar{\boldsymbol \tau}}]}
\right]^{\bar{\rho}_0(\lambda_0,\mu_0)\delta_0[{\boldsymbol \tau}_0]},
\label{eq:rel-unit-second-quantized-isospin}
\end{aligned}
\end{equation}
where $\check{\Phi}[\dots]$ and $\cev{\Phi}[\dots]$ are $\grpsu{4}$ interchange
and reversal phase matrices analogous to those defined for $\grpsu{3}$.  For
two-body operators, the coupling $[\bar{\boldsymbol\tau}]\times[{\boldsymbol
    \tau}_0]\rightarrow [\bar{\boldsymbol\tau}']$ is always multiplicity free
and thus these phase matrices reduce to a phase factors, which are given in, e.g., Ref.~\cite{jmp-10-1969-1571-Hecht}.

 \section{Conclusions}

Solving the nuclear many-body problem in a basis that directly encodes symmetries associated with nuclear collectivity, such as the $\grpsptr$ symplectic symmetry, has the potential to significantly reduce the size of the many-body basis required to describe highly collective nuclei. 
However, carrying out calculations in such a basis is challenging. 
In this work, we presented the symplectic no-core configuration interaction (SpNCCI) framework, which makes extensive use of group-theoretical machinery to simplify calculations in a symplectic many-body basis.

In the SpNCCI framework, we solve the nuclear many-body problem in a basis that explicitly encodes $\grpsptr$ and $\grpu{3}$ symmetries. 
The many-body Hilbert space is organized into irreducible representations (irreps) of $\grpsptr$. 
Each irrep is obtained by starting from a unique lowest-grade $\grpu{3}$ irrep (LGI) and repeatedly acting on this LGI with symplectic raising polynomials, thereby generating a set of $\grpu{3}$ irreps with differing numbers of excitation quanta belonging to the same $\grpsptr$ irrep. 
More specifically, we construct $\grpsptr$ irreps with no center-of-mass excitation, i.e., center-of-mass–free (CMF) irreps. 
We obtain these CMF states by first constructing CMF LGIs and then generating the remaining $\grpu{3}$ irreps in the $\grpsptr$ irrep by acting on the LGI with intrinsic symplectic raising polynomials. 
Because the intrinsic operators do not act on the center-of-mass coordinates, the resulting excited $\grpu{3}$ irreps also have no center-of-mass excitation.

Constructing the basis in this way enables us to compute matrix elements of operators—such as the Hamiltonian—directly in the symplectic basis, without expanding all $\grpsptr$ states in terms of an $\grpsu{3}$-coupled configuration basis. 
Matrix elements are evaluated via a recurrence relation that expresses a given reduced matrix element (RME) in terms of RMEs between states of the same $\grpsptr$ irrep but with fewer oscillator quanta. 
In this work, we derive recurrence relations for computing the matrix elements of relative operators.

Central to the derivation of these recurrences is the decomposition of any relative operator into $\grpsu{3}\times\grpsuS$ tensors. 
Specifically, we express relative two-body operators in terms of relative unit tensors, which are special tensors defined to have a single non-zero RME in the relative harmonic-oscillator basis. 
The expansion of a relative operator in terms of relative unit tensors is given in terms of known matrix elements of the operator in the relative harmonic-oscillator basis and standard $\grpsu{3}$ and $\grpsu{2}$ coupling and recoupling coefficients.

The recurrence is then defined in terms of RMEs of the unit tensors in the symplectic many-body basis. 
As shown, the coefficients of the recurrence involve $\grpsu{3}$ and $\grpsu{2}$ recoupling coefficients, dimension factors, and analytically known RMEs of the $\grpsptr$ generators. 
The seeds of the recurrence are the RMEs of unit tensors between LGIs, which can be computed by expanding the LGIs in terms of $\grpsu{3}$-coupled configurations.

The formalism presented in this paper applies equally when working in a symplectic basis where protons and neutrons are treated as distinguishable particles (proton–neutron scheme) or in a basis where nucleons are treated as indistinguishable (isospin scheme).
Only the seeds of the recurrence—i.e., the RMEs of unit tensors between the LGIs for each $\grpsptr$ irrep—are scheme dependent.
When working in the isospin scheme, one does also have the option to further decompose the operators in terms of good $\grpsu{4}$ tensors, e.g., so as to apply the $\grpsu{4}$ Wigner-Eckart theorem to further factorize calculation of the many-body matrix elements.
Such simplification requires the use of $\grpsu{4}$ coupling and recoupling coefficients.
Although code is readily available for computing $\grpsu{4}$ coupling coefficients in bases reducing the canonical $\grpsu{4}\supset\grpsu{3}\times\grpu{1}$ group chain~\cite{cpc-179-2008-733-Kuhn}, the present application---involving bases reducing the subgroup chain~\eqref{su3_so3_su4_chain}---requires evaluation of coupling coefficients for the spin-isospin subgroup chain $\grpsu{4}\supset\grpsuS\times\grpsuT$~\cite{epjplus-139-2024-933-Dang}.

 \appendix
\section{Decomposing nuclear interactions into $\grpsu{3}$ tensors
\label{appendix:upcoupling} 
}
In this appendix, we derive the relation given in Section~\ref{section:recurrence} for decomposing a general relative operator in terms of $\grpsu{3}\times\grpsuS$ unit tensors.  
We also provide details on how this expression may be adapted for decomposing a relative operator in terms of $\grpsu{3}\times\grpsu{4}$ unit tensors for use in the isospin scheme. 

The basic idea underpinning this decomposition is that any operator is uniquely defined by its complete set of matrix elements in a given basis.  
Thus, if another operator,  e.g., a linear combination of unit tensors, has all the same matrix elements, then the two operators are equivalent.  
In this work, we want to identify a linear combination of $\grpsu{3}\times\grpsuS$ unit tensors [defined in \eqref{eq:relative_unit_tensors_u3s}], which have the same matrix elements as a given operator $V$ in the relative harmonic oscillator basis. 

However, before we dive into decomposing $V$ in terms of $\grpsu{3}\times\grpsuS$ unit tensors, we will first consider a much simpler case, which is decomposing an operator in terms of $\grpsu{2}$ unit tensors. 
Here, we illustrate this case for an operator with good $M_T$, such as a relative proton-neutron operator, which has $M_T=0$, but which is not a good $\grpsuT$ tensor.  
Note that, while we illustrate the decomposition method for this specific case of isospin,  the expressions derived are equally valid for any realization of $\grpsu{2}$ --- such as $\grpsuJ$ or $\grpsuS$. 

\subsection{Decomposing operators in terms of $\grpsu{2}$ unit tensors}
To decompose an operator with definite $M_T$ in terms of $\grpsuT$ tensors, we start with the matrix elements of this operator $V_{M_T}$ in the relative harmonic oscillator basis with states labeled by $\ket{[f] NL SJ MTM_T}$.  
For the purpose of decomposing the operator in terms of $\grpsuT$ unit tensors, most of these labels are additional labels serving to distinguish states of the same $\grpsuT$ quantum numbers.  
Thus, to simplify notation in this  first example, we label the states by $\ket{\Gamma TM_T}$, where  $\Gamma=[f]NLSJM$.
Our goal is then to write the operator $V_{M_T}$ in terms of $\grpsuT$ unit tensors, which are defined with respect to this basis by
\begin{equation}
  \label{eq:su2t_unit_tensor_defn}
\rme{\Gamma'T'}{\mathcal{U}^{T_0}(\bar{\Gamma}'\bar{T}',\bar{\Gamma}\bar{T})}{\Gamma T}
=\delta_{\bar{\Gamma},\Gamma}\delta_{\bar{T},T}
\delta_{\bar{\Gamma}',\Gamma'}\delta_{\bar{T}',T'}.
\end{equation}

We begin by writing the operator in terms of dyads.  Such an expression can be obtained by inserting resolutions of the identity around the operator: 
\begin{equation}
V_{M_{T_0}}
=\sum_{\substack{
	\bar{\Gamma}\bar{T}\bar{M}_T\\
	\bar{\Gamma}\bar{T}'\bar{M}_T'
	}}
\me{\bar{\Gamma}'\bar{T}'\bar{M}_T'}{V_{M_{T_0}}}{\bar{\Gamma}\bar{T}\bar{M}_T}
\left[\ket{\bar{\Gamma}'\bar{T}'\bar{M}_T'}\bra{\bar{\Gamma}\bar{T}\bar{M}_T}\right]. 
\label{su2t_dyad_expansion}
\end{equation}
We then want to couple the bra and ket of the dyad to yield an operator of definite isospin tensor character $T_0$.  
However, before we can do that, we must first take into account that the bra  does not transform as a member of a standard basis for the dual space of an $\grpsu{2}$ irrep~\cite{Hall2015, Wybourne1974}.  
It can, however, be related to the covariant adjoint of a standard basis ket, which does have the desired transformation properties, by the conjugation relation
\begin{equation}
\bra{\bar{\Gamma}\bar{T}\bar{M}_T} = (-)T^{\bar{T}+\bar{M}_T}\bra{\widetilde{\bar{\Gamma}\bar{T}\,-\bar{M}_T}}.
\label{eq:su2_conj_rel}
\end{equation}
The tilde on the bra on the right-hand side of the equation indicates this bra is the covariant adjoint. 
After substituting the relationship \eqref{eq:su2_conj_rel} between the standard basis state and adjoint state into \eqref{su2t_dyad_expansion}, we can then couple the bra and ket of the dyad using standard coupling coefficients.  The expansion of the operator in terms of coupled dyads is given by 
\begin{equation}
V_{M_{T_0}}
=\sum_{\substack{
	\bar{\Gamma}\bar{T}\bar{M}_T\\
	\bar{\Gamma}\bar{T}'\bar{M}_T'
	}}
\me{\bar{\Gamma}'\bar{T}'\bar{M}_T'}{V_{M_{T_0}}}{\bar{\Gamma}\bar{T}\bar{M}_T}
\sum_{T_0}(-)^{\bar{T}+\bar{M}_T}(\bar{T}'\bar{M}_T';\bar{T}\,-\bar{M}_T|T_0M_{T_0})
\left[\ket{\bar{\Gamma}'\bar{T}'}\bra{\widetilde{\bar{\Gamma}\bar{T}}}\right]^{T_0}_{M_{T_0}}.
\label{su2t_coupled_dyad_expansion}
\end{equation}

The expression above for the operator in terms of coupled dyads can also be used to obtain the relation between $\grpsuT$ unit tensors and the coupled dyad.  If the operator in \eqref{su2t_coupled_dyad_expansion} is a component of an $\grpsuT$ tensor 
$V_{M_{T_0}}=V_{M_{T_0}}^{T_0}$, then the above expression can be simplified by first using the Wigner-Eckart theorem to factor the matrix element into a coupling coefficient and an RME:
\begin{multline}
V_{M_{T_0}}^{T_0}
=\sum_{\substack{
	\bar{\Gamma}\bar{T}\bar{M}_T\\
	\bar{\Gamma}\bar{T}'\bar{M}_T'
	}}
(\bar{T}\bar{M}_T; T_0M_{T_0}|\bar{T}'\bar{M}_T')
\rme{\bar{\Gamma}'\bar{T}'}{V^{T_0}}{\bar{\Gamma}\bar{T}}\sum_{T_0'}
(-)^{\bar{T}+\bar{M}_T}
(\bar{T}'\bar{M}_T';\bar{T}\,-\bar{M}_T|T_0'M_{T_0})
\left[\ket{\bar{\Gamma}'\bar{T}'}\bra{\widetilde{\bar{\Gamma}\bar{T}}}\right]^{T_0'}_{M_{T_0}}.
\label{su2t_tensor_coupled_dyad_expansion}
\end{multline}
By using interchange identities and orthogonality relations~\cite{Varshalovich1988} for the coupling coefficients, we can simplify the expression to 
\begin{equation}
V_{M_{T_0}}^{T_0}
=\sum_{\substack{
	\bar{\Gamma}\bar{T}\\
	\bar{\Gamma}\bar{T}'
	}}
\rme{\bar{\Gamma}'\bar{T}'}{V^{T_0}}{\bar{\Gamma}\bar{T}}
\frac{\hat{\bar{T}}'}{\hat{T}_0}
\left[\ket{\bar{\Gamma}'\bar{T}'}\bra{\widetilde{\bar{\Gamma}\bar{T}}}\right]^{T_0}_{M_{T_0}}.
\label{su2t_tensor_coupled_dyad_expansion2}
\end{equation}
In the special case that $V_{M_{T_0}}^{T_0}$ is an $\grpsuT$ unit tensor, so that the RME is given by Kronecker $\delta$ symbols as in~\eqref{eq:su2t_unit_tensor_defn}, we have 
\begin{equation}
\scrU^{T_0}_{M_{T_0}}(\bar{\Gamma}'\bar{T}',\bar{\Gamma} \bar{T})
=
\frac{\hat{\bar{T}}'}{\hat{T}_0}
\left[\ket{\bar{\Gamma}'\bar{T}'}\bra{\widetilde{\bar{\Gamma}\bar{T}}}\right]^{T_0}_{M_{T_0}}.
\label{su2t_unit_tensor_coupled_dyad_expansion}
\end{equation}
Combining \eqref{su2t_unit_tensor_coupled_dyad_expansion} with \eqref{su2t_tensor_coupled_dyad_expansion2}, we obtain an expression for an  $\grpsuT$ tensor expanded in terms of $\grpsuT$ unit tensors:
\begin{equation}
V_{M_{T_0}}^{T_0}
=\sum_{\substack{
	\bar{\Gamma}\bar{T}\\
	\bar{\Gamma}\bar{T}'
	}}
\rme{\bar{\Gamma}'\bar{T}'}{V^{T_0}}{\bar{\Gamma}\bar{T}}
\scrU^{T_0}_{M_{T_0}}(\bar{\Gamma}'\bar{T}',\bar{\Gamma} \bar{T}).
\label{su2t_tensor_unit_expansion}
\end{equation}

Finally, by combining  \eqref{su2t_unit_tensor_coupled_dyad_expansion} with \eqref{su2t_coupled_dyad_expansion}, we obtain the desired expression for an operator which has definite $M_{T_0}$  in terms of $\grpsuT$ unit tensors (or, more specifically, their $M_{T_0}$ components):
\begin{equation}
V_{M_{T_0}}
=
\sum_{T_0}
\sum_{\substack{
	\bar{\Gamma}\bar{T}\bar{M}_T\\
	\bar{\Gamma}\bar{T}'\bar{M}_T'
	}}
\me{\bar{\Gamma}'\bar{T}'\bar{M}_T'}{V_{M_{T_0}}}{\bar{\Gamma}\bar{T}\bar{M}_T}
(-)^{\bar{T}+\bar{M}_T}
(\bar{T}'\bar{M}_T';\bar{T}\,-\bar{M}_T|T_0M_{T_0})
\frac{\hat{T}_0}{\hat{T}'}
\scrU^{T_0}_{M_{T_0}}(\bar{\Gamma}'\bar{T}',\bar{\Gamma}\bar{T}).
\label{mtop_unit_tensor_expansion}
\end{equation}

\subsection{Decomposing operators in terms of $\grpsu{3}\times\grpsuS$ unit tensors}
We now turn to the more complicated case, which is decomposing a given operator in terms of $\grpsu{3}\times\grpsuS$ relative unit tensors. 
Our goal now is to expand the operator of interest in terms of $\grpsu{3}\times\grpsuS$ unit tensors defined with respect to the relative harmonic oscillator basis with states $\ket{NL\Sigma SJM_J}$.  
We take the operator to already be an angular momentum tensor, with tensor character $J_0$.
Consequently, we start from the $\grpsuJ$ RMEs (rather than simply matrix elements) of this operator $V_{J_0}$ in the relative harmonic oscillator basis.

The derivation of the expansion \eqref{eq:suJ-su3suS-unit-tensors} of the operator in terms of $\grpsu{3}\times\grpsuS$ tensor components follows a similar process as above.  
We start from the expression for $V_{J_0}$ in terms of $\grpsuJ$ coupled dyads~\eqref{su2t_tensor_coupled_dyad_expansion2}, and then recouple the bra and ket of the dyad to obtain $\grpsu{3}\times\grpsuS$ coupled dyads.  
Then, we then relate the $\grpsu{3}\times\grpsuS$ coupled dyads to $\grpsu{3}\times\grpsuS$ unit tensors to obtain the desired relation.  

We begin by expressing the operator in terms of $\grpsuJ$ coupled dyads, as in \eqref{su2t_tensor_coupled_dyad_expansion2}, as 
\begin{equation}
V_{J_0 M_0}=\sum_{\substack{
\bar{N}'\bar{L}'\bar{\Sigma}'\bar{S}'\bar{J}'\\
\bar{N}\bar{L}\bar{\Sigma}\bar{S}\bar{J}
}}
\rme{\bar{N}'\bar{L}'\bar{\Sigma}'\bar{S}'\bar{J}'}{V_{J_0}}{\bar{N}\bar{L}\bar{\Sigma}\bar{S}\bar{J}}\frac{\hat{\bar{J}}'}{\hat{J}_0}
\left[\ket{\bar{N}'\bar{L}'\bar{\Sigma}'\bar{S}'\bar{J}'}\bra{\widetilde{\bar{N}\bar{L}\bar{\Sigma}\bar{S}\bar{J}}}\right]_{J_0 M_0}.
\label{su2j_tensor_coupled_dyad_expansion}
\end{equation}
To relate the $\grpsuJ$ coupled dyads to $\grpsu{3}\times\grpsuS$ coupled dyads,  we first need to ensure that the bra of the dyad transforms as a member of a standard basis for the $\grpsu{3}\times\grpsuS$ group, and not just as a member of a standard basis for $\grpsuJ$. 
For the relative harmonic oscillator basis the relation [corresponding to \eqref{su3_tensor_conj}] between the bras obtained as adjoints to the standard basis kets and the standard basis for the adjoint irrep is given by 
\begin{equation}
\bra{\widetilde{(0,N)L\Sigma SJ-M}}=\tilde{\Phi}[(N,0)LSJM]\bra{(N,0)L\Sigma SJM},
\end{equation}
where the conjugation phase, as defined in footnote~\ref{fn:sp_ho_conjugation_phase_pn}, is given by $\tilde{\Phi}[(N,0)LSJM]= (-)^N(-)^{J+M}$.
The $\grpsuJ$ coupled dyads are then related to the $\grpsu{3}\times\grpsuS$ coupled dyads by 
\begin{multline}
\left[\ket{\bar{N}'\bar{L}'\bar{\Sigma}'\bar{S}'\bar{J}'}\bra{\widetilde{\bar{N}\bar{L}\bar{\Sigma}\bar{S}\bar{J}}}\right]_{J_0}
 = \sum_{\substack{(\lambda_0\mu_0)\kappa_0\\L_0S_0}}
(-)^{\bar{N}}
\rcc{(\bar{N}',0)}{\bar{L}'}{(0,\bar{N})}{\bar{L}}{(\lambda_0\mu_0)}{\kappa_0L_0}
\uninej{\bar{L}'}{\bar{S}'}{\bar{J}'}{\bar{L}}{\bar{S}}{\bar{J}}{L_0}{S_0}{J_0}
\left[\ket{\bar{N}'\bar{\Sigma}'\bar{S}'}\bra{\widetilde{\bar{N}\bar{\Sigma}\bar{S}}}\right]^{(\lambda_0\mu_0)S_0}_{\kappa_0L_0J_0}.
\label{eq:suj-dyads-su3sus-coupled-dyads}
\end{multline}
Substituting \eqref{eq:suj-dyads-su3sus-coupled-dyads} into \eqref{su2j_tensor_coupled_dyad_expansion}, we obtain an expression for $V_{J_0}$ in terms of $\grpsu{3}\times\grpsuS$ coupled dyads:
\begin{multline}
V_{J_0}=\sum_{\substack{
\bar{N}'\bar{L}'\bar{\Sigma}'\bar{S}'\bar{J}'\\
\bar{N}\bar{L}\bar{\Sigma}\bar{S}\bar{J}
}}
\rme{\bar{N}'\bar{L}'\bar{\Sigma}'\bar{S}'\bar{J}'}{V_{J_0}}{\bar{N}\bar{L}\bar{\Sigma}\bar{S}\bar{J}}\frac{\hat{\bar{J}}'}{\hat{J}_0}\\
\times 
 \sum_{\substack{(\lambda_0\mu_0)\kappa_0\\L_0S_0}}
(-)^{\bar{N}}
\rcc{(\bar{N}',0)}{\bar{L}'}{(0,\bar{N})}{\bar{L}}{(\lambda_0\mu_0)}{\kappa_0L_0}
\uninej{\bar{L}'}{\bar{S}'}{\bar{J}'}{\bar{L}}{\bar{S}}{\bar{J}}{L_0}{S_0}{J_0}
\left[\ket{\bar{N}'\bar{\Sigma}'\bar{S}'}\bra{\widetilde{\bar{N}\bar{\Sigma}\bar{S}}}\right]^{(\lambda_0\mu_0)S_0}_{\kappa_0L_0J_0}.
\label{eq:suJ-su3suS-dyads}
\end{multline}

In the special case that $V_{J_0}$ is already a component of an $\grpsu{3}\times\grpsuS$ tensor, i.e., $V_{J_0}=V^{(\lambda_0\mu_0)S_0}_{\kappa_0L_0J_0}$, then the above expression can be simplified by using the Wigner-Eckart theorem to factor the $J$-reduced matrix element as the product of a reduced coupling coefficient and an $\grpsu{3}\times\grpsuS$ RME as 
\begin{equation}
\rme{\bar{N}'\bar{L}'\bar{\Sigma}'\bar{S}'\bar{J}'}{V^{(\lambda_0\mu_0)S_0}_{\kappa_0L_0J_0}}{\bar{N}\bar{L}\bar{\Sigma}\bar{S}\bar{J}}
=\rcc{\bar{N},0)}{\bar{L}}{(\lambda_0\mu_0)}{\kappa_0L_0}{(\bar{N}',0)}{\bar{L}'}
\uninej{\bar{L}}{\bar{S}}{\bar{J}}{L_0}{S_0}{J_0}{\bar{L}'}{\bar{S}'}{\bar{J}'}
\rme{\bar{N}'\bar{\Sigma}'\bar{S}'}{V^{(\lambda_0\mu_0)S_0}}{\bar{N}\bar{\Sigma}\bar{S}}. 
\label{eq:we_rel_u3s}
\end{equation} 
Substituting \eqref{eq:we_rel_u3s} into \eqref{eq:suJ-su3suS-dyads}, and applying various coupling coefficient identities and orthogonality relations~\cite{McCoy2018}, gives 
\begin{equation}
V^{(\lambda_0\mu_0)S_0}_{\kappa_0L_0J_0}
=\sum_{\substack{
\bar{N}'\bar{\Sigma}'\bar{S}'\\
\bar{N}\bar{\Sigma}\bar{S}
}}
\rme{\bar{N}'\bar{\Sigma}'\bar{S}'}{V^{(\lambda_0\mu_0)S_0}}{\bar{N}\bar{\Sigma}\bar{S}}
\frac{\hat{\bar{S}}'}{\hat{S}_0}
\left[\frac{\dim(\bar{N}',0)}{\dim(\lambda_0\mu_0)}\right]^{1/2}
\left[\ket{\bar{N}'\bar{\Sigma}'\bar{S}'}\bra{\widetilde{\bar{N}\bar{\Sigma}\bar{S}}}\right]^{(\lambda_0\mu_0)S_0}_{\kappa_0L_0J_0}.
\end{equation}
The relation between an $\grpsu{3}\times\grpsu{2}$ unit tensor and a coupled dyad is a special case of the above expression, where the RME is defined in terms of Kronecker $\delta$ symbols as in~\eqref{eq:relative_unit_tensors_u3s}, giving 
\begin{equation}
\scrU^{(\lambda_0\mu_0)S_0}_{\kappa_0L_0J_0}(\bar{N}'\bar{\Sigma}'\bar{S}',\bar{N}\bar{\Sigma}\bar{S})
=\frac{\hat{\bar{S}}'}{\hat{S}_0}
\left[\frac{\dim(\bar{N}',0)}{\dim(\lambda_0\mu_0)}\right]^{1/2}
\left[\ket{\bar{N}'\bar{\Sigma}'\bar{S}'}\bra{\widetilde{\bar{N}\bar{\Sigma}\bar{S}}}\right]^{(\lambda_0\mu_0)S_0}_{\kappa_0L_0J_0}.
\label{eq:su3xsu2_relative_unit_tensor_dyads}
\end{equation}
Finally, combining \eqref{eq:su3xsu2_relative_unit_tensor_dyads} with \eqref{eq:suJ-su3suS-dyads}, we obtain the expansion given in \eqref{eq:suJ-su3suS-unit-tensors}--\eqref{eq:suJ-su3suS-unit-tensors-coef} for $V_{J_0}$ in terms of $\grpsu{3}\times\grpsuS$ unit tensors.

\subsection{Decomposing operators in terms of $\grpsu{3}\times\grpsu{4}$ unit tensors}
One may follow a similar derivation when working in an isospin scheme, to decompose the relative operator in terms of $\grpsu{3}\times\grpsu{4}$ unit tensors~\eqref{eq:u3u4_unit_tensor_def}.
Here we assume the operator is already a  good  angular momentum and isospin tensor, i.e., a good $\grpsuJ\times\grpsuT$ tensor, and we start from $\grpsuJ\times\grpsuT$ RMEs of this operator $V_{J_0 T_0}$.
Note that, if we have an operator which is not a good isospin tensor, we may first use \eqref{mtop_unit_tensor_expansion} to express it in terms of $\grpsuT$ unit tensors, and then proceed as below.
We start with the operator expressed in terms of $\grpsuJ\times\grpsuT$-coupled dyads
\begin{equation}
V_{J_0 T_0}=\sum_{\substack{
{}[{\bar{\bf f}}']\bar{N}'\bar{L}'\bar{S}'\bar{J}'\bar{T}'\\
{}[{\bar{\bf f}}]\bar{N}\bar{L}\bar{S}\bar{J}\bar{T}
}}
\rme{\bar{N}'\bar{L}'[{\bar{\bf f}}']\bar{S}'\bar{T}'\bar{J}'}{V_{J_0 T_0}}{\bar{N}\bar{L}[{\bar{\bf f}}]\bar{S}\bar{T}\bar{J}}
\frac{\hat{\bar{J}}'}{\hat{J}_0}
\left[\ket{\bar{N}'\bar{L}'[{\bar{\bf f}}']\bar{S}'\bar{T}'\bar{J}'}\bra{\widetilde{\bar{N}\bar{L}[{\bar{\bf f}}]\bar{S}\bar{J}\bar{T}}}\right]_{J_0 T_0},
\label{su2j_tensor_coupled_dyad_expansion-isospin}
\end{equation}
We then need to apply the appropriate conjugation relation to relate the standard basis bra to the covariant adjoint of a standard basis ket for $\grpu{3}\times\grpu{4}$.  
For $\grpsu{3}\times\grpsu{4}$, the conjugation relation~\cite{jmp-10-1969-1571-Hecht}  is given by 
\begin{equation}
\bra{\widetilde{\bar{N}\bar{L}[{\tilde{\bar{\bf f}}}]\bar{S}\bar{T}\,-\bar{M}_T\bar{J}\,-\bar{M}}}
=(-)^{\phi([\bar{\bf f}])+\bar{N}+\bar{J}+\bar{M}+\bar{T}+\bar{M}_t}
\bra{[\bar{\bf f}]\bar{N}\bar{L}\bar{S}\bar{J}\bar{M}\bar{T}\bar{M}_T}.
\end{equation}
After applying the conjugation relation, we can recouple the dyads to have definite $\grpsu{3}\times\grpsu{4}$ tensor character, using the  coupling coefficient \eqref{eq:su3su4cc}, and then relate the coupled dyad to the $\grpsu{3}\times\grpsu{4}$ unit tensor defined in~\eqref{eq:u3u4_unit_tensor_def}.
The resulting expression is given by 
\begin{multline}
V_{J_0T_0}=\sum_{
\substack{
{}[{\bar{\bf f}}']\bar{N}'\bar{L}'\bar{S}'\bar{J}'\bar{T}'\\
{}[{\bar{\bf f}}]\bar{N}\bar{L}\bar{S}\bar{J}\bar{T}}
}
(-)^{\phi([\bar{\bf f}])+\bar{N}}
\rme{\bar{N}'\bar{L}'[{\bar{\bf f}}']\bar{S}'\bar{T}'\bar{J}'}{V_{J_0 T_0}}{\bar{N}\bar{L}[{\bar{\bf f}}]\bar{S}\bar{T}\bar{J}}\frac{\hat{\bar{J}}'}{\hat{J}_0}
\uninej{\bar{L}'}{\bar{S}'}{\bar{J}'}
{\bar{L}}{\bar{S}}{\bar{J}}
{L_0}{S_0}{J_0}
 \rcc{(\bar{N}',0)}{\bar{L}'}{(0,\bar{N})}{\bar{L}}{(\lambda_0\mu_0)}{\kappa_0 L_0}\\
\times
\rcc{[\bar{\bf f}]}{\bar{S}'\bar{T}'}
{[\tilde{\bar{\bf f}}]}{\bar{S}\bar{T}}
{[{\bf f}_0]}{\beta_0S_0T_0}
\check{\Phi}\big([{\bf f}_0][\bar{\bf f}][\bar{\bf f}']\big)
\cev{\Phi}\big([\bar{\bf f}'][\tilde{\bar{{\bf f}}}][{\bf f}_0]
\big)
\left[
\frac{
\dim(\lambda_0\mu_0)
\dim([{\bf f}_0])}
{
\dim(\bar{N}',0)
\dim([\bar{\bf f}'])
}\right]^{1/2}
\scrU^{(\lambda_0\mu_0)[{\bf f}_0]}_{\kappa_0L_0\beta_0S_0J_0T_0}\big(\bar{N}'[\bar{\bf f}'],\bar{N}[\bar{\bf f}]\big).
\end{multline}
Here $\check{\Phi}\big([{\bf f}_0][\bar{\bf f}][\bar{\bf f}']\big)$ and 
$\cev{\Phi}\big([\bar{\bf f}'][\tilde{\bar{{\bf f}}}][{\bf f}_0])$ are the phase factors that arise from  interchanging the coupling order of the irreps in the Kronecker product $[{\bf f}_0]\times [\bar{\bf f}]\rightarrow[\bar{\bf f}']$ and from reversing the coupling order the three $\grpu{4}$ irreps, respectively. See Ref.~\cite{McCoy2018} for more details on the definitions and properties of these phase factors for a general group and Ref.~\cite{jmp-10-1969-1571-Hecht} for specific phase factors for $\grpu{4}$.
 
\section{Coupled commutator for $\grpsu{3}$}
\label{sec:app-commutator}

\newcommand{\nocheck}{}  

The coupled commutator~\cite{French1966,npa-554-1993-61-Chen} of spherical tensor operators provides a convenient mechanism for reordering operators within an angular momentum coupled product, circumventing the tedious process of first uncoupling the product, which introduces summations over angular momentum projection quantum numbers and a proliferation of coupling coefficients, carrying out the commutations, and then recoupling the results.
The idea of the coupled commutator may be extended to coupled products of tensors with respect to $\grpsu{3}$ or $\grpu{3}$.
In the case of multiplicity-free products, as already noted by Chen \textit{et al.}~\cite{npa-554-1993-61-Chen}, the extension is trivial.
In this appendix, we set out more general relations allowing for outer multiplicities in the couplings.
These relations generalize in a straightforward fashion to tensors with respect to product groups $\grpsu{3}\times\grpsu{2}$, $\grpsu{3}\times\grpsu{2}\times\grpsu{2}$, \textit{etc.}

The ordinary angular momentum  or $\grpsu{2}$-coupled commutator, of two spherical tensor operators $A^{S_a}$ and $B^{S_b}$, is defined by
\begin{equation}
  \label{eqn:coupled-commutator-su2-defn}
        [A^{S_a},B^{S_b}]^{S_c}_{M_{S c}}
        =\sum_{M_{S a} M_{S b}}\tcg{S_a}{M_{S a}}{S_b}{M_{S b}}{S_c}{M_{S c}} [A^{S_a}_{M_{S a}},B^{S_b}_{M_{S b}}].
\end{equation}
As a linear combination (in this case, simply a sum or difference) of
$\grpsu{2}$-coupled products,
\begin{equation}
  \label{eqn:coupled-commutator-su2-difference}
  [A^{S_a},B^{S_b}]^{S_c}= (A^{S_a}\times B^{S_b})^{S_c} - (-)^{S_c-S_a-S_b}  (B^{S_b}\times A^{S_a})^{S_c},
\end{equation}
the resulting operator is itself manifestly an $\grpsu{2}$ tensor.
The phase factor in the second term arises from transposition of coupling order.
The coupled commutator obeys the symmetry relation
\begin{equation}
  \label{eqn:coupled-commutator-su2-symmetry}
    [A^{S_a},B^{S_b}]^{S_c} = - (-)^{S_c-S_a-S_b} [B^{S_b}, A^{S_a}]^{S_c}.
\end{equation}

The coupled commutator of two $\grpu{3}$ tensor operators $A^{\omega_a}$ and $B^{\omega_b}$, is defined, analogously, by
\begin{equation}
  \label{eqn:coupled-commutator-su3-defn}
        [A^{\omega_a},B^{\omega_b}]^{\rho\omega_c}_{\kappa_c L_c M_c}
        =\sum_{\substack{\kappa_a L_a M_a \\ \kappa_b L_b M_b}}
         \ccg{\omega_a}{\kappa_a L_a}{M_a}{\omega_b}{\kappa_b L_b}{M_b}{\omega_c}{\kappa_c L_c}{M_c}_\rho
                    [A^{\omega_a}_{\kappa_a L_a M_a},B^{\omega_b}_{\kappa_b L_b M_b}],
\end{equation}
where we have used the $\kappa L M$ labels of the $\grpsu{3}\supset\grpso{3}$ subgroup chain~(\ref{su3_chain}) to label the basis for a $\grpu{3}$ irrep, but the coupled commutator could equivalently be defined in terms of a summation over any standard basis for the irrep, \textit{e.g.}, one reducing the canonical $\grpsu{3}\supset\grpsu{2}\times\grpu{1}$ subgroup chain.
We have used $\grpu{3}$ labels $\omega$ for the irreps in~(\ref{eqn:coupled-commutator-su3-defn}), but the corresponding expressions for $\grpsu{3}$ tensors are obtained by substitution of $\grpsu{3}$ labels $(\lambda,\mu)$.
The coupled commutator may be written as a linear combination of $\grpsu{3}$-coupled products as
\begin{equation}
  \label{eqn:coupled-commutator-su3-difference}
        [A^{\omega_a},B^{\omega_b}]^{\rho\omega_c}
        =
          (A^{\omega_a}\times B^{\omega_b})^{\rho\omega_c}
        -\sum_{\rho'} \nocheck{\Phi}_{\rho\rho'}[\omega_a\omega_b\omega_c] (B^{\omega_b} \times A^{\omega_a})^{\rho'\omega_c},
\end{equation}
where the transposition of coupling order in the second term now introduces a phase matrix $\nocheck{\Phi}$ and summation over an outer multiplicity index.
The coupled commutator obeys the symmetry relation
\begin{equation}
  \label{eqn:coupled-commutator-su3-symmetry}
  [A^{\omega_a},B^{\omega_b}]^{\rho\omega_c} = -\sum_{\rho'} \nocheck{\Phi}_{\rho\rho'}[\omega_a\omega_b\omega_c] [B^{\omega_b},A^{\omega_a}]^{\rho'\omega_c}.
\end{equation}

Ordinary commutators obey the well-known product rules
\begin{math}
  [AB,C]=A[B,C]+[A,C]B
\end{math}
and
\begin{math}
  [A,BC]=B[A,C]+[A,B]C.
\end{math}
The corresponding product rules for the $\grpsu{2}$-coupled commutator~\cite{npa-554-1993-61-Chen} (see also Appendix of Ref.~\cite{jpa-44-2011-075303-Caprio}) involve recoupling coefficients for products of three angular momenta, \textit{i.e.}, of the $6$-$j$ type.
The expressions take on their simplest form, and most directly generalize to $\grpu{3}$, when written in terms of the unitary recoupling coefficients, \textit{i.e.}, transformation brackets between $(12)3$ and $1(23)$ couplings or between $(12)3$ and $(13)2$ couplings. The product rules may then be written as
\begin{equation}
  \label{eqn:coupled-commutator-su2-product}
  \begin{gathered}
    \begin{multlined}
      [(A^{S_a}\times B^{S_b})^{S_e},C^{S_c}]^{S_d}
      =
      \sum_{S_f} U(S_aS_bS_dS_c;S_eS_f) (A^{S_a}\times[B^{S_b},C^{S_c}]^{S_f})^{S_d}
      \\
      +\sum_{S_f} Z(S_bS_aS_dS_c;S_eS_f) ([A^{S_a},C^{S_c}]^{S_f}\times B^{S_b})^{S_d}
    \end{multlined}
    \\
    \begin{multlined}
      [A^{S_a}, (B^{S_b}\times C^{S_c})^{S_e}]^{S_d}
      =
      \sum_{S_f} Z(S_cS_aS_dS_b;S_fS_e) (B^{S_b}\times[A^{S_a},C^{S_c}]^{S_f})^{S_d}
      \\
      +\sum_{S_f} U(S_aS_bS_dS_c;S_fS_e) ([A^{S_a},B^{S_b}]^{S_f}\times C^{S_c})^{S_d}.
    \end{multlined}
  \end{gathered}
\end{equation}
The unitary recoupling coefficients for $(12)3$--$1(23)$ and $(12)3$--$(13)2$ are denoted here by $U$ and $Z$ coefficients, defined as\footnote{Note that the $U$ coefficient here is that of Jahn~\cite{prsla-205-1951-192-Jahn}, to which the $\grpsu{3}$ $U$ coefficient of Hecht~\cite{np-62-1965-1-Hecht} is defined in direct analogy.
However, the $Z$ coefficient is defined here for parallelism with the $\grpsu{3}$ $Z$ coefficient of Millener~\cite{jmp-19-1978-1513-Millener}, and is therefore \textit{not} to be conflated with the $Z$ coefficient of Biedenharn, Blatt, and Rose~\cite{rmp-24-1952-249-Biedenharn}.
See also (6.2.13)--(6.2.15) of Edmonds~\cite{Edmonds1960}.}
\begin{equation}
  \label{eqn:u-z-6j}
  \begin{aligned}
    U(J_1J_2JJ_3;J_{12}J_{23})
&\equiv\toverlap{\wick{\c1J_1\c1J_2J_3;J_{12}J}}{\wick{J_1\c1J_2\c1J_3;J_{23}J}}
    \\
    &=(-)^{J_1+J_2+J_3+J} \hat{J}_{12}\hat{J}_{23}\sixj{J_1}{J_2}{J_{12}}{J_3}{J}{J_{23}}
    \\
    Z(J_2J_1JJ_3;J_{12}J_{13})
&\equiv\toverlap{\wick{\c1J_1\c1J_2J_3;J_{12}J}}{\wick{\c1J_1J_2\c1J_3;J_{13}J}}
    \\
    &=(-)^{J_1+J_2+J_3+J} \hat{J}_{12}\hat{J}_{23}\sixj{J_1}{J_2}{J_{12}}{J_3}{J}{J_{23}},
  \end{aligned}
\end{equation}
where the ``contraction'' lines simply serve to indicate which angular momenta are subject to the intermediate coupling in each bra and ket.

The corresponding product rule for $\grpu{3}$ commutators, in the presence of outer multiplicities, may be written as
\begin{equation}
  \label{eqn:coupled-commutator-su3-product}
  \begin{gathered}
    \begin{multlined}
      [(A^{\omega_a}\times B^{\omega_b})^{\rho_e\omega_e},C^{\omega_c}]^{\rho_d\omega_d}
      =
      \sum_{\rho'_f\omega_f\rho'_d} U(\omega_a\omega_b\omega_d\omega_c;\omega_e\rho_e\rho_d\omega_f\rho'_f\rho'_d) (A^{\omega_a}\times[B^{\omega_b},C^{\omega_c}]^{\rho'_f\omega_f})^{\rho'_d\omega_d}
      \\
      +\sum_{\rho'_f\omega_f\rho'_d} Z(\omega_b\omega_a\omega_d\omega_c;\omega_e\rho_e\rho_d\omega_f\rho'_f\rho'_d) ([A^{\omega_a},C^{\omega_c}]^{\rho'_f\omega_f}\times B^{\omega_b})^{\rho'_d\omega_d}
    \end{multlined}
    \\
    \begin{multlined}
      [A^{\omega_a}, (B^{\omega_b}\times C^{\omega_c})^{\rho_e\omega_e}]^{\rho_d\omega_d}
      =
      \sum_{\rho'_f\omega_f\rho'_d} Z(\omega_c\omega_a\omega_d\omega_b;\omega_f\rho'_f\rho'_d\omega_e\rho_e\rho_d) (B^{\omega_b}\times[A^{\omega_a},C^{\omega_c}]^{\rho'_f\omega_f})^{\rho'_d\omega_d}
      \\
      +\sum_{\rho'_f\omega_f\rho'_d} U(\omega_a\omega_b\omega_d\omega_c;\omega_f\rho'_f\rho'_d\omega_e\rho_e\rho_d) ([A^{\omega_a},B^{\omega_b}]^{\rho'_f\omega_f}\times C^{\omega_c})^{\rho'_d\omega_d},
    \end{multlined}
  \end{gathered}
\end{equation}
in terms of the $U$ and $Z$ unitary recoupling coefficients for $\grpsu{3}$~\cite{np-62-1965-1-Hecht,jmp-14-1973-1904-Draayer,jmp-19-1978-1513-Millener,jmp-39-1998-5123-Escher}.
Alternatively, each of these expressions may be expressed entirely in terms of $U$ coefficients or entirely in terms of $Z$ coefficients, but eliminating either of these coefficients in favor of the other introduces additional phase matrices and summations over multiplicity indices.

The extension of these relations to product groups $\grpsu{3}\times\grpsu{2}$, $\grpsu{3}\times\grpsu{2}\times\grpsu{2}$, \textit{etc.}, is straightforward, since the couplings are carried out independently.
Thus, \textit{e.g.}, for $\grpsu{3}\times\grpsu{2}$, combining~(\ref{eqn:coupled-commutator-su2-defn}) and~(\ref{eqn:coupled-commutator-su3-defn}) gives the definition
\begin{multline}
  \label{eqn:coupled-commutator-su3-su2-defn}
        [A^{\omega_a S_a},B^{\omega_b S_b}]^{\rho\omega_c S_c}_{\kappa_c L_c M_c M_{S c}}
        =\sum_{\substack{\kappa_a L_a M_a M_{S a}\\ \kappa_b L_b M_b M_{S b}}}
        \ccg{\omega_a}{\kappa_a L_a}{M_a}{\omega_b}{\kappa_b L_b}{M_b}{\omega_c}{\kappa_c L_c}{M_c}_\rho \tcg{S_a}{M_{S a}}{S_b}{M_{S b}}{S_c}{M_{S c}}
        \\\times
            [A^{\omega_a S_a}_{\kappa_a L_a M_a M_{S a}},B^{\omega_b S_b}_{\kappa_b L_b M_b M_{S b}}].
\end{multline}
The extensions of subsequent relations follow by combining the summations and prefactors arising from each group, \textit{e.g.}, (\ref{eqn:coupled-commutator-su2-difference}) and (\ref{eqn:coupled-commutator-su3-difference}) combine to give
\begin{equation}
  \label{eqn:coupled-commutator-su3-su2-difference}
        [A^{\omega_a S_a},B^{\omega_b S_b}]^{\rho\omega_c S_c}
        =
          (A^{\omega_a S_a}\times B^{\omega_b S_b})^{\rho\omega_c S_c}
        -\sum_{\rho'} \nocheck{\Phi}_{\rho\rho'}[\omega_a\omega_b\omega_c] (-)^{S_c-S_a-S_b}  (B^{\omega_b S_b} \times A^{\omega_a S_a})^{\rho'\omega_c S_c}.
\end{equation}
 \section{Coupled commutator of unit tensors
\label{sec:app-coupled-commutator-unit-tensors}
}

In this section, we derive the coupled commutator of two relative unit tensors, which is used in the derivation of the recurrence in Section~\ref{section:recurrence}.  
This coupled commutator can be expanded again in terms of relative unit tensors.   

To evaluate the coupled commutator, we first rewrite the coupled product of two unit tensors as a single relative unit tensor.  
To obtain this expression, we use the relation between unit tensors and coupled dyads~\eqref{eq:su3xsu2_relative_unit_tensor_dyads} to rewrite the coupled product of two unit tensors in terms of coupled dyads as 
\begin{equation}
\begin{split}
&\big[
\scrU^{(\lambda_1\mu_1)S_1}(\bar{N}_1'\bar{S}_1',\bar{N}_1\bar{S}_1)
\times\scrU^{(\lambda_2\mu_2)S_2}(\bar{N}_2'\bar{S}_2',\bar{N}_2\bar{S}_2)
\big]^{\rho_0(\lambda_0\mu_0)S_0}\\
&=
\frac{\hat{\bar{S}}_1\hat{\bar{S}}_2}{\hat{S}_1\hat{S}_2}
\left[\frac
	{\dim(\bar{N}_1',0)\dim(\bar{N}_2',0)}
	{\dim(\lambda_1\mu_1)\dim(\lambda_2\mu_2)}
\right]^{\frac12}
\bigg[
	\Big[
	\ket{(\bar{N}_1',0)\bar{S}_1'}\bra{\widetilde{(0,\bar{N}_1)\bar{S}_1}}
	\Big]^{(\lambda_1\mu_1)S_1}
\times
	\Big[
	\ket{(\bar{N}_2',0)\bar{S}_2'}\bra{\widetilde{(0,\bar{N}_2)\bar{S}_2}}
	\Big]^{(\lambda_2\mu_2)S_2}
\bigg]^{\rho_0(\lambda_0\mu_0)S_0},
\label{coupled_unit_tensors1}
\end{split}
\end{equation}
If we recouple so that the inner bra and ket first couple to each other, then the coupled product of these two states is just a coupled overlap, or  inner product, which reduces to
\begin{equation}
\big[\bra{\widetilde{(0,\bar{N}_1)\bar{S}_1}}\times \ket{(\bar{N}_2',0)\bar{S}_2'}\big]^{(\bar{\lambda}'\bar{\mu}')\bar{S}'}=[\dim(\bar{N}_1,0)]^{1/2}\hat{\bar{S}}_1\delta_{\bar{N}_2',\bar{N}_1}\delta_{\bar{S}_2'\bar{S}_1}\delta_{(\bar{\lambda}\bar{\mu}),(0,0)}\delta_{\bar{S}',0}.
\label{eq:coupled_product_overlap}
\end{equation}

Then, the coupled product of two coupled dyads reduces to 
\begin{equation}
\begin{split}
\bigg[
	\Big[
	\ket{(\bar{N}_1',0)\bar{S}_1'}&\bra{\widetilde{(0,\bar{N}_1)\bar{S}_1}}
	\Big]^{(\lambda_1\mu_1)S_1}
\times
	\Big[
	\ket{(\bar{N}_2',))\bar{S}_2'}\bra{\widetilde{(0,\bar{N}_2)\bar{S}_2}}
	\Big]^{(\lambda_2\mu_2)S_2}
\bigg]^{\rho_0(\lambda_0\mu_0)S_0}\\
&=
(-)^{\bar{N}_1+\bar{N}_2+\lambda_2+\mu_2}
(-)^{\bar{S}_1+\bar{S}_2+S_2}
\left[
\frac{\dim(\lambda_2\mu_2)}
	{
	\dim(0,\bar{N}_2)}
\right]^{\frac12} \frac{
	\hat{S}_2}
	{
	\hat{\bar{S}}_2
	}\\
&\qquad\times U\Big(
	(\bar{N}_1',0)(0,\bar{N}_1)(\lambda_0\mu_0)(\lambda_2\mu_2);
	(\lambda_1\mu_1)\_\,\rho_0; 
	(0,\bar{N}_2)\_\,\_
\Big)
U\big(\bar{S}_1'\bar{S}_1S_0S_2;S_1;\bar{S}_2\big)\\
&\qquad\qquad\times
\Bigg[
\ket{(\bar{N}_1',0)\bar{S}_1'}
\bra{\widetilde{(0,\bar{N}_2)\bar{S}_2}}
\Bigg]^{(\lambda_0\mu_0)S_0
}.
\end{split}
\end{equation}
Finally, we invert the relation between unit tensors and coupled dyads \eqref{eq:su3xsu2_relative_unit_tensor_dyads} to rewrite the coupled dyads in terms of unit tensors and obtain an expression for the product of two $\grpsu{3}\times\grpsuS$ relative unit tenors as 
\begin{equation}
\begin{split}
\big[
&\scrU^{(\lambda_1\mu_1)S_1}(\bar{N}_1'\bar{S}_1',\bar{N}_1\bar{S}_1)
\times\scrU^{(\lambda_2\mu_2)S_2}(\bar{N}_2'\bar{S}_2',\bar{N}_2\bar{S}_2)
\big]^{\rho_0(\lambda_0\mu_0)S_0}\\
&=
(-)^{\bar{N}_1+\bar{N}_2+\lambda_2+\mu_2}
(-)^{\bar{S}_1+\bar{S}_2+S_2}
\frac{\hat{\bar{S}}_1\hat{S}_0}{\hat{S}_1\hat{\bar{S}}_2}
\left[
\frac
{\dim(\bar{N}_1,0)\dim(\lambda_0\mu_0)}
{\dim(\lambda_1\mu_1)\dim(0,\bar{N}_2)}
\right]^{\frac12}\\
&\qquad\times U\Big(
	(\bar{N}_1',0)(0,\bar{N}_1)(\lambda_0\mu_0)(\lambda_2\mu_2);
	(\lambda_1\mu_1)\_\,\rho_0; 
	(0\bar{N}_2)\_\,\_
\Big)
U\big(\bar{S}_1'\bar{S}_1S_0S_2;S_1;\bar{S}_2\big)\,
\scrU^{(\lambda_0\mu_0)S_0}(\bar{N}_1'\bar{S}_1',\bar{N}_2\bar{S}_2).
\end{split}
\label{eq:coupled_unit_tensor_expansion}
\end{equation}

The expansion of the coupled commutator of two relative unit tensors in terms  of unit tensors immediately follows.  From the expression~\eqref{eqn:coupled-commutator-su3-su2-difference} for an $\grpsu{3}\times\grpsu{2}$ coupled commutator, we obtain
\begin{equation}
\begin{split}
\big[
	\scrU^{(\lambda_1\mu_1)S_1}&(\bar{N}_1'\bar{S}_1',\bar{N}_1\bar{S}_1),\,
	\scrU^{(\lambda_2\mu_2)S_2}(\bar{N}_2'\bar{S}_2',\bar{N}_2\bar{S}_2)
\big]^{(\lambda_0\mu_0)S_0}\\
&=
\big[
	\scrU^{(\lambda_1\mu_1)S_1}(\bar{N}_1'\bar{S}_1',\bar{N}_1\bar{S}_1)
	\times
	\scrU^{(\lambda_2\mu_2)S_2}(\bar{N}_2'\bar{S}_2',\bar{N}_2\bar{S}_2)
\big]^{\rho_0(\lambda_0\mu_0)S_0}\\
&-
\quad\sum_{\rho_0\rho_0'}[(\lambda_1\mu_1)(\lambda_2\mu_2)(\lambda_0\mu_0)]
(-)^{S_0-S_1-S_2}\\
&\qquad\times
\big[
	\scrU^{(\lambda_2\mu_2)S_2}(\bar{N}_2'\bar{S}_2',\bar{N}_2\bar{S}_2)
	\times
	\scrU^{(\lambda_1\mu_1)S_1}(\bar{N}_1'\bar{S}_1',\bar{N}_1\bar{S}_1)	
\big]^{\rho_0'(\lambda_0\mu_0)S_0}.
\label{eq:unit_coupled_commutator}
\end{split}
\end{equation}
Substituting the expression \eqref{eq:coupled_unit_tensor_expansion} for the coupled product of two relative unit tensors  into \eqref{eq:unit_coupled_commutator},  we obtain
\begin{equation}
\begin{split}
\big[&
	\scrU^{(\lambda_1\mu_1)S_1}(\bar{N}_1'\bar{S}_1',\bar{N}_1\bar{S}_1),\,
	\scrU^{(\lambda_2\mu_2)S_2}(\bar{N}_2'\bar{S}_2',\bar{N}_2\bar{S}_2)
\big]^{(\lambda_0\mu_0)S_0}\\
&=
(-)^{\bar{N}_1+\bar{N}_2+\lambda_2+\mu_2}
(-)^{\bar{S}_1+\bar{S}_2+S_2}
\frac{\hat{\bar{S}}_1\hat{S}_0}{\hat{S}_1\hat{\bar{S}}_2}
\left[
\frac
{\dim(\bar{N}_1,0)\dim(\lambda_0\mu_0)}
{\dim(\lambda_1\mu_1)\dim(0,\bar{N}_2)}
\right]^{\frac12}\\
&\qquad\times U\Big(
	(\bar{N}_1',0)(0,\bar{N}_1)(\lambda_0\mu_0)(\lambda_2\mu_2);
	(\lambda_1\mu_1)\_\,\rho_0; 
	(0\bar{N}_2)\_\,\_
\Big)
U\big(\bar{S}_1'\bar{S}_1S_0S_2;S_1;\bar{S}_2\big)\,
\scrU^{(\lambda_0\mu_0)S_0}(\bar{N}_1'\bar{S}_1',\bar{N}_2\bar{S}_2)\\[2ex]
&\quad-
\sum_{\rho_0\rho_0'}[(\lambda_1\mu_1)(\lambda_2\mu_2)(\lambda_0\mu_0)]
(-)^{\bar{N}_1+\bar{N}_2+\lambda_1+\mu_1}
(-)^{\bar{S}_1+\bar{S}_2+S_1}
\frac{\hat{\bar{S}}_2\hat{S}_0}{\hat{S}_2\hat{\bar{S}}_1}
\left[
\frac
{\dim(\bar{N}_2,0)\dim(\lambda_0\mu_0)}
{\dim(\lambda_2\mu_2)\dim(0,\bar{N}_1)}
\right]^{\frac12}\\
&\quad\qquad\times U\Big(
	(\bar{N}_2',0)(0,\bar{N}_2)(\lambda_0\mu_0)(\lambda_1\mu_1);
	(\lambda_2\mu_2)\_\,\rho_0; 
	(0\bar{N}_1)\_\,\_
\Big)
U\big(\bar{S}_2'\bar{S}_2S_0S_1;S_2;\bar{S}_1\big)\,
\scrU^{(\lambda_0\mu_0)S_0}(\bar{N}_2'\bar{S}_2',\bar{N}_1\bar{S}_1).
\end{split}
\label{eq:unit_tensor_commutator}
\end{equation}

\begin{acknowledgments}
We thank David J.~Rowe for indispensable assistance with the $\grpsptr$ formalism and Daniel Lay for helpful feedback on the manuscript.
This material is based upon work supported by the U.S.~Department of Energy,
Office of Science, Office of Nuclear Physics, under Award Numbers
DE-FG02-95ER-40934 and DE-AC02-06CH11357, the Office of Science SCGSR program, under Award Numbers DE-AC05-06OR23100 and DE‐SC0014664, and the Research Corporation for
Science Advancement, under a Cottrell Scholar Award.
\end{acknowledgments}

\bibliographystyle{apsrev4-2}

\begin{thebibliography}{193}\makeatletter
\providecommand \@ifxundefined [1]{\@ifx{#1\undefined}
}\providecommand \@ifnum [1]{\ifnum #1\expandafter \@firstoftwo
 \else \expandafter \@secondoftwo
 \fi
}\providecommand \@ifx [1]{\ifx #1\expandafter \@firstoftwo
 \else \expandafter \@secondoftwo
 \fi
}\providecommand \natexlab [1]{#1}\providecommand \enquote  [1]{``#1''}\providecommand \bibnamefont  [1]{#1}\providecommand \bibfnamefont [1]{#1}\providecommand \citenamefont [1]{#1}\providecommand \href@noop [0]{\@secondoftwo}\providecommand \href [0]{\begingroup \@sanitize@url \@href}\providecommand \@href[1]{\@@startlink{#1}\@@href}\providecommand \@@href[1]{\endgroup#1\@@endlink}\providecommand \@sanitize@url [0]{\catcode `\\12\catcode `\$12\catcode
  `\&12\catcode `\#12\catcode `\^12\catcode `\_12\catcode `\%12\relax}\providecommand \@@startlink[1]{}\providecommand \@@endlink[0]{}\providecommand \url  [0]{\begingroup\@sanitize@url \@url }\providecommand \@url [1]{\endgroup\@href {#1}{\urlprefix }}\providecommand \urlprefix  [0]{URL }\providecommand \Eprint [0]{\href }\providecommand \doibase [0]{https://doi.org/}\providecommand \selectlanguage [0]{\@gobble}\providecommand \bibinfo  [0]{\@secondoftwo}\providecommand \bibfield  [0]{\@secondoftwo}\providecommand \translation [1]{[#1]}\providecommand \BibitemOpen [0]{}\providecommand \bibitemStop [0]{}\providecommand \bibitemNoStop [0]{.\EOS\space}\providecommand \EOS [0]{\spacefactor3000\relax}\providecommand \BibitemShut  [1]{\csname bibitem#1\endcsname}\let\auto@bib@innerbib\@empty
\bibitem [{\citenamefont {Morris}\ \emph {et~al.}(2018)\citenamefont {Morris},
  \citenamefont {Simonis}, \citenamefont {Stroberg}, \citenamefont {Strumpf},
  \citenamefont {Hagen}, \citenamefont {Holt}, \citenamefont {Jansen},
  \citenamefont {Papenbrock}, \citenamefont {Roth},\ and\ \citenamefont
  {Schwenk}}]{prl-120-2018-152503-Morris}\BibitemOpen
  \bibfield  {author} {\bibinfo {author} {\bibfnamefont {T.~D.}\ \bibnamefont
  {Morris}}, \bibinfo {author} {\bibfnamefont {J.}~\bibnamefont {Simonis}},
  \bibinfo {author} {\bibfnamefont {S.~R.}\ \bibnamefont {Stroberg}}, \bibinfo
  {author} {\bibfnamefont {C.}~\bibnamefont {Strumpf}}, \bibinfo {author}
  {\bibfnamefont {G.}~\bibnamefont {Hagen}}, \bibinfo {author} {\bibfnamefont
  {J.~D.}\ \bibnamefont {Holt}}, \bibinfo {author} {\bibfnamefont {G.~R.}\
  \bibnamefont {Jansen}}, \bibinfo {author} {\bibfnamefont {T.}~\bibnamefont
  {Papenbrock}}, \bibinfo {author} {\bibfnamefont {R.}~\bibnamefont {Roth}},\
  and\ \bibinfo {author} {\bibfnamefont {A.}~\bibnamefont {Schwenk}},\
  }\href@noop {} {\bibfield  {journal} {\bibinfo  {journal} {Phys. Rev. Lett.}\
  }\textbf {\bibinfo {volume} {120}},\ \bibinfo {pages} {152503} (\bibinfo
  {year} {2018})}\BibitemShut {NoStop}\bibitem [{\citenamefont {Gysbers}\ \emph {et~al.}(2019)\citenamefont
  {Gysbers}, \citenamefont {Hagen}, \citenamefont {Holt}, \citenamefont
  {Jansen}, \citenamefont {Morris}, \citenamefont {Navr{\'a}til}, \citenamefont
  {Papenbrock}, \citenamefont {Quaglioni}, \citenamefont {Schwenk},
  \citenamefont {Stroberg},\ and\ \citenamefont
  {Wendt}}]{np-15-2019-428-Gysbers}\BibitemOpen
  \bibfield  {author} {\bibinfo {author} {\bibfnamefont {P.}~\bibnamefont
  {Gysbers}}, \bibinfo {author} {\bibfnamefont {G.}~\bibnamefont {Hagen}},
  \bibinfo {author} {\bibfnamefont {J.~D.}\ \bibnamefont {Holt}}, \bibinfo
  {author} {\bibfnamefont {G.~R.}\ \bibnamefont {Jansen}}, \bibinfo {author}
  {\bibfnamefont {T.~D.}\ \bibnamefont {Morris}}, \bibinfo {author}
  {\bibfnamefont {P.}~\bibnamefont {Navr{\'a}til}}, \bibinfo {author}
  {\bibfnamefont {T.}~\bibnamefont {Papenbrock}}, \bibinfo {author}
  {\bibfnamefont {S.}~\bibnamefont {Quaglioni}}, \bibinfo {author}
  {\bibfnamefont {A.}~\bibnamefont {Schwenk}}, \bibinfo {author} {\bibfnamefont
  {S.~R.}\ \bibnamefont {Stroberg}},\ and\ \bibinfo {author} {\bibfnamefont
  {K.~A.}\ \bibnamefont {Wendt}},\ }\href
  {https://doi.org/10.1038/s41567-019-0450-7} {\bibfield  {journal} {\bibinfo
  {journal} {Nature Physics}\ }\textbf {\bibinfo {volume} {15}},\ \bibinfo
  {pages} {428} (\bibinfo {year} {2019})}\BibitemShut {NoStop}\bibitem [{\citenamefont {Yao}\ \emph {et~al.}(2020)\citenamefont {Yao},
  \citenamefont {Bally}, \citenamefont {Engel}, \citenamefont {Wirth},
  \citenamefont {Rodr\'{\i}guez},\ and\ \citenamefont
  {Hergert}}]{prl-124-2020-232501-Yao}\BibitemOpen
  \bibfield  {author} {\bibinfo {author} {\bibfnamefont {J.~M.}\ \bibnamefont
  {Yao}}, \bibinfo {author} {\bibfnamefont {B.}~\bibnamefont {Bally}}, \bibinfo
  {author} {\bibfnamefont {J.}~\bibnamefont {Engel}}, \bibinfo {author}
  {\bibfnamefont {R.}~\bibnamefont {Wirth}}, \bibinfo {author} {\bibfnamefont
  {T.~R.}\ \bibnamefont {Rodr\'{\i}guez}},\ and\ \bibinfo {author}
  {\bibfnamefont {H.}~\bibnamefont {Hergert}},\ }\href
  {https://doi.org/10.1103/PhysRevLett.124.232501} {\bibfield  {journal}
  {\bibinfo  {journal} {Phys. Rev. Lett.}\ }\textbf {\bibinfo {volume} {124}},\
  \bibinfo {pages} {232501} (\bibinfo {year} {2020})}\BibitemShut {NoStop}\bibitem [{\citenamefont {Stroberg}\ \emph {et~al.}(2021)\citenamefont
  {Stroberg}, \citenamefont {Holt}, \citenamefont {Schwenk},\ and\
  \citenamefont {Simonis}}]{prl-126-2021-022501-Stroberg}\BibitemOpen
  \bibfield  {author} {\bibinfo {author} {\bibfnamefont {S.~R.}\ \bibnamefont
  {Stroberg}}, \bibinfo {author} {\bibfnamefont {J.~D.}\ \bibnamefont {Holt}},
  \bibinfo {author} {\bibfnamefont {A.}~\bibnamefont {Schwenk}},\ and\ \bibinfo
  {author} {\bibfnamefont {J.}~\bibnamefont {Simonis}},\ }\href
  {https://doi.org/10.1103/PhysRevLett.126.022501} {\bibfield  {journal}
  {\bibinfo  {journal} {Phys. Rev. Lett.}\ }\textbf {\bibinfo {volume} {126}},\
  \bibinfo {pages} {022501} (\bibinfo {year} {2021})}\BibitemShut {NoStop}\bibitem [{\citenamefont {Hu}\ \emph {et~al.}(2022)\citenamefont {Hu},
  \citenamefont {Jiang}, \citenamefont {Miyagi}, \citenamefont {Sun},
  \citenamefont {Ekstrom}, \citenamefont {Forss{\'e}n}, \citenamefont {Hagen},
  \citenamefont {Holt}, \citenamefont {Papenbrock}, \citenamefont {Stroberg},\
  and\ \citenamefont {Vernon}}]{np-18-2022-1196-Hu}\BibitemOpen
  \bibfield  {author} {\bibinfo {author} {\bibfnamefont {B.}~\bibnamefont
  {Hu}}, \bibinfo {author} {\bibfnamefont {W.}~\bibnamefont {Jiang}}, \bibinfo
  {author} {\bibfnamefont {T.}~\bibnamefont {Miyagi}}, \bibinfo {author}
  {\bibfnamefont {Z.}~\bibnamefont {Sun}}, \bibinfo {author} {\bibfnamefont
  {A.}~\bibnamefont {Ekstrom}}, \bibinfo {author} {\bibfnamefont
  {C.}~\bibnamefont {Forss{\'e}n}}, \bibinfo {author} {\bibfnamefont
  {G.}~\bibnamefont {Hagen}}, \bibinfo {author} {\bibfnamefont {J.~D.}\
  \bibnamefont {Holt}}, \bibinfo {author} {\bibfnamefont {T.}~\bibnamefont
  {Papenbrock}}, \bibinfo {author} {\bibfnamefont {S.~R.}\ \bibnamefont
  {Stroberg}},\ and\ \bibinfo {author} {\bibfnamefont {I.}~\bibnamefont
  {Vernon}},\ }\href {https://doi.org/10.1038/s41567-022-01715-8} {\bibfield
  {journal} {\bibinfo  {journal} {Nature Physics}\ }\textbf {\bibinfo {volume}
  {18}},\ \bibinfo {pages} {1196} (\bibinfo {year} {2022})}\BibitemShut
  {NoStop}\bibitem [{\citenamefont {Hergert}(2020)}]{fp-8-2020-1-Hergert}\BibitemOpen
  \bibfield  {author} {\bibinfo {author} {\bibfnamefont {H.}~\bibnamefont
  {Hergert}},\ }\href {https://doi.org/10.3389/fphy.2020.00379} {\bibfield
  {journal} {\bibinfo  {journal} {Frontiers in Physics}\ }\textbf {\bibinfo
  {volume} {8}},\ \bibinfo {pages} {1} (\bibinfo {year} {2020})}\BibitemShut
  {NoStop}\bibitem [{\citenamefont {Belley}\ \emph {et~al.}(2021)\citenamefont {Belley},
  \citenamefont {Payne}, \citenamefont {Stroberg}, \citenamefont {Miyagi},\
  and\ \citenamefont {Holt}}]{prl-126-2021-042502-Belley}\BibitemOpen
  \bibfield  {author} {\bibinfo {author} {\bibfnamefont {A.}~\bibnamefont
  {Belley}}, \bibinfo {author} {\bibfnamefont {C.~G.}\ \bibnamefont {Payne}},
  \bibinfo {author} {\bibfnamefont {S.~R.}\ \bibnamefont {Stroberg}}, \bibinfo
  {author} {\bibfnamefont {T.}~\bibnamefont {Miyagi}},\ and\ \bibinfo {author}
  {\bibfnamefont {J.~D.}\ \bibnamefont {Holt}},\ }\href
  {https://doi.org/10.1103/PhysRevLett.126.042502} {\bibfield  {journal}
  {\bibinfo  {journal} {Phys. Rev. Lett.}\ }\textbf {\bibinfo {volume} {126}},\
  \bibinfo {pages} {042502} (\bibinfo {year} {2021})}\BibitemShut {NoStop}\bibitem [{\citenamefont {Belley}\ \emph {et~al.}(2024)\citenamefont {Belley},
  \citenamefont {Yao}, \citenamefont {Bally}, \citenamefont {Pitcher},
  \citenamefont {Engel}, \citenamefont {Hergert}, \citenamefont {Holt},
  \citenamefont {Miyagi}, \citenamefont {Rodr\'{\i}guez}, \citenamefont
  {Romero}, \citenamefont {Stroberg},\ and\ \citenamefont
  {Zhang}}]{prl-132-2024-182502-Belley}\BibitemOpen
  \bibfield  {author} {\bibinfo {author} {\bibfnamefont {A.}~\bibnamefont
  {Belley}}, \bibinfo {author} {\bibfnamefont {J.~M.}\ \bibnamefont {Yao}},
  \bibinfo {author} {\bibfnamefont {B.}~\bibnamefont {Bally}}, \bibinfo
  {author} {\bibfnamefont {J.}~\bibnamefont {Pitcher}}, \bibinfo {author}
  {\bibfnamefont {J.}~\bibnamefont {Engel}}, \bibinfo {author} {\bibfnamefont
  {H.}~\bibnamefont {Hergert}}, \bibinfo {author} {\bibfnamefont {J.~D.}\
  \bibnamefont {Holt}}, \bibinfo {author} {\bibfnamefont {T.}~\bibnamefont
  {Miyagi}}, \bibinfo {author} {\bibfnamefont {T.~R.}\ \bibnamefont
  {Rodr\'{\i}guez}}, \bibinfo {author} {\bibfnamefont {A.~M.}\ \bibnamefont
  {Romero}}, \bibinfo {author} {\bibfnamefont {S.~R.}\ \bibnamefont
  {Stroberg}},\ and\ \bibinfo {author} {\bibfnamefont {X.}~\bibnamefont
  {Zhang}},\ }\href {https://doi.org/10.1103/PhysRevLett.132.182502} {\bibfield
   {journal} {\bibinfo  {journal} {Phys. Rev. Lett.}\ }\textbf {\bibinfo
  {volume} {132}},\ \bibinfo {pages} {182502} (\bibinfo {year}
  {2024})}\BibitemShut {NoStop}\bibitem [{\citenamefont {Hu}\ \emph {et~al.}(2024)\citenamefont {Hu},
  \citenamefont {Sun}, \citenamefont {Hagen},\ and\ \citenamefont
  {Papenbrock}}]{prc-110-2024-L011302-Hu}\BibitemOpen
  \bibfield  {author} {\bibinfo {author} {\bibfnamefont {B.~S.}\ \bibnamefont
  {Hu}}, \bibinfo {author} {\bibfnamefont {Z.~H.}\ \bibnamefont {Sun}},
  \bibinfo {author} {\bibfnamefont {G.}~\bibnamefont {Hagen}},\ and\ \bibinfo
  {author} {\bibfnamefont {T.}~\bibnamefont {Papenbrock}},\ }\href@noop {}
  {\bibfield  {journal} {\bibinfo  {journal} {Phys. Rev. C}\ }\textbf {\bibinfo
  {volume} {110}},\ \bibinfo {pages} {L011302} (\bibinfo {year}
  {2024})}\BibitemShut {NoStop}\bibitem [{\citenamefont {Caprio}\ \emph {et~al.}(2013)\citenamefont {Caprio},
  \citenamefont {Maris},\ and\ \citenamefont {Vary}}]{plb-719-2013-179-Caprio}\BibitemOpen
  \bibfield  {author} {\bibinfo {author} {\bibfnamefont {M.~A.}\ \bibnamefont
  {Caprio}}, \bibinfo {author} {\bibfnamefont {P.}~\bibnamefont {Maris}},\ and\
  \bibinfo {author} {\bibfnamefont {J.~P.}\ \bibnamefont {Vary}},\ }\href
  {https://doi.org/10.1016/j.physletb.2012.12.064} {\bibfield  {journal}
  {\bibinfo  {journal} {Phys. Lett. B}\ }\textbf {\bibinfo {volume} {719}},\
  \bibinfo {pages} {179} (\bibinfo {year} {2013})}\BibitemShut {NoStop}\bibitem [{\citenamefont {Caprio}\ \emph {et~al.}(2019)\citenamefont {Caprio},
  \citenamefont {Fasano}, \citenamefont {McCoy}, \citenamefont {Maris},\ and\
  \citenamefont {Vary}}]{bjp-46-2019-445-Caprio}\BibitemOpen
  \bibfield  {author} {\bibinfo {author} {\bibfnamefont {M.~A.}\ \bibnamefont
  {Caprio}}, \bibinfo {author} {\bibfnamefont {P.~J.}\ \bibnamefont {Fasano}},
  \bibinfo {author} {\bibfnamefont {A.~E.}\ \bibnamefont {McCoy}}, \bibinfo
  {author} {\bibfnamefont {P.}~\bibnamefont {Maris}},\ and\ \bibinfo {author}
  {\bibfnamefont {J.~P.}\ \bibnamefont {Vary}},\ }\href
  {https://doi.org/10.55318/bgjp.2022.49.1.057} {\bibfield  {journal} {\bibinfo
   {journal} {Bulg. J. Phys.}\ }\textbf {\bibinfo {volume} {46}},\ \bibinfo
  {pages} {445} (\bibinfo {year} {2019})}\BibitemShut {NoStop}\bibitem [{\citenamefont {Caprio}\ \emph {et~al.}(2022)\citenamefont {Caprio},
  \citenamefont {McCoy}, \citenamefont {Fasano},\ and\ \citenamefont
  {Dytrych}}]{bjp-49-2022-057066-Caprio}\BibitemOpen
  \bibfield  {author} {\bibinfo {author} {\bibfnamefont {M.~A.}\ \bibnamefont
  {Caprio}}, \bibinfo {author} {\bibfnamefont {A.~E.}\ \bibnamefont {McCoy}},
  \bibinfo {author} {\bibfnamefont {P.~J.}\ \bibnamefont {Fasano}},\ and\
  \bibinfo {author} {\bibfnamefont {T.}~\bibnamefont {Dytrych}},\ }\href
  {https://doi.org/10.55318/bgjp.2022.49.1.057} {\bibfield  {journal} {\bibinfo
   {journal} {Bulg. J. Phys.}\ }\textbf {\bibinfo {volume} {49}},\ \bibinfo
  {pages} {057066} (\bibinfo {year} {2022})}\BibitemShut {NoStop}\bibitem [{\citenamefont {Caprio}\ \emph
  {et~al.}(2020{\natexlab{a}})\citenamefont {Caprio}, \citenamefont {Fasano},
  \citenamefont {Maris}, \citenamefont {McCoy},\ and\ \citenamefont
  {Vary}}]{epja-56-2020-120-Caprio}\BibitemOpen
  \bibfield  {author} {\bibinfo {author} {\bibfnamefont {M.~A.}\ \bibnamefont
  {Caprio}}, \bibinfo {author} {\bibfnamefont {P.~J.}\ \bibnamefont {Fasano}},
  \bibinfo {author} {\bibfnamefont {P.}~\bibnamefont {Maris}}, \bibinfo
  {author} {\bibfnamefont {A.~E.}\ \bibnamefont {McCoy}},\ and\ \bibinfo
  {author} {\bibfnamefont {J.~P.}\ \bibnamefont {Vary}},\ }\href
  {https://doi.org/10.1140/epja/s10050-020-00112-0} {\bibfield  {journal}
  {\bibinfo  {journal} {Eur. Phys. J. A}\ }\textbf {\bibinfo {volume} {56}},\
  \bibinfo {pages} {120} (\bibinfo {year} {2020}{\natexlab{a}})}\BibitemShut
  {NoStop}\bibitem [{\citenamefont {Caprio}\ \emph {et~al.}(2010)\citenamefont {Caprio},
  \citenamefont {Maris}, \citenamefont {Vary},\ and\ \citenamefont
  {Smith}}]{ijmpe-24-2015-1541002-Caprio}\BibitemOpen
  \bibfield  {author} {\bibinfo {author} {\bibfnamefont {M.~A.}\ \bibnamefont
  {Caprio}}, \bibinfo {author} {\bibfnamefont {P.}~\bibnamefont {Maris}},
  \bibinfo {author} {\bibfnamefont {J.~P.}\ \bibnamefont {Vary}},\ and\
  \bibinfo {author} {\bibfnamefont {R.}~\bibnamefont {Smith}},\ }\href
  {https://doi.org/10.1142/S0218301315410025} {\bibfield  {journal} {\bibinfo
  {journal} {Int. J. Mod. Phys.}\ }\textbf {\bibinfo {volume} {24}},\ \bibinfo
  {pages} {1541002} (\bibinfo {year} {2010})}\BibitemShut {NoStop}\bibitem [{\citenamefont {Zbikowski}\ \emph {et~al.}(2021)\citenamefont
  {Zbikowski}, \citenamefont {Johnson}, \citenamefont {McCoy}, \citenamefont
  {Caprio},\ and\ \citenamefont {Fasano}}]{jpg-48-2021-075102-Zbikowski}\BibitemOpen
  \bibfield  {author} {\bibinfo {author} {\bibfnamefont {R.}~\bibnamefont
  {Zbikowski}}, \bibinfo {author} {\bibfnamefont {C.~W.}\ \bibnamefont
  {Johnson}}, \bibinfo {author} {\bibfnamefont {A.~E.}\ \bibnamefont {McCoy}},
  \bibinfo {author} {\bibfnamefont {M.~A.}\ \bibnamefont {Caprio}},\ and\
  \bibinfo {author} {\bibfnamefont {P.~J.}\ \bibnamefont {Fasano}},\ }\href
  {https://doi.org/10.1088/1361-6471/abdd8e} {\bibfield  {journal} {\bibinfo
  {journal} {J. Phys. G}\ }\textbf {\bibinfo {volume} {48}},\ \bibinfo {pages}
  {075102} (\bibinfo {year} {2021})}\BibitemShut {NoStop}\bibitem [{\citenamefont {McCoy}\ \emph {et~al.}(2024)\citenamefont {McCoy},
  \citenamefont {Caprio}, \citenamefont {Maris},\ and\ \citenamefont
  {Fasano}}]{plb-856-2024-138870-McCoy}\BibitemOpen
  \bibfield  {author} {\bibinfo {author} {\bibfnamefont {A.~E.}\ \bibnamefont
  {McCoy}}, \bibinfo {author} {\bibfnamefont {M.~A.}\ \bibnamefont {Caprio}},
  \bibinfo {author} {\bibfnamefont {P.}~\bibnamefont {Maris}},\ and\ \bibinfo
  {author} {\bibfnamefont {P.~J.}\ \bibnamefont {Fasano}},\ }\href
  {https://doi.org/10.1016/j.physletb.2024.138870} {\bibfield  {journal}
  {\bibinfo  {journal} {Phys. Lett. B}\ }\textbf {\bibinfo {volume} {856}},\
  \bibinfo {pages} {138870} (\bibinfo {year} {2024})}\BibitemShut {NoStop}\bibitem [{\citenamefont {Maris}\ \emph {et~al.}(2015)\citenamefont {Maris},
  \citenamefont {Caprio},\ and\ \citenamefont
  {Vary}}]{prc-91-2015-014310-Maris}\BibitemOpen
  \bibfield  {author} {\bibinfo {author} {\bibfnamefont {P.}~\bibnamefont
  {Maris}}, \bibinfo {author} {\bibfnamefont {M.~A.}\ \bibnamefont {Caprio}},\
  and\ \bibinfo {author} {\bibfnamefont {J.~P.}\ \bibnamefont {Vary}},\ }\href
  {https://doi.org/10.1103/PhysRevC.91.014310} {\bibfield  {journal} {\bibinfo
  {journal} {Phys. Rev. C}\ }\textbf {\bibinfo {volume} {91}},\ \bibinfo
  {pages} {014310} (\bibinfo {year} {2015})}\BibitemShut {NoStop}\bibitem [{\citenamefont {Maris}\ \emph {et~al.}(2019)\citenamefont {Maris},
  \citenamefont {Caprio},\ and\ \citenamefont
  {Vary}}]{prc-99-2019-029902-Maris-ERRATUM}\BibitemOpen
  \bibfield  {author} {\bibinfo {author} {\bibfnamefont {P.}~\bibnamefont
  {Maris}}, \bibinfo {author} {\bibfnamefont {M.~A.}\ \bibnamefont {Caprio}},\
  and\ \bibinfo {author} {\bibfnamefont {J.~P.}\ \bibnamefont {Vary}},\ }\href
  {https://doi.org/10.1103/PhysRevC.99.029902} {\bibfield  {journal} {\bibinfo
  {journal} {Phys. Rev. C}\ }\textbf {\bibinfo {volume} {99}},\ \bibinfo
  {pages} {029902(E)} (\bibinfo {year} {2019})}\BibitemShut {NoStop}\bibitem [{\citenamefont {Stroberg}\ \emph {et~al.}(2016)\citenamefont
  {Stroberg}, \citenamefont {Hergert}, \citenamefont {Holt}, \citenamefont
  {Bogner},\ and\ \citenamefont {Schwenk}}]{prc-93-2016-051301-Stroberg}\BibitemOpen
  \bibfield  {author} {\bibinfo {author} {\bibfnamefont {S.~R.}\ \bibnamefont
  {Stroberg}}, \bibinfo {author} {\bibfnamefont {H.}~\bibnamefont {Hergert}},
  \bibinfo {author} {\bibfnamefont {J.~D.}\ \bibnamefont {Holt}}, \bibinfo
  {author} {\bibfnamefont {S.~K.}\ \bibnamefont {Bogner}},\ and\ \bibinfo
  {author} {\bibfnamefont {A.}~\bibnamefont {Schwenk}},\ }\href
  {https://doi.org/10.1103/PhysRevC.93.051301} {\bibfield  {journal} {\bibinfo
  {journal} {Phys. Rev. C}\ }\textbf {\bibinfo {volume} {93}},\ \bibinfo
  {pages} {051301(R)} (\bibinfo {year} {2016})}\BibitemShut {NoStop}\bibitem [{\citenamefont {Dytrych}\ \emph {et~al.}(2013)\citenamefont
  {Dytrych}, \citenamefont {Launey}, \citenamefont {Draayer}, \citenamefont
  {Maris}, \citenamefont {Vary}, \citenamefont {Saule}, \citenamefont
  {Catalyurek}, \citenamefont {Sosonkina}, \citenamefont {Langr},\ and\
  \citenamefont {Caprio}}]{prl-111-2013-252501-Dytrych}\BibitemOpen
  \bibfield  {author} {\bibinfo {author} {\bibfnamefont {T.}~\bibnamefont
  {Dytrych}}, \bibinfo {author} {\bibfnamefont {K.~D.}\ \bibnamefont {Launey}},
  \bibinfo {author} {\bibfnamefont {J.~P.}\ \bibnamefont {Draayer}}, \bibinfo
  {author} {\bibfnamefont {P.}~\bibnamefont {Maris}}, \bibinfo {author}
  {\bibfnamefont {J.~P.}\ \bibnamefont {Vary}}, \bibinfo {author}
  {\bibfnamefont {E.}~\bibnamefont {Saule}}, \bibinfo {author} {\bibfnamefont
  {U.}~\bibnamefont {Catalyurek}}, \bibinfo {author} {\bibfnamefont
  {M.}~\bibnamefont {Sosonkina}}, \bibinfo {author} {\bibfnamefont
  {D.}~\bibnamefont {Langr}},\ and\ \bibinfo {author} {\bibfnamefont {M.~A.}\
  \bibnamefont {Caprio}},\ }\href
  {https://doi.org/10.1103/PhysRevLett.111.252501} {\bibfield  {journal}
  {\bibinfo  {journal} {Phys. Rev. Lett.}\ }\textbf {\bibinfo {volume} {111}},\
  \bibinfo {pages} {252501} (\bibinfo {year} {2013})}\BibitemShut {NoStop}\bibitem [{\citenamefont {Jansen}\ \emph {et~al.}(2016)\citenamefont {Jansen},
  \citenamefont {Schuster}, \citenamefont {Signoracci}, \citenamefont {Hagen},\
  and\ \citenamefont {Navr\'atil}}]{prc-94-2016-011301-Jansen}\BibitemOpen
  \bibfield  {author} {\bibinfo {author} {\bibfnamefont {G.~R.}\ \bibnamefont
  {Jansen}}, \bibinfo {author} {\bibfnamefont {M.~D.}\ \bibnamefont
  {Schuster}}, \bibinfo {author} {\bibfnamefont {A.}~\bibnamefont
  {Signoracci}}, \bibinfo {author} {\bibfnamefont {G.}~\bibnamefont {Hagen}},\
  and\ \bibinfo {author} {\bibfnamefont {P.}~\bibnamefont {Navr\'atil}},\
  }\href {https://doi.org/10.1103/PhysRevC.94.011301} {\bibfield  {journal}
  {\bibinfo  {journal} {Phys. Rev. C}\ }\textbf {\bibinfo {volume} {94}},\
  \bibinfo {pages} {011301 (R)} (\bibinfo {year} {2016})}\BibitemShut {NoStop}\bibitem [{\citenamefont {McCoy}\ \emph {et~al.}(2020)\citenamefont {McCoy},
  \citenamefont {Caprio},\ and\ \citenamefont
  {Dytrych}}]{prl-125-2020-102505-McCoy}\BibitemOpen
  \bibfield  {author} {\bibinfo {author} {\bibfnamefont {A.~E.}\ \bibnamefont
  {McCoy}}, \bibinfo {author} {\bibfnamefont {M.~A.}\ \bibnamefont {Caprio}},\
  and\ \bibinfo {author} {\bibfnamefont {T.}~\bibnamefont {Dytrych}},\ }\href
  {https://doi.org/10.1103/PhysRevLett.125.102505} {\bibfield  {journal}
  {\bibinfo  {journal} {Phys. Rev. Lett.}\ }\textbf {\bibinfo {volume} {125}},\
  \bibinfo {pages} {102505} (\bibinfo {year} {2020})}\BibitemShut {NoStop}\bibitem [{\citenamefont {Pieper}\ \emph {et~al.}(2004)\citenamefont {Pieper},
  \citenamefont {Wiringa},\ and\ \citenamefont
  {Carlson}}]{prc-70-2004-054325-Pieper}\BibitemOpen
  \bibfield  {author} {\bibinfo {author} {\bibfnamefont {S.~C.}\ \bibnamefont
  {Pieper}}, \bibinfo {author} {\bibfnamefont {R.~B.}\ \bibnamefont
  {Wiringa}},\ and\ \bibinfo {author} {\bibfnamefont {J.}~\bibnamefont
  {Carlson}},\ }\href {https://doi.org/10.1103/PhysRevC.70.054325} {\bibfield
  {journal} {\bibinfo  {journal} {Phys. Rev. C}\ }\textbf {\bibinfo {volume}
  {70}},\ \bibinfo {pages} {054325} (\bibinfo {year} {2004})}\BibitemShut
  {NoStop}\bibitem [{\citenamefont {Neff}\ and\ \citenamefont
  {Feldmeier}(2004)}]{npa-738-2004-357-Neff}\BibitemOpen
  \bibfield  {author} {\bibinfo {author} {\bibfnamefont {T.}~\bibnamefont
  {Neff}}\ and\ \bibinfo {author} {\bibfnamefont {H.}~\bibnamefont
  {Feldmeier}},\ }\href {https://doi.org/10.1016/j.nuclphysa.2004.04.061}
  {\bibfield  {journal} {\bibinfo  {journal} {Nucl. Phys. A}\ }\textbf
  {\bibinfo {volume} {738}},\ \bibinfo {pages} {357} (\bibinfo {year}
  {2004})}\BibitemShut {NoStop}\bibitem [{\citenamefont {Maris}(2012)}]{jpcs-402-2012-012031-Maris}\BibitemOpen
  \bibfield  {author} {\bibinfo {author} {\bibfnamefont {P.}~\bibnamefont
  {Maris}},\ }\href {https://doi.org/10.1088/1742-6596/402/1/012031} {\bibfield
   {journal} {\bibinfo  {journal} {J. Phys. Conf. Ser.}\ }\textbf {\bibinfo
  {volume} {402}},\ \bibinfo {pages} {012031} (\bibinfo {year}
  {2012})}\BibitemShut {NoStop}\bibitem [{\citenamefont {Yoshida}\ \emph {et~al.}(2013)\citenamefont
  {Yoshida}, \citenamefont {Shimizu}, \citenamefont {Abe},\ and\ \citenamefont
  {Otsuka}}]{fbs-54-2013-1465-Yoshida}\BibitemOpen
  \bibfield  {author} {\bibinfo {author} {\bibfnamefont {T.}~\bibnamefont
  {Yoshida}}, \bibinfo {author} {\bibfnamefont {N.}~\bibnamefont {Shimizu}},
  \bibinfo {author} {\bibfnamefont {T.}~\bibnamefont {Abe}},\ and\ \bibinfo
  {author} {\bibfnamefont {T.}~\bibnamefont {Otsuka}},\ }\href
  {https://doi.org/10.1007/s00601-013-0680-7} {\bibfield  {journal} {\bibinfo
  {journal} {Few-Body Syst.}\ }\textbf {\bibinfo {volume} {54}},\ \bibinfo
  {pages} {1465} (\bibinfo {year} {2013})}\BibitemShut {NoStop}\bibitem [{\citenamefont {Romero-Redondo}\ \emph {et~al.}(2016)\citenamefont
  {Romero-Redondo}, \citenamefont {Quaglioni}, \citenamefont {Navr\'{a}til},\
  and\ \citenamefont {Hupin}}]{prl-117-2016-2225-Romero-Redondo}\BibitemOpen
  \bibfield  {author} {\bibinfo {author} {\bibfnamefont {C.}~\bibnamefont
  {Romero-Redondo}}, \bibinfo {author} {\bibfnamefont {S.}~\bibnamefont
  {Quaglioni}}, \bibinfo {author} {\bibfnamefont {P.}~\bibnamefont
  {Navr\'{a}til}},\ and\ \bibinfo {author} {\bibfnamefont {G.}~\bibnamefont
  {Hupin}},\ }\href {https://doi.org/10.1103/PhysRevLett.117.222501} {\bibfield
   {journal} {\bibinfo  {journal} {Phys. Rev. Lett.}\ }\textbf {\bibinfo
  {volume} {117}},\ \bibinfo {pages} {222501} (\bibinfo {year}
  {2016})}\BibitemShut {NoStop}\bibitem [{\citenamefont {Navr\'{a}til}\ \emph {et~al.}(2016)\citenamefont
  {Navr\'{a}til}, \citenamefont {Quaglioni}, \citenamefont {Hupin},
  \citenamefont {Romero-Redondo},\ and\ \citenamefont
  {Calci}}]{ps-91-2016-053002-Navratil}\BibitemOpen
  \bibfield  {author} {\bibinfo {author} {\bibfnamefont {P.}~\bibnamefont
  {Navr\'{a}til}}, \bibinfo {author} {\bibfnamefont {S.}~\bibnamefont
  {Quaglioni}}, \bibinfo {author} {\bibfnamefont {G.}~\bibnamefont {Hupin}},
  \bibinfo {author} {\bibfnamefont {C.}~\bibnamefont {Romero-Redondo}},\ and\
  \bibinfo {author} {\bibfnamefont {A.}~\bibnamefont {Calci}},\ }\href
  {https://doi.org/10.1088/0031-8949/91/5/053002} {\bibfield  {journal}
  {\bibinfo  {journal} {Phys. Scr.}\ }\textbf {\bibinfo {volume} {91}},\
  \bibinfo {pages} {053002} (\bibinfo {year} {2016})}\BibitemShut {NoStop}\bibitem [{\citenamefont {Shen}\ \emph {et~al.}(2025)\citenamefont {Shen},
  \citenamefont {Elhatisari}, \citenamefont {Lee}, \citenamefont
  {Mei{\ss{}}ner},\ and\ \citenamefont {Ren}}]{prl-134-2025-162503-Shen}\BibitemOpen
  \bibfield  {author} {\bibinfo {author} {\bibfnamefont {S.}~\bibnamefont
  {Shen}}, \bibinfo {author} {\bibfnamefont {S.}~\bibnamefont {Elhatisari}},
  \bibinfo {author} {\bibfnamefont {D.}~\bibnamefont {Lee}}, \bibinfo {author}
  {\bibfnamefont {U.}~\bibnamefont {Mei{\ss{}}ner}},\ and\ \bibinfo {author}
  {\bibfnamefont {Z.}~\bibnamefont {Ren}},\ }\href
  {https://doi.org/10.1103/PhysRevLett.134.162503} {\bibfield  {journal}
  {\bibinfo  {journal} {Phys. Rev. Lett.}\ }\textbf {\bibinfo {volume} {134}},\
  \bibinfo {pages} {162503} (\bibinfo {year} {2025})}\BibitemShut {NoStop}\bibitem [{\citenamefont {Novario}\ \emph {et~al.}(2020)\citenamefont
  {Novario}, \citenamefont {Hagen}, \citenamefont {Jansen},\ and\ \citenamefont
  {Papenbrock}}]{prc-102-2020-051303R-Novario}\BibitemOpen
  \bibfield  {author} {\bibinfo {author} {\bibfnamefont {S.~J.}\ \bibnamefont
  {Novario}}, \bibinfo {author} {\bibfnamefont {G.}~\bibnamefont {Hagen}},
  \bibinfo {author} {\bibfnamefont {G.~R.}\ \bibnamefont {Jansen}},\ and\
  \bibinfo {author} {\bibfnamefont {T.}~\bibnamefont {Papenbrock}},\ }\href
  {https://doi.org/DOI: https://doi.org/10.1103/PhysRevC.102.051303} {\bibfield
   {journal} {\bibinfo  {journal} {Phys. Rev. C}\ }\textbf {\bibinfo {volume}
  {102}},\ \bibinfo {pages} {051303R} (\bibinfo {year} {2020})}\BibitemShut
  {NoStop}\bibitem [{\citenamefont {Sun}\ \emph {et~al.}(2025{\natexlab{a}})\citenamefont
  {Sun}, \citenamefont {Dj\"arv}, \citenamefont {Hagen}, \citenamefont
  {Jansen},\ and\ \citenamefont {Papenbrock}}]{prc-111-2025-044304-Sun}\BibitemOpen
  \bibfield  {author} {\bibinfo {author} {\bibfnamefont {Z.~H.}\ \bibnamefont
  {Sun}}, \bibinfo {author} {\bibfnamefont {T.~R.}\ \bibnamefont {Dj\"arv}},
  \bibinfo {author} {\bibfnamefont {G.}~\bibnamefont {Hagen}}, \bibinfo
  {author} {\bibfnamefont {G.~R.}\ \bibnamefont {Jansen}},\ and\ \bibinfo
  {author} {\bibfnamefont {T.}~\bibnamefont {Papenbrock}},\ }\href
  {https://doi.org/10.1103/PhysRevC.111.044304} {\bibfield  {journal} {\bibinfo
   {journal} {Phys. Rev. C}\ }\textbf {\bibinfo {volume} {111}},\ \bibinfo
  {pages} {044304} (\bibinfo {year} {2025}{\natexlab{a}})}\BibitemShut
  {NoStop}\bibitem [{\citenamefont {Sun}\ \emph {et~al.}(2025{\natexlab{b}})\citenamefont
  {Sun}, \citenamefont {Ekstr\"om}, \citenamefont {Forss\'en}, \citenamefont
  {Hagen}, \citenamefont {Jansen},\ and\ \citenamefont
  {Papenbrock}}]{prx-15-2025-011028-Sun}\BibitemOpen
  \bibfield  {author} {\bibinfo {author} {\bibfnamefont {Z.~H.}\ \bibnamefont
  {Sun}}, \bibinfo {author} {\bibfnamefont {A.}~\bibnamefont {Ekstr\"om}},
  \bibinfo {author} {\bibfnamefont {C.}~\bibnamefont {Forss\'en}}, \bibinfo
  {author} {\bibfnamefont {G.}~\bibnamefont {Hagen}}, \bibinfo {author}
  {\bibfnamefont {G.~R.}\ \bibnamefont {Jansen}},\ and\ \bibinfo {author}
  {\bibfnamefont {T.}~\bibnamefont {Papenbrock}},\ }\href
  {https://doi.org/10.1103/PhysRevX.15.011028} {\bibfield  {journal} {\bibinfo
  {journal} {Phys. Rev. X}\ }\textbf {\bibinfo {volume} {15}},\ \bibinfo
  {pages} {011028} (\bibinfo {year} {2025}{\natexlab{b}})}\BibitemShut
  {NoStop}\bibitem [{\citenamefont {Stroberg}\ \emph {et~al.}(2017)\citenamefont
  {Stroberg}, \citenamefont {Calci}, \citenamefont {Hergert}, \citenamefont
  {Holt}, \citenamefont {Bogner}, \citenamefont {Roth},\ and\ \citenamefont
  {Schwenk}}]{prl-118-2017-032502-Stroberg}\BibitemOpen
  \bibfield  {author} {\bibinfo {author} {\bibfnamefont {S.~R.}\ \bibnamefont
  {Stroberg}}, \bibinfo {author} {\bibfnamefont {A.}~\bibnamefont {Calci}},
  \bibinfo {author} {\bibfnamefont {H.}~\bibnamefont {Hergert}}, \bibinfo
  {author} {\bibfnamefont {J.~D.}\ \bibnamefont {Holt}}, \bibinfo {author}
  {\bibfnamefont {S.~K.}\ \bibnamefont {Bogner}}, \bibinfo {author}
  {\bibfnamefont {R.}~\bibnamefont {Roth}},\ and\ \bibinfo {author}
  {\bibfnamefont {A.}~\bibnamefont {Schwenk}},\ }\href
  {https://doi.org/10.1103/PhysRevLett.118.032502} {\bibfield  {journal}
  {\bibinfo  {journal} {Phys. Rev. Lett.}\ }\textbf {\bibinfo {volume} {118}},\
  \bibinfo {pages} {032502} (\bibinfo {year} {2017})}\BibitemShut {NoStop}\bibitem [{\citenamefont {Hagen}\ \emph {et~al.}(2014)\citenamefont {Hagen},
  \citenamefont {Papenbrock}, \citenamefont {Hjorth-Jensen},\ and\
  \citenamefont {Dean}}]{rpp-77-2014-096302-Hagen}\BibitemOpen
  \bibfield  {author} {\bibinfo {author} {\bibfnamefont {G.}~\bibnamefont
  {Hagen}}, \bibinfo {author} {\bibfnamefont {T.}~\bibnamefont {Papenbrock}},
  \bibinfo {author} {\bibfnamefont {M.}~\bibnamefont {Hjorth-Jensen}},\ and\
  \bibinfo {author} {\bibfnamefont {D.~J.}\ \bibnamefont {Dean}},\ }\href@noop
  {} {\bibfield  {journal} {\bibinfo  {journal} {Rep. Prog. Phys.}\ }\textbf
  {\bibinfo {volume} {77}},\ \bibinfo {pages} {096302} (\bibinfo {year}
  {2014})}\BibitemShut {NoStop}\bibitem [{\citenamefont {Barrett}\ \emph {et~al.}(2013)\citenamefont
  {Barrett}, \citenamefont {Navr\'{a}til},\ and\ \citenamefont
  {Vary}}]{ppnp-69-2013-131-Barrett}\BibitemOpen
  \bibfield  {author} {\bibinfo {author} {\bibfnamefont {B.~R.}\ \bibnamefont
  {Barrett}}, \bibinfo {author} {\bibfnamefont {P.}~\bibnamefont
  {Navr\'{a}til}},\ and\ \bibinfo {author} {\bibfnamefont {J.~P.}\ \bibnamefont
  {Vary}},\ }\href@noop {} {\bibfield  {journal} {\bibinfo  {journal} {Prog.
  Part. Nucl. Phys.}\ }\textbf {\bibinfo {volume} {69}},\ \bibinfo {pages}
  {131} (\bibinfo {year} {2013})}\BibitemShut {NoStop}\bibitem [{\citenamefont {Pieper}\ \emph {et~al.}(2002)\citenamefont {Pieper},
  \citenamefont {Varga},\ and\ \citenamefont
  {Wiringa}}]{prc-66-2002-044310-Pieper}\BibitemOpen
  \bibfield  {author} {\bibinfo {author} {\bibfnamefont {S.~C.}\ \bibnamefont
  {Pieper}}, \bibinfo {author} {\bibfnamefont {K.}~\bibnamefont {Varga}},\ and\
  \bibinfo {author} {\bibfnamefont {R.~B.}\ \bibnamefont {Wiringa}},\
  }\href@noop {} {\bibfield  {journal} {\bibinfo  {journal} {Phys. Rev. C}\
  }\textbf {\bibinfo {volume} {66}},\ \bibinfo {pages} {044310} (\bibinfo
  {year} {2002})}\BibitemShut {NoStop}\bibitem [{\citenamefont {Whitehead}\ \emph {et~al.}(1977)\citenamefont
  {Whitehead}, \citenamefont {Watt}, \citenamefont {Cole},\ and\ \citenamefont
  {Morrison}}]{anp-9-1977-123-Whitehead}\BibitemOpen
  \bibfield  {author} {\bibinfo {author} {\bibfnamefont {R.~R.}\ \bibnamefont
  {Whitehead}}, \bibinfo {author} {\bibfnamefont {A.}~\bibnamefont {Watt}},
  \bibinfo {author} {\bibfnamefont {B.~J.}\ \bibnamefont {Cole}},\ and\
  \bibinfo {author} {\bibfnamefont {I.}~\bibnamefont {Morrison}},\ }\href@noop
  {} {\bibfield  {journal} {\bibinfo  {journal} {Adv. Nucl. Phys.}\ }\textbf
  {\bibinfo {volume} {9}},\ \bibinfo {pages} {123} (\bibinfo {year}
  {1977})}\BibitemShut {NoStop}\bibitem [{\citenamefont {Brussard}\ and\ \citenamefont
  {Glaudemans}(1977)}]{Brussard1977}\BibitemOpen
  \bibfield  {author} {\bibinfo {author} {\bibfnamefont {P.}~\bibnamefont
  {Brussard}}\ and\ \bibinfo {author} {\bibfnamefont {P.}~\bibnamefont
  {Glaudemans}},\ }\href@noop {} {\emph {\bibinfo {title} {Shell-model
  applications in nuclear spectroscopy}}}\ (\bibinfo  {publisher}
  {North-Holland Publishing Co., Amsterdam},\ \bibinfo {year}
  {1977})\BibitemShut {NoStop}\bibitem [{\citenamefont {Caprio}\ \emph {et~al.}(2012)\citenamefont {Caprio},
  \citenamefont {Maris},\ and\ \citenamefont
  {Vary}}]{prc-86-2012-034312-Caprio}\BibitemOpen
  \bibfield  {author} {\bibinfo {author} {\bibfnamefont {M.~A.}\ \bibnamefont
  {Caprio}}, \bibinfo {author} {\bibfnamefont {P.}~\bibnamefont {Maris}},\ and\
  \bibinfo {author} {\bibfnamefont {J.~P.}\ \bibnamefont {Vary}},\ }\href@noop
  {} {\bibfield  {journal} {\bibinfo  {journal} {Phys. Rev. C}\ }\textbf
  {\bibinfo {volume} {86}},\ \bibinfo {pages} {034312} (\bibinfo {year}
  {2012})}\BibitemShut {NoStop}\bibitem [{\citenamefont {Caprio}\ \emph {et~al.}(2014)\citenamefont {Caprio},
  \citenamefont {Maris},\ and\ \citenamefont
  {Vary}}]{prc-90-2014-034305-Caprio}\BibitemOpen
  \bibfield  {author} {\bibinfo {author} {\bibfnamefont {M.~A.}\ \bibnamefont
  {Caprio}}, \bibinfo {author} {\bibfnamefont {P.}~\bibnamefont {Maris}},\ and\
  \bibinfo {author} {\bibfnamefont {J.~P.}\ \bibnamefont {Vary}},\ }\href@noop
  {} {\bibfield  {journal} {\bibinfo  {journal} {Phys. Rev. C}\ }\textbf
  {\bibinfo {volume} {90}},\ \bibinfo {pages} {034305} (\bibinfo {year}
  {2014})}\BibitemShut {NoStop}\bibitem [{\citenamefont {Caprio}\ \emph {et~al.}(2021)\citenamefont {Caprio},
  \citenamefont {Fasano}, \citenamefont {Maris},\ and\ \citenamefont
  {McCoy}}]{prc-104-2021-034319-Caprio}\BibitemOpen
  \bibfield  {author} {\bibinfo {author} {\bibfnamefont {M.~A.}\ \bibnamefont
  {Caprio}}, \bibinfo {author} {\bibfnamefont {P.~J.}\ \bibnamefont {Fasano}},
  \bibinfo {author} {\bibfnamefont {P.}~\bibnamefont {Maris}},\ and\ \bibinfo
  {author} {\bibfnamefont {A.~E.}\ \bibnamefont {McCoy}},\ }\href
  {https://doi.org/10.1103/PhysRevC.104.034319} {\bibfield  {journal} {\bibinfo
   {journal} {Phys. Rev. C}\ }\textbf {\bibinfo {volume} {104}},\ \bibinfo
  {pages} {034319} (\bibinfo {year} {2021})}\BibitemShut {NoStop}\bibitem [{\citenamefont {Fasano}\ \emph {et~al.}(2022)\citenamefont {Fasano},
  \citenamefont {Constantinou}, \citenamefont {Caprio}, \citenamefont {Maris},\
  and\ \citenamefont {Vary}}]{prc-105-2022-054301-Fasano}\BibitemOpen
  \bibfield  {author} {\bibinfo {author} {\bibfnamefont {P.~J.}\ \bibnamefont
  {Fasano}}, \bibinfo {author} {\bibfnamefont {C.}~\bibnamefont
  {Constantinou}}, \bibinfo {author} {\bibfnamefont {M.~A.}\ \bibnamefont
  {Caprio}}, \bibinfo {author} {\bibfnamefont {P.}~\bibnamefont {Maris}},\ and\
  \bibinfo {author} {\bibfnamefont {J.~P.}\ \bibnamefont {Vary}},\ }\href
  {https://doi.org/10.1103/PhysRevC.105.054301} {\bibfield  {journal} {\bibinfo
   {journal} {Phys. Rev. C}\ }\textbf {\bibinfo {volume} {105}},\ \bibinfo
  {pages} {054301} (\bibinfo {year} {2022})}\BibitemShut {NoStop}\bibitem [{\citenamefont {Caprio}(2022)}]{prc-105-2022-L061302-Caprio}\BibitemOpen
  \bibfield  {author} {\bibinfo {author} {\bibfnamefont {M.~A.}\ \bibnamefont
  {Caprio}},\ }\href {https://doi.org/10.1103/PhysRevC.105.L061302} {\bibfield
  {journal} {\bibinfo  {journal} {Phys. Rev. C}\ }\textbf {\bibinfo {volume}
  {105}},\ \bibinfo {pages} {L061302} (\bibinfo {year} {2022})}\BibitemShut
  {NoStop}\bibitem [{\citenamefont {Caprio}\ and\ \citenamefont
  {Fasano}(2022)}]{prc-106-2022-034320-Caprio}\BibitemOpen
  \bibfield  {author} {\bibinfo {author} {\bibfnamefont {M.~A.}\ \bibnamefont
  {Caprio}}\ and\ \bibinfo {author} {\bibfnamefont {P.~J.}\ \bibnamefont
  {Fasano}},\ }\href {https://doi.org/10.1103/PhysRevC.106.034320} {\bibfield
  {journal} {\bibinfo  {journal} {Phys. Rev. C}\ }\textbf {\bibinfo {volume}
  {106}},\ \bibinfo {pages} {034320} (\bibinfo {year} {2022})}\BibitemShut
  {NoStop}\bibitem [{\citenamefont {Pervin}\ \emph {et~al.}(2007)\citenamefont {Pervin},
  \citenamefont {Pieper},\ and\ \citenamefont
  {Wiringa}}]{prc-76-2007-064319-Pervin}\BibitemOpen
  \bibfield  {author} {\bibinfo {author} {\bibfnamefont {M.}~\bibnamefont
  {Pervin}}, \bibinfo {author} {\bibfnamefont {S.~C.}\ \bibnamefont {Pieper}},\
  and\ \bibinfo {author} {\bibfnamefont {R.~B.}\ \bibnamefont {Wiringa}},\
  }\href {https://doi.org/10.1103/PhysRevC.76.064319} {\bibfield  {journal}
  {\bibinfo  {journal} {Phys. Rev. C}\ }\textbf {\bibinfo {volume} {76}},\
  \bibinfo {pages} {064319} (\bibinfo {year} {2007})}\BibitemShut {NoStop}\bibitem [{\citenamefont {Bogner}\ \emph {et~al.}(2008)\citenamefont {Bogner}
  \emph {et~al.}}]{npa-801-2008-21-Bogner}\BibitemOpen
  \bibfield  {author} {\bibinfo {author} {\bibfnamefont {S.~K.}\ \bibnamefont
  {Bogner}} \emph {et~al.},\ }\href@noop {} {\bibfield  {journal} {\bibinfo
  {journal} {Nucl. Phys. A}\ }\textbf {\bibinfo {volume} {801}},\ \bibinfo
  {pages} {21} (\bibinfo {year} {2008})}\BibitemShut {NoStop}\bibitem [{\citenamefont {Maris}\ and\ \citenamefont
  {Vary}(2013)}]{ijmpe-22-2013-1330016-Maris}\BibitemOpen
  \bibfield  {author} {\bibinfo {author} {\bibfnamefont {P.}~\bibnamefont
  {Maris}}\ and\ \bibinfo {author} {\bibfnamefont {J.~P.}\ \bibnamefont
  {Vary}},\ }\href@noop {} {\bibfield  {journal} {\bibinfo  {journal} {Int. J.
  Mod. Phys. E}\ }\textbf {\bibinfo {volume} {22}},\ \bibinfo {pages} {1330016}
  (\bibinfo {year} {2013})}\BibitemShut {NoStop}\bibitem [{\citenamefont {Carlson}\ \emph {et~al.}(2015)\citenamefont
  {Carlson}, \citenamefont {Gandolfi}, \citenamefont {Pederiva}, \citenamefont
  {Pieper}, \citenamefont {Schiavilla}, \citenamefont {Schmidt},\ and\
  \citenamefont {Wiringa}}]{rmp-87-2015-1067-Carlson}\BibitemOpen
  \bibfield  {author} {\bibinfo {author} {\bibfnamefont {J.}~\bibnamefont
  {Carlson}}, \bibinfo {author} {\bibfnamefont {S.}~\bibnamefont {Gandolfi}},
  \bibinfo {author} {\bibfnamefont {F.}~\bibnamefont {Pederiva}}, \bibinfo
  {author} {\bibfnamefont {S.~C.}\ \bibnamefont {Pieper}}, \bibinfo {author}
  {\bibfnamefont {R.}~\bibnamefont {Schiavilla}}, \bibinfo {author}
  {\bibfnamefont {K.~E.}\ \bibnamefont {Schmidt}},\ and\ \bibinfo {author}
  {\bibfnamefont {R.~B.}\ \bibnamefont {Wiringa}},\ }\href
  {https://doi.org/10.1103/RevModPhys.87.1067} {\bibfield  {journal} {\bibinfo
  {journal} {Rev. Mod. Phys.}\ }\textbf {\bibinfo {volume} {87}},\ \bibinfo
  {pages} {1067} (\bibinfo {year} {2015})}\BibitemShut {NoStop}\bibitem [{\citenamefont {Odell}\ \emph {et~al.}(2016)\citenamefont {Odell},
  \citenamefont {Papenbrock},\ and\ \citenamefont
  {Platter}}]{prc-93-2016-044331-Odell}\BibitemOpen
  \bibfield  {author} {\bibinfo {author} {\bibfnamefont {D.}~\bibnamefont
  {Odell}}, \bibinfo {author} {\bibfnamefont {T.}~\bibnamefont {Papenbrock}},\
  and\ \bibinfo {author} {\bibfnamefont {L.}~\bibnamefont {Platter}},\ }\href
  {https://doi.org/10.1103/PhysRevC.93.044331} {\bibfield  {journal} {\bibinfo
  {journal} {Phys. Rev. C}\ }\textbf {\bibinfo {volume} {93}},\ \bibinfo
  {pages} {044331} (\bibinfo {year} {2016})}\BibitemShut {NoStop}\bibitem [{\citenamefont {Roth}\ and\ \citenamefont
  {Petri}(2024)}]{ptrsla-382-2024-20230119-Roth}\BibitemOpen
  \bibfield  {author} {\bibinfo {author} {\bibfnamefont {R.}~\bibnamefont
  {Roth}}\ and\ \bibinfo {author} {\bibfnamefont {M.}~\bibnamefont {Petri}},\
  }\href {https://doi.org/10.1098/rsta.2023.0119} {\bibfield  {journal}
  {\bibinfo  {journal} {Phil. Trans. Roy. Soc. (London) A}\ }\textbf {\bibinfo
  {volume} {382}},\ \bibinfo {pages} {20230119} (\bibinfo {year}
  {2024})}\BibitemShut {NoStop}\bibitem [{\citenamefont {Scalesi}\ \emph {et~al.}(2025)\citenamefont
  {Scalesi}, \citenamefont {Duguet}, \citenamefont {Frosini},\ and\
  \citenamefont {Soma}}]{epja-61-2025-1-Scalesi}\BibitemOpen
  \bibfield  {author} {\bibinfo {author} {\bibfnamefont {A.}~\bibnamefont
  {Scalesi}}, \bibinfo {author} {\bibfnamefont {T.}~\bibnamefont {Duguet}},
  \bibinfo {author} {\bibfnamefont {M.}~\bibnamefont {Frosini}},\ and\ \bibinfo
  {author} {\bibfnamefont {V.}~\bibnamefont {Soma}},\ }\href
  {https://doi.org/10.1140/epja/s10050-024-01466-5} {\bibfield  {journal}
  {\bibinfo  {journal} {Eur. Phys. J. A}\ }\textbf {\bibinfo {volume} {61}},\
  \bibinfo {pages} {1} (\bibinfo {year} {2025})}\BibitemShut {NoStop}\bibitem [{\citenamefont {Tichai}\ \emph {et~al.}(2019)\citenamefont {Tichai},
  \citenamefont {Muller}, \citenamefont {Vobig},\ and\ \citenamefont
  {Roth}}]{prc-99-2019-034321-Tichai}\BibitemOpen
  \bibfield  {author} {\bibinfo {author} {\bibfnamefont {A.}~\bibnamefont
  {Tichai}}, \bibinfo {author} {\bibfnamefont {J.}~\bibnamefont {Muller}},
  \bibinfo {author} {\bibfnamefont {K.}~\bibnamefont {Vobig}},\ and\ \bibinfo
  {author} {\bibfnamefont {R.}~\bibnamefont {Roth}},\ }\href
  {https://doi.org/DOI: https://doi.org/10.1103/PhysRevC.99.034321} {\bibfield
  {journal} {\bibinfo  {journal} {Phys. Rev. C}\ }\textbf {\bibinfo {volume}
  {99}},\ \bibinfo {pages} {034321} (\bibinfo {year} {2019})}\BibitemShut
  {NoStop}\bibitem [{\citenamefont {Hoppe}\ \emph {et~al.}(2021)\citenamefont {Hoppe},
  \citenamefont {Tichai}, \citenamefont {Heinz}, \citenamefont {Hebeler},\ and\
  \citenamefont {Schwenk}}]{prc-103-2021-014321-Hoppe}\BibitemOpen
  \bibfield  {author} {\bibinfo {author} {\bibfnamefont {J.}~\bibnamefont
  {Hoppe}}, \bibinfo {author} {\bibfnamefont {A.}~\bibnamefont {Tichai}},
  \bibinfo {author} {\bibfnamefont {M.}~\bibnamefont {Heinz}}, \bibinfo
  {author} {\bibfnamefont {K.}~\bibnamefont {Hebeler}},\ and\ \bibinfo {author}
  {\bibfnamefont {A.}~\bibnamefont {Schwenk}},\ }\href@noop {} {\bibfield
  {journal} {\bibinfo  {journal} {Phys. Rev. C}\ }\textbf {\bibinfo {volume}
  {103}},\ \bibinfo {pages} {014321} (\bibinfo {year} {2021})}\BibitemShut
  {NoStop}\bibitem [{\citenamefont {Dytrych}\ \emph
  {et~al.}(2008{\natexlab{a}})\citenamefont {Dytrych}, \citenamefont
  {Sviratcheva}, \citenamefont {Draayer}, \citenamefont {Bahri},\ and\
  \citenamefont {Vary}}]{jpg-35-2008-123101-Dytrych}\BibitemOpen
  \bibfield  {author} {\bibinfo {author} {\bibfnamefont {T.}~\bibnamefont
  {Dytrych}}, \bibinfo {author} {\bibfnamefont {K.~D.}\ \bibnamefont
  {Sviratcheva}}, \bibinfo {author} {\bibfnamefont {J.~P.}\ \bibnamefont
  {Draayer}}, \bibinfo {author} {\bibfnamefont {C.}~\bibnamefont {Bahri}},\
  and\ \bibinfo {author} {\bibfnamefont {J.~P.}\ \bibnamefont {Vary}},\
  }\href@noop {} {\bibfield  {journal} {\bibinfo  {journal} {J. Phys. G}\
  }\textbf {\bibinfo {volume} {35}},\ \bibinfo {pages} {123101} (\bibinfo
  {year} {2008}{\natexlab{a}})}\BibitemShut {NoStop}\bibitem [{\citenamefont {Dytrych}\ \emph {et~al.}(2016)\citenamefont
  {Dytrych}, \citenamefont {Maris}, \citenamefont {Launey}, \citenamefont
  {Draayer}, \citenamefont {Vary}, \citenamefont {Langr}, \citenamefont
  {Saule}, \citenamefont {Caprio}, \citenamefont {Catalyurek},\ and\
  \citenamefont {Sosonkina}}]{cpc-207-2016-202-Dytrych}\BibitemOpen
  \bibfield  {author} {\bibinfo {author} {\bibfnamefont {T.}~\bibnamefont
  {Dytrych}}, \bibinfo {author} {\bibfnamefont {P.}~\bibnamefont {Maris}},
  \bibinfo {author} {\bibfnamefont {K.~D.}\ \bibnamefont {Launey}}, \bibinfo
  {author} {\bibfnamefont {J.~P.}\ \bibnamefont {Draayer}}, \bibinfo {author}
  {\bibfnamefont {J.~P.}\ \bibnamefont {Vary}}, \bibinfo {author}
  {\bibfnamefont {D.}~\bibnamefont {Langr}}, \bibinfo {author} {\bibfnamefont
  {E.}~\bibnamefont {Saule}}, \bibinfo {author} {\bibfnamefont {M.~A.}\
  \bibnamefont {Caprio}}, \bibinfo {author} {\bibfnamefont {U.}~\bibnamefont
  {Catalyurek}},\ and\ \bibinfo {author} {\bibfnamefont {M.}~\bibnamefont
  {Sosonkina}},\ }\href@noop {} {\bibfield  {journal} {\bibinfo  {journal}
  {Comput. Phys. Commun.}\ }\textbf {\bibinfo {volume} {207}},\ \bibinfo
  {pages} {202} (\bibinfo {year} {2016})}\BibitemShut {NoStop}\bibitem [{\citenamefont {Launey}\ \emph {et~al.}(2016)\citenamefont {Launey},
  \citenamefont {Dytrych},\ and\ \citenamefont
  {Draayer}}]{ppnp-89-2016-101-Launey}\BibitemOpen
  \bibfield  {author} {\bibinfo {author} {\bibfnamefont {K.~D.}\ \bibnamefont
  {Launey}}, \bibinfo {author} {\bibfnamefont {T.}~\bibnamefont {Dytrych}},\
  and\ \bibinfo {author} {\bibfnamefont {J.~P.}\ \bibnamefont {Draayer}},\
  }\href@noop {} {\bibfield  {journal} {\bibinfo  {journal} {Prog. Part. Nucl.
  Phys.}\ }\textbf {\bibinfo {volume} {89}},\ \bibinfo {pages} {101} (\bibinfo
  {year} {2016})}\BibitemShut {NoStop}\bibitem [{\citenamefont {McCoy}\ \emph {et~al.}(2018)\citenamefont {McCoy},
  \citenamefont {Caprio},\ and\ \citenamefont
  {Dytrych}}]{aarsscps-3-2018-17-McCoy}\BibitemOpen
  \bibfield  {author} {\bibinfo {author} {\bibfnamefont {A.~E.}\ \bibnamefont
  {McCoy}}, \bibinfo {author} {\bibfnamefont {M.~A.}\ \bibnamefont {Caprio}},\
  and\ \bibinfo {author} {\bibfnamefont {T.}~\bibnamefont {Dytrych}},\ }\href
  {http://www.aos.ro/wp-content/anale/PCVol3Nr1Art.2.pdf} {\bibfield  {journal}
  {\bibinfo  {journal} {Ann. Acad. Rom. Sci. Ser. Chem. Phys. Sci.}\ }\textbf
  {\bibinfo {volume} {3}},\ \bibinfo {pages} {17} (\bibinfo {year}
  {2018})}\BibitemShut {NoStop}\bibitem [{\citenamefont {McCoy}(2018)}]{McCoy2018}\BibitemOpen
  \bibfield  {author} {\bibinfo {author} {\bibfnamefont {A.~E.}\ \bibnamefont
  {McCoy}},\ }\emph {\bibinfo {title} {\textit{Ab initio} multi-irrep
  symplectic no-core configuration interaction calculations}},\ \href
  {https://curate.nd.edu/show/pz50gt57p16} {Ph.D. thesis},\ \bibinfo  {school}
  {University of Notre Dame} (\bibinfo {year} {2018})\BibitemShut {NoStop}\bibitem [{\citenamefont
  {Elliott}(1958{\natexlab{a}})}]{prsla-245-1958-128-Elliott}\BibitemOpen
  \bibfield  {author} {\bibinfo {author} {\bibfnamefont {J.~P.}\ \bibnamefont
  {Elliott}},\ }\href {https://doi.org/10.1098/rspa.1958.0072} {\bibfield
  {journal} {\bibinfo  {journal} {Proc. R. Soc. A}\ }\textbf {\bibinfo {volume}
  {245}},\ \bibinfo {pages} {128} (\bibinfo {year}
  {1958}{\natexlab{a}})}\BibitemShut {NoStop}\bibitem [{\citenamefont
  {Elliott}(1958{\natexlab{b}})}]{prsla-245-1958-562-Elliott}\BibitemOpen
  \bibfield  {author} {\bibinfo {author} {\bibfnamefont {J.~P.}\ \bibnamefont
  {Elliott}},\ }\href {https://doi.org/10.1098/rspa.1958.0101} {\bibfield
  {journal} {\bibinfo  {journal} {Proc. R. Soc. A}\ }\textbf {\bibinfo {volume}
  {245}},\ \bibinfo {pages} {562} (\bibinfo {year}
  {1958}{\natexlab{b}})}\BibitemShut {NoStop}\bibitem [{\citenamefont {Elliott}\ and\ \citenamefont
  {Harvey}(1963)}]{prsla-272-1963-557-Elliott}\BibitemOpen
  \bibfield  {author} {\bibinfo {author} {\bibfnamefont {J.~P.}\ \bibnamefont
  {Elliott}}\ and\ \bibinfo {author} {\bibfnamefont {M.}~\bibnamefont
  {Harvey}},\ }\href {https://doi.org/10.1098/rspa.1963.0071} {\bibfield
  {journal} {\bibinfo  {journal} {Proc. R. Soc. A}\ }\textbf {\bibinfo {volume}
  {272}},\ \bibinfo {pages} {557} (\bibinfo {year} {1963})}\BibitemShut
  {NoStop}\bibitem [{\citenamefont {Harvey}(1968)}]{anp-1-67-1968-Harvey}\BibitemOpen
  \bibfield  {author} {\bibinfo {author} {\bibfnamefont {M.}~\bibnamefont
  {Harvey}},\ }\bibinfo {title} {The nuclear $\mathit{SU}_3$ model},\ in\ \href
  {https://doi.org/10.1007/978-1-4757-0103-6_2} {\emph {\bibinfo {booktitle}
  {Adv. Nucl. Phys.}}},\ Vol.~\bibinfo {volume} {1},\ \bibinfo {editor} {edited
  by\ \bibinfo {editor} {\bibfnamefont {M.}~\bibnamefont {Baranger}}\ and\
  \bibinfo {editor} {\bibfnamefont {E.}~\bibnamefont {Vogt}}}\ (\bibinfo
  {publisher} {Plenum},\ \bibinfo {address} {New York},\ \bibinfo {year}
  {1968})\ p.~\bibinfo {pages} {67}\BibitemShut {NoStop}\bibitem [{\citenamefont {Rosensteel}\ and\ \citenamefont
  {Rowe}(1977{\natexlab{a}})}]{prl-38-1977-10-Rosensteel}\BibitemOpen
  \bibfield  {author} {\bibinfo {author} {\bibfnamefont {G.}~\bibnamefont
  {Rosensteel}}\ and\ \bibinfo {author} {\bibfnamefont {D.~J.}\ \bibnamefont
  {Rowe}},\ }\href {https://doi.org/10.1103/PhysRevLett.38.10} {\bibfield
  {journal} {\bibinfo  {journal} {Phys. Rev. Lett.}\ }\textbf {\bibinfo
  {volume} {38}},\ \bibinfo {pages} {10} (\bibinfo {year}
  {1977}{\natexlab{a}})}\BibitemShut {NoStop}\bibitem [{\citenamefont {Rosensteel}\ and\ \citenamefont
  {Rowe}(1980)}]{ap-126-1980-343-Rosensteel}\BibitemOpen
  \bibfield  {author} {\bibinfo {author} {\bibfnamefont {G.}~\bibnamefont
  {Rosensteel}}\ and\ \bibinfo {author} {\bibfnamefont {D.~J.}\ \bibnamefont
  {Rowe}},\ }\href@noop {} {\bibfield  {journal} {\bibinfo  {journal} {Ann.
  Phys. (N. Y.)}\ }\textbf {\bibinfo {volume} {126}},\ \bibinfo {pages} {343}
  (\bibinfo {year} {1980})}\BibitemShut {NoStop}\bibitem [{\citenamefont {Rowe}(1985)}]{rpp-48-1985-1419-Rowe}\BibitemOpen
  \bibfield  {author} {\bibinfo {author} {\bibfnamefont {D.~J.}\ \bibnamefont
  {Rowe}},\ }\href@noop {} {\bibfield  {journal} {\bibinfo  {journal} {Rep.
  Prog. Phys.}\ }\textbf {\bibinfo {volume} {48}},\ \bibinfo {pages} {1419}
  (\bibinfo {year} {1985})}\BibitemShut {NoStop}\bibitem [{\citenamefont {Rosensteel}\ and\ \citenamefont
  {Rowe}(1981)}]{prl-47-1981-223-Rosensteel}\BibitemOpen
  \bibfield  {author} {\bibinfo {author} {\bibfnamefont {G.}~\bibnamefont
  {Rosensteel}}\ and\ \bibinfo {author} {\bibfnamefont {D.~J.}\ \bibnamefont
  {Rowe}},\ }\href@noop {} {\bibfield  {journal} {\bibinfo  {journal} {Phys.
  Rev. Lett.}\ }\textbf {\bibinfo {volume} {47}},\ \bibinfo {pages} {223}
  (\bibinfo {year} {1981})}\BibitemShut {NoStop}\bibitem [{\citenamefont {{Le Blanc}}\ \emph {et~al.}(1984)\citenamefont {{Le
  Blanc}}, \citenamefont {Carvalho},\ and\ \citenamefont
  {Rowe}}]{plb-140-1984-155-Le_Blanc}\BibitemOpen
  \bibfield  {author} {\bibinfo {author} {\bibfnamefont {R.}~\bibnamefont {{Le
  Blanc}}}, \bibinfo {author} {\bibfnamefont {J.}~\bibnamefont {Carvalho}},\
  and\ \bibinfo {author} {\bibfnamefont {D.~J.}\ \bibnamefont {Rowe}},\
  }\href@noop {} {\bibfield  {journal} {\bibinfo  {journal} {Phys. Lett. B}\
  }\textbf {\bibinfo {volume} {140}},\ \bibinfo {pages} {155} (\bibinfo {year}
  {1984})}\BibitemShut {NoStop}\bibitem [{\citenamefont {Launey}\ \emph {et~al.}(2020)\citenamefont {Launey},
  \citenamefont {Dytrych}, \citenamefont {Sargsyan}, \citenamefont {Baker},\
  and\ \citenamefont {Draayer}}]{epj-229-2429-2020-Launey}\BibitemOpen
  \bibfield  {author} {\bibinfo {author} {\bibfnamefont {K.~D.}\ \bibnamefont
  {Launey}}, \bibinfo {author} {\bibfnamefont {T.}~\bibnamefont {Dytrych}},
  \bibinfo {author} {\bibfnamefont {G.~H.}\ \bibnamefont {Sargsyan}}, \bibinfo
  {author} {\bibfnamefont {R.~B.}\ \bibnamefont {Baker}},\ and\ \bibinfo
  {author} {\bibfnamefont {J.~P.}\ \bibnamefont {Draayer}},\ }\href@noop {}
  {\bibfield  {journal} {\bibinfo  {journal} {Eur. Phys. J. Special Topics}\
  }\textbf {\bibinfo {volume} {229}},\ \bibinfo {pages} {2429} (\bibinfo {year}
  {2020})}\BibitemShut {NoStop}\bibitem [{\citenamefont {Becker}\ \emph {et~al.}(2023)\citenamefont {Becker},
  \citenamefont {Launey}, \citenamefont {Erkstrom},\ and\ \citenamefont
  {Dytrych}}]{fp-11-2023-1-Becker}\BibitemOpen
  \bibfield  {author} {\bibinfo {author} {\bibfnamefont {K.~S.}\ \bibnamefont
  {Becker}}, \bibinfo {author} {\bibfnamefont {K.~D.}\ \bibnamefont {Launey}},
  \bibinfo {author} {\bibfnamefont {A.}~\bibnamefont {Erkstrom}},\ and\
  \bibinfo {author} {\bibfnamefont {T.}~\bibnamefont {Dytrych}},\ }\href
  {https://doi.org/10.3389/fphy.2023.1064601} {\bibfield  {journal} {\bibinfo
  {journal} {Front. Phys.}\ }\textbf {\bibinfo {volume} {11}},\ \bibinfo
  {pages} {1} (\bibinfo {year} {2023})}\BibitemShut {NoStop}\bibitem [{\citenamefont {Dytrych}\ \emph
  {et~al.}(2008{\natexlab{b}})\citenamefont {Dytrych}, \citenamefont
  {Sviratcheva}, \citenamefont {Bahri}, \citenamefont {Draayer},\ and\
  \citenamefont {Vary}}]{jpg-35-2008-095101-Dytrych}\BibitemOpen
  \bibfield  {author} {\bibinfo {author} {\bibfnamefont {T.}~\bibnamefont
  {Dytrych}}, \bibinfo {author} {\bibfnamefont {K.~D.}\ \bibnamefont
  {Sviratcheva}}, \bibinfo {author} {\bibfnamefont {C.}~\bibnamefont {Bahri}},
  \bibinfo {author} {\bibfnamefont {J.~P.}\ \bibnamefont {Draayer}},\ and\
  \bibinfo {author} {\bibfnamefont {J.~P.}\ \bibnamefont {Vary}},\ }\href@noop
  {} {\bibfield  {journal} {\bibinfo  {journal} {J. Phys. G}\ }\textbf
  {\bibinfo {volume} {35}},\ \bibinfo {pages} {095101} (\bibinfo {year}
  {2008}{\natexlab{b}})}\BibitemShut {NoStop}\bibitem [{\citenamefont {Dytrych}\ \emph
  {et~al.}(2007{\natexlab{a}})\citenamefont {Dytrych}, \citenamefont
  {Sviratcheva}, \citenamefont {Bahri},\ and\ \citenamefont
  {Draayer}}]{prc-76-2007-014315-Dytrych}\BibitemOpen
  \bibfield  {author} {\bibinfo {author} {\bibfnamefont {T.}~\bibnamefont
  {Dytrych}}, \bibinfo {author} {\bibfnamefont {K.~D.}\ \bibnamefont
  {Sviratcheva}}, \bibinfo {author} {\bibfnamefont {C.}~\bibnamefont {Bahri}},\
  and\ \bibinfo {author} {\bibfnamefont {J.~P.}\ \bibnamefont {Draayer}},\
  }\href@noop {} {\bibfield  {journal} {\bibinfo  {journal} {Phys. Rev. C}\
  }\textbf {\bibinfo {volume} {76}},\ \bibinfo {pages} {014315} (\bibinfo
  {year} {2007}{\natexlab{a}})}\BibitemShut {NoStop}\bibitem [{\citenamefont {Molchanov}\ \emph {et~al.}(2022)\citenamefont
  {Molchanov}, \citenamefont {Launey}, \citenamefont {Mercenne}, \citenamefont
  {Sargsyan}, \citenamefont {Dytrych},\ and\ \citenamefont
  {Draayer}}]{prc-105-2022-034306-Molchanov}\BibitemOpen
  \bibfield  {author} {\bibinfo {author} {\bibfnamefont {O.~M.}\ \bibnamefont
  {Molchanov}}, \bibinfo {author} {\bibfnamefont {K.~D.}\ \bibnamefont
  {Launey}}, \bibinfo {author} {\bibfnamefont {A.}~\bibnamefont {Mercenne}},
  \bibinfo {author} {\bibfnamefont {G.~H.}\ \bibnamefont {Sargsyan}}, \bibinfo
  {author} {\bibfnamefont {T.}~\bibnamefont {Dytrych}},\ and\ \bibinfo {author}
  {\bibfnamefont {J.~P.}\ \bibnamefont {Draayer}},\ }\href
  {https://doi.org/10.1103/PhysRevC.105.034306} {\bibfield  {journal} {\bibinfo
   {journal} {Phys. Rev. C}\ }\textbf {\bibinfo {volume} {105}},\ \bibinfo
  {pages} {034306} (\bibinfo {year} {2022})}\BibitemShut {NoStop}\bibitem [{\citenamefont {Dytrych}\ \emph
  {et~al.}(2007{\natexlab{b}})\citenamefont {Dytrych}, \citenamefont
  {Sviratcheva}, \citenamefont {Bahri}, \citenamefont {Draayer},\ and\
  \citenamefont {Vary}}]{prl-98-2007-162503-Dytrych}\BibitemOpen
  \bibfield  {author} {\bibinfo {author} {\bibfnamefont {T.}~\bibnamefont
  {Dytrych}}, \bibinfo {author} {\bibfnamefont {K.~D.}\ \bibnamefont
  {Sviratcheva}}, \bibinfo {author} {\bibfnamefont {C.}~\bibnamefont {Bahri}},
  \bibinfo {author} {\bibfnamefont {J.~P.}\ \bibnamefont {Draayer}},\ and\
  \bibinfo {author} {\bibfnamefont {J.~P.}\ \bibnamefont {Vary}},\ }\href@noop
  {} {\bibfield  {journal} {\bibinfo  {journal} {Phys. Rev. Lett.}\ }\textbf
  {\bibinfo {volume} {98}},\ \bibinfo {pages} {162503} (\bibinfo {year}
  {2007}{\natexlab{b}})}\BibitemShut {NoStop}\bibitem [{\citenamefont {Dytrych}\ \emph {et~al.}(2020)\citenamefont
  {Dytrych}, \citenamefont {Launey}, \citenamefont {Draayer}, \citenamefont
  {Rowe}, \citenamefont {Wood}, \citenamefont {Rosensteel}, \citenamefont
  {Bahri}, \citenamefont {Langr},\ and\ \citenamefont
  {Baker}}]{prl-124-2020-042501-Dytrych}\BibitemOpen
  \bibfield  {author} {\bibinfo {author} {\bibfnamefont {T.}~\bibnamefont
  {Dytrych}}, \bibinfo {author} {\bibfnamefont {K.~D.}\ \bibnamefont {Launey}},
  \bibinfo {author} {\bibfnamefont {J.~P.}\ \bibnamefont {Draayer}}, \bibinfo
  {author} {\bibfnamefont {D.~J.}\ \bibnamefont {Rowe}}, \bibinfo {author}
  {\bibfnamefont {J.~L.}\ \bibnamefont {Wood}}, \bibinfo {author}
  {\bibfnamefont {G.}~\bibnamefont {Rosensteel}}, \bibinfo {author}
  {\bibfnamefont {C.}~\bibnamefont {Bahri}}, \bibinfo {author} {\bibfnamefont
  {D.}~\bibnamefont {Langr}},\ and\ \bibinfo {author} {\bibfnamefont {R.~B.}\
  \bibnamefont {Baker}},\ }\href@noop {} {\bibfield  {journal} {\bibinfo
  {journal} {Phys. Rev. Lett.}\ }\textbf {\bibinfo {volume} {124}},\ \bibinfo
  {pages} {042501} (\bibinfo {year} {2020})}\BibitemShut {NoStop}\bibitem [{\citenamefont {Rosensteel}\ and\ \citenamefont
  {Rowe}(1977{\natexlab{b}})}]{ap-104-1977-134-Rosensteel}\BibitemOpen
  \bibfield  {author} {\bibinfo {author} {\bibfnamefont {G.}~\bibnamefont
  {Rosensteel}}\ and\ \bibinfo {author} {\bibfnamefont {D.~J.}\ \bibnamefont
  {Rowe}},\ }\href@noop {} {\bibfield  {journal} {\bibinfo  {journal} {Ann. of
  Phys.}\ }\textbf {\bibinfo {volume} {104}},\ \bibinfo {pages} {134} (\bibinfo
  {year} {1977}{\natexlab{b}})}\BibitemShut {NoStop}\bibitem [{\citenamefont {Rosensteel}\ and\ \citenamefont
  {Rowe}(1979)}]{ap-123-1979-36-Rosensteel}\BibitemOpen
  \bibfield  {author} {\bibinfo {author} {\bibfnamefont {G.}~\bibnamefont
  {Rosensteel}}\ and\ \bibinfo {author} {\bibfnamefont {D.~J.}\ \bibnamefont
  {Rowe}},\ }\href@noop {} {\bibfield  {journal} {\bibinfo  {journal} {Ann.
  Phys. (N. Y.)}\ }\textbf {\bibinfo {volume} {123}},\ \bibinfo {pages} {36}
  (\bibinfo {year} {1979})}\BibitemShut {NoStop}\bibitem [{\citenamefont {Rowe}\ and\ \citenamefont
  {Rosensteel}(1980)}]{ap-126-1980-198-Rowe}\BibitemOpen
  \bibfield  {author} {\bibinfo {author} {\bibfnamefont {D.~J.}\ \bibnamefont
  {Rowe}}\ and\ \bibinfo {author} {\bibfnamefont {G.}~\bibnamefont
  {Rosensteel}},\ }\href@noop {} {\bibfield  {journal} {\bibinfo  {journal}
  {Ann. Phys. (N. Y.)}\ }\textbf {\bibinfo {volume} {126}},\ \bibinfo {pages}
  {198} (\bibinfo {year} {1980})}\BibitemShut {NoStop}\bibitem [{\citenamefont {Elliott}\ and\ \citenamefont
  {Wilsdon}(1968)}]{prsla-302-1968-509-Elliott}\BibitemOpen
  \bibfield  {author} {\bibinfo {author} {\bibfnamefont {J.~P.}\ \bibnamefont
  {Elliott}}\ and\ \bibinfo {author} {\bibfnamefont {C.~E.}\ \bibnamefont
  {Wilsdon}},\ }\href {https://doi.org/10.1098/rspa.1968.0033} {\bibfield
  {journal} {\bibinfo  {journal} {Proc. R. Soc. A}\ }\textbf {\bibinfo {volume}
  {302}},\ \bibinfo {pages} {509} (\bibinfo {year} {1968})}\BibitemShut
  {NoStop}\bibitem [{\citenamefont {Fu}\ \emph {et~al.}(2021)\citenamefont {Fu},
  \citenamefont {Johnson}, \citenamefont {Isacker},\ and\ \citenamefont
  {Ren}}]{prc-103-2021-L021302-Fu}\BibitemOpen
  \bibfield  {author} {\bibinfo {author} {\bibfnamefont {G.~J.}\ \bibnamefont
  {Fu}}, \bibinfo {author} {\bibfnamefont {C.~W.}\ \bibnamefont {Johnson}},
  \bibinfo {author} {\bibfnamefont {P.~F.}\ \bibnamefont {Isacker}},\ and\
  \bibinfo {author} {\bibfnamefont {Z.}~\bibnamefont {Ren}},\ }\href
  {https://doi.org/10.1103/PhysRevC.103.L021302} {\bibfield  {journal}
  {\bibinfo  {journal} {Phys. Rev. C}\ }\textbf {\bibinfo {volume} {103}},\
  \bibinfo {pages} {L021302} (\bibinfo {year} {2021})}\BibitemShut {NoStop}\bibitem [{\citenamefont {Parikh}(1978)}]{Parikh1978}\BibitemOpen
  \bibfield  {author} {\bibinfo {author} {\bibfnamefont {J.~C.}\ \bibnamefont
  {Parikh}},\ }\bibinfo {title} {Su(3) symmetry},\ in\ \href
  {https://doi.org/10.1007/978-1-4684-2376-1_10} {\emph {\bibinfo {booktitle}
  {Group Symmetries in Nuclear Structure}}}\ (\bibinfo  {publisher} {Springer
  US},\ \bibinfo {address} {Boston, MA},\ \bibinfo {year} {1978})\ pp.\
  \bibinfo {pages} {195--234}\BibitemShut {NoStop}\bibitem [{\citenamefont {Hirsch}\ \emph {et~al.}(1999)\citenamefont {Hirsch},
  \citenamefont {Hess}, \citenamefont {Hernandez}, \citenamefont {Vargas},
  \citenamefont {Beuschel},\ and\ \citenamefont
  {Draayer}}]{rmf-45-1999-86-Hirsch}\BibitemOpen
  \bibfield  {author} {\bibinfo {author} {\bibfnamefont {J.~G.}\ \bibnamefont
  {Hirsch}}, \bibinfo {author} {\bibfnamefont {P.~O.}\ \bibnamefont {Hess}},
  \bibinfo {author} {\bibfnamefont {L.}~\bibnamefont {Hernandez}}, \bibinfo
  {author} {\bibfnamefont {C.}~\bibnamefont {Vargas}}, \bibinfo {author}
  {\bibfnamefont {T.}~\bibnamefont {Beuschel}},\ and\ \bibinfo {author}
  {\bibfnamefont {J.~P.}\ \bibnamefont {Draayer}},\ }\href@noop {} {\bibfield
  {journal} {\bibinfo  {journal} {Rev. Mex. Fis.}\ }\textbf {\bibinfo {volume}
  {45}},\ \bibinfo {pages} {86} (\bibinfo {year} {1999})}\BibitemShut {NoStop}\bibitem [{\citenamefont {Wilsdon}(1965)}]{Wilsdon1965}\BibitemOpen
  \bibfield  {author} {\bibinfo {author} {\bibfnamefont {C.~E.}\ \bibnamefont
  {Wilsdon}},\ }\emph {\bibinfo {title} {A survey of the nuclear $s-d$ shell,
  using the su(3) coupling scheme.}},\ \href@noop {} {Ph.D. thesis},\ \bibinfo
  {school} {University of Sussex} (\bibinfo {year} {1965})\BibitemShut
  {NoStop}\bibitem [{\citenamefont {Raju}\ \emph {et~al.}(1973)\citenamefont {Raju},
  \citenamefont {Draayer},\ and\ \citenamefont
  {Hecht}}]{npa-202-1973-433-Raju}\BibitemOpen
  \bibfield  {author} {\bibinfo {author} {\bibfnamefont {R.}~\bibnamefont
  {Raju}}, \bibinfo {author} {\bibfnamefont {J.}~\bibnamefont {Draayer}},\ and\
  \bibinfo {author} {\bibfnamefont {K.}~\bibnamefont {Hecht}},\ }\href
  {https://doi.org/10.1016/0375-9474(73)90635-0} {\bibfield  {journal}
  {\bibinfo  {journal} {Nuclear Physics A}\ }\textbf {\bibinfo {volume}
  {202}},\ \bibinfo {pages} {433} (\bibinfo {year} {1973})}\BibitemShut
  {NoStop}\bibitem [{\citenamefont {Draayer}\ \emph {et~al.}(1982)\citenamefont
  {Draayer}, \citenamefont {Weeks},\ and\ \citenamefont
  {Hecht}}]{npa-381-1982-1-Draayer}\BibitemOpen
  \bibfield  {author} {\bibinfo {author} {\bibfnamefont {J.}~\bibnamefont
  {Draayer}}, \bibinfo {author} {\bibfnamefont {K.}~\bibnamefont {Weeks}},\
  and\ \bibinfo {author} {\bibfnamefont {K.}~\bibnamefont {Hecht}},\ }\href
  {https://doi.org/10.1016/0375-9474(82)90497-3} {\bibfield  {journal}
  {\bibinfo  {journal} {Nuclear Physics A}\ }\textbf {\bibinfo {volume}
  {381}},\ \bibinfo {pages} {1} (\bibinfo {year} {1982})}\BibitemShut {NoStop}\bibitem [{\citenamefont {Draayer}\ and\ \citenamefont
  {Weeks}(1984)}]{apny-156-1984-41-Draayer}\BibitemOpen
  \bibfield  {author} {\bibinfo {author} {\bibfnamefont {J.}~\bibnamefont
  {Draayer}}\ and\ \bibinfo {author} {\bibfnamefont {K.}~\bibnamefont
  {Weeks}},\ }\href {https://doi.org/10.1016/0003-4916(84)90210-0} {\bibfield
  {journal} {\bibinfo  {journal} {Annals of Physics}\ }\textbf {\bibinfo
  {volume} {156}},\ \bibinfo {pages} {41} (\bibinfo {year} {1984})}\BibitemShut
  {NoStop}\bibitem [{\citenamefont {Ginocchio}(1997)}]{prl-78-1997-436-Ginocchio}\BibitemOpen
  \bibfield  {author} {\bibinfo {author} {\bibfnamefont {J.~N.}\ \bibnamefont
  {Ginocchio}},\ }\href {https://doi.org/10.1103/PhysRevLett.78.436} {\bibfield
   {journal} {\bibinfo  {journal} {Phys. Rev. Lett.}\ }\textbf {\bibinfo
  {volume} {78}},\ \bibinfo {pages} {436} (\bibinfo {year} {1997})}\BibitemShut
  {NoStop}\bibitem [{\citenamefont {Bonatsos}\ \emph {et~al.}(2017)\citenamefont
  {Bonatsos}, \citenamefont {Assimakis}, \citenamefont {Minkov}, \citenamefont
  {Martinou}, \citenamefont {Cakirli}, \citenamefont {Casten},\ and\
  \citenamefont {Blaum}}]{prc-95-2017-064325-Bonatsos}\BibitemOpen
  \bibfield  {author} {\bibinfo {author} {\bibfnamefont {D.}~\bibnamefont
  {Bonatsos}}, \bibinfo {author} {\bibfnamefont {I.~E.}\ \bibnamefont
  {Assimakis}}, \bibinfo {author} {\bibfnamefont {N.}~\bibnamefont {Minkov}},
  \bibinfo {author} {\bibfnamefont {A.}~\bibnamefont {Martinou}}, \bibinfo
  {author} {\bibfnamefont {R.~B.}\ \bibnamefont {Cakirli}}, \bibinfo {author}
  {\bibfnamefont {R.~F.}\ \bibnamefont {Casten}},\ and\ \bibinfo {author}
  {\bibfnamefont {K.}~\bibnamefont {Blaum}},\ }\href
  {https://doi.org/10.1103/PhysRevC.95.064325} {\bibfield  {journal} {\bibinfo
  {journal} {Phys. Rev. C}\ }\textbf {\bibinfo {volume} {95}},\ \bibinfo
  {pages} {064325} (\bibinfo {year} {2017})}\BibitemShut {NoStop}\bibitem [{\citenamefont {Bonatsos}(2017)}]{epja-53-2017-148-Bonatsos}\BibitemOpen
  \bibfield  {author} {\bibinfo {author} {\bibfnamefont {D.}~\bibnamefont
  {Bonatsos}},\ }\href {https://doi.org/10.1140/epja/i2017-12346-x} {\bibfield
  {journal} {\bibinfo  {journal} {Eur. Phys. J. A}\ }\textbf {\bibinfo {volume}
  {53}},\ \bibinfo {pages} {148} (\bibinfo {year} {2017})}\BibitemShut
  {NoStop}\bibitem [{\citenamefont {Martinou}\ \emph {et~al.}(2020)\citenamefont
  {Martinou}, \citenamefont {Bonatsos}, \citenamefont {Minkov}, \citenamefont
  {Assimakis}, \citenamefont {Peroulis}, \citenamefont {Sarantopoulou},\ and\
  \citenamefont {Cseh}}]{epja-56-2020-239-Martinou}\BibitemOpen
  \bibfield  {author} {\bibinfo {author} {\bibfnamefont {A.}~\bibnamefont
  {Martinou}}, \bibinfo {author} {\bibfnamefont {D.}~\bibnamefont {Bonatsos}},
  \bibinfo {author} {\bibfnamefont {N.}~\bibnamefont {Minkov}}, \bibinfo
  {author} {\bibfnamefont {I.~E.}\ \bibnamefont {Assimakis}}, \bibinfo {author}
  {\bibfnamefont {S.~K.}\ \bibnamefont {Peroulis}}, \bibinfo {author}
  {\bibfnamefont {S.}~\bibnamefont {Sarantopoulou}},\ and\ \bibinfo {author}
  {\bibfnamefont {J.}~\bibnamefont {Cseh}},\ }\href
  {https://doi.org/10.1140/epja/s10050-020-00239-0} {\bibfield  {journal}
  {\bibinfo  {journal} {The European Physical Journal A}\ }\textbf {\bibinfo
  {volume} {56}},\ \bibinfo {pages} {239} (\bibinfo {year} {2020})}\BibitemShut
  {NoStop}\bibitem [{\citenamefont {Martinou}\ \emph {et~al.}(2021)\citenamefont
  {Martinou}, \citenamefont {Bonatsos}, \citenamefont {Mertzimekis},
  \citenamefont {Karakatsanis}, \citenamefont {Assimakis}, \citenamefont
  {Peroulis}, \citenamefont {Sarantopoulou},\ and\ \citenamefont
  {Minkov}}]{epja-57-2021-84-Martinou}\BibitemOpen
  \bibfield  {author} {\bibinfo {author} {\bibfnamefont {A.}~\bibnamefont
  {Martinou}}, \bibinfo {author} {\bibfnamefont {D.}~\bibnamefont {Bonatsos}},
  \bibinfo {author} {\bibfnamefont {T.~J.}\ \bibnamefont {Mertzimekis}},
  \bibinfo {author} {\bibfnamefont {K.~E.}\ \bibnamefont {Karakatsanis}},
  \bibinfo {author} {\bibfnamefont {I.~E.}\ \bibnamefont {Assimakis}}, \bibinfo
  {author} {\bibfnamefont {S.~K.}\ \bibnamefont {Peroulis}}, \bibinfo {author}
  {\bibfnamefont {S.}~\bibnamefont {Sarantopoulou}},\ and\ \bibinfo {author}
  {\bibfnamefont {N.}~\bibnamefont {Minkov}},\ }\href
  {https://doi.org/10.1140/epja/s10050-021-00396-w} {\bibfield  {journal}
  {\bibinfo  {journal} {The European Physical Journal A}\ }\textbf {\bibinfo
  {volume} {57}},\ \bibinfo {pages} {84} (\bibinfo {year} {2021})}\BibitemShut
  {NoStop}\bibitem [{\citenamefont {Rosensteel}\ \emph {et~al.}(1984)\citenamefont
  {Rosensteel}, \citenamefont {Draayer},\ and\ \citenamefont
  {Weeks}}]{npa-419-1984-1-Rosensteel}\BibitemOpen
  \bibfield  {author} {\bibinfo {author} {\bibfnamefont {G.}~\bibnamefont
  {Rosensteel}}, \bibinfo {author} {\bibfnamefont {J.~P.}\ \bibnamefont
  {Draayer}},\ and\ \bibinfo {author} {\bibfnamefont {K.~J.}\ \bibnamefont
  {Weeks}},\ }\href@noop {} {\bibfield  {journal} {\bibinfo  {journal} {Nucl.
  Phys. A}\ }\textbf {\bibinfo {volume} {419}},\ \bibinfo {pages} {1} (\bibinfo
  {year} {1984})}\BibitemShut {NoStop}\bibitem [{\citenamefont {Draayer}\ \emph {et~al.}(1984)\citenamefont
  {Draayer}, \citenamefont {Weeks},\ and\ \citenamefont
  {Rosensteel}}]{npa-413-1984-215-Draayer}\BibitemOpen
  \bibfield  {author} {\bibinfo {author} {\bibfnamefont {J.~P.}\ \bibnamefont
  {Draayer}}, \bibinfo {author} {\bibfnamefont {K.~J.}\ \bibnamefont {Weeks}},\
  and\ \bibinfo {author} {\bibfnamefont {G.}~\bibnamefont {Rosensteel}},\
  }\href@noop {} {\bibfield  {journal} {\bibinfo  {journal} {Nucl. Phys. A}\
  }\textbf {\bibinfo {volume} {413}},\ \bibinfo {pages} {215} (\bibinfo {year}
  {1984})}\BibitemShut {NoStop}\bibitem [{\citenamefont {Hecht}\ and\ \citenamefont
  {Braunschweig}(1978)}]{npa-295-1978-34-Hecht}\BibitemOpen
  \bibfield  {author} {\bibinfo {author} {\bibfnamefont {K.~T.}\ \bibnamefont
  {Hecht}}\ and\ \bibinfo {author} {\bibfnamefont {D.}~\bibnamefont
  {Braunschweig}},\ }\href {https://doi.org/DOI: 10.1016/0375-9474(78)90018-0}
  {\bibfield  {journal} {\bibinfo  {journal} {Nucl. Phys. A}\ }\textbf
  {\bibinfo {volume} {295}},\ \bibinfo {pages} {34} (\bibinfo {year}
  {1978})}\BibitemShut {NoStop}\bibitem [{\citenamefont {Carvalho}\ \emph {et~al.}(1982)\citenamefont
  {Carvalho}, \citenamefont {Park},\ and\ \citenamefont
  {Rowe}}]{plb-119-1982-249-Carvalho}\BibitemOpen
  \bibfield  {author} {\bibinfo {author} {\bibfnamefont {J.}~\bibnamefont
  {Carvalho}}, \bibinfo {author} {\bibfnamefont {P.}~\bibnamefont {Park}},\
  and\ \bibinfo {author} {\bibfnamefont {D.~J.}\ \bibnamefont {Rowe}},\
  }\href@noop {} {\bibfield  {journal} {\bibinfo  {journal} {Phys. Lett. B}\
  }\textbf {\bibinfo {volume} {119}},\ \bibinfo {pages} {249} (\bibinfo {year}
  {1982})}\BibitemShut {NoStop}\bibitem [{\citenamefont {Draayer}\ and\ \citenamefont
  {Rosensteel}(1983{\natexlab{a}})}]{plb-124-1983-281-Draayer}\BibitemOpen
  \bibfield  {author} {\bibinfo {author} {\bibfnamefont {J.~P.}\ \bibnamefont
  {Draayer}}\ and\ \bibinfo {author} {\bibfnamefont {G.}~\bibnamefont
  {Rosensteel}},\ }\href@noop {} {\bibfield  {journal} {\bibinfo  {journal}
  {Phys. Lett. B}\ }\textbf {\bibinfo {volume} {124}},\ \bibinfo {pages} {281}
  (\bibinfo {year} {1983}{\natexlab{a}})}\BibitemShut {NoStop}\bibitem [{\citenamefont {Draayer}\ and\ \citenamefont
  {Rosensteel}(1983{\natexlab{b}})}]{plb-125-1983-237-Draayer}\BibitemOpen
  \bibfield  {author} {\bibinfo {author} {\bibfnamefont {J.~P.}\ \bibnamefont
  {Draayer}}\ and\ \bibinfo {author} {\bibfnamefont {G.}~\bibnamefont
  {Rosensteel}},\ }\href@noop {} {\bibfield  {journal} {\bibinfo  {journal}
  {Phys. Lett. B}\ }\textbf {\bibinfo {volume} {125}},\ \bibinfo {pages} {237}
  (\bibinfo {year} {1983}{\natexlab{b}})}\BibitemShut {NoStop}\bibitem [{\citenamefont {Park}\ \emph {et~al.}(1984)\citenamefont {Park},
  \citenamefont {Carvalho}, \citenamefont {Vassanji},\ and\ \citenamefont
  {Rowe}}]{npa-414-1984-93-Park}\BibitemOpen
  \bibfield  {author} {\bibinfo {author} {\bibfnamefont {P.}~\bibnamefont
  {Park}}, \bibinfo {author} {\bibfnamefont {J.}~\bibnamefont {Carvalho}},
  \bibinfo {author} {\bibfnamefont {M.}~\bibnamefont {Vassanji}},\ and\
  \bibinfo {author} {\bibfnamefont {D.~J.}\ \bibnamefont {Rowe}},\ }\href@noop
  {} {\bibfield  {journal} {\bibinfo  {journal} {Nucl. Phys. A}\ }\textbf
  {\bibinfo {volume} {414}},\ \bibinfo {pages} {93} (\bibinfo {year}
  {1984})}\BibitemShut {NoStop}\bibitem [{\citenamefont {Carvalho}\ \emph
  {et~al.}(1986{\natexlab{a}})\citenamefont {Carvalho}, \citenamefont {{Le
  Blanc}}, \citenamefont {Vassanji},\ and\ \citenamefont
  {Rowe}}]{npa-452-1986-240-Carvalho}\BibitemOpen
  \bibfield  {author} {\bibinfo {author} {\bibfnamefont {J.}~\bibnamefont
  {Carvalho}}, \bibinfo {author} {\bibfnamefont {R.}~\bibnamefont {{Le
  Blanc}}}, \bibinfo {author} {\bibfnamefont {M.~G.}\ \bibnamefont
  {Vassanji}},\ and\ \bibinfo {author} {\bibfnamefont {D.~J.}\ \bibnamefont
  {Rowe}},\ }\href@noop {} {\bibfield  {journal} {\bibinfo  {journal} {Nucl.
  Phys. A}\ }\textbf {\bibinfo {volume} {452}},\ \bibinfo {pages} {240}
  (\bibinfo {year} {1986}{\natexlab{a}})}\BibitemShut {NoStop}\bibitem [{\citenamefont {{Le Blanc}}\ \emph {et~al.}(1986)\citenamefont {{Le
  Blanc}}, \citenamefont {Carvalho}, \citenamefont {Vassanji},\ and\
  \citenamefont {Rowe}}]{npa-452-1986-263-Le_Blanc}\BibitemOpen
  \bibfield  {author} {\bibinfo {author} {\bibfnamefont {R.}~\bibnamefont {{Le
  Blanc}}}, \bibinfo {author} {\bibfnamefont {J.}~\bibnamefont {Carvalho}},
  \bibinfo {author} {\bibfnamefont {M.}~\bibnamefont {Vassanji}},\ and\
  \bibinfo {author} {\bibfnamefont {D.~J.}\ \bibnamefont {Rowe}},\ }\href@noop
  {} {\bibfield  {journal} {\bibinfo  {journal} {Nucl. Phys. A}\ }\textbf
  {\bibinfo {volume} {452}},\ \bibinfo {pages} {263} (\bibinfo {year}
  {1986})}\BibitemShut {NoStop}\bibitem [{\citenamefont {Vassanji}\ and\ \citenamefont
  {Rowe}(1986)}]{npa-454-1986-288-Vassanji}\BibitemOpen
  \bibfield  {author} {\bibinfo {author} {\bibfnamefont {M.~G.}\ \bibnamefont
  {Vassanji}}\ and\ \bibinfo {author} {\bibfnamefont {D.~J.}\ \bibnamefont
  {Rowe}},\ }\href@noop {} {\bibfield  {journal} {\bibinfo  {journal} {Nucl.
  Phys. A}\ }\textbf {\bibinfo {volume} {454}},\ \bibinfo {pages} {288}
  (\bibinfo {year} {1986})}\BibitemShut {NoStop}\bibitem [{\citenamefont {Reske}(1984)}]{Reske1984}\BibitemOpen
  \bibfield  {author} {\bibinfo {author} {\bibfnamefont {E.~J.}\ \bibnamefont
  {Reske}},\ }\emph {\bibinfo {title} {$\mathrm{Sp}(6,R)$ symmetry and the
  giant quadrupole resonance in $^{24}\mathrm{Mg}$}},\ \href@noop {} {Ph.D.
  thesis},\ \bibinfo  {school} {University of Michigan} (\bibinfo {year}
  {1984})\BibitemShut {NoStop}\bibitem [{\citenamefont {Suzuki}\ and\ \citenamefont
  {Hecht}(1986)}]{npa-455-1986-315-Suzuki}\BibitemOpen
  \bibfield  {author} {\bibinfo {author} {\bibfnamefont {Y.}~\bibnamefont
  {Suzuki}}\ and\ \bibinfo {author} {\bibfnamefont {K.~T.}\ \bibnamefont
  {Hecht}},\ }\href@noop {} {\bibfield  {journal} {\bibinfo  {journal} {Nucl.
  Phys. A}\ }\textbf {\bibinfo {volume} {455}},\ \bibinfo {pages} {315}
  (\bibinfo {year} {1986})}\BibitemShut {NoStop}\bibitem [{\citenamefont {Suzuki}\ and\ \citenamefont
  {Hecht}(1987)}]{ptp-77-1987-190-Suzuki}\BibitemOpen
  \bibfield  {author} {\bibinfo {author} {\bibfnamefont {Y.}~\bibnamefont
  {Suzuki}}\ and\ \bibinfo {author} {\bibfnamefont {K.~T.}\ \bibnamefont
  {Hecht}},\ }\href@noop {} {\bibfield  {journal} {\bibinfo  {journal} {Prog.
  Theor. Phys.}\ }\textbf {\bibinfo {volume} {77}},\ \bibinfo {pages} {190}
  (\bibinfo {year} {1987})}\BibitemShut {NoStop}\bibitem [{\citenamefont {Carvalho}\ \emph
  {et~al.}(1986{\natexlab{b}})\citenamefont {Carvalho}, \citenamefont
  {Vassanji},\ and\ \citenamefont {Rowe}}]{npa-465-1987-265-Carvalho}\BibitemOpen
  \bibfield  {author} {\bibinfo {author} {\bibfnamefont {J.}~\bibnamefont
  {Carvalho}}, \bibinfo {author} {\bibfnamefont {M.~G.}\ \bibnamefont
  {Vassanji}},\ and\ \bibinfo {author} {\bibfnamefont {D.~J.}\ \bibnamefont
  {Rowe}},\ }\href@noop {} {\bibfield  {journal} {\bibinfo  {journal} {Nucl.
  Phys. A}\ }\textbf {\bibinfo {volume} {465}},\ \bibinfo {pages} {265}
  (\bibinfo {year} {1986}{\natexlab{b}})}\BibitemShut {NoStop}\bibitem [{\citenamefont {Rowe}\ \emph {et~al.}(2006)\citenamefont {Rowe},
  \citenamefont {Thiamova},\ and\ \citenamefont
  {Wood}}]{prl-97-2006-202501-Rowe}\BibitemOpen
  \bibfield  {author} {\bibinfo {author} {\bibfnamefont {D.~J.}\ \bibnamefont
  {Rowe}}, \bibinfo {author} {\bibfnamefont {G.}~\bibnamefont {Thiamova}},\
  and\ \bibinfo {author} {\bibfnamefont {J.~L.}\ \bibnamefont {Wood}},\
  }\href@noop {} {\bibfield  {journal} {\bibinfo  {journal} {Phys. Rev. Lett.}\
  }\textbf {\bibinfo {volume} {97}},\ \bibinfo {pages} {202501} (\bibinfo
  {year} {2006})}\BibitemShut {NoStop}\bibitem [{\citenamefont {Escher}(1997)}]{Escher1997}\BibitemOpen
  \bibfield  {author} {\bibinfo {author} {\bibfnamefont {J.}~\bibnamefont
  {Escher}},\ }\emph {\bibinfo {title} {Electron Scattering Studies in the
  Framework of the Symplectic Shell Model}},\ \href@noop {} {Ph.D. thesis},\
  \bibinfo  {school} {Louisiana State University} (\bibinfo {year}
  {1997})\BibitemShut {NoStop}\bibitem [{\citenamefont {Escher}\ and\ \citenamefont
  {Draayer}(1998)}]{jmp-39-1998-5123-Escher}\BibitemOpen
  \bibfield  {author} {\bibinfo {author} {\bibfnamefont {J.}~\bibnamefont
  {Escher}}\ and\ \bibinfo {author} {\bibfnamefont {J.~P.}\ \bibnamefont
  {Draayer}},\ }\href@noop {} {\bibfield  {journal} {\bibinfo  {journal} {J.
  Math. Phys. (N.Y.)}\ }\textbf {\bibinfo {volume} {39}},\ \bibinfo {pages}
  {5123} (\bibinfo {year} {1998})}\BibitemShut {NoStop}\bibitem [{\citenamefont {Dytrych}(2008)}]{Dytrych2001}\BibitemOpen
  \bibfield  {author} {\bibinfo {author} {\bibfnamefont {T.}~\bibnamefont
  {Dytrych}},\ }\emph {\bibinfo {title} {Evidence for symplectic symmetry in ab
  initio no-core shell model results}},\ \href@noop {} {Ph.D. thesis},\
  \bibinfo  {school} {Louisiana State University} (\bibinfo {year}
  {2008})\BibitemShut {NoStop}\bibitem [{\citenamefont {Wybourne}(1974)}]{Wybourne1974}\BibitemOpen
  \bibfield  {author} {\bibinfo {author} {\bibfnamefont {B.~G.}\ \bibnamefont
  {Wybourne}},\ }\href@noop {} {\emph {\bibinfo {title} {Classical Groups for
  Physicists}}}\ (\bibinfo  {publisher} {Wiley},\ \bibinfo {address} {New
  York},\ \bibinfo {year} {1974})\BibitemShut {NoStop}\bibitem [{\citenamefont {Klimyk}(1972)}]{tmp-13-1972-1171-Klimyk}\BibitemOpen
  \bibfield  {author} {\bibinfo {author} {\bibfnamefont {A.~U.}\ \bibnamefont
  {Klimyk}},\ }\href@noop {} {\bibfield  {journal} {\bibinfo  {journal}
  {Teoret. Mat. Fiz}\ }\textbf {\bibinfo {volume} {13}},\ \bibinfo {pages}
  {1171} (\bibinfo {year} {1972})}\BibitemShut {NoStop}\bibitem [{\citenamefont {Klimyk}(1975)}]{rmp-7-1975-153-Klimyk}\BibitemOpen
  \bibfield  {author} {\bibinfo {author} {\bibfnamefont {A.~U.}\ \bibnamefont
  {Klimyk}},\ }\href@noop {} {\bibfield  {journal} {\bibinfo  {journal} {Rep.
  Math. Phys.}\ }\textbf {\bibinfo {volume} {7}},\ \bibinfo {pages} {153}
  (\bibinfo {year} {1975})}\BibitemShut {NoStop}\bibitem [{\citenamefont {Butler}(1974)}]{ptrsla-277-1975-545-Butler}\BibitemOpen
  \bibfield  {author} {\bibinfo {author} {\bibfnamefont {P.~H.}\ \bibnamefont
  {Butler}},\ }\href@noop {} {\bibfield  {journal} {\bibinfo  {journal} {Phil.
  Trans. Roy. Soc. (London) A}\ }\textbf {\bibinfo {volume} {277}},\ \bibinfo
  {pages} {545} (\bibinfo {year} {1974})}\BibitemShut {NoStop}\bibitem [{\citenamefont {Draayer}\ \emph {et~al.}(2012)\citenamefont
  {Draayer}, \citenamefont {Dytrych}, \citenamefont {Launey},\ and\
  \citenamefont {Langr}}]{ppnp-67-2012-516-Draayer}\BibitemOpen
  \bibfield  {author} {\bibinfo {author} {\bibfnamefont {J.~P.}\ \bibnamefont
  {Draayer}}, \bibinfo {author} {\bibfnamefont {T.}~\bibnamefont {Dytrych}},
  \bibinfo {author} {\bibfnamefont {K.~D.}\ \bibnamefont {Launey}},\ and\
  \bibinfo {author} {\bibfnamefont {D.}~\bibnamefont {Langr}},\ }\href@noop {}
  {\bibfield  {journal} {\bibinfo  {journal} {Prog. Part. Nucl. Phys.}\
  }\textbf {\bibinfo {volume} {67}},\ \bibinfo {pages} {516} (\bibinfo {year}
  {2012})}\BibitemShut {NoStop}\bibitem [{\citenamefont {Launey}\ \emph {et~al.}(2021)\citenamefont {Launey},
  \citenamefont {Mercenne},\ and\ \citenamefont
  {Dytrych}}]{arnps-71-2021-253-Launey}\BibitemOpen
  \bibfield  {author} {\bibinfo {author} {\bibfnamefont {K.~D.}\ \bibnamefont
  {Launey}}, \bibinfo {author} {\bibfnamefont {A.}~\bibnamefont {Mercenne}},\
  and\ \bibinfo {author} {\bibfnamefont {T.}~\bibnamefont {Dytrych}},\ }\href
  {https://doi.org/10.1146/annurev-nucl-102419-033316} {\bibfield  {journal}
  {\bibinfo  {journal} {Ann. Rev. Nucl. Part. Sci.}\ }\textbf {\bibinfo
  {volume} {71}},\ \bibinfo {pages} {253} (\bibinfo {year} {2021})}\BibitemShut
  {NoStop}\bibitem [{\citenamefont {Sargsyan}\ \emph {et~al.}(2023)\citenamefont
  {Sargsyan}, \citenamefont {Launey}, \citenamefont {Shaffer}, \citenamefont
  {Marley}, \citenamefont {Dudeck}, \citenamefont {Mercenne}, \citenamefont
  {Dytrych},\ and\ \citenamefont {Draayer}}]{prc-108-2023-054303-Sargsyan}\BibitemOpen
  \bibfield  {author} {\bibinfo {author} {\bibfnamefont {G.~H.}\ \bibnamefont
  {Sargsyan}}, \bibinfo {author} {\bibfnamefont {K.~D.}\ \bibnamefont
  {Launey}}, \bibinfo {author} {\bibfnamefont {R.~M.}\ \bibnamefont {Shaffer}},
  \bibinfo {author} {\bibfnamefont {S.~T.}\ \bibnamefont {Marley}}, \bibinfo
  {author} {\bibfnamefont {N.}~\bibnamefont {Dudeck}}, \bibinfo {author}
  {\bibfnamefont {A.}~\bibnamefont {Mercenne}}, \bibinfo {author}
  {\bibfnamefont {T.}~\bibnamefont {Dytrych}},\ and\ \bibinfo {author}
  {\bibfnamefont {J.~P.}\ \bibnamefont {Draayer}},\ }\href
  {https://doi.org/10.1103/PhysRevC.108.054303} {\bibfield  {journal} {\bibinfo
   {journal} {Phys. Rev. C}\ }\textbf {\bibinfo {volume} {108}},\ \bibinfo
  {pages} {054303} (\bibinfo {year} {2023})}\BibitemShut {NoStop}\bibitem [{\citenamefont {Mercenne}\ \emph {et~al.}(2022)\citenamefont
  {Mercenne}, \citenamefont {Launey}, \citenamefont {Dytrych}, \citenamefont
  {Escher}, \citenamefont {Quaglioni}, \citenamefont {Sargsyan}, \citenamefont
  {Langr},\ and\ \citenamefont {Draayer}}]{cpc-280-2022-108476-Mercenne}\BibitemOpen
  \bibfield  {author} {\bibinfo {author} {\bibfnamefont {A.}~\bibnamefont
  {Mercenne}}, \bibinfo {author} {\bibfnamefont {K.}~\bibnamefont {Launey}},
  \bibinfo {author} {\bibfnamefont {T.}~\bibnamefont {Dytrych}}, \bibinfo
  {author} {\bibfnamefont {J.}~\bibnamefont {Escher}}, \bibinfo {author}
  {\bibfnamefont {S.}~\bibnamefont {Quaglioni}}, \bibinfo {author}
  {\bibfnamefont {G.}~\bibnamefont {Sargsyan}}, \bibinfo {author}
  {\bibfnamefont {D.}~\bibnamefont {Langr}},\ and\ \bibinfo {author}
  {\bibfnamefont {J.}~\bibnamefont {Draayer}},\ }\href
  {https://doi.org/10.1016/j.cpc.2022.108476} {\bibfield  {journal} {\bibinfo
  {journal} {Computer Physics Communications}\ }\textbf {\bibinfo {volume}
  {280}},\ \bibinfo {pages} {108476} (\bibinfo {year} {2022})}\BibitemShut
  {NoStop}\bibitem [{\citenamefont {Sargsyan}\ \emph {et~al.}(2022)\citenamefont
  {Sargsyan}, \citenamefont {Launey}, \citenamefont {Burkey}, \citenamefont
  {Gallant}, \citenamefont {Scielzo}, \citenamefont {Savard}, \citenamefont
  {Mercenne}, \citenamefont {Dytrych}, \citenamefont {Langr}, \citenamefont
  {Varriano}, \citenamefont {Longfellow}, \citenamefont {Hirsh},\ and\
  \citenamefont {Draayer}}]{prl-128-2022-202503-Sargsyan}\BibitemOpen
  \bibfield  {author} {\bibinfo {author} {\bibfnamefont {G.~H.}\ \bibnamefont
  {Sargsyan}}, \bibinfo {author} {\bibfnamefont {K.~D.}\ \bibnamefont
  {Launey}}, \bibinfo {author} {\bibfnamefont {M.~T.}\ \bibnamefont {Burkey}},
  \bibinfo {author} {\bibfnamefont {A.~T.}\ \bibnamefont {Gallant}}, \bibinfo
  {author} {\bibfnamefont {N.~D.}\ \bibnamefont {Scielzo}}, \bibinfo {author}
  {\bibfnamefont {G.}~\bibnamefont {Savard}}, \bibinfo {author} {\bibfnamefont
  {A.}~\bibnamefont {Mercenne}}, \bibinfo {author} {\bibfnamefont
  {T.}~\bibnamefont {Dytrych}}, \bibinfo {author} {\bibfnamefont
  {D.}~\bibnamefont {Langr}}, \bibinfo {author} {\bibfnamefont
  {L.}~\bibnamefont {Varriano}}, \bibinfo {author} {\bibfnamefont
  {B.}~\bibnamefont {Longfellow}}, \bibinfo {author} {\bibfnamefont {T.~Y.}\
  \bibnamefont {Hirsh}},\ and\ \bibinfo {author} {\bibfnamefont {J.~P.}\
  \bibnamefont {Draayer}},\ }\href
  {https://doi.org/10.1103/PhysRevLett.128.202503} {\bibfield  {journal}
  {\bibinfo  {journal} {Phys. Rev. Lett.}\ }\textbf {\bibinfo {volume} {128}},\
  \bibinfo {pages} {202503} (\bibinfo {year} {2022})}\BibitemShut {NoStop}\bibitem [{\citenamefont {Heller}\ \emph {et~al.}(2023)\citenamefont {Heller},
  \citenamefont {Sargsyan}, \citenamefont {Launey}, \citenamefont {Johnson},
  \citenamefont {Dytrych},\ and\ \citenamefont
  {Draayer}}]{prc-108-2023-024304-Heller}\BibitemOpen
  \bibfield  {author} {\bibinfo {author} {\bibfnamefont {N.~D.}\ \bibnamefont
  {Heller}}, \bibinfo {author} {\bibfnamefont {G.~H.}\ \bibnamefont
  {Sargsyan}}, \bibinfo {author} {\bibfnamefont {K.~D.}\ \bibnamefont
  {Launey}}, \bibinfo {author} {\bibfnamefont {C.~W.}\ \bibnamefont {Johnson}},
  \bibinfo {author} {\bibfnamefont {T.}~\bibnamefont {Dytrych}},\ and\ \bibinfo
  {author} {\bibfnamefont {J.~P.}\ \bibnamefont {Draayer}},\ }\href
  {https://doi.org/10.1103/PhysRevC.108.024304} {\bibfield  {journal} {\bibinfo
   {journal} {Phys. Rev. C}\ }\textbf {\bibinfo {volume} {108}},\ \bibinfo
  {pages} {024304} (\bibinfo {year} {2023})}\BibitemShut {NoStop}\bibitem [{\citenamefont {Becker}\ \emph {et~al.}()\citenamefont {Becker},
  \citenamefont {Launey}, \citenamefont {Ekstrom}, \citenamefont {Dytrych},
  \citenamefont {Langr}, \citenamefont {Sargsyan},\ and\ \citenamefont
  {Draayer}}]{arxiv:2501.00682-Becker}\BibitemOpen
  \bibinfo {author} {\bibfnamefont {K.~S.}\ \bibnamefont {Becker}}, \bibinfo
  {author} {\bibfnamefont {K.~D.}\ \bibnamefont {Launey}}, \bibinfo {author}
  {\bibfnamefont {A.}~\bibnamefont {Ekstrom}}, \bibinfo {author} {\bibfnamefont
  {T.}~\bibnamefont {Dytrych}}, \bibinfo {author} {\bibfnamefont
  {D.}~\bibnamefont {Langr}}, \bibinfo {author} {\bibfnamefont {G.~H.}\
  \bibnamefont {Sargsyan}},\ and\ \bibinfo {author} {\bibfnamefont {J.~P.}\
  \bibnamefont {Draayer}}\BibitemShut {NoStop}\bibitem [{\citenamefont {Sargsyan}\ \emph {et~al.}(2025)\citenamefont
  {Sargsyan}, \citenamefont {Yoshida}, \citenamefont {Ogata}, \citenamefont
  {Launey}, \citenamefont {Escher}, \citenamefont {Langr},\ and\ \citenamefont
  {Dytrych}}]{plb-866-2025-139563-Sargsyan}\BibitemOpen
\bibfield  {author} {  }\bibfield  {author} {\bibinfo {author} {\bibfnamefont
  {G.~H.}\ \bibnamefont {Sargsyan}}, \bibinfo {author} {\bibfnamefont
  {K.}~\bibnamefont {Yoshida}}, \bibinfo {author} {\bibfnamefont
  {K.}~\bibnamefont {Ogata}}, \bibinfo {author} {\bibfnamefont {K.~D.}\
  \bibnamefont {Launey}}, \bibinfo {author} {\bibfnamefont {J.~E.}\
  \bibnamefont {Escher}}, \bibinfo {author} {\bibfnamefont {D.}~\bibnamefont
  {Langr}},\ and\ \bibinfo {author} {\bibfnamefont {T.}~\bibnamefont
  {Dytrych}},\ }\href {https://doi.org/10.1016/j.physletb.2025.139563}
  {\bibfield  {journal} {\bibinfo  {journal} {Phys. Lett. B}\ }\textbf
  {\bibinfo {volume} {866}},\ \bibinfo {pages} {139563} (\bibinfo {year}
  {2025})}\BibitemShut {NoStop}\bibitem [{\citenamefont {Launey}\ \emph {et~al.}(2018)\citenamefont {Launey},
  \citenamefont {Mercenne}, \citenamefont {Sargsyan}, \citenamefont {Shows},
  \citenamefont {Baker}, \citenamefont {Miora}, \citenamefont {Dytrych},\ and\
  \citenamefont {Draayer}}]{psotancp-2038-2018-020004-Launey}\BibitemOpen
  \bibfield  {author} {\bibinfo {author} {\bibfnamefont {K.~D.}\ \bibnamefont
  {Launey}}, \bibinfo {author} {\bibfnamefont {A.}~\bibnamefont {Mercenne}},
  \bibinfo {author} {\bibfnamefont {G.~H.}\ \bibnamefont {Sargsyan}}, \bibinfo
  {author} {\bibfnamefont {H.}~\bibnamefont {Shows}}, \bibinfo {author}
  {\bibfnamefont {R.~B.}\ \bibnamefont {Baker}}, \bibinfo {author}
  {\bibfnamefont {M.~E.}\ \bibnamefont {Miora}}, \bibinfo {author}
  {\bibfnamefont {T.}~\bibnamefont {Dytrych}},\ and\ \bibinfo {author}
  {\bibfnamefont {J.~P.}\ \bibnamefont {Draayer}},\ }in\ \href
  {https://doi.org/10.1063/1.5078823} {\emph {\bibinfo {booktitle} {Proceedings
  of the 4th International Workshop on ``State of the Art in Nuclear Cluster
  Physics'' (SOTANCP4)}}},\ \bibinfo {series and number} {\bibinfo {series}
  {AIP Conf. Proc.}\ No.\ \bibinfo {number} {2038}},\ \bibinfo {editor} {edited
  by\ \bibinfo {editor} {\bibfnamefont {M.}~\bibnamefont {Barbui}}, \bibinfo
  {editor} {\bibfnamefont {C.~M.}\ \bibnamefont {Folden}, \bibfnamefont {III}},
  \bibinfo {editor} {\bibfnamefont {V.~Z.}\ \bibnamefont {Goldberg}},\ and\
  \bibinfo {editor} {\bibfnamefont {G.~V.}\ \bibnamefont {Rogachev}}}\
  (\bibinfo  {publisher} {AIP},\ \bibinfo {address} {New York},\ \bibinfo
  {year} {2018})\ p.\ \bibinfo {pages} {020004}\BibitemShut {NoStop}\bibitem [{\citenamefont {Herko}(2024)}]{Herko2024}\BibitemOpen
  \bibfield  {author} {\bibinfo {author} {\bibfnamefont {J.}~\bibnamefont
  {Herko}},\ }\emph {\bibinfo {title} {Ab initio Description of Nuclear
  Structure and Reactions in a Unified Framework}},\ \href
  {https://doi.org/10.7274/26288734.v1} {Ph.D. thesis},\ \bibinfo  {school}
  {University of Notre Dame} (\bibinfo {year} {2024})\BibitemShut {NoStop}\bibitem [{\citenamefont {Moshinsky}(1962)}]{rmp-34-1962-813-Moshinsky}\BibitemOpen
  \bibfield  {author} {\bibinfo {author} {\bibfnamefont {M.}~\bibnamefont
  {Moshinsky}},\ }\href {https://doi.org/10.1103/RevModPhys.34.813} {\bibfield
  {journal} {\bibinfo  {journal} {Rev. Mod. Phys.}\ }\textbf {\bibinfo {volume}
  {34}},\ \bibinfo {pages} {813} (\bibinfo {year} {1962})}\BibitemShut
  {NoStop}\bibitem [{\citenamefont {Hecht}(1965)}]{np-62-1965-1-Hecht}\BibitemOpen
  \bibfield  {author} {\bibinfo {author} {\bibfnamefont {K.~T.}\ \bibnamefont
  {Hecht}},\ }\href {https://doi.org/10.1016/0029-5582(65)90068-4} {\bibfield
  {journal} {\bibinfo  {journal} {Nucl. Phys.}\ }\textbf {\bibinfo {volume}
  {62}},\ \bibinfo {pages} {1} (\bibinfo {year} {1965})}\BibitemShut {NoStop}\bibitem [{\citenamefont {Kaufman}\ and\ \citenamefont
  {Noack}(1965)}]{jmp-6-1965-142-Kaufman}\BibitemOpen
  \bibfield  {author} {\bibinfo {author} {\bibfnamefont {B.}~\bibnamefont
  {Kaufman}}\ and\ \bibinfo {author} {\bibfnamefont {C.}~\bibnamefont
  {Noack}},\ }\href@noop {} {\bibfield  {journal} {\bibinfo  {journal} {J.
  Math. Phys.}\ }\textbf {\bibinfo {volume} {6}},\ \bibinfo {pages} {142}
  (\bibinfo {year} {1965})}\BibitemShut {NoStop}\bibitem [{\citenamefont {Derome}\ and\ \citenamefont
  {Sharp}(1965)}]{jmp-6-1965-1584-Derome}\BibitemOpen
  \bibfield  {author} {\bibinfo {author} {\bibfnamefont {J.}~\bibnamefont
  {Derome}}\ and\ \bibinfo {author} {\bibfnamefont {W.~T.}\ \bibnamefont
  {Sharp}},\ }\href@noop {} {\bibfield  {journal} {\bibinfo  {journal} {J.
  Math. Phys.}\ }\textbf {\bibinfo {volume} {6}},\ \bibinfo {pages} {1584}
  (\bibinfo {year} {1965})}\BibitemShut {NoStop}\bibitem [{\citenamefont {Derome}(1966)}]{jmp-7-1966-612-Derome}\BibitemOpen
  \bibfield  {author} {\bibinfo {author} {\bibfnamefont {J.}~\bibnamefont
  {Derome}},\ }\href@noop {} {\bibfield  {journal} {\bibinfo  {journal} {J.
  Math. Phys.}\ }\textbf {\bibinfo {volume} {7}},\ \bibinfo {pages} {612}
  (\bibinfo {year} {1966})}\BibitemShut {NoStop}\bibitem [{\citenamefont {Derome}(1967)}]{jmp-8-1967-714-Derome}\BibitemOpen
  \bibfield  {author} {\bibinfo {author} {\bibfnamefont {J.}~\bibnamefont
  {Derome}},\ }\href@noop {} {\bibfield  {journal} {\bibinfo  {journal} {J.
  Math. Phys.}\ }\textbf {\bibinfo {volume} {8}},\ \bibinfo {pages} {714}
  (\bibinfo {year} {1967})}\BibitemShut {NoStop}\bibitem [{\citenamefont {Resnikoff}(1967)}]{jmp-8-1967-63-Resnikoff}\BibitemOpen
  \bibfield  {author} {\bibinfo {author} {\bibfnamefont {M.}~\bibnamefont
  {Resnikoff}},\ }\href@noop {} {\bibfield  {journal} {\bibinfo  {journal} {J.
  Math. Phys.}\ }\textbf {\bibinfo {volume} {8}},\ \bibinfo {pages} {63}
  (\bibinfo {year} {1967})}\BibitemShut {NoStop}\bibitem [{\citenamefont {Vergados}(1968)}]{npa-111-1968-681-Vergados}\BibitemOpen
  \bibfield  {author} {\bibinfo {author} {\bibfnamefont {J.~D.}\ \bibnamefont
  {Vergados}},\ }\href@noop {} {\bibfield  {journal} {\bibinfo  {journal}
  {Nucl. Phys. A}\ }\textbf {\bibinfo {volume} {111}},\ \bibinfo {pages} {681}
  (\bibinfo {year} {1968})}\BibitemShut {NoStop}\bibitem [{\citenamefont {Biedenharn}\ and\ \citenamefont
  {Louck}(1972)}]{jmp-13-1972-1985-Biedenharn}\BibitemOpen
  \bibfield  {author} {\bibinfo {author} {\bibfnamefont {L.~C.}\ \bibnamefont
  {Biedenharn}}\ and\ \bibinfo {author} {\bibfnamefont {J.~D.}\ \bibnamefont
  {Louck}},\ }\href {https://doi.org/10.1063/1.1665941} {\bibfield  {journal}
  {\bibinfo  {journal} {J. Math. Phys.}\ }\textbf {\bibinfo {volume} {13}},\
  \bibinfo {pages} {1985} (\bibinfo {year} {1972})}\BibitemShut {NoStop}\bibitem [{\citenamefont {Draayer}\ and\ \citenamefont
  {Akiyama}(1973)}]{jmp-14-1973-1904-Draayer}\BibitemOpen
  \bibfield  {author} {\bibinfo {author} {\bibfnamefont {J.~P.}\ \bibnamefont
  {Draayer}}\ and\ \bibinfo {author} {\bibfnamefont {Y.}~\bibnamefont
  {Akiyama}},\ }\href@noop {} {\bibfield  {journal} {\bibinfo  {journal} {J.
  Math. Phys.}\ }\textbf {\bibinfo {volume} {14}},\ \bibinfo {pages} {1904}
  (\bibinfo {year} {1973})}\BibitemShut {NoStop}\bibitem [{\citenamefont {Akiyama}\ and\ \citenamefont
  {Draayer}(1973)}]{cpc-5-1973-405-Akiyama}\BibitemOpen
  \bibfield  {author} {\bibinfo {author} {\bibfnamefont {Y.}~\bibnamefont
  {Akiyama}}\ and\ \bibinfo {author} {\bibfnamefont {J.~P.}\ \bibnamefont
  {Draayer}},\ }\href@noop {} {\bibfield  {journal} {\bibinfo  {journal}
  {Comput. Phys. Commun.}\ }\textbf {\bibinfo {volume} {5}},\ \bibinfo {pages}
  {405} (\bibinfo {year} {1973})}\BibitemShut {NoStop}\bibitem [{\citenamefont {Hecht}\ and\ \citenamefont
  {Braunschweig}(1975)}]{npa-244-1975-365-Hecht}\BibitemOpen
  \bibfield  {author} {\bibinfo {author} {\bibfnamefont {K.~T.}\ \bibnamefont
  {Hecht}}\ and\ \bibinfo {author} {\bibfnamefont {D.}~\bibnamefont
  {Braunschweig}},\ }\href@noop {} {\bibfield  {journal} {\bibinfo  {journal}
  {Nucl. Phys. A}\ }\textbf {\bibinfo {volume} {244}},\ \bibinfo {pages} {365}
  (\bibinfo {year} {1975})}\BibitemShut {NoStop}\bibitem [{\citenamefont {Moshinsky}\ \emph {et~al.}(1975)\citenamefont
  {Moshinsky}, \citenamefont {Patera}, \citenamefont {Sharp},\ and\
  \citenamefont {Winternitz}}]{apny-95-1975-139-Moshinsky}\BibitemOpen
  \bibfield  {author} {\bibinfo {author} {\bibfnamefont {M.}~\bibnamefont
  {Moshinsky}}, \bibinfo {author} {\bibfnamefont {J.}~\bibnamefont {Patera}},
  \bibinfo {author} {\bibfnamefont {R.~T.}\ \bibnamefont {Sharp}},\ and\
  \bibinfo {author} {\bibfnamefont {P.}~\bibnamefont {Winternitz}},\
  }\href@noop {} {\bibfield  {journal} {\bibinfo  {journal} {Ann. Phys. (N.
  Y.)}\ }\textbf {\bibinfo {volume} {95}},\ \bibinfo {pages} {139} (\bibinfo
  {year} {1975})}\BibitemShut {NoStop}\bibitem [{\citenamefont {Millener}(1978)}]{jmp-19-1978-1513-Millener}\BibitemOpen
  \bibfield  {author} {\bibinfo {author} {\bibfnamefont {D.~J.}\ \bibnamefont
  {Millener}},\ }\href@noop {} {\bibfield  {journal} {\bibinfo  {journal} {J.
  Math. Phys.}\ }\textbf {\bibinfo {volume} {19}},\ \bibinfo {pages} {1513}
  (\bibinfo {year} {1978})}\BibitemShut {NoStop}\bibitem [{\citenamefont {O'Reilly}(1982)}]{jmp-23-1982-2022-OReilly}\BibitemOpen
  \bibfield  {author} {\bibinfo {author} {\bibfnamefont {M.~F.}\ \bibnamefont
  {O'Reilly}},\ }\href@noop {} {\bibfield  {journal} {\bibinfo  {journal} {J.
  Math. Phys. (N.Y.)}\ }\textbf {\bibinfo {volume} {23}},\ \bibinfo {pages}
  {2022} (\bibinfo {year} {1982})}\BibitemShut {NoStop}\bibitem [{\citenamefont {Hecht}(1987)}]{Hecht1987}\BibitemOpen
  \bibfield  {author} {\bibinfo {author} {\bibfnamefont {K.}~\bibnamefont
  {Hecht}},\ }\href@noop {} {\emph {\bibinfo {title} {The Vector Coherent State
  Method and Its Application to Problems of Higher Symmetries}}}\ (\bibinfo
  {publisher} {Springer-Verlag},\ \bibinfo {year} {1987})\BibitemShut {NoStop}\bibitem [{\citenamefont {Hecht}(1990)}]{jpa-23-1990-407-Hecht}\BibitemOpen
  \bibfield  {author} {\bibinfo {author} {\bibfnamefont {K.~T.}\ \bibnamefont
  {Hecht}},\ }\href@noop {} {\bibfield  {journal} {\bibinfo  {journal} {J.
  Phys. A}\ }\textbf {\bibinfo {volume} {23}},\ \bibinfo {pages} {407}
  (\bibinfo {year} {1990})}\BibitemShut {NoStop}\bibitem [{\citenamefont {Chen}\ \emph {et~al.}(2002)\citenamefont {Chen},
  \citenamefont {Ping},\ and\ \citenamefont {Wang}}]{Chen2002}\BibitemOpen
  \bibfield  {author} {\bibinfo {author} {\bibfnamefont {J.-Q.}\ \bibnamefont
  {Chen}}, \bibinfo {author} {\bibfnamefont {J.}~\bibnamefont {Ping}},\ and\
  \bibinfo {author} {\bibfnamefont {F.}~\bibnamefont {Wang}},\ }\href@noop {}
  {\emph {\bibinfo {title} {Group Representation Theory for Physicists}}},\
  \bibinfo {edition} {2nd}\ ed.\ (\bibinfo  {publisher} {World Scientific
  Publishing Co Pte Ltd.},\ \bibinfo {year} {2002})\BibitemShut {NoStop}\bibitem [{\citenamefont {Bahri}\ \emph {et~al.}(2004)\citenamefont {Bahri},
  \citenamefont {Rowe},\ and\ \citenamefont
  {Draayer}}]{cpc-159-2004-121-Bahri}\BibitemOpen
  \bibfield  {author} {\bibinfo {author} {\bibfnamefont {C.}~\bibnamefont
  {Bahri}}, \bibinfo {author} {\bibfnamefont {D.~J.}\ \bibnamefont {Rowe}},\
  and\ \bibinfo {author} {\bibfnamefont {J.~P.}\ \bibnamefont {Draayer}},\
  }\href@noop {} {\bibfield  {journal} {\bibinfo  {journal} {Comput. Phys.
  Commun.}\ }\textbf {\bibinfo {volume} {159}},\ \bibinfo {pages} {121}
  (\bibinfo {year} {2004})}\BibitemShut {NoStop}\bibitem [{\citenamefont {Rowe}\ and\ \citenamefont
  {Thiamova}(2008)}]{jpa-41-2008-065206-Rowe}\BibitemOpen
  \bibfield  {author} {\bibinfo {author} {\bibfnamefont {D.~J.}\ \bibnamefont
  {Rowe}}\ and\ \bibinfo {author} {\bibfnamefont {G.}~\bibnamefont
  {Thiamova}},\ }\href {https://doi.org/10.1088/1751-8113/41/6/065206}
  {\bibfield  {journal} {\bibinfo  {journal} {J. Phys. A}\ }\textbf {\bibinfo
  {volume} {41}},\ \bibinfo {pages} {065206} (\bibinfo {year}
  {2008})}\BibitemShut {NoStop}\bibitem [{\citenamefont {Rowe}(2016)}]{privcom-Rowe-2016}\BibitemOpen
  \bibfield  {author} {\bibinfo {author} {\bibfnamefont {D.~J.}\ \bibnamefont
  {Rowe}},\ }\href@noop {} {}\bibinfo {howpublished} {private communication}
  (\bibinfo {year} {2016})\BibitemShut {NoStop}\bibitem [{\citenamefont {Langr}\ \emph {et~al.}(2019)\citenamefont {Langr},
  \citenamefont {Dytrych}, \citenamefont {Draayer}, \citenamefont {Launey},\
  and\ \citenamefont {Tvrd\'{i}k}}]{cpc-244-2019-442-Langr}\BibitemOpen
  \bibfield  {author} {\bibinfo {author} {\bibfnamefont {D.}~\bibnamefont
  {Langr}}, \bibinfo {author} {\bibfnamefont {T.}~\bibnamefont {Dytrych}},
  \bibinfo {author} {\bibfnamefont {J.~P.}\ \bibnamefont {Draayer}}, \bibinfo
  {author} {\bibfnamefont {K.~D.}\ \bibnamefont {Launey}},\ and\ \bibinfo
  {author} {\bibfnamefont {P.}~\bibnamefont {Tvrd\'{i}k}},\ }\href
  {https://doi.org/10.1016/j.cpc.2019.05.018} {\bibfield  {journal} {\bibinfo
  {journal} {Comput. Phys. Commun.}\ }\textbf {\bibinfo {volume} {244}},\
  \bibinfo {pages} {442} (\bibinfo {year} {2019})}\BibitemShut {NoStop}\bibitem [{\citenamefont {Dytrych}\ \emph {et~al.}(2021)\citenamefont
  {Dytrych}, \citenamefont {Langr}, \citenamefont {Draayer}, \citenamefont
  {Launey},\ and\ \citenamefont {Gazda}}]{cpc-269-2021-108137-Dytrych}\BibitemOpen
  \bibfield  {author} {\bibinfo {author} {\bibfnamefont {T.}~\bibnamefont
  {Dytrych}}, \bibinfo {author} {\bibfnamefont {D.}~\bibnamefont {Langr}},
  \bibinfo {author} {\bibfnamefont {J.~P.}\ \bibnamefont {Draayer}}, \bibinfo
  {author} {\bibfnamefont {K.~D.}\ \bibnamefont {Launey}},\ and\ \bibinfo
  {author} {\bibfnamefont {D.}~\bibnamefont {Gazda}},\ }\href
  {https://doi.org/10.1016/j.cpc.2021.108137} {\bibfield  {journal} {\bibinfo
  {journal} {Comput. Phys. Commun.}\ }\textbf {\bibinfo {volume} {269}},\
  \bibinfo {pages} {108137} (\bibinfo {year} {2021})}\BibitemShut {NoStop}\bibitem [{\citenamefont {Herko}\ \emph {et~al.}(ress)\citenamefont {Herko},
  \citenamefont {Caprio}, \citenamefont {McCoy},\ and\ \citenamefont
  {Fasano}}]{arxiv-Herko}\BibitemOpen
  \bibfield  {author} {\bibinfo {author} {\bibfnamefont {J.}~\bibnamefont
  {Herko}}, \bibinfo {author} {\bibfnamefont {M.~A.}\ \bibnamefont {Caprio}},
  \bibinfo {author} {\bibfnamefont {A.~E.}\ \bibnamefont {McCoy}},\ and\
  \bibinfo {author} {\bibfnamefont {P.~J.}\ \bibnamefont {Fasano}},\
  }\href@noop {} {\bibfield  {journal} {\bibinfo  {journal} {Eur. Phys. J. A}\
  } (\bibinfo {year} {in press})}\BibitemShut {NoStop}\bibitem [{\citenamefont {Rowe}\ \emph {et~al.}(2016)\citenamefont {Rowe},
  \citenamefont {McCoy},\ and\ \citenamefont
  {Caprio}}]{ps-91-2016-033003-Rowe}\BibitemOpen
  \bibfield  {author} {\bibinfo {author} {\bibfnamefont {D.~J.}\ \bibnamefont
  {Rowe}}, \bibinfo {author} {\bibfnamefont {A.~E.}\ \bibnamefont {McCoy}},\
  and\ \bibinfo {author} {\bibfnamefont {M.~A.}\ \bibnamefont {Caprio}},\
  }\href {https://doi.org/10.1088/0031-8949/91/3/033003} {\bibfield  {journal}
  {\bibinfo  {journal} {Phys. Scr.}\ }\textbf {\bibinfo {volume} {91}},\
  \bibinfo {pages} {033003} (\bibinfo {year} {2016})}\BibitemShut {NoStop}\bibitem [{\citenamefont {Castanos}\ \emph {et~al.}(1988)\citenamefont
  {Castanos}, \citenamefont {Draayer},\ and\ \citenamefont
  {Leschber}}]{zpa-329-1988-33-Castanos}\BibitemOpen
  \bibfield  {author} {\bibinfo {author} {\bibfnamefont {O.}~\bibnamefont
  {Castanos}}, \bibinfo {author} {\bibfnamefont {J.~P.}\ \bibnamefont
  {Draayer}},\ and\ \bibinfo {author} {\bibfnamefont {Y.}~\bibnamefont
  {Leschber}},\ }\href@noop {} {\bibfield  {journal} {\bibinfo  {journal} {Z.
  Phys. A}\ }\textbf {\bibinfo {volume} {329}},\ \bibinfo {pages} {33}
  (\bibinfo {year} {1988})}\BibitemShut {NoStop}\bibitem [{\citenamefont {Peierls}\ and\ \citenamefont
  {Yoccoz}(1957)}]{ppsla-70-1957-381-Peierls}\BibitemOpen
  \bibfield  {author} {\bibinfo {author} {\bibfnamefont {R.~E.}\ \bibnamefont
  {Peierls}}\ and\ \bibinfo {author} {\bibfnamefont {J.}~\bibnamefont
  {Yoccoz}},\ }\href@noop {} {\bibfield  {journal} {\bibinfo  {journal} {Proc.
  Phys. Society A}\ }\textbf {\bibinfo {volume} {70}},\ \bibinfo {pages} {381}
  (\bibinfo {year} {1957})}\BibitemShut {NoStop}\bibitem [{\citenamefont {Bohr}\ and\ \citenamefont
  {Mottelson}(1998)}]{Bohr1998:v1}\BibitemOpen
  \bibfield  {author} {\bibinfo {author} {\bibfnamefont {A.}~\bibnamefont
  {Bohr}}\ and\ \bibinfo {author} {\bibfnamefont {B.~R.}\ \bibnamefont
  {Mottelson}},\ }\href {https://doi.org/10.1142/3530} {\emph {\bibinfo {title}
  {Nuclear Structure}}},\ Vol.~\bibinfo {volume} {1}\ (\bibinfo  {publisher}
  {World Scientific},\ \bibinfo {address} {Singapore},\ \bibinfo {year}
  {1998})\BibitemShut {NoStop}\bibitem [{\citenamefont {Rowe}(2010)}]{Rowe2010}\BibitemOpen
  \bibfield  {author} {\bibinfo {author} {\bibfnamefont {D.~J.}\ \bibnamefont
  {Rowe}},\ }\href {https://doi.org/10.1142/6721} {\emph {\bibinfo {title}
  {Nuclear Collective Motion: Models and Theory}}}\ (\bibinfo  {publisher}
  {World Scientific},\ \bibinfo {address} {Singapore},\ \bibinfo {year}
  {2010})\BibitemShut {NoStop}\bibitem [{\citenamefont {Casten}(2000)}]{Casten2000}\BibitemOpen
  \bibfield  {author} {\bibinfo {author} {\bibfnamefont {R.~F.}\ \bibnamefont
  {Casten}},\ }\href@noop {} {\emph {\bibinfo {title} {Nuclear Structure from a
  Simple Perspective}}},\ \bibinfo {edition} {2nd}\ ed.,\ Vol.~\bibinfo
  {volume} {23}\ (\bibinfo  {publisher} {Oxford University Press},\ \bibinfo
  {address} {Oxford, New York},\ \bibinfo {year} {2000})\BibitemShut {NoStop}\bibitem [{\citenamefont {Rowe}\ and\ \citenamefont
  {Bahri}(2000)}]{jmp-41-2000-6544-Rowe}\BibitemOpen
  \bibfield  {author} {\bibinfo {author} {\bibfnamefont {D.~J.}\ \bibnamefont
  {Rowe}}\ and\ \bibinfo {author} {\bibfnamefont {C.}~\bibnamefont {Bahri}},\
  }\href@noop {} {\bibfield  {journal} {\bibinfo  {journal} {J. Math. Phys.}\
  }\textbf {\bibinfo {volume} {41}},\ \bibinfo {pages} {6544} (\bibinfo {year}
  {2000})}\BibitemShut {NoStop}\bibitem [{\citenamefont {Racah}(1949)}]{rmp-21-1949-494-Racah}\BibitemOpen
  \bibfield  {author} {\bibinfo {author} {\bibfnamefont {G.}~\bibnamefont
  {Racah}},\ }\href@noop {} {\bibfield  {journal} {\bibinfo  {journal} {Rev.
  Mod. Phys.}\ }\textbf {\bibinfo {volume} {21}},\ \bibinfo {pages} {494}
  (\bibinfo {year} {1949})}\BibitemShut {NoStop}\bibitem [{\citenamefont {Braunschweig}\ and\ \citenamefont
  {Hecht}(1978)}]{jmp-19-1978-720-Braunschweig}\BibitemOpen
  \bibfield  {author} {\bibinfo {author} {\bibfnamefont {D.}~\bibnamefont
  {Braunschweig}}\ and\ \bibinfo {author} {\bibfnamefont {K.~T.}\ \bibnamefont
  {Hecht}},\ }\href@noop {} {\bibfield  {journal} {\bibinfo  {journal} {J.
  Math. Phys.}\ }\textbf {\bibinfo {volume} {19}},\ \bibinfo {pages} {720}
  (\bibinfo {year} {1978})}\BibitemShut {NoStop}\bibitem [{\citenamefont {Hecht}(1971)}]{npa-170-1971-34-Hecht}\BibitemOpen
  \bibfield  {author} {\bibinfo {author} {\bibfnamefont {K.~T.}\ \bibnamefont
  {Hecht}},\ }\href@noop {} {\bibfield  {journal} {\bibinfo  {journal} {Nucl.
  Phys. A}\ }\textbf {\bibinfo {volume} {170}},\ \bibinfo {pages} {34}
  (\bibinfo {year} {1971})}\BibitemShut {NoStop}\bibitem [{\citenamefont {Rowe}\ and\ \citenamefont
  {Wood}(2010)}]{RoweWood2010}\BibitemOpen
  \bibfield  {author} {\bibinfo {author} {\bibfnamefont {D.~J.}\ \bibnamefont
  {Rowe}}\ and\ \bibinfo {author} {\bibfnamefont {J.~L.}\ \bibnamefont
  {Wood}},\ }\href@noop {} {\emph {\bibinfo {title} {Fundamentals of Nuclear
  Models: Foundational Models}}}\ (\bibinfo  {publisher} {World Scientific},\
  \bibinfo {address} {Singapore},\ \bibinfo {year} {2010})\BibitemShut
  {NoStop}\bibitem [{\citenamefont {Caprio}\ \emph
  {et~al.}(2020{\natexlab{b}})\citenamefont {Caprio}, \citenamefont {McCoy},\
  and\ \citenamefont {Fasano}}]{jpg-47-2020-122001-Caprio}\BibitemOpen
  \bibfield  {author} {\bibinfo {author} {\bibfnamefont {M.~A.}\ \bibnamefont
  {Caprio}}, \bibinfo {author} {\bibfnamefont {A.~E.}\ \bibnamefont {McCoy}},\
  and\ \bibinfo {author} {\bibfnamefont {P.~J.}\ \bibnamefont {Fasano}},\
  }\href {https://doi.org/10.1088/1361-6471/ab9d38} {\bibfield  {journal}
  {\bibinfo  {journal} {J. Phys. G}\ }\textbf {\bibinfo {volume} {47}},\
  \bibinfo {pages} {122001} (\bibinfo {year} {2020}{\natexlab{b}})}\BibitemShut
  {NoStop}\bibitem [{\citenamefont {Rowe}(2012)}]{privcom-Rowe-2012}\BibitemOpen
  \bibfield  {author} {\bibinfo {author} {\bibfnamefont {D.~J.}\ \bibnamefont
  {Rowe}},\ }\href@noop {} {}\bibinfo {howpublished} {private communication}
  (\bibinfo {year} {2012})\BibitemShut {NoStop}\bibitem [{\citenamefont {Anyas-Weiss}\ \emph {et~al.}(1974)\citenamefont
  {Anyas-Weiss}, \citenamefont {Cornell}, \citenamefont {Fisher}, \citenamefont
  {Hudson}, \citenamefont {Menchaca-Rocha}, \citenamefont {Millener},
  \citenamefont {Panagiotou}, \citenamefont {Scott},\ and\ \citenamefont
  {Strottman}}]{pr-12-1974-201-Anyas-Weiss}\BibitemOpen
  \bibfield  {author} {\bibinfo {author} {\bibfnamefont {N.}~\bibnamefont
  {Anyas-Weiss}}, \bibinfo {author} {\bibfnamefont {J.~C.}\ \bibnamefont
  {Cornell}}, \bibinfo {author} {\bibfnamefont {P.~S.}\ \bibnamefont {Fisher}},
  \bibinfo {author} {\bibfnamefont {P.~N.}\ \bibnamefont {Hudson}}, \bibinfo
  {author} {\bibfnamefont {A.}~\bibnamefont {Menchaca-Rocha}}, \bibinfo
  {author} {\bibfnamefont {D.~J.}\ \bibnamefont {Millener}}, \bibinfo {author}
  {\bibfnamefont {A.~D.}\ \bibnamefont {Panagiotou}}, \bibinfo {author}
  {\bibfnamefont {D.~K.}\ \bibnamefont {Scott}},\ and\ \bibinfo {author}
  {\bibfnamefont {D.}~\bibnamefont {Strottman}},\ }\href@noop {} {\bibfield
  {journal} {\bibinfo  {journal} {Phys. Rep.}\ }\textbf {\bibinfo {volume}
  {12}},\ \bibinfo {pages} {201} (\bibinfo {year} {1974})}\BibitemShut
  {NoStop}\bibitem [{\citenamefont {Luo}\ \emph {et~al.}(2013)\citenamefont {Luo},
  \citenamefont {Caprio},\ and\ \citenamefont
  {Dytrych}}]{npa-897-2013-109-Luo}\BibitemOpen
  \bibfield  {author} {\bibinfo {author} {\bibfnamefont {F.~Q.}\ \bibnamefont
  {Luo}}, \bibinfo {author} {\bibfnamefont {M.~A.}\ \bibnamefont {Caprio}},\
  and\ \bibinfo {author} {\bibfnamefont {T.}~\bibnamefont {Dytrych}},\ }\href
  {https://doi.org/10.1016/j.nuclphysa.2012.11.003} {\bibfield  {journal}
  {\bibinfo  {journal} {Nucl. Phys. A}\ }\textbf {\bibinfo {volume} {897}},\
  \bibinfo {pages} {109} (\bibinfo {year} {2013})}\BibitemShut {NoStop}\bibitem [{\citenamefont {Wigner}(1937)}]{pr-51-1937-106-Wigner}\BibitemOpen
  \bibfield  {author} {\bibinfo {author} {\bibfnamefont {E.}~\bibnamefont
  {Wigner}},\ }\href {https://doi.org/10.1103/PhysRev.51.106} {\bibfield
  {journal} {\bibinfo  {journal} {Phys. Rev.}\ }\textbf {\bibinfo {volume}
  {51}},\ \bibinfo {pages} {106} (\bibinfo {year} {1937})}\BibitemShut
  {NoStop}\bibitem [{\citenamefont {Hecht}\ and\ \citenamefont
  {Pang}(1969)}]{jmp-10-1969-1571-Hecht}\BibitemOpen
  \bibfield  {author} {\bibinfo {author} {\bibfnamefont {K.~T.}\ \bibnamefont
  {Hecht}}\ and\ \bibinfo {author} {\bibfnamefont {S.~C.}\ \bibnamefont
  {Pang}},\ }\href@noop {} {\bibfield  {journal} {\bibinfo  {journal} {J. Math.
  Phys.}\ }\textbf {\bibinfo {volume} {10}},\ \bibinfo {pages} {1571} (\bibinfo
  {year} {1969})}\BibitemShut {NoStop}\bibitem [{\citenamefont {Littlewood}(1943)}]{trsa-239-1943-305-Littlewood}\BibitemOpen
  \bibfield  {author} {\bibinfo {author} {\bibfnamefont {D.~E.}\ \bibnamefont
  {Littlewood}},\ }\href@noop {} {\bibfield  {journal} {\bibinfo  {journal}
  {Trans. R. Sco. A}\ }\textbf {\bibinfo {volume} {239}},\ \bibinfo {pages}
  {305} (\bibinfo {year} {1943})}\BibitemShut {NoStop}\bibitem [{\citenamefont {Bahri}\ and\ \citenamefont
  {Draayer}(1994)}]{cpc-83-1994-59-Bahri}\BibitemOpen
  \bibfield  {author} {\bibinfo {author} {\bibfnamefont {C.}~\bibnamefont
  {Bahri}}\ and\ \bibinfo {author} {\bibfnamefont {J.~P.}\ \bibnamefont
  {Draayer}},\ }\href@noop {} {\bibfield  {journal} {\bibinfo  {journal}
  {Comput. Phys. Commun.}\ }\textbf {\bibinfo {volume} {83}},\ \bibinfo {pages}
  {59} (\bibinfo {year} {1994})}\BibitemShut {NoStop}\bibitem [{\citenamefont {Borel}\ and\ \citenamefont
  {Wallach}(2000)}]{Borel2000}\BibitemOpen
  \bibfield  {author} {\bibinfo {author} {\bibfnamefont {A.}~\bibnamefont
  {Borel}}\ and\ \bibinfo {author} {\bibfnamefont {N.}~\bibnamefont
  {Wallach}},\ }\href@noop {} {\emph {\bibinfo {title} {Continuous Cohomology,
  Discrete Subgroups, and Representations of Reductive Groups}}},\ \bibinfo
  {edition} {2nd}\ ed.,\ edited by\ \bibinfo {editor} {\bibfnamefont
  {A.}~\bibnamefont {Borel}}\ (\bibinfo  {publisher} {American Mathematical
  Society},\ \bibinfo {year} {2000})\BibitemShut {NoStop}\bibitem [{\citenamefont {Draayer}\ \emph {et~al.}(1989)\citenamefont
  {Draayer}, \citenamefont {Leschber}, \citenamefont {Park},\ and\
  \citenamefont {Lopez}}]{cpc-56-1989-279-Draayer}\BibitemOpen
  \bibfield  {author} {\bibinfo {author} {\bibfnamefont {J.~P.}\ \bibnamefont
  {Draayer}}, \bibinfo {author} {\bibfnamefont {Y.}~\bibnamefont {Leschber}},
  \bibinfo {author} {\bibfnamefont {S.~C.}\ \bibnamefont {Park}},\ and\
  \bibinfo {author} {\bibfnamefont {R.}~\bibnamefont {Lopez}},\ }\href@noop {}
  {\bibfield  {journal} {\bibinfo  {journal} {Comput. Phys. Commun.}\ }\textbf
  {\bibinfo {volume} {56}},\ \bibinfo {pages} {279} (\bibinfo {year}
  {1989})}\BibitemShut {NoStop}\bibitem [{\citenamefont
  {Rosensteel}(1992)}]{rosensteel1992:sp3r-tensors-gtssnp91}\BibitemOpen
  \bibfield  {author} {\bibinfo {author} {\bibfnamefont {G.}~\bibnamefont
  {Rosensteel}},\ }in\ \href@noop {} {\emph {\bibinfo {booktitle} {Group Theory
  and Special Symmetries in Nuclear Physics}}},\ \bibinfo {editor} {edited by\
  \bibinfo {editor} {\bibfnamefont {J.~P.}\ \bibnamefont {Draayer}}\ and\
  \bibinfo {editor} {\bibfnamefont {J.}~\bibnamefont {J{\"a}necke}}}\ (\bibinfo
   {publisher} {World Scientific},\ \bibinfo {address} {Singapore},\ \bibinfo
  {year} {1992})\ p.\ \bibinfo {pages} {332}\BibitemShut {NoStop}\bibitem [{\citenamefont {Escher}\ and\ \citenamefont
  {Leviatan}(2002)}]{prc-65-2002-054309-Escher}\BibitemOpen
  \bibfield  {author} {\bibinfo {author} {\bibfnamefont {J.}~\bibnamefont
  {Escher}}\ and\ \bibinfo {author} {\bibfnamefont {A.}~\bibnamefont
  {Leviatan}},\ }\href {https://doi.org/10.1103/PhysRevC.65.054309} {\bibfield
  {journal} {\bibinfo  {journal} {Phys. Rev. C}\ }\textbf {\bibinfo {volume}
  {65}},\ \bibinfo {pages} {054309} (\bibinfo {year} {2002})}\BibitemShut
  {NoStop}\bibitem [{\citenamefont {Dreyfuss}\ \emph {et~al.}(2013)\citenamefont
  {Dreyfuss}, \citenamefont {Launey}, \citenamefont {Dytrych},\ and\
  \citenamefont {Draayer}}]{plb-727-2013-511-Dreyfuss}\BibitemOpen
  \bibfield  {author} {\bibinfo {author} {\bibfnamefont {A.~C.}\ \bibnamefont
  {Dreyfuss}}, \bibinfo {author} {\bibfnamefont {K.~D.}\ \bibnamefont
  {Launey}}, \bibinfo {author} {\bibfnamefont {T.}~\bibnamefont {Dytrych}},\
  and\ \bibinfo {author} {\bibfnamefont {J.~P.}\ \bibnamefont {Draayer}},\
  }\href@noop {} {\bibfield  {journal} {\bibinfo  {journal} {Phys. Lett. B}\
  }\textbf {\bibinfo {volume} {727}},\ \bibinfo {pages} {511} (\bibinfo {year}
  {2013})}\BibitemShut {NoStop}\bibitem [{\citenamefont {Dreyfuss}\ \emph {et~al.}(2017)\citenamefont
  {Dreyfuss}, \citenamefont {Launey}, \citenamefont {Dytrych}, \citenamefont
  {Draayer}, \citenamefont {Baker}, \citenamefont {Deibel},\ and\ \citenamefont
  {Bahri}}]{prc-95-2017-044312-Dreyfuss}\BibitemOpen
  \bibfield  {author} {\bibinfo {author} {\bibfnamefont {A.~C.}\ \bibnamefont
  {Dreyfuss}}, \bibinfo {author} {\bibfnamefont {K.~D.}\ \bibnamefont
  {Launey}}, \bibinfo {author} {\bibfnamefont {T.}~\bibnamefont {Dytrych}},
  \bibinfo {author} {\bibfnamefont {J.~P.}\ \bibnamefont {Draayer}}, \bibinfo
  {author} {\bibfnamefont {R.~B.}\ \bibnamefont {Baker}}, \bibinfo {author}
  {\bibfnamefont {C.~M.}\ \bibnamefont {Deibel}},\ and\ \bibinfo {author}
  {\bibfnamefont {C.}~\bibnamefont {Bahri}},\ }\href@noop {} {\bibfield
  {journal} {\bibinfo  {journal} {Phys. Rev. C}\ }\textbf {\bibinfo {volume}
  {95}},\ \bibinfo {pages} {044312} (\bibinfo {year} {2017})}\BibitemShut
  {NoStop}\bibitem [{\citenamefont {Rosensteel}\ and\ \citenamefont
  {Rowe}(1983)}]{jmp-24-1983-2461-Rosensteel}\BibitemOpen
  \bibfield  {author} {\bibinfo {author} {\bibfnamefont {G.}~\bibnamefont
  {Rosensteel}}\ and\ \bibinfo {author} {\bibfnamefont {D.~J.}\ \bibnamefont
  {Rowe}},\ }\href@noop {} {\bibfield  {journal} {\bibinfo  {journal} {J. Math.
  Phys.}\ }\textbf {\bibinfo {volume} {24}},\ \bibinfo {pages} {2461} (\bibinfo
  {year} {1983})}\BibitemShut {NoStop}\bibitem [{\citenamefont {Rowe}(1984)}]{jmp-25-1984-2662-Rowe}\BibitemOpen
  \bibfield  {author} {\bibinfo {author} {\bibfnamefont {D.~J.}\ \bibnamefont
  {Rowe}},\ }\href@noop {} {\bibfield  {journal} {\bibinfo  {journal} {J. Math.
  Phys.}\ }\textbf {\bibinfo {volume} {25}},\ \bibinfo {pages} {2662} (\bibinfo
  {year} {1984})}\BibitemShut {NoStop}\bibitem [{\citenamefont {Rowe}(1995)}]{jmp-36-1995-1520-Rowe}\BibitemOpen
  \bibfield  {author} {\bibinfo {author} {\bibfnamefont {D.~J.}\ \bibnamefont
  {Rowe}},\ }\href@noop {} {\bibfield  {journal} {\bibinfo  {journal} {J. Math.
  Phys.}\ }\textbf {\bibinfo {volume} {36}},\ \bibinfo {pages} {1520} (\bibinfo
  {year} {1995})}\BibitemShut {NoStop}\bibitem [{\citenamefont {Kretzschmar}(1960)}]{zp-158-1960-284-Kretzschmar}\BibitemOpen
  \bibfield  {author} {\bibinfo {author} {\bibfnamefont {M.}~\bibnamefont
  {Kretzschmar}},\ }\href {https://doi.org/10.1007/BF01340561} {\bibfield
  {journal} {\bibinfo  {journal} {Z. Phys.}\ }\textbf {\bibinfo {volume}
  {158}},\ \bibinfo {pages} {284} (\bibinfo {year} {1960})}\BibitemShut
  {NoStop}\bibitem [{\citenamefont {Verhaar}(1960)}]{np-21-1960-508-Verhaar}\BibitemOpen
  \bibfield  {author} {\bibinfo {author} {\bibfnamefont {B.~J.}\ \bibnamefont
  {Verhaar}},\ }\href@noop {} {\bibfield  {journal} {\bibinfo  {journal} {Nucl.
  Phys.}\ }\textbf {\bibinfo {volume} {21}},\ \bibinfo {pages} {508} (\bibinfo
  {year} {1960})}\BibitemShut {NoStop}\bibitem [{\citenamefont {Millener}\ and\ \citenamefont
  {Kurath}(1975)}]{npa-255-1975-315-Millener}\BibitemOpen
  \bibfield  {author} {\bibinfo {author} {\bibfnamefont {D.~J.}\ \bibnamefont
  {Millener}}\ and\ \bibinfo {author} {\bibfnamefont {D.}~\bibnamefont
  {Kurath}},\ }\href {https://doi.org/10.1016/0375-9474(75)90683-1} {\bibfield
  {journal} {\bibinfo  {journal} {Nucl. Phys. A}\ }\textbf {\bibinfo {volume}
  {255}},\ \bibinfo {pages} {315} (\bibinfo {year} {1975})}\BibitemShut
  {NoStop}\bibitem [{\citenamefont {Draayer}\ and\ \citenamefont
  {Williams}(1969)}]{npa-129-1969-647-Draayer}\BibitemOpen
  \bibfield  {author} {\bibinfo {author} {\bibfnamefont {J.~P.}\ \bibnamefont
  {Draayer}}\ and\ \bibinfo {author} {\bibfnamefont {S.~A.}\ \bibnamefont
  {Williams}},\ }\href@noop {} {\bibfield  {journal} {\bibinfo  {journal}
  {Nucl. Phys. A}\ }\textbf {\bibinfo {volume} {129}},\ \bibinfo {pages} {647}
  (\bibinfo {year} {1969})}\BibitemShut {NoStop}\bibitem [{\citenamefont {Hecht}\ \emph {et~al.}(1981)\citenamefont {Hecht},
  \citenamefont {Reske}, \citenamefont {Seligman},\ and\ \citenamefont
  {Zahn}}]{npa-356-1981-146-Hecht}\BibitemOpen
  \bibfield  {author} {\bibinfo {author} {\bibfnamefont {K.~T.}\ \bibnamefont
  {Hecht}}, \bibinfo {author} {\bibfnamefont {E.~J.}\ \bibnamefont {Reske}},
  \bibinfo {author} {\bibfnamefont {T.~H.}\ \bibnamefont {Seligman}},\ and\
  \bibinfo {author} {\bibfnamefont {W.}~\bibnamefont {Zahn}},\ }\href@noop {}
  {\bibfield  {journal} {\bibinfo  {journal} {Nucl. Phys. A}\ }\textbf
  {\bibinfo {volume} {356}},\ \bibinfo {pages} {146} (\bibinfo {year}
  {1981})}\BibitemShut {NoStop}\bibitem [{\citenamefont {Hecht}\ and\ \citenamefont
  {Suzuki}(1983)}]{jmp-24-1983-785-Hecht}\BibitemOpen
  \bibfield  {author} {\bibinfo {author} {\bibfnamefont {K.}~\bibnamefont
  {Hecht}}\ and\ \bibinfo {author} {\bibfnamefont {Y.}~\bibnamefont {Suzuki}},\
  }\href@noop {} {\bibfield  {journal} {\bibinfo  {journal} {J. Math. Phys.}\
  }\textbf {\bibinfo {volume} {24}},\ \bibinfo {pages} {785} (\bibinfo {year}
  {1983})}\BibitemShut {NoStop}\bibitem [{\citenamefont {Hecht}\ and\ \citenamefont
  {Chen}(1990)}]{jpa-23-1990-Hecht}\BibitemOpen
  \bibfield  {author} {\bibinfo {author} {\bibfnamefont {K.~T.}\ \bibnamefont
  {Hecht}}\ and\ \bibinfo {author} {\bibfnamefont {J.~Q.}\ \bibnamefont
  {Chen}},\ }\href@noop {} {\bibfield  {journal} {\bibinfo  {journal} {Nucl.
  Phys. A}\ }\textbf {\bibinfo {volume} {512}},\ \bibinfo {pages} {365}
  (\bibinfo {year} {1990})}\BibitemShut {NoStop}\bibitem [{\citenamefont {Hecht}\ and\ \citenamefont
  {Zahn}(1979)}]{npa-318-1979-1-Hecht}\BibitemOpen
  \bibfield  {author} {\bibinfo {author} {\bibfnamefont {K.~T.}\ \bibnamefont
  {Hecht}}\ and\ \bibinfo {author} {\bibfnamefont {W.}~\bibnamefont {Zahn}},\
  }\href@noop {} {\bibfield  {journal} {\bibinfo  {journal} {Nucl. Phys. A}\
  }\textbf {\bibinfo {volume} {318}},\ \bibinfo {pages} {1} (\bibinfo {year}
  {1979})}\BibitemShut {NoStop}\bibitem [{\citenamefont {Moshinsky}\ and\ \citenamefont
  {Brody}(1960)}]{rmf-9-1960-181-moshinsky}\BibitemOpen
  \bibfield  {author} {\bibinfo {author} {\bibfnamefont {M.}~\bibnamefont
  {Moshinsky}}\ and\ \bibinfo {author} {\bibfnamefont {T.~A.}\ \bibnamefont
  {Brody}},\ }\href@noop {} {\bibfield  {journal} {\bibinfo  {journal} {Rev.
  Mex. Fis.}\ }\textbf {\bibinfo {volume} {9}},\ \bibinfo {pages} {181}
  (\bibinfo {year} {1960})}\BibitemShut {NoStop}\bibitem [{\citenamefont {Edmonds}(1960)}]{Edmonds1960}\BibitemOpen
  \bibfield  {author} {\bibinfo {author} {\bibfnamefont {A.~R.}\ \bibnamefont
  {Edmonds}},\ }\href@noop {} {\emph {\bibinfo {title} {Angular momentum in
  quantum mechanics}}}\ (\bibinfo  {publisher} {Princeton University Press},\
  \bibinfo {year} {1960})\BibitemShut {NoStop}\bibitem [{\citenamefont {Kuhn}\ and\ \citenamefont
  {Walliser}(2008)}]{cpc-179-2008-733-Kuhn}\BibitemOpen
  \bibfield  {author} {\bibinfo {author} {\bibfnamefont {M.}~\bibnamefont
  {Kuhn}}\ and\ \bibinfo {author} {\bibfnamefont {H.}~\bibnamefont
  {Walliser}},\ }\href {https://doi.org/10.1016/j.cpc.2008.06.009} {\bibfield
  {journal} {\bibinfo  {journal} {Comput. Phys. Commun.}\ }\textbf {\bibinfo
  {volume} {179}},\ \bibinfo {pages} {733} (\bibinfo {year}
  {2008})}\BibitemShut {NoStop}\bibitem [{\citenamefont {Dang}\ \emph {et~al.}(2024)\citenamefont {Dang},
  \citenamefont {Draayer}, \citenamefont {Pan}, \citenamefont {Dytrych},
  \citenamefont {Langr}, \citenamefont {Kekejian}, \citenamefont {Becker},\
  and\ \citenamefont {Thompson}}]{epjplus-139-2024-933-Dang}\BibitemOpen
  \bibfield  {author} {\bibinfo {author} {\bibfnamefont {P.}~\bibnamefont
  {Dang}}, \bibinfo {author} {\bibfnamefont {J.~P.}\ \bibnamefont {Draayer}},
  \bibinfo {author} {\bibfnamefont {F.}~\bibnamefont {Pan}}, \bibinfo {author}
  {\bibfnamefont {T.}~\bibnamefont {Dytrych}}, \bibinfo {author} {\bibfnamefont
  {D.}~\bibnamefont {Langr}}, \bibinfo {author} {\bibfnamefont
  {D.}~\bibnamefont {Kekejian}}, \bibinfo {author} {\bibfnamefont {K.~S.}\
  \bibnamefont {Becker}},\ and\ \bibinfo {author} {\bibfnamefont
  {N.}~\bibnamefont {Thompson}},\ }\href
  {https://doi.org/10.1140/epjp/s13360-024-05581-6} {\bibfield  {journal}
  {\bibinfo  {journal} {Eur. Phys. J. Plus}\ }\textbf {\bibinfo {volume}
  {139}},\ \bibinfo {pages} {933} (\bibinfo {year} {2024})}\BibitemShut
  {NoStop}\bibitem [{\citenamefont {Hall}(2015)}]{Hall2015}\BibitemOpen
  \bibfield  {author} {\bibinfo {author} {\bibfnamefont {B.~C.}\ \bibnamefont
  {Hall}},\ }\href {https://doi.org/10.1007/978-3-319-13467-3} {\emph {\bibinfo
  {title} {Lie Groups, Lie Algebras and Representations}}},\ \bibinfo {edition}
  {2nd}\ ed.\ (\bibinfo  {publisher} {Springer International},\ \bibinfo
  {address} {Switzerland},\ \bibinfo {year} {2015})\BibitemShut {NoStop}\bibitem [{\citenamefont {Varshalovich}\ \emph {et~al.}(1988)\citenamefont
  {Varshalovich}, \citenamefont {Moskalev},\ and\ \citenamefont
  {Khersonskii}}]{Varshalovich1988}\BibitemOpen
  \bibfield  {author} {\bibinfo {author} {\bibfnamefont {D.~A.}\ \bibnamefont
  {Varshalovich}}, \bibinfo {author} {\bibfnamefont {A.~N.}\ \bibnamefont
  {Moskalev}},\ and\ \bibinfo {author} {\bibfnamefont {V.~K.}\ \bibnamefont
  {Khersonskii}},\ }\href@noop {} {\emph {\bibinfo {title} {Quantum Theory of
  Angular Momentum}}}\ (\bibinfo  {publisher} {World Scientific Publishing Co
  Pte Ltd.},\ \bibinfo {year} {1988})\BibitemShut {NoStop}\bibitem [{\citenamefont {French}(1966)}]{French1966}\BibitemOpen
  \bibfield  {author} {\bibinfo {author} {\bibfnamefont {J.~B.}\ \bibnamefont
  {French}},\ }in\ \href@noop {} {\emph {\bibinfo {booktitle} {Proceedings of
  the International School of Physics ``Enrico Fermi'', Course XXXVI}}},\
  \bibinfo {editor} {edited by\ \bibinfo {editor} {\bibfnamefont
  {C.}~\bibnamefont {Bloch}}}\ (\bibinfo  {publisher} {Academic Press},\
  \bibinfo {address} {New York},\ \bibinfo {year} {1966})\ p.\ \bibinfo {pages}
  {278}\BibitemShut {NoStop}\bibitem [{\citenamefont {Chen}\ \emph {et~al.}(1993)\citenamefont {Chen},
  \citenamefont {Chen},\ and\ \citenamefont {Klein}}]{npa-554-1993-61-Chen}\BibitemOpen
  \bibfield  {author} {\bibinfo {author} {\bibfnamefont {J.~Q.}\ \bibnamefont
  {Chen}}, \bibinfo {author} {\bibfnamefont {B.~Q.}\ \bibnamefont {Chen}},\
  and\ \bibinfo {author} {\bibfnamefont {A.}~\bibnamefont {Klein}},\
  }\href@noop {} {\bibfield  {journal} {\bibinfo  {journal} {Nucl. Phys. A}\
  }\textbf {\bibinfo {volume} {554}},\ \bibinfo {pages} {61} (\bibinfo {year}
  {1993})}\BibitemShut {NoStop}\bibitem [{\citenamefont {Caprio}\ \emph {et~al.}(2011)\citenamefont {Caprio},
  \citenamefont {Skrabacz},\ and\ \citenamefont
  {Iachello}}]{jpa-44-2011-075303-Caprio}\BibitemOpen
  \bibfield  {author} {\bibinfo {author} {\bibfnamefont {M.~A.}\ \bibnamefont
  {Caprio}}, \bibinfo {author} {\bibfnamefont {J.~H.}\ \bibnamefont
  {Skrabacz}},\ and\ \bibinfo {author} {\bibfnamefont {F.}~\bibnamefont
  {Iachello}},\ }\href@noop {} {\bibfield  {journal} {\bibinfo  {journal} {J.
  Phys. A}\ }\textbf {\bibinfo {volume} {44}},\ \bibinfo {pages} {075303}
  (\bibinfo {year} {2011})}\BibitemShut {NoStop}\bibitem [{\citenamefont {Jahn}(1951)}]{prsla-205-1951-192-Jahn}\BibitemOpen
  \bibfield  {author} {\bibinfo {author} {\bibfnamefont {H.~A.}\ \bibnamefont
  {Jahn}},\ }\href {https://doi.org/10.1098/rspa.1951.0026} {\bibfield
  {journal} {\bibinfo  {journal} {Proc. R. Soc. A}\ }\textbf {\bibinfo {volume}
  {205}},\ \bibinfo {pages} {192} (\bibinfo {year} {1951})}\BibitemShut
  {NoStop}\bibitem [{\citenamefont {Biedenharn}\ \emph {et~al.}(1952)\citenamefont
  {Biedenharn}, \citenamefont {Blatt},\ and\ \citenamefont
  {Rose}}]{rmp-24-1952-249-Biedenharn}\BibitemOpen
  \bibfield  {author} {\bibinfo {author} {\bibfnamefont {L.~C.}\ \bibnamefont
  {Biedenharn}}, \bibinfo {author} {\bibfnamefont {J.~M.}\ \bibnamefont
  {Blatt}},\ and\ \bibinfo {author} {\bibfnamefont {M.~E.}\ \bibnamefont
  {Rose}},\ }\href {https://doi.org/10.1103/RevModPhys.24.249} {\bibfield
  {journal} {\bibinfo  {journal} {Rev. Mod. Phys.}\ }\textbf {\bibinfo {volume}
  {24}},\ \bibinfo {pages} {249} (\bibinfo {year} {1952})}\BibitemShut
  {NoStop}\end{thebibliography}
\nocite{control:spncci-recurrence}

\end{document}